\documentclass[10pt,latterpaper,openany]{book}
\usepackage{times}
\usepackage[utf8]{inputenc}
\usepackage[english]{babel}
\usepackage{amsmath}
\usepackage{amsfonts}
\usepackage{amssymb}
\usepackage{graphicx}
\usepackage{bm}
\usepackage{fancyhdr}
\usepackage{array}
\usepackage{transparent}
\usepackage{eso-pic}
\usepackage{multirow}
\usepackage{framed}
\usepackage{color}
\usepackage{wrapfig}\definecolor{shadecolor}{RGB}{224,238,238}
\usepackage{multicol}
\usepackage{pifont}
\usepackage{hyperref}
\usepackage{wrapfig}
\usepackage{cancel} 
\usepackage{booktabs}
\usepackage{float}
\usepackage{subfigure}
\usepackage{physics}
\usepackage{braket}
\usepackage{longtable}
\usepackage{verbatim}
\usepackage{longtable}
\usepackage{tikz-feynman}
\usepackage{pgffor}
\usepackage{ulem}
\usepackage{csquotes}
\usepackage{cancel}
\usepackage{blindtext}
\usepackage{lipsum}
\usepackage{CJK}
\usepackage{ulem}
\usepackage{epsfig}
\usepackage{enumerate}
\usepackage{tikz}
\usepackage{cite} 

\newcommand{\nn}{\nonumber}

\usetikzlibrary{arrows,shapes}
\usetikzlibrary{trees}
\usetikzlibrary{matrix,arrows} 				
\usetikzlibrary{positioning}				
\usetikzlibrary{calc,through}				
\usetikzlibrary{decorations.pathreplacing}  
\usepackage{pgffor}							

\usetikzlibrary{decorations.pathmorphing}	
\usetikzlibrary{decorations.markings}

\tikzset{
arrow/.style = { very thick, color=black, ->, >=Triangle},
ar/.style = { very thick, color=red, ->, >=Triangle},
}
\tikzset{
    vector/.style={decorate, decoration={snake}, draw},
provector/.style={decorate, decoration={snake,amplitude=2.5pt}, draw},
antivector/.style={decorate, decoration={snake,amplitude=-2.5pt}, draw},
    fermion/.style={draw=black, postaction={decorate},
        decoration={markings,mark=at position .55 with {\arrow[draw=black]{>}}}},
    fermionbar/.style={draw=black, postaction={decorate},
        decoration={markings,mark=at position .55 with {\arrow[draw=black]{<}}}},
    fermionnoarrow/.style={draw=black},
    gluon/.style={decorate, draw=black,
        decoration={coil,amplitude=4pt, segment length=5pt}},
    scalar/.style={dashed,draw=black, postaction={decorate},
        decoration={markings,mark=at position .55 with {\arrow[draw=black]{>}}}},
    scalarbar/.style={dashed,draw=black, postaction={decorate},
        decoration={markings,mark=at position .55 with {\arrow[draw=black]{<}}}},
    scalarnoarrow/.style={dashed,draw=black},
    electron/.style={draw=black, postaction={decorate},
        decoration={markings,mark=at position .55 with {\arrow[draw=black]{>}}}},
bigvector/.style={decorate, decoration={snake,amplitude=4pt}, draw},
    line/.style={draw=black},
}\usetikzlibrary{decorations.markings}

\usepackage{anyfontsize}

\usepackage[left=3.5cm,right=2cm,top=2.5cm,bottom=2.5cm]{geometry}
\providecommand{\abs}[1]{\lvert#1\rvert}
\usepackage{fancyhdr}
\fancypagestyle{plain}{
\fancyhead[L]{}
\fancyhead[C]{}
\fancyhead[R]{}
\fancyfoot[L]{}
\fancyfoot[C]{\thepage}
\fancyfoot[R]{}

}

\begin{document}

\begin{titlepage}
\begin{center}
\vspace*{0.3\baselineskip}
{
\bf\fontsize{16}{0}{\selectfont{UNIVERSIDAD T\'ECNICA FEDERICO SANTA MAR\'IA}}\\[2mm]
\fontsize{14}{0}{\selectfont{DEPARTAMENTO DE F\'ISICA}}\\[2mm]
\fontsize{14}{0}{\selectfont{VALPARASO - CHILE}}
}

\vspace*{2.5\baselineskip}
\includegraphics[scale=0.9]{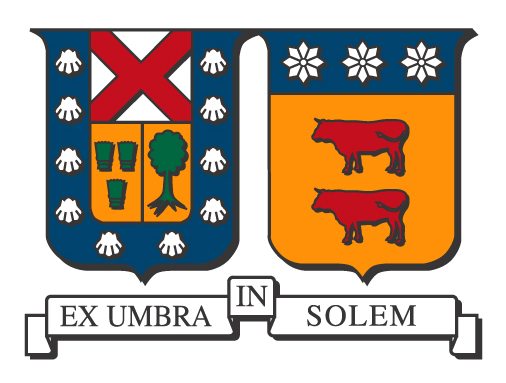} \\[10PT]
\vspace*{2.5\baselineskip}

{
\bf\fontsize{16}{0}{\selectfont{PHENOMENOLOGICAL ASPECTS OF MODELS WITH
LOW SCALE SEESAW.}}
}

\vspace*{3.5\baselineskip}

{
\bf\fontsize{14}{0}{\selectfont{JUAN MANUEL MARCHANT GONZ\'ALEZ}}
}\\[30pt]

{
\bf\fontsize{12}{0}{\selectfont{TESIS PARA OPTAR AL GRADO DE DOCTOR EN CIENCIAS F\'ISICAS.}}
}\\[30pt]

\begin{tabular}{rl}
{\bf\fontsize{12}{0}{\selectfont{DIRECTOR:}}} & {\bf\fontsize{12}{0}{\selectfont{Ph.D. ANTONIO C\'ARCAMO HERN\'ANDEZ,}}}\\
{\bf\fontsize{12}{0}{\selectfont{Co-DIRECTOR:}}} & {\bf\fontsize{12}{0}{\selectfont{Ph.D. CESAR BONILLA.}}}
\end{tabular}

\vspace*{3\baselineskip}

{
\bf\fontsize{14}{0}{\selectfont{2024}}
}
\end{center}
\end{titlepage}
\newpage
$\ $
\thispagestyle{empty} 

\chapter*{}
\pagenumbering{roman} 
\begin{flushright}
\textit{\fontsize{14}{0}{\selectfont{''No te preocupes por lo que otras personas piensan. Haz lo que te haga feliz.''\\[5pt]
''Don't worry about what other people think. Do what makes you happy.''}}}\\[1.5cm]

\textit{\fontsize{14}{0}{\selectfont{Richard P. Feynman}}}
\end{flushright}

\newpage
$\ $
\thispagestyle{empty} 
\chapter*{Acknowledgments} 
\addcontentsline{toc}{chapter}{Acknowledgments} 

Quiero iniciar mis agradecimiento con la persona más importante en mi vida, quiero agradecer a mi madre, es gracias a ella que he logrado llegar hasta esta instancia de mi vida, desde que tengo uso de la razón ella ha realizado ciento de sacrificios para que a mi no me faltara nada y pudiera dedicarme y enfocarme en mis estudios, se que si ella lee esto diría que ``No fueron sacrificios'' que es una mujer fuerte capaz de muchas cosas, pero el hecho es que, trabajar de domingo a domingo, más de diez horas diarias, quitarse el pan de la boca para dármelo a mi, caminar pueblos enteros vendiendo puerta a puerta para que no nos falte la comida en la casa, entre otras muchas cosas, no son cosas fáciles de hacer y menos de mantener durante años, es algo que por lo menos requiere un esfuerzo enorme. También se que son cosas que ha hecho no para que yo le este agradeciendo, simplemente lo hizo por que es mi madre y movería cielo, mar y tierra para que no me falte nada, pero no puedo dejar pasar esta oportunidad para expresar mi agradecimiento a ella, pero me faltan las palabras para poder expresar lo muy agradecido que estoy, siento que no importa lo que haga en mi vida, nunca podría retribuirla y darle todo lo que se pueda merecer. Muchas gracias mamá, gracias por estar siempre a mi lado y apoyarme, gracias por todo los que has hecho y haces por mi, te quiero mucho. Gracias.\\

El Taekwondo ha sido una disciplina que me ha entregado mucho, pero no sólo hablo de las habilidades y conocimientos en las artes marciales. También me ha entregado a grandes personas, donde quiero agradecer a dos en especifico, primero a quien inicialmente lo conocí como mi maestro, mi ``profe'', pero que en realidad ha sido un verdadero padre para mi, estando ahí cuando lo he necesitado, ayudándome y apoyándome en diversas circunstancias, siento que sin él, haber llegado hasta acá hubiera sido aún más complicado. Papá, eres quien me dio la fortaleza y me enseño que rendirse no es una opción, que por muy complicado se vea el camino, el trabajo duro siempre dará frutos, que quizás no sea lo que uno quería inicialmente, pero ese esfuerzo llevará tarde o temprano a la meta que nos propusimos inicialmente, muchas gracias papá, gracias por transmitirme tu fortaleza. También quiero agradecer a quien ha sido un hermano, brindándome su apoyo y energía, confiando en mi a ciegas, compartiendo y celebrando los buenos momentos, pero también dándome su apoyo en los momentos malos, estando ahí con toda su energía y positivismo, que siempre me sorprende, muchas gracias Javier, eres mi hermano y no sólo de palabra, me lo has demostrado en un sin número de ocasiones, que al igual que el ``profe'', no compartimos lasos sanguíneos, pero ustedes han sido una verdadera familia para mi, tu sabes que no soy de expresar mucho mis sentimiento y que rara vez lo digo, pero ustedes saben que los quiero y estoy muy agradecido por todo lo que hacen por mi.\\

Quiero aprovechar esta oportunidad para agradecer a mis tutores Antonio y Cesar, por ayudarme y tener paciencia conmigo, por estar siempre dispuestos a atender mis dudas y brindarme toda la ayuda que he necesitado en nuestras investigaciones. Muchas gracias por todo.\\

Por último, quiero agradecer a mis amigos del programa de postgrado de la USM, Pata, Rocio, Daniel, Nico y Bob. Puede ser algo normal para la mayoría, pero compartir para ir a comer unos completos, juntarnos a jugar catan, ponernos hablar de varias cosas en la sala de postgrado y salir de la rutina del doctorado, es algo que yo agradezco mucho. Sin ustedes el doctorado no hubiera sido lo mismo.

\newpage
$\ $
\thispagestyle{empty} 
\chapter*{Abstract} 
\addcontentsline{toc}{chapter}{Abstract} 
\markboth{ABSTRACT}{ABSTRACT} 

Various phenomenological consequences of seesaw theories for the generation of the fermion mass hierarchy of the Standard Model have been analyzed, with an emphasis on models in which the light-active neutrino masses are derived from low-scale seesaw mechanisms. In particular, fermion masses and lepton flavor-violating decay processes, the flavor-changing neutral current, have been studied, and the implications of these theories for the observed dark matter relic density in the Universe have been determined. From the analysis of these phenomenological aspects, it was possible to determine the allowed parameter spaces of these theories and to obtain a parameter fit consistent with the currently measured experimental values. In this way, correlations between the different observables of the fermionic sector could be obtained, where all values were within the experimental ranges at $3\sigma$. Predictions for new physics consistent with cosmological limits were also obtained.


\newpage
$\ $
\thispagestyle{empty} 
\chapter*{List of publications} 
\addcontentsline{toc}{chapter}{List of publications} 
\markboth{LIST OF PUBLICATIONS}{LIST OF PUBLICATIONS} 

This doctoral thesis is presented as a compendium of the following publications.
\begin{enumerate}[1)]
\item \textbf{Dark Matter from a Radiative Inverse Seesaw Majoron Model}\\
Cesar Bonilla, Antonio Cárcamo Hernández, Bastián Díaz Sáez, Sergey Kovalenko, Juan Marchant González.\\
\href{https://www.sciencedirect.com/science/article/pii/S0370269323006160}{arXiv:2306.08453/Physics Letters B, 847, (2023)}\label{paper:DM}

\item \textbf{Phenomenology of extended multiHiggs doublet models with $S_4$ family symmetry.}\\
Antonio Cárcamo Hernández, Catalina Espinoza, Juan Carlos Gómez-Izquierdo, Juan Marchant González, Myriam Mondragón.\\
\href{https://doi.org/10.1140/epjc/s10052-024-13633-5}{arXiv:2212.12000/Eur. Phys. J. C, 1239, (2024)}\label{paper:S4}

\item \textbf{Phenomenological aspects of the fermion and scalar sectors of a $S_4$ flavored 3-3-1 model}\\
Antonio Cárcamo Hernández, Juan Marchant González, María Luisa Morra-Urritia, Daniel Salinas-Arizmendi.\\
\href{https://www.sciencedirect.com/science/article/pii/S0550321324001548}{arXiv:2305.13441/Nucl. Phys. B, 1005, (2024)}\label{paper:331}
\end{enumerate}

\newpage
$\ $
\thispagestyle{empty} 
\tableofcontents 

\cleardoublepage
\addcontentsline{toc}{chapter}{List of figures} 
\listoffigures 

\cleardoublepage
\addcontentsline{toc}{chapter}{List of tables} 
\listoftables 
\newpage
$\ $
\thispagestyle{empty} 
\chapter{Introduction}\label{cap.introduccion}
\markboth{INTRODUCTION}{INTRODUCTION}
\pagenumbering{arabic} 

The Standard Model of particle physics (SM) is a quantum field theory that unifies the electromagnetic, weak and strong interactions under the gauge symmetry group $SU(3)_C\otimes SU(2)_L \otimes U(1)_Y$. The symmetry group $SU(3)_C$ is associated with the strong interaction, i.e. the interaction between the different colors of quarks, where the mediator particles are the gluons and each quark flavor is a triplet of the color of the group $SU(3)_C$. On the other hand, the electromagnetic and weak interactions are unified under the electroweak theory associated with the group $SU(2)_L \otimes U(1)_Y$, where left-handed fermions are doublets and right-handed fermions are singlets of $SU(2)_L$, and the mediating particles of this interaction are four gauge bosons (photon, $Z$, and $W^{\pm}$). In addition to having the representations of each symmetry group, the content of fermionic matter is classified into three families of quarks and three families of leptons, where asignment under the SM gauge group are show in table. \ref{tab:asignacion}
\begin{table}[!h]
\begin{center}
\begin{tabular}{c|c|c|c|c}
\toprule[0.13em] & $SU(3)_{C}\otimes SU(2)_{L}\otimes U(1)_{Y}$ & I & II & III \\

\hline
 & $\left(3,2,\frac{1}{6}\right)$ & ${u_{L} \choose d_{L}}$ & ${c_{L} \choose s_{L}}$ & ${t_{L} \choose b_{L}}$ \\

\cline{2-5}
Quarks & $\left(3,1,\frac{2}{3}\right)$ & $u_{R}$ & $c_{R}$ & $t_{R}$ \\

\cline{2-5}
 & $\left(3,1,-\frac{1}{3}\right)$ & $d_{R}$ & $s_{R}$ & $b_{R}$ \\

\hline
 & $\left(1,2,-\frac{1}{2}\right)$ & ${\nu_{e_{L}} \choose e_{L}}$ & ${\nu_{\mu_{L}} \choose \mu_{L}}$ & ${\nu_{\tau_{L}} \choose \tau_{L}}$ \\
 
\cline{2-5}
Leptones & $\left(1,1,-1\right)$ & $e_{R}$ & $\mu_{R}$ & $\tau_{R}$ \\

\cline{2-5}
 & $\left(1,1,0\right)$ & $\nu_{e_{R}}$ & $\mu_{\mu_{R}}$ & $\nu_{\tau_{R}}$ \\
\bottomrule[0.13em]
\end{tabular}
\end{center}
\caption{Multiplets of Standard Model fields.}
\label{tab:asignacion}
\end{table}
%
%
%
%
%
%

However, despite the great experimental success of the SM, it does not explain why there are three families of fermions, it does not naturally explain the hierarchy present in the fermionic sector, which extends over a range of thirteen orders of magnitude, from the active neutrino mass scale up to the top quark mass. Furthermore, the SM cannot describe the pattern of fermion masses and mixings, with very small mixing angles for the quarks, giving us a mixing matrix known as Cabibbo-Kobayashi-Maskawa (CKM), which is close to the identity, but for the leptonic sector, on the other hand, there are two large and one small mixing angle, giving us the Pontecorvo-Maki-Nakagawa-Sakata (PMNS) matrix, which is far from the identity. This whole puzzle is known as the "flavor problem".\\

In addition to the points mentioned above, we must also add the mass of the neutrinos, since the SM predicts massless neutrinos, which we know to be incorrect, because neutrino oscillation has been observed experimentally, starting with the observation of solar neutrinos~\cite{Kamiokande-II:1990wrs}, then with atmospheric neutrinos, and finally with reactor experiments~\cite{Davis:1968cp,Super-Kamiokande:1998kpq,Kamiokande-II:1992hns,Super-Kamiokande:2001ljr,KamLAND:2002uet,KamLAND:2008dgz,T2K:2014ghj,RENO:2012mkc}. However, although these experiments show us the mixtures of neutrino eigenstates ($\nu_{e,\mu,\tau}$) as they propagate as their mass eigenstates, they do not give us a clear answer as to what is the absolute mass scale of the neutrinos, which is their dynamical origin, or whether they are Dirac or Majorana fermions. data from neutrino oscillation experiments and from other experiments limit the range of mass values to be between $m_{\nu}\sim 0.01 - 1\; \text{eV}$~\cite{deSalas:2020pgw}.\\

Because of these drawbacks, there is a need to search for new physics beyond the Standard Model, where several proposals can be found in the literature, such as theries with enlarged particle spectrum, extended symmetries ans low scale seesaw mechanisms. In the BSM theories with extended symmetries, the spontaneous breaking of these symmetries yields the observed pattern of SM fermion masses and mixing parameters.\\

This thesis is organized as follows. In chapter \ref{cap.marco} we will discuss the main features of the SM and look at some of the problems it cannot explain. Then, in chapter \ref{cap.neutrino}, we will analyze neutrino physics and see a brief discussion of seesaw mechanisms for neutrino mass production. In chapter \ref{cap.modelo1}, we will look at the first paper on which this thesis is based, where we will study the implementation of a radiative inverse seesaw mechanism and the implications for dark matter candidates and leptonic flavor violating processes. Then, in chapter \ref{cap.modelo3HDM}, we will present the analyses of the second paper on which this thesis is based, where we will study an extension of the theories known as 3HDM and 4HDM, analyzing the implementation of a radiative seesaw mechanism, analyzing the masses and mixtures of fermions, the scalar sector and dark matter, in addition, we will see the implication of the model in the oscillation of mesons and the corrections obtained for the oblique parameters. In chapter \ref{cap.modelo331} we will present the last paper on which this work is based, where we will analyze the extension of a 331 model with discrete and cyclic symmetries, studying the masses and mixtures of the fermions, where an inverse seesaw mechanism is implemented for the neutrino sector and the scalar sector will be analyzed, we will also see an analysis of the oscillation of mesons and the corrections obtained for the oblique parameters will be studied. Finally, in chapter \ref{cap.conclu} the conclusions of this work will be presented.

\newpage
$\ $
\thispagestyle{empty} 
\chapter{Important aspects }\label{cap.marco}
\markboth{IMPORTANT ASPECTS}{IMPORTANT ASPECTS}
\section{The Standard Model}
\lhead[\thepage]{\thesection. The Standard Model} 

The Standard Model (SM) is a gauge theory based on the group of local symmetries $SU(3)_{C}\otimes SU(2)_{L}\otimes U(1)_{Y}$, which describes the strong, weak, and electromagnetic interactions mediated by the corresponding spin-1 fields (gauge bosons): 8 massless gluons and 1 massless $(\gamma)$ photon for the strong and electromagnetic interactions, respectively, and 3 massive bosons ($W^{\pm}$ and $Z$) for the weak interaction\cite{string2007model} (Table 1.2). The fermionic matter consists of three families of quarks and three families of leptons. Each family consists of two spin 1/2 particles, f and f', with electric charges $Q_{f}=Q_{f'}+1$ multiples of the charge of the proton (Table \ref{tab:fermions}), and their corresponding antiparticles. Quarks appear in three possible color states (conventionally red, green, and blue).\\

The fields are grouped into multiplets (irreducible representations) under the group transformations (table \ref{tab:asignacion}). Quarks are triplets and leptons are singlets under the color group $SU(3)_{C}$. Under the $SU(2)_{L}$ group, the levorotatory (left) components are transformed differently from the dextrogyro (right) ones: the left fields are doublets, and the right ones are singlets of weak isospin $T$. The $Y$ index refers to the hypercharge. The electric charge, isospin and hypercharge of the fields are related by $Q = T_{3} + Y$. The three families of quarks and leptons have the same properties (gauge interactions), differing only in the masses and flavor quantum numbers of their fields.\\

The gauge symmetry is spontaneously broken, which requires introducing a scalar field (the Higgs field) and allows weak bosons and fermions to be massive, as we observe them in nature. In the following, we will construct the SM Lagrangian of the electromagnetic and weak interactions for a single family of quarks and leptons. We will ignore the flavor independent interactions.
\begin{table}[!h]
\begin{center}
\scalebox{1.2}[1.1]{
\begin{tabular}{c|c|c|c|c|c|c}
\toprule[0.13em]\multicolumn{7}{c}{Bosons}\\

\hline
&  \multicolumn{1}{|c}{Photon} ($\gamma$) & \multicolumn{5}{|c}{Electromagnetic interaction}\\

\cline{2-7}
spin $1$ & \multicolumn{1}{|c}{$W_{^{\pm}}$, $Z$} & \multicolumn{5}{|c}{Weak interaction}\\

\cline{2-7}
 & \multicolumn{1}{|c}{8 Gluons ($G$)} & \multicolumn{5}{|c}{Strong interaction}\\
 
\hline
spin $0$ & \multicolumn{1}{|c}{Higgs} & \multicolumn{5}{|c}{Origin of the masses}\\
\bottomrule[0.13em]
\end{tabular}
}
\end{center}
\caption{Standard Model Interactions}
\label{tabla1.2}
\end{table}
\begin{table}[!h]
\begin{center}
\scalebox{1.4}[1.2]{
\begin{tabular}{cc|c|c|c|c|c}
\toprule[0.13em]\multicolumn{3}{c|}{Fermions} & I & II & II & Q \\

\hline
 & \multicolumn{1}{|c|}{Quarks} & f & \textcolor{red}{u}\textcolor{green}{u}\textcolor{blue}{u} & \textcolor{red}{c}\textcolor{green}{c}\textcolor{blue}{c} & \textcolor{red}{t}\textcolor{green}{t}\textcolor{blue}{t} & $2/3$ \\

\cline{3-7}
spin $\frac{1}{2}$ & \multicolumn{1}{|c|}{} & f' & \textcolor{red}{d}\textcolor{green}{d}\textcolor{blue}{d} & \textcolor{red}{s}\textcolor{green}{s}\textcolor{blue}{s} & \textcolor{red}{b}\textcolor{green}{b}\textcolor{blue}{b} & $-1/3$ \\

\cline{2-7}
 & \multicolumn{1}{|c|}{Leptons} & f & $\nu_{e}$ & $\nu_{\mu}$ & $\nu_{\tau}$ & 0 \\

\cline{3-7}
 & \multicolumn{1}{|c|}{} & f' & $e$ & $\mu$ & $\tau$ & $-1$ \\
\bottomrule[0.13em]
\end{tabular}
}
\caption{Standard Model Particles}
\end{center}
\label{tab:fermions}
\end{table}

\subsection{Abelian gauge symmetry}

Consider just one family of free and massless spin 1/2 $\psi$ and $\overline{\psi}$ fermions (quarks or leptons), described by the fields $\psi(x)$ and $\overline{\psi}(x)$, respectively. The Dirac Lagrangian will describe them~\cite{illana2017modelo}:
\begin{equation}
\mathcal{L}_{0}=\overline{\psi}(x)\left(i\gamma^{\mu}\partial_{\mu} - m\right) \psi(x)\;,
\label{eq2:ldirac}
\end{equation}

with $\partial_{\mu}=\frac{\partial}{\partial x^{\mu}}$ and $\gamma^{\mu}=\left(\gamma^{0},\gamma^{j}\right)$, are linear operators described by $4 \times 4$ matrices, known as Dirac matrices~\cite{maggiore2005modern, peskin2018introduction}:
\begin{equation}
\gamma^{0} = \sigma_{3}\otimes 1 \quad \text{y} \quad \gamma^{j} = \gamma^{0}\left(\sigma_{3}\otimes \sigma_{j}\right)\;.
\end{equation}

Now we want to see if $\mathcal{L}_{0}$ remains invariant under local gauge transformations $U(1)$~\cite{Pich:2007vu}, which is unitary, so: $UU^{dagger}=U^{\dagger}U=1$.
\begin{equation}
\psi(x):\hspace{5mm} \underrightarrow{U(1)}\hspace{5mm} \psi'(x)=\exp\left\lbrace iQ\theta\right\rbrace \psi(x)\;,
\end{equation}
\noindent
where $Q$ is the generator of the Abelian group $U(1)$\cite{peskin2018introduction, schwartz2014quantum} and $\theta = \theta(x)$ is a spacetime dependent parameter. In such a case, the Lagrangian is not invariant under local transformations,
\begin{align}
\mathcal{L}_0^{\prime}&=i\overline{\psi}(x)^{\prime}\gamma^{\mu}\partial_{\mu}\psi(x)^{\prime}-m\overline{\psi}(x)^{\prime}\psi(x)^{\prime} \notag\\
&=i\overline{\psi}(x)e^{-i\theta}\gamma^{\mu}\partial_{\mu}\left(e^{i\theta}\psi(x)\right)-m\overline{\psi}(x)e^{-i\theta}e^{i\theta}\psi(x) \notag\\
&=i\overline{\psi}(x)\gamma^{\mu}\partial_{\mu}\psi(x)-\overline{\psi}(x)\gamma^{\mu}\psi(x)\partial_{\mu}\theta-m\overline{\psi}(x)\psi(x)\notag\\
\mathcal{L}_0^{\prime}&=\mathcal{L}-\overline{\psi}(x)\gamma^{\mu}\psi(x)\partial_{\mu}\theta.\label{eq2:lang.noinv}
\end{align}

The kinetic term is not invariant due to the partial derivative, so if we want to ensure that the phase invariance of $U(1)$ is preserved locally, we have to define a new derivative which must have the following transformation,
\begin{equation}
\left(D_{\mu}\psi\right)^{\prime}= D_{\mu}^{\prime}\left(\exp{iQ\theta}\psi\right)=\exp{iQ\theta}\left(D_{\mu}\psi\right).\label{eq2:derv-co}
\end{equation}
\noindent
$D_{\mu}$ is known as the covariant derivative, which we can define as,
\begin{equation}
D_{\mu}= \partial_{\mu}-iQA_{\mu},
\label{eq2:dcovariante}
\end{equation}
\noindent
where $A_{\mu}$ is a 4-vector, whose transformation we can obtain by replacing \eqref{eq2:dcovariante} in \eqref{eq2:derv-co}, so,
\begin{align}
\left(D_{\mu}\psi\right)^{\prime}&= \left(\partial_{\mu}-iQA_{\mu}^{\prime}\right)\left(e^{iQ\theta}\psi\right)\notag \\
&= e^{iQ\theta}\left(\partial_{\mu}+ iQ\partial_{\mu}\theta -iQA_{\mu}^{\prime}\right)\psi,
\end{align}
\noindent
but,
\begin{align}
\left(D_{\mu}\psi\right)^{\prime}&=e^{iQ\theta}\left(D_{\mu}\psi\right) \notag\\
e^{iQ\theta}\left(\partial_{\mu}+ iQ\partial_{\mu}\theta -iQA_{\mu}^{\prime}\right)\psi &= e^{iQ\theta}\left(\partial_{\mu}-iQA_{\mu}\right)\psi \notag\\
A_{\mu}^{\prime}&= A_{\mu}+\partial_{\mu}\theta.\label{eq2:Amu}
\end{align}
\noindent
Here we can see $A_{\mu}$ as the 4-vector potential of the gauge field. Consequently,
\begin{equation}
\mathcal{L}=i\overline{\psi}(x)\gamma^{\mu}D_{\mu}\psi(x) - m\overline{\psi}(x) \psi(x)= \mathcal{L}_{0}-Q\overline{\psi}(x)\gamma^{\mu}\psi A_{\mu}\;,
\label{eq2:Linvariante}
\end{equation}

\noindent
the Dirac Lagrangian invariant under local $U(1)$ transformations. Thus the gauge principle has generated the interaction between the Dirac spinor $\psi$ and the gauge field $A_{\mu}$.\\

\noindent
Now let us define the following tensor \cite{peskin2018introduction, schwartz2014quantum}:
\begin{equation}
F_{\mu\nu} = \partial_{\mu}A_{\nu}-\partial_{\nu}A_{\mu}\;,
\label{eq2:tensorEL}
\end{equation}
\noindent
which, as we can see, is invariant under the gauge transformation described in the equation \eqref{eq2:Amu}. Hence the kinetic term of the Lagrangian for the gauge field $A_{\mu}$ is
\begin{equation}
\mathcal{L}_{A}=-\frac{1}{4}F_{\mu\nu}F^{\mu\nu}\;.
\label{eq2:Lcinetico}
\end{equation}

If we associate the field $A_{\mu}$ with the photon and $Q$ with the charge generator, the equation \eqref{eq2:tensorEL} is the electromagnetic tensor, and the equation \eqref{eq2:Linvariante} allows us to fully describe quantum electrodynamics (QED)~\cite{Pich:2007vu} by the QED Lagrangian.
\begin{equation}
\mathcal{L}_{QED}=-\frac{1}{4}F_{\mu\nu}F^{\mu\nu}+\mathcal{L}_{0}-J_{\text{EM}}^{\mu}A_{\mu}\;,
\label{eq:LQED}
\end{equation}
\noindent
where $J_{\text{EM}}^{\mu}=Q\overline{\psi}\gamma^{\mu}\psi$ is the 4-electromagnetic current.


\subsection{Non-Abelian gauge symmetry}

As in the Abelian case, we want to impose that the Dirac Lagrangian for the free particle is invariant under local $G$ transformations~\cite{Pich:2007vu}.
\begin{equation}
\psi(x)\hspace{5mm} \underrightarrow{G}\hspace{5mm} \psi'(x)=U \psi(x)\;.
\end{equation}

\noindent
So $UU^{\dagger}=U^{\dagger}U=1$ and $\det|U|=1$, where $G=SU(N)$ represents a non-abelian Lie group~\cite{georgi2018lie}, then:
\begin{equation}
\psi(x)\hspace{5mm} \underrightarrow{G}\hspace{5mm} \psi'(x)=\exp\left\lbrace iT_{a}\theta^{a}(x)\right\rbrace \psi(x)\;,
\end{equation}

\noindent
with $T_{a}=\lambda_{a}/2$, where $\theta^{a}(x)$ is an arbitrary parameter and is also a function of spacetime, while $\lambda_{a}(\text{with} \hspace{3mm} a=1,2,...,N^{2}-1 )$ are the generators of the Lie group representation $SU(N)$, which satisfies the Lie algebra \cite{schwartz2014quantum}
\begin{equation}
\left[\frac{\lambda^{a}}{2},\frac{\lambda^{b}}{2}\right]=if_{abc}\frac{\lambda^{c}}{2}\;,
\label{eq2:alLie}
\end{equation}

\noindent
where $f_{abc}$ is the structure constant of the Lie algebra~\cite{georgi2018lie}. Again, one must resort to the gauge principle, since the Lagrangian is not invariant under $G$ gauge transformations. Therefore $\left(N^{2}-1\right)$ transforming fields are introduced as:
\begin{equation}
G_{\mu}^{a}:\hspace{5mm} \underrightarrow{G}\hspace{5mm} \left(G_{\mu}^{a}\right)'= G_{\mu}^{a}-\frac{1}{g}\left(\partial_{\mu}\theta^{a}\right)+f_{abc}\theta^{b}G_{\mu}^{c}\;.
\end{equation}

\noindent
The covariant derivative is defined as:
\begin{equation}
D_{\mu} = \partial_{\mu}+igT_{a}G_{\mu}^{a}\;.
\label{eq2:covarianteD}
\end{equation}

A gauge field will be introduced for each group generator associated with the interactions; therefore, we can associate a symmetry group and a set of gauge fields with each interaction.\\

In addition, $g$ is an arbitrary coupling constant that characterizes the intensity of the interaction. Then the Lagrangian is a function of the covariant derivative:
\begin{equation}
\mathcal{L}=\overline{\psi}(x)\left(i\gamma^{\mu}D_{\mu}-m\right)\psi(x)\;.
\label{eq2:Lcovariante}
\end{equation}

This Lagrangian is invariant under local gauge transformations $G$. If a theory is both non-Abelian and locally invariant, it is called a Yang-Mills\cite{schwartz2014quantum} theory.\\

\noindent
By generalizing the field strenght tensor to a Yang-Mills theory, we obtain
\begin{equation}
F_{\mu\nu}^{a} = \partial_{\mu}G_{\nu}^{a}-\partial_{\nu}G_{\mu}^{a}-gf_{abc}G_{\mu}^{b}G_{\nu}^{c}\;,
\label{eq2:F no-abe}
\end{equation}

\noindent
whose transformation is:
\begin{equation}
F_{\mu\nu}^{a}\hspace{5mm} \underrightarrow{G}\hspace{5mm} \left(F_{\mu\nu}^{a}\right)'= F_{\mu\nu}^{a}+f_{abc}\theta^{b}F_{\mu\nu}^{c}\;.
\label{eq2:trans F}
\end{equation}

\noindent
Then we can write the kinetic term of $G_{\mu}^{a}$ as,
\begin{equation}
\mathcal{L}_{G} = -\frac{1}{4}F_{\mu\nu}^{a}F_{a}^{\mu\nu}\;.
\label{eq2:FG cinetico}
\end{equation}

\noindent
Thus, the invariant Lagrangian under $SU(N)$ is:
\begin{equation}
\mathcal{L} = -\frac{1}{4}F_{\mu\nu}^{a}F_{a}^{\mu\nu}+\overline{\psi}(x)\left(i\gamma^{\mu}D_{\mu}-m\right)\psi(x)\;.
\label{eq2:LG invariante}
\end{equation}

\subsection{Gauge symmetry in the Standard Model}

Let us take the electroweak sector, i.e., the symmetry group $SU(2)_{L}\times U(1)_{Y}$ and for simplicity consider only a family of free and massless $1/2$ spin fermions (quarks or leptons) described by the fields $f(x)$ and $f'(x)$, respectively. As seen above, one has the following Dirac lagrangian:
\begin{equation}
\mathcal{L}_{0}= i\overline{f}(x)\cancel{\partial}f(x)+i\overline{f}'(x)\cancel{\partial}f'(x) = i\sum_{j=1}^{3}i\overline{\psi_{j}}(x)\cancel{\partial}\psi_{j}(x)\;,
\end{equation}

\noindent
with $ \cancel{\partial} = \gamma^{\mu}\partial_{\mu}$ and we have grouped the \textit{left} components into one doublet and the \textit{right} components into two singlets for convenience.
\begin{equation}
\psi_{1} = {f_{L} \choose f'_{L}} \quad , \quad \psi_{2}=f_{R} \quad , \quad \psi_{3}=f'_{R}\;.
\end{equation}

If we want the Lagrangian to be invariant under gauge transformations of the group $G=SU(2)_{L} \otimes U(1)_{Y}$, we consider abelian and non-abelian gauge transformations simultaneously, so that $\psi(x)$ transforms to,
\begin{equation}
\psi_{j}(x):\hspace{5mm} \underrightarrow{G}\hspace{5mm} \psi'_{j}(x)= U_{Y}U_{L}\psi_{j}(x)\;,
\end{equation}

\noindent
where
\begin{equation}
U_{L}=\exp\left\lbrace i\frac{\sigma^{j}}{2}\alpha^{j}(x)\right\rbrace \quad \text{;} \quad U_{Y}=\exp\left\lbrace iY_{j}\beta(x)\right\rbrace \hspace{1cm} j=(1,2,3)\;,
\end{equation}

\noindent
$\beta=\beta(x)$ and $\alpha^{j}=\theta^{j}(x)$ are arbitrary parameters depending on the space-time coordinates. The parameter $Y$ is the hypercharge and $\sigma_{j}$ are the Pauli matrices.\\

The Lagrangian described by the equation \eqref{eq2:LG invariante} is invariant under global transformations of the group $G$, i.e:
\begin{align}
\psi_{1}(x)\hspace{5mm} \underrightarrow{G}\hspace{5mm} \psi'_{1}(x) &= \exp\left\lbrace iY_{1}\beta(x)\right\rbrace U_{L}\psi_{1}(x)\;, \nonumber \\[1pt]
\psi_{2}(x)\hspace{5mm} \underrightarrow{G}\hspace{5mm} \psi'_{2}(x) &= \exp\left\lbrace iY_{2}\beta(x)\right\rbrace \psi_{2}(x)\;, \\[1pt]
\psi_{3}(x)\hspace{5mm} \underrightarrow{G}\hspace{5mm} \psi'_{3}(x) &= \exp\left\lbrace iY_{3}\beta(x)\right\rbrace \psi_{3}(x)\;. \nonumber
\end{align}

If we want the Lagrangian to be invariant under local transformations of the group $G$, the covariant derivative must have the following form~\cite{Pich:2007vu},
\begin{equation}
D_{\mu} = \partial_{\mu}+ig\frac{\sigma^{j}}{2}W_{\mu}^{j}(x)+ig'YB_{\mu}(x)\;,
\end{equation}

where $g$ and $g'$ are the coupling constants associated with the $SU(2)_{L}$ and $U(1)_{Y}$ groups, respectively. $B_{\mu}$ is the gauge field required to maintain invariance under the $U(1)_{Y}$ transformation, and $W_{\mu}^{j}$ are the 3 associated gauge fields ($W^{\pm}$ and $Z$)\cite{Pich:2007vu} to maintain invariance under the $SU(2)_{L}$ group transformation.\\

Therefore, since we want $D_{\mu}\psi_{j}(x)$ to transform the same as $\psi(x)$, since the properties of the gauge field transformations remain fixed, then:
\begin{align}
B_{\mu}(x)\hspace{5mm} \underrightarrow{G}\hspace{5mm} B'_{\mu}(x) &= B_{\mu}(x)-\frac{1}{g'}\partial_{\mu}\beta(x)\;, \\[1pt] 
W_{\mu}^{j}(x)\hspace{5mm} \underrightarrow{G}\hspace{5mm} \left(W_{\mu}^{j}(x)\right)'&= W_{\mu}^{j}(x)-\frac{1}{g}\left(\partial_{\mu}\alpha^{j}\right)+\epsilon_{ijk}\alpha^{k}W_{\mu}^{l}\;.
\end{align}

For the introduction of the invariant kinetic terms for the gauge fields, we introduce the corresponding intensity tensors (abelian and non-abelian).
\begin{align}
B_{\mu\nu} &= \partial_{\mu}B_{\nu}-\partial_{\nu}B_{\mu}\;, \\[1pt]
W_{\mu\nu}^{j} &= \partial_{\mu}W_{\nu}^{j}-\partial_{\nu}W_{\mu}^{j}+g\epsilon^{jik}W_{\mu}^{i}W_{\nu}^{k}\;.
\end{align}

We can observe that $B_{\mu\nu}$ remains invariant under transformations of the group $G$, while $W_{\mu\nu}^{j}$ transforms covariantly according to~\cite{Pich:2007vu}:
\begin{equation}
B_{\mu\nu}:\hspace{5mm} \underrightarrow{G}\hspace{5mm} B_{\mu\nu} \quad \text{;} \quad W_{\mu\nu}^{j}\hspace{5mm} \underrightarrow{G}\hspace{5mm} U_{L}W_{\mu\nu}^{j}U_{L}^{\dagger}\;.
\end{equation}

\noindent
Therefore, the kinetic terms for the gauge fields are given by:
\begin{equation}
\mathcal{L}_{G} = -\frac{1}{4}B_{\mu\nu}B^{\mu\nu}-\frac{1}{4}W_{\mu\nu}^{j}W_{j}^{\mu\nu}\;.
\end{equation}

\subsection{Spontaneous symmetry breaking}

The gauge symmetry $SU(2)_{L}\times U(1)_{Y}$ forbids mass terms for gauge bosons. Mass terms for fermions are also impossible.\\

Spontaneous symmetry breaking (SSB) occurs when the vacuum of the system (minimum energy state) is degenerate. The physical vacuum is one of the possible minimum energy states connected by the symmetries of the Lagrangian. When nature chooses it, the symmetry of the physical states is broken, although the symmetry of the Lagrangian is preserved.\\

The result of the SSB depends on the type of symmetries. If the lagrangian is invariant under a continuous group of symmetries G, but the vacuum is invariant only under a subgroup $H\subset G$, then as many massless, spin-0 states (Goldstone bosons) appear as $G$ generators that are not those of $H$, i.e., the number of symmetries that have been broken (Goldstone theorem~\cite{goldstone1961field}). If the symmetries of the Lagrangian are local (gauge), these Goldstone bosons are absorbed by the gauge bosons associated with the broken symmetries, giving them mass (Higgs-Kibble mechanism~\cite{englert1964broken}).\\

Let us illustrate the SSB with the simplest example: a complex scalar field $\phi(x)$, whose Lagrangian is described by~\cite{langacker2017standard,peskin2018introduction}:
\begin{equation}
\mathcal{L}_{\phi}= \partial_{\mu}\phi^{\dagger}\partial^{\mu}\phi - V(\Phi) \quad \text{with} \quad V(\Phi)=\mu^{2}\Phi^{\dagger}\Phi + \lambda(\Phi^{\dagger}\Phi)^{2}\;,
\label{eq2:La escalar}
\end{equation}

\noindent
with real and positive $\lambda$, where $\phi$ was said before and $\Phi$ is represented by a doublet of $SU(2)_{L}$ with hypercharge $Y = 1$:
\begin{equation}
\Phi = \left( 
\begin{array}{c}
\phi^{\pm} \\
\phi^{0}
\end{array}
\right)
= 
\left(
\begin{array}{c}
\phi^{1}+i\phi^{2}\\
\sigma+i\chi
\end{array} 
\right)\;.
\end{equation}

This definition ensures that the scalar Lagrangian is invariant under $SU(2)_{L}\otimes U(1)_{Y}$. We can get the vacuum from the Hamiltonian:
\begin{equation}
H= \frac{1}{2}\left[\left(\partial_{0}\Phi_{0}\right)^{2}+(\nabla\Phi)^{2}\right]+V(\Phi^{\dagger}\Phi)\;.
\end{equation}

The minimum of $H$ is obtained for $\langle\Phi\rangle_0= \text{constant}$. Due to the symmetry $SU(2)_{L}$, it can be chosen without loss of generality.
\begin{equation}
\langle\Phi\rangle_{0}=\frac{1}{\sqrt{2}} {0 \choose v}\;,
\label{eq2:Phi0}
\end{equation}

\noindent
so that $v$ corresponds to the minimum of the potential $V(\Phi^{\dagger}\Phi)$ and thus of the energy. So the equation for this minimum is
\begin{equation}
\mu^{2}+\lambda\langle\Phi\rangle_{0}^{2}=0\;,
\end{equation}

\noindent
where it is obtained:
\begin{equation}
v=\sqrt{\frac{-\mu^{2}}{\lambda}}\;.
\end{equation}\\

For a minimum energy state (the vacuum) to exist, the parameter $\lambda$ must satisfy $\lambda > 0$. With respect to $\mu^{2}$ there are two possibilities:
\begin{itemize}
\item[i)] If $\mu^{2} > 0$, the potential has only a trivial minimum. It is then a scalar field with mass $\mu$ and quartic coupling $\lambda$\cite{illana2007modelo,langacker2017standard}.

\item[ii)] If $\mu^{2} < 0$, we do not have a unique vacuum state, since there is a continuous set of degenerate vacua, which can be clearly seen in the figure \ref{fig:pote}b of the potential. The minimum corresponds to satisfying field configurations:
\begin{figure}
\centering
\includegraphics[scale=0.8]{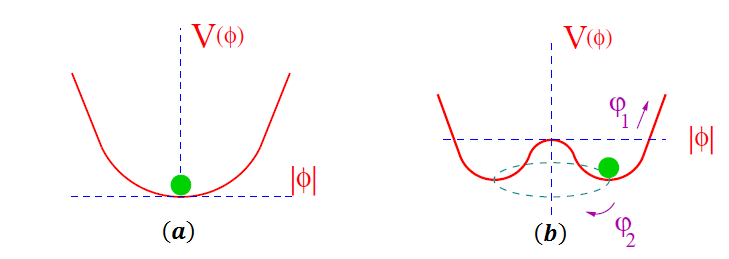}
\caption{(a) Form of the scalar potential for $\mu^{2}>0$ and (b) $\mu^{2} < 0$. In the second case, there is a continuum of degenerate vacuums corresponding to different phases, connected by a massless field excitation $\phi_{2}$.}
\label{fig:pote}
\end{figure}
\begin{align}
\abs{\bra{0}\phi(x)\ket{0}}= \abs{\phi_{0}(x)} &= \sqrt{\frac{-\mu^{2}}{2\lambda}}= \frac{v}{\sqrt{2}}>0\;, \nonumber \\[1pt]
V(\psi_{0}) &= -\frac{\lambda}{4}v^{4}\;.
\end{align}
\end{itemize}
Since the hypercharge of the $\Phi$ field is 1, it is obtained for the vacuum state:
\begin{equation}
\left(T_{3}+\frac{Y}{2}\right)\langle\Phi\rangle_{0}=0\;.
\end{equation}

Then the vacuum is annihilated by the $T_{3} +\frac{Y}{2}$ operator, and recalling the relationship between electric charge, isospin, and hypercharge, we conclude that the $U(1)$ symmetry associated with the charge operator is not broken. So we have the following symmetry breaking pattern,
\begin{equation}
SU(2)_{L}\otimes U(1)_{Y}\hspace{5mm} \underrightarrow{SSB}\hspace{5mm} U(1)_{Q}\;.
\end{equation}

The photon, the generator of $U(1)_{Q}$, remains massless. Instead, we will see that the gauge bosons associated with the generators $T_1$; $T_2$ and $\left(T_3+\frac{Y}{2}\right)$ acquire mass. If we express $\phi^{0}$ in terms of $\langle\phi\rangle_0$, we get
\begin{equation}
\sigma+i\chi=\frac{v}{\sqrt{2}}+\frac{H}{\sqrt{2}}+i\chi\;,
\label{phi 0}
\end{equation}

\noindent
where the $H$ field is called the Higgs boson, we can write the Lagrangian (we consider only the terms associated with $\sigma$ as:
\begin{equation}
\mathcal{L}_{\phi}= \left\vert \left(\partial_\mu+igT^{a}W_{\mu}^{a}+i\frac{g'}{2}B_{\mu}\right)\frac{v+H}{\sqrt{2}}{0 \choose 1}\right\vert^{2} -\mu^{2}\frac{(v+H)^{2}}{2}-\lambda\frac{(v+H)^{4}}{4}\;,
\end{equation}

\noindent
where the quadratic terms for the vector fields are:
\begin{equation}
\frac{g^{2}(v+H)^{2}}{4}W_{\mu}^{+}W^{-\mu}+\frac{g^{2}(v+H)^{2}}{8\cos^{2}\theta_{W}}Z_{\mu}Z^{\mu}\;.
\end{equation}

It is important to note that the terms with $v^{2}$ represent mass terms for the $W^{\pm}$ and $Z$ bosons. Therefore, the masses of the $W^{\pm}$ and $Z$ bosons are given by:
\begin{equation}
M_{W}=\frac{gv}{2} \quad \text{y} \quad M_{Z}=\frac{gv}{2\cos\theta_{W}}=\frac{M_{W}}{\cos\theta_{W}}\;.
\label{masa W Y Z}
\end{equation}

And so, the Higgs mechanism has endowed the $W$ and $Z$ bosons with mass starting from a gauge invariant Lagrangian, through the spontaneous breaking of the electroweak symmetry $SU(2)_{L}\otimes U(1)_{Y}$.


\subsection{Yukawa Coupling}

The fermion masses are generated by the Yukawa interactions between the Higgs doublet and fermions, represented through the spontaneous symmetry breaking of $SU(2){L}\times U(1){Y}$. The most general gauge-invariant Yukawa Lagrangian for the $3$ families of fermions takes the following form: \cite{langacker2017standard}:
\begin{equation}
-\mathcal{L}_{\text{y}}= \sum_{jk=1}^{3}\left[ y_{jk}^{(d)}\overline{q}_{jL}^{0}\Phi d_{kR}^{0}+ y_{jk}^{(u)}\overline{q}_{jL}^{0}\widetilde{\Phi}u_{kR}^{0} +y_{jk}^{(l)}\overline{l}_{jk}^{0}\Phi e_{kR}^{0}+y_{jk}^{(\nu)}\overline{l}_{jL}^{0}\widetilde{\Phi}\nu_{kR}^{0}\right]+h.c\;.
\label{eq2:L yukawa}
\end{equation}

\noindent
where the superscript $^{0}$ refers to the fact that these fields are weak eigenstates, that is, they have defined gauge transformation properties, with the element of each doublet transforming into each other under $SU(2)_{L} $ and $y_{jk}^{(d)}$, $y_{jk}^{(u)}$, $y_{jk}^{(l)}$ $(j,k= 1,2,3)$ are arbitrary coupling constants. On the other hand, we also have:
\begin{equation}
\Phi={\phi^{+} \choose \phi^{-}} \quad ; \quad \widetilde{\Phi}=i\tau^{2}\Phi^{\dagger}={\phi^{0\dagger} \choose -\phi^{-}}\;,
\end{equation}

After SSB, and as we know from equation \eqref{eq2:Phi0}, in the unitary gauge, the Yukawa Lagrangian can be written as \cite{illana2007modelo}:
\begin{equation}
-\mathcal{L}_{Yk}= \left(1+\frac{H}{v}\right)\left\lbrace \overline{d}_{L}^{0}M_{d}d_{R}^{0}+\overline{u}_{L}^{0}M_{u}u_{R}^{0}+\overline{l}_{L}^{0}M_{l}l_{R}^{0}+h.c. \right\rbrace\;.
\end{equation}

Here $d^{0}$, $u^{0}$ and $l^{0}$ denote vectors in the three-dimensional flavour space. The mass matrices are given by
\begin{equation}
\left({\bf M}_{d}^{'}\right)_{ij}=y_{ij}^{(d)}\frac{v}{\sqrt{2}}\quad ; \quad \left({\bf M}_{u}^{'}\right)_{ij}=y_{ij}^{(u)}\frac{v}{\sqrt{2}} \quad ; \quad \left({\bf M}_{l}^{'}\right)_{ij}=y_{ij}^{(l)}\frac{v}{\sqrt{2}}\;. \quad (i,j= 1,2,3)
\label{eq2:masa q}
\end{equation}

The diagonalization of these matrices determines the mass eigenstates $d_{j}$, $u_{j}$, and $l_{j}$, (with $j=1,2,3$). The three matrices ${\bf M}{f}$ can be written as
\begin{equation}
{\bf M}_{f}={\bf H}_{f}\mathcal{U}_{f}=R_{f}^{\dagger}\mathcal{M}_{f}R_{f}\mathcal{U}_{f}\;,
\end{equation}

\noindent
where ${\bf H}{f}=\sqrt{{\bf M}_{f}{\bf M}_{f}^{\dagger}}$ is a positive definite Hermitian matrix, and $\mathcal{U}_{f}$ is a unitary matrix. Each ${\bf H}_{f}$ can be diagonalized by a unitary matrix $R_{f}$. The resulting matrix $\mathcal{M}_{f}$ is diagonal and positive-definite. Thus, in terms of the diagonal matrices, we have:
\begin{equation}
\mathcal{M}_{d}=\text{diag}(m_{d},m_{s},m_{b})\;,\quad \mathcal{M}_{u}=\text{diag}(m_{u},m_{c},m_{t})\;, \quad \mathcal{M}_{l}=\text{diag}(m_{e},m_{\mu},m_{\tau})\;.
\end{equation}
\noindent
The Yukawa Lagrangian takes the following form \cite{illana2007modelo}
\begin{equation}
\mathcal{L}_{Yk}= -\left(1+\frac{H}{v}\right)\left\lbrace \overline{d}\mathcal{M}_{d}d+\overline{u}\mathcal{M}_{u}u+\overline{l}\mathcal{M}_{l}l\right\rbrace\;,
\end{equation}

\noindent
where the mass eigenstates are defined by the relations:
\begin{align}
d_{L}&=R_{d}d_{L}^{0}\;, & u_{L}&=R_{u}u_{L}^{0}\;, & l_{L}&=R_{l}l_{L}^{0}\;, \nonumber \\
d_{R}&=R_{d}\mathcal{U}_{d} d_{R}^{0}\;, & u_{R}&=R_{u}\mathcal{U}_{u} u_{R}^{0}\;, & l_{R}&=R_{l}\mathcal{U}_{l} l_{R}^{0}\;.
\end{align}
\subsection{Cabibbo-Kobayashi-Maskawa (CKM) Matrix and CP Violating}

As mentioned in chapter \ref{cap.introduccion}, in the Standard Model, the Cabibbo-Kobayashi-Maskawa (CKM) matrix is a unitary matrix containing information about the strength of weak interactions that change flavor. That is, it specifies the difference between the quantum states of quarks when they propagate freely and when they participate in weak interactions.\\

However, since $\overline{u}_{L}d_{L}=\overline{u}_{L}R_{u}S_{d}^{\dagger}d_{L}$, as in general $R_{u}\neq R_{d}$. The CKM matrix is defined \cite{Pich:2018njk},
\begin{equation}
V=R_{u}R_{d}^{\dagger} \quad \Rightarrow \quad \overline{u}_{L}^{'}d_{L}^{'}=\overline{u}_{L}Vd_{L}\;.
\end{equation}

The $3\times 3$ CKM matrix is unitary and appears in the charged current interactions of quarks:
\begin{equation}
\mathcal{L}_{CC}= \frac{g}{2\sqrt{2}}\left\lbrace Q_{\mu}^{+}\left[\sum_{ij}^{3}\overline{u}_{i}\gamma^{\mu}(1-\gamma_{5})V_{ij}d_{j}+\sum_{l=e,\mu,\tau}\overline{\nu}_{l}\gamma^{\mu}1-\gamma_{5})l \right]+h.c. \right\rbrace\;.
\label{Lcc}
\end{equation}

\noindent
The matrix $V$ couples each $up$ type quark to all $down$ type quarks.\\

We have assumed that neutrinos are massless. In that case, we can always redefine the flavors of neutrinos so that we eliminate the analogous mixing in the leptonic sector: $\overline{\nu}_{L}^{0}l_{L}^{0}=\overline{\nu}_{L}S_{l}l_{L}=\overline{\nu}_{L}l_{L}$, and we have flavor conservation. Note that if the $u_{i}$ or $d_{j}$ had degenerate masses, we could also redefine the fields, and there would also be flavor conservation in the quark sector. If $\nu_{R}$ fields are included, Yukawa couplings for neutrinos could be introduced, leading to a mass matrix $({\bf M}_{\nu})_{ij}=y_{ij}^{(\nu)}v/\sqrt{2}$ and we would obtain leptonic flavor violation through a mixing matrix analogous to the CKM. The next chapter will cover the phenomenon of neutrino oscillations, which indicates that neutrinos have masses, albeit very small.\\

The masses of the fermions and the mixing matrix $V$ of the quarks are determined by the corresponding Yukawa coupling matrices $y_{ij}^{(f)}$, which are free parameters. A general $n\times n$ unitary matrix is characterized by $n^{2}$ real parameters: $n(n-1)/2$ moduli and $n(n+1)/2$ phases. Several of these phases are irrelevant, as one can redefine the phases of the fields (they are not physical): $u_{i} \rightarrow e^{i\phi_{i}} u_{i}$ and $d_{j} \rightarrow e^{i\theta_{j}} d_{j}$, such that $V_{ij} \rightarrow V_{ij} e^{i(\theta_{j}-\phi_{i})}$. This means there are $2n-1$ unobservable phases. Therefore, the number of physical free parameters is reduced to $(n- 1)^{2}; n(n-1)/2$ moduli and $(n-1)(n-2)/2$ phases.\\

Thus, if only two generations are mixed, $V$ is determined by a single parameter, the Cabibbo angle,
\begin{equation}
V=\left(
\begin{array}{cc}
\cos\theta_{C} & \sin\theta_{C}\\
-\sin\theta_{C} & \cos\theta_{C}
\end{array}
\right)\;.
\end{equation}

\noindent
If we now consider a mix of 3 generations, the matrix is defined as:
\begin{align}
V &=\left(
\begin{array}{ccc}
V_{ud} & V_{us} & V_{ub}\\
V_{cd} & V_{cs} & V_{cb}\\
V_{td} & V_{ts} & V_{tb}
\end{array}
\right)\;.
\end{align}

It's important to know how many independent parameters the CKM matrix contains; for three families of quarks, it is a complex $3\times 3$ matrix with $2\cdot3^{2}=18$ real parameters and from the unitarity condition:
\begin{align}
V^{\dagger}V =\left(
\begin{array}{ccc}
V_{ud}^{*} & V_{us}^{*} & V_{ub}^{*}\\
V_{cd}^{*} & V_{cs}^{*} & V_{cb}^{*}\\
V_{td}^{*} & V_{ts}^{*} & V_{tb}^{*}
\end{array}
\right)^{T}
\left(
\begin{array}{ccc}
V_{ud} & V_{us} & V_{ub}\\
V_{cd} & V_{cs} & V_{cb}\\
V_{td} & V_{ts} & V_{tb}
\end{array}
\right)
= 1\;,
\end{align}

we obtain 9 conditions that eliminate 9 of the 18 real parameters. Of the remaining 9 parameters, we can parameterise the rotation angles with 3 real numbers; the remaining 6 parameters can be associated with the phases of the quark fields, of which 5 phases are removable~\cite{Zupan:2019uoi}, and $V$ can be parameterised in terms of 3 real rotation angles $\theta_{ij}$ and one phase $\delta_{13}$. There are different (but equivalent) representations in the literature. The so-called standard parameterisation is given by~\cite{illana2017modelo,Pich:2007vu}:
\begin{align}
V_{CKM}&=\left(
\begin{array}{ccc}
1 & 0 & 0 \\
0 & c_{23} & s_{23} \\
0 & -s_{23} & c_{23}
\end{array}
\right)
\left(
\begin{array}{ccc}
c_{13} & 0 & s_{13}e^{-i\delta_{13}} \\
0 & 1 & 0 \\
-s_{13}e^{i\delta_{13}} & 0 & c_{13} \\
\end{array}
\right)
\left(
\begin{array}{ccc}
c_{12} & s_{12} & 0 \\
-s_{12} & c_{12} & 0 \\
0 & 0 & 1 \\
\end{array}
\right)\;, \nonumber \\[1pt]
&=
\left(
\begin{array}{ccc}
c_{12}c_{13} & s_{12}c_{13} & s_{13}e^{-i\delta_{13}} \\
-s_{12}c_{23}-c_{12}s_{23}s_{13}e^{i\delta_{13}} & c_{12}c_{23}-s_{12}s_{23}s_{13}e^{i\delta_{13}} & s_{23}c_{13} \\
s_{12}s_{23}-c_{12}c_{23}s_{13}e^{i\delta_{13}} & -c_{12}s_{23}-s_{12}c_{23}s_{13}e^{i\delta_{13}} & c_{23}c_{13} \\
\end{array}
\right)\;.
\end{align}

\noindent
where $c_{ij} = \cos\theta_{ij}$ and $s_{ij} = \sin\theta_{ij}$ $(i, j = 1, 2, 3)$. The angles $\theta_{12}$, $\theta_{13}$, and $\theta_{23}$ can all be made to lie in the first quadrant, through an appropriate redefinition of the phases of the quark fields. Thus, $c_{ij} \geq 0$, $s_{ij} \geq 0$ and $0 \leq \delta_{13} \leq 2\pi$. It should be noted that $\delta_{13}$ is the only physical phase in the SM Lagrangian. Therefore, it is the only possible source of CP violation~\cite{illana2017modelo}.\\

Experimentally, only the moduli of $V_{ij}$ can be accessed. The table \ref{tab:elementos ckm} shows the values that have been determined directly.
\begin{table}[t]
\centering
\begin{tabular}{ccc|c|c}
\toprule[0.13em]& CKM & & Value & Source \\
\hline
& $|V_{ud}|$ & & $0.97377 \pm 0.00027$ & Disintegration $\beta$ nuclear \\
& & & $0.9746 \pm 0.0019$ & $n \rightarrow pe^{-}\overline{\nu}_{e}$ \\ 
& & & $0.9728 \pm 0.0030$ & $\pi^{+}\rightarrow \pi^{0}e^{+}\nu_{e}$ \\ 
& & & $0.97378 \pm 0.00027$ & average \\
\hline
& $|V_{us}|$ & & $0.2234 \pm 0.0024$ & $K\rightarrow \pi l^{+}\nu_{l}$ \\ 
& & & $0.2220 \pm 0.0033$ & decay of $\tau$ \\ 
& & & $0.2226_{-0.0014}^{+0.0026}$ & $K^{+}/\pi^{+}\rightarrow \mu^{+}\nu_{\mu}, \quad V_{ud}$ \\ 
& & & $0.226 \pm 0.005$ & hyperion decay \\ 
 &  &  & $0.2230 \pm 0.0015$ & average \\ 
\hline
 & $|V_{cd}|$ &  & $0.213 \pm 0.022$ & $D \rightarrow\pi l\overline{\nu}_{l}$ \\ 
 &  &  & $0.230 \pm 0.011$ & $\nu d \rightarrow cX$ \\ 
 &  &  & $0.227 \pm 0.010$ & average \\ 
\hline
 & $|V_{cs}|$ &  & $0.957 \pm 0.095$ & $D\rightarrow Kl\overline{\nu}_{l}$ \\ 
 &  &  & $0.94_{-0.29}^{+0.35}$ & $W^{+}\rightarrow c\overline{s}$ \\ 
 &  &  & $0.974 \pm 0.013$ & $W^{+}\rightarrow$had., $V_{uj}$, $V_{cd}$, $V_{cb}$ \\ 
\hline
 & $|V_{cb}|$ &  & $0.0392 \pm 0.0016$ & $B\rightarrow D^{*}l\overline{\nu}_{l}$ \\ 
 &  &  & $0.0417 \pm 0.0007$ & $b\rightarrow cl\overline{\nu}_{l}$ \\ 
 &  &  & $0.0413 \pm 0.0006$ & average \\ 
\hline
 & $|V_{ub}|$ &  & $0.0039 \pm 0.0006$ & $B\rightarrow \pi l\overline{\nu}_{l}$ \\ 
 &  &  & $0.0045 \pm 0.0003$ & $b\rightarrow ul\overline{\nu}_{l}$ \\ 
 &  &  & $0.0044 \pm 0.0003$ & average \\ 
\hline
 & $|V_{tb}|/\sqrt{\sum_{q}|V_{tq}|^{2}}$ &  & $>0.78$ & $t\rightarrow bW/qW$ \\ 
 & $|V_{tb}|$ &  & $>0.68 \quad ; \quad \le 1$ & $p\overline{p}\rightarrow tb + X$ \\ 
\bottomrule[0.13em]
\end{tabular} 
\caption{Direct determination of the elements of the CKM matrix $V_{ij}$. Taken from~\cite{Pich:2007vu}.}
\label{tab:elementos ckm} 
\end{table}


\section{Problem of Standard Model}
\lhead[\thepage]{\thesection. Problem of Standard Model}

The SM is one of the most successful scientific theories and has proven to be very useful for particle physics as it explains many known phenomena. However, it has difficulties in explaining several problems, leaving some aspects unanswered, such as:
\begin{itemize}
\item The origin and hierarchy of the fermion mass spectrum.
\item The number of generations (flavor) in the theory is arbitrary.
\item The origin of CP violation is a mystery.
\item The asymmetry between matter and antimatter in the universe.
\item It does not predict mass for neutrinos.
\item It does not contemplate dark matter.
\end{itemize}

In this section, we will discuss only some of the SM problems: the hierarchy problem, the flavor problem, and dark matter. In the case of neutrinos, we will discuss the general aspects since, in chapter \ref{cap.neutrino}, we will discuss neutrinos in more detail.

\subsection{Hierarchy problem}

In physics, we expect macroscopic behavior to follow from a microscopic theory. The microscopic theory is unlikely to contain several free parameters that are carefully tuned to give the macroscopic system some special properties.\\

What we want is for the theory on a low energy scale $\mu_{1}$ to follow the properties of a much higher energy scale $\mu_{2}$ without the need to tune the various parameters in the high energy theory to a precision of order $\mu_{1}/\mu_{2}$. However, one parameter is allowed to be very small (of order $\mu_{1}/\mu_{2}$), provided that this property is not affected by higher order effects~\cite{jegerlehner2013hierarchy}.\\

As mentioned above, the SM is a very successful description of particle physics, however this is at the weak scale ($\sim$ 100 GeV). Therefore, if we have TeV-scale physics, we run into the hierarchy problem, where the following question arises: why is the electroweak scale so much smaller than the Planck scale ($\sim 10^{19}$ GeV)? The Higgs vacuum expectation value $v\sim$ 246 GeV gives the electroweak scale, the only dimensional parameter in the SM \cite{burdman2007new}. However, the Higgs VEV is not naturally stable under radiative corrections. Suppose we consider radiative corrections to the Higgs mass arising from its gauge boson couplings, Yukawa fermion couplings, and scalar self-couplings. In that case, these corrections exhibit a quadratic sensitivity to the ultraviolet cutoff. Therefore, if the SM were valid up to the Planck scale, the radiative corrections would push $m_{h}$ and the minimum of the Higgs potential, $v$, to the Planck scale. To avoid this, one has to adjust the Higgs vacuum mass into the SM Lagrangian by one part in $10^{17}$\cite{burdman2007new}, which is quite unnatural and is known as the hierarchy problem.

\subsection{The flavor problem}

\begin{figure}[!ht]
\centering
\includegraphics[scale=0.6]{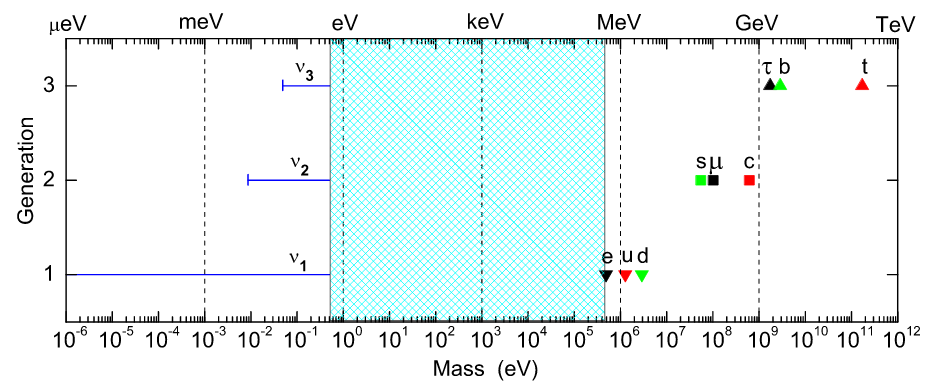}
\caption{Schematic of the flavor ''hierarchy'' problem and the ''desert'' in the SM mass spectrum at an electroweak scale. From \cite{Xing:2014sja}. }
\label{fig:jerarquia-masas}
\end{figure}

If we focus on the electroweak sector of the SM, we see that it has thirteen free parameters: three charged lepton masses, six quark masses, three mixing angles, and one CP violation phase. Since all three neutrinos must be massive beyond the SM \cite{Xing:2014sja}, one has to introduce seven more free parameters to describe the flavor properties: three neutrino masses, three mixing angles, and one CP violation phase if we assume Dirac neutrinos. However, if the neutrinos were Majorana neutrinos, we would have to introduce nine extra parameters: three neutrino masses, three mixing angles, three complex phases, the two Majorana phases, and the lepton CP violation phase. Therefore, at least 18 degrees of freedom are in the low-energy flavor, raising the question: Why is the number of degrees of freedom so large in the flavor sector?\\

In addition to this question, other questions arise related to the flavor problem, such as: What is the number of fermion families? Why are there three fermion families and not another number? Why are the masses of the fermions so hierarchical and not of the same order? This can be seen if we express the fermion masses as a function of the top quark mass and the Cabibbo angle \cite{Xing:2014sja,Feruglio:2015jfa,Fritzsch:1999ee}.
\begin{align}
\sqrt{|\Delta m_{13}^{2}|}\sim \lambda^{20}m_{t} \quad ; \quad \sqrt{|\Delta m_{12}^{2}|}\sim \lambda^{21}m_{t}\;, \nonumber \\[1pt]
m_{e}\sim \lambda^{9}m_{t} \quad ; \quad m_{u}\sim m_{d}\sim \lambda^{8}m_{t}\;, \nonumber \\[1pt]
m_{s}\sim m_{\mu}\sim\lambda^{5}m_{t}\;, \nonumber \\[1pt]
m_{c}\sim \lambda^{4}m_{t} \quad ; \quad m_{b}\sim m_{\tau}\sim \lambda^{3}m_{t}\;, \\[1pt]
\sin\theta_{12}^{(q)}\sim \lambda \quad ; \quad \sin\theta_{23}^{(q)}\sim \lambda^{2} \quad ; \quad \sin\theta_{13}^{(q)}\sim \lambda^{4}\;, \nonumber \\[1pt]
\sin\theta_{12}^{(l)}\sim \sqrt{\frac{1}{3}} \quad ; \quad \sin\theta_{23}^{(l)}\sim \sqrt{\frac{1}{2}} \quad ; \quad \sin\theta_{13}^{(l)}\sim \frac{\lambda}{\sqrt{2}}\;,\nonumber
\label{eq:masas-fermiones}
\end{align}
\noindent
where $\lambda=\sin\theta_{C}=0.225$\cite{Xing:2014sja} and $m_{t}=173.8 \pm 5.2$ GeV \cite{Fritzsch:1999ee}.\\

To better understand the problem of the fermion mass hierarchy, a schematic of the six-quark, six-lepton mass spectrum on the electroweak scale is shown (see Fig.~\ref{fig:mass-hierarchy}), where a normal ordering in the neutrino mass spectrum has been assumed. It can be seen that the difference between the neutrino masses $m_{i}$ and the top quark mass $m_{t}$ is at least twelve orders of magnitude. Moreover, the ''desert'' between the heaviest neutral fermion $\nu_{3}$ and the lightest charged fermion $e^{-}$ spans at least six orders of magnitude.

\subsection{Neutrino masses}

For many years, neutrinos have been considered massless, however, through hard theoretical and experimental work, it has been confirmed that neutrinos are massive, with a mass of the order of $10^{-10}$ GeV. Thus confirming Bruno Pontecorvo's hypothesis \cite{osti_4343073} between 1998 and 2002 with the atmospheric and solar neutrino oscillations experiments of the Super-Kamioka Neutrino Detection Experiment (Super-Kamiokande)\cite{Super-Kamiokande:1998kpq}, the Sudbury Neutrino Observatory (SNO)\cite{SNO:2002tuh} and the Kamioka Liquid Scintillator Antineutrino Detector (KamLAND) \cite{KamLAND:2002uet}. But this leads to the question, why are neutrino masses so small compared to those of charged leptons? This suggests that the origin of neutrino masses should be studied.\\

However, unlike the origin of the masses of charged leptons, the nature of the neutrinos mass is unknown, since, being neutral fermions, their masses can be Dirac or Majorana. These two mass terms are phenomenologically very different because, in the case of Dirac, the lepton number is conserved. In contrast, in the case of Majorana, it is broken into two units \cite{King:2014nza}.

\subsection{Dark matter}

One of the main evidences that motivated the existence of dark matter was the speed of rotation of galaxies, since according to the mass present in the galaxy, it will have a defined speed based on the radius according to:
\begin{align}
G\frac{Mm}{r^2}&=m\frac{v^2}{r} \notag\\
v&	=\sqrt{\frac{GM}{r}}.
\end{align}

\begin{figure}
\centering
\subfigure[]{\includegraphics[scale=0.4]{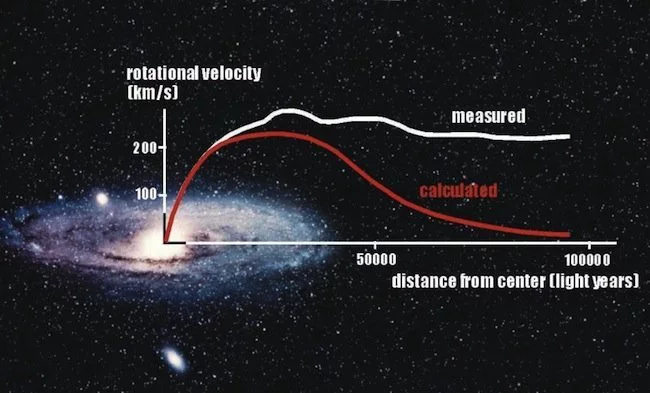}\label{fig:DM-velocidad}} \quad
\subfigure[]{\includegraphics[scale=0.4]{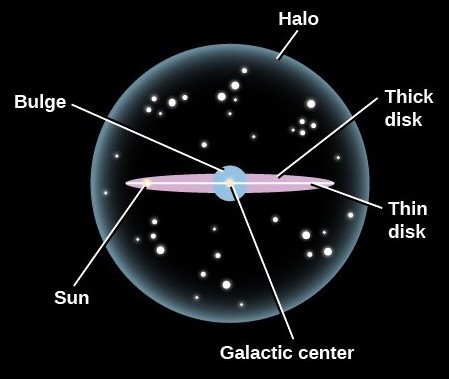}\label{fig:DM-halo}}
\caption{a) The relationship between rotation speed and distance is shown, where the red curve is the one expected by theory and the white curve is the one observed. b) The dark halo surrounding a galaxy is observed, showing that most of the matter in a galaxy is dark matter.}
\end{figure}

Where we can see that the speed is proportional to $v\sim\frac{1}{\sqrt{r}}$, however, in the 70s, Vera Rubin among others~\cite{Rubin:1970zza} found that the rotation speed of galaxies did not behave in the form $\frac{1}{\sqrt{r}}$, but as the radius increased, the speed tended to a constant, as can be seen in Fig.~\ref{fig:DM-velocidad}, that is, $M\sim r$, this means that the mass increases as the radius increases, even if there is no light, so there is something with gravitational effects that does not emit or absorb light, that is, dark matter (DM). This DM surrounds our galaxy, forming a dark halo (see Fig.\ref{fig:DM-halo}) and containing approximately 80\% of the total mass of the galaxy.
\begin{figure}[h]
\centering
\subfigure[]{\includegraphics[scale=0.25]{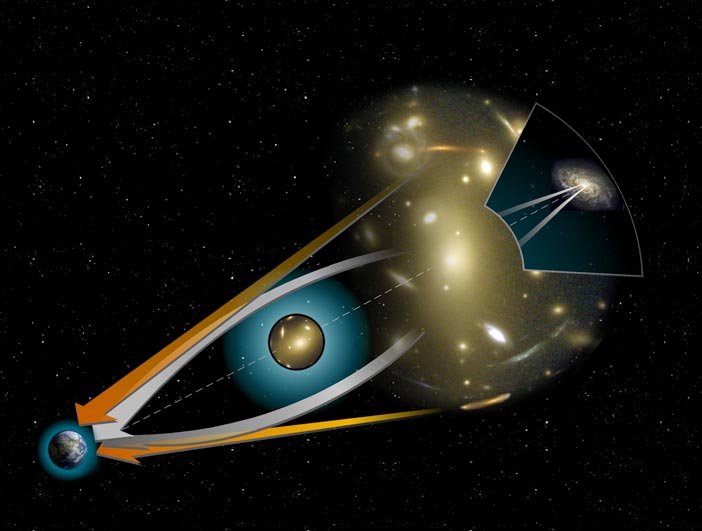}\label{fig:DM-anillo}} \quad
\subfigure[]{\includegraphics[scale=0.25]{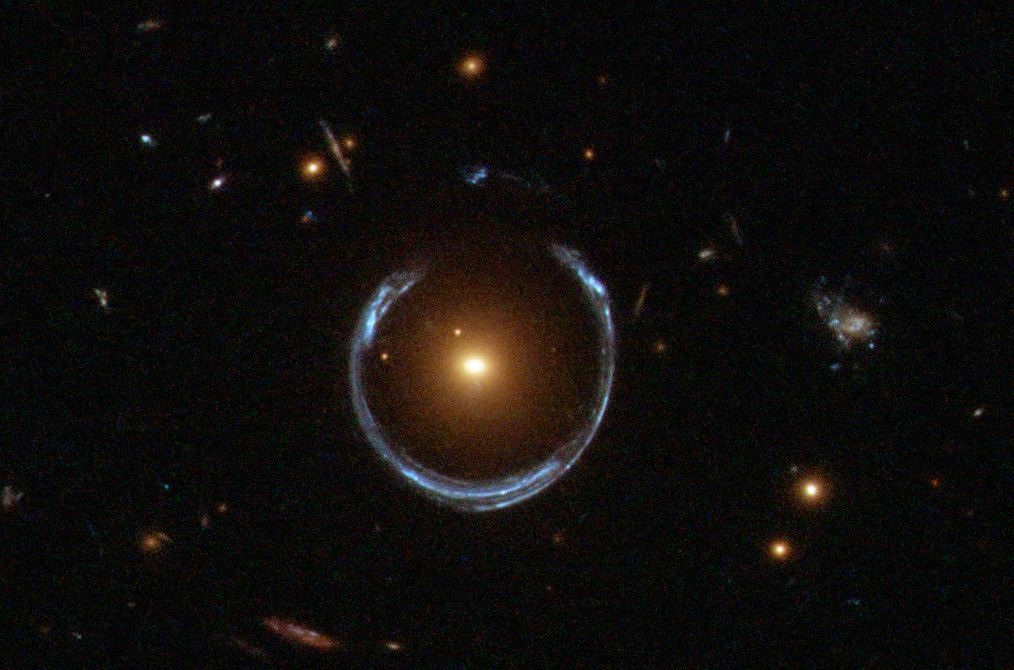}\label{fig:DM-anillo2}}
\caption{The figure shows the gravitational lensing effect, in which a) it shows how this effect occurs and b) it shows when light coming from distant stars passes through several paths and a ring is observed.}
\end{figure}
\begin{figure}[h]
\centering
\includegraphics[scale=0.4]{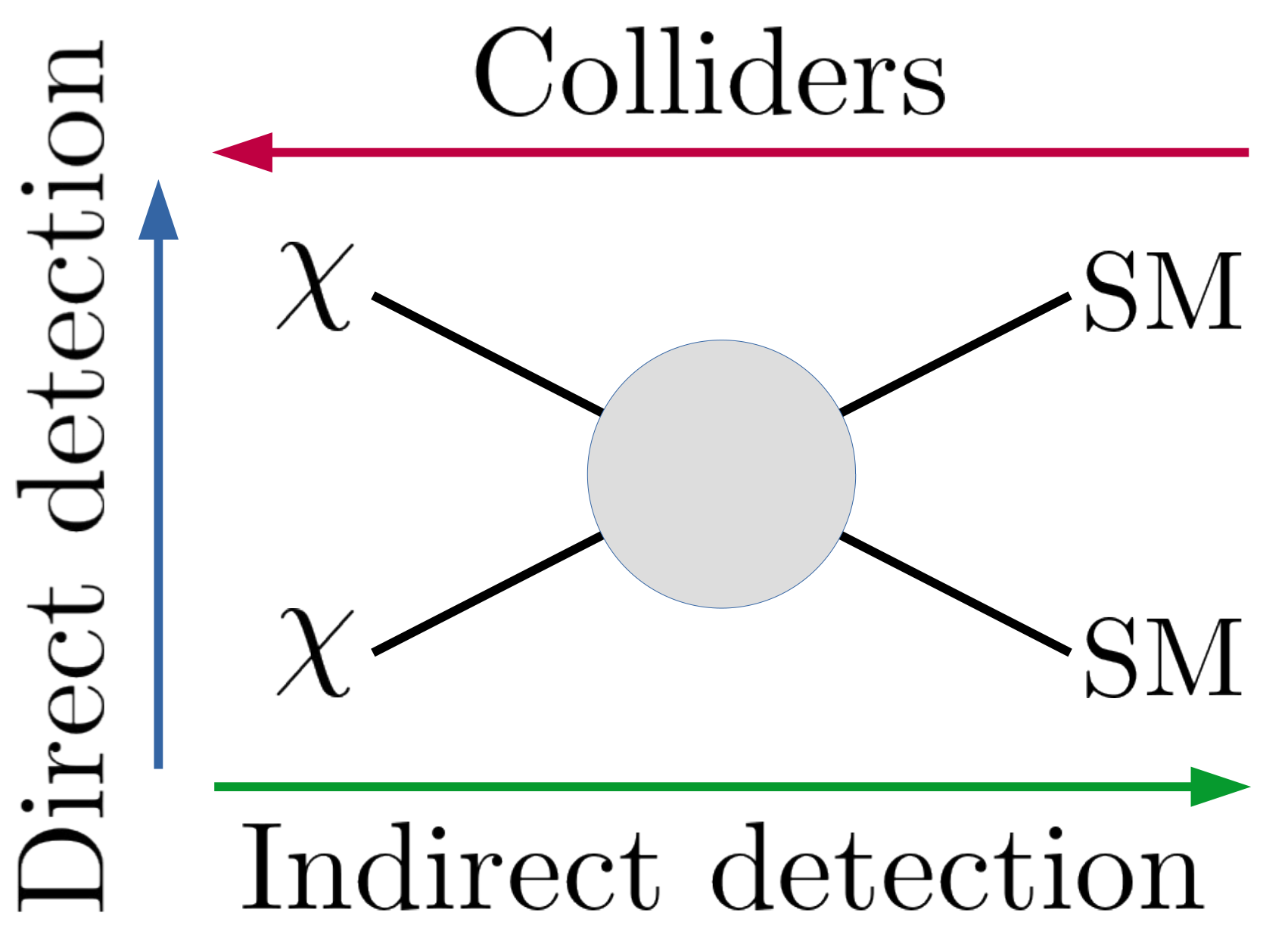}
\caption{Different methods of searching for dark matter.}
\label{fig:DM-busqueda}
\end{figure}
The above discussion is made by comparing the rotation of the solar system according to its mass distribution, so it can be said that the mass distribution of the galaxy does not have to be like that of the solar system. Therefore, the behavior of the galaxy does not have to be like that of the solar system. However, the existence of dark matter can also be inferred by the deflection of light from distant galaxies, since the light from these galaxies is deflected due to the gravitational field of the dark matter cluster between the Earth and the galaxy, being able to generate two images of the same galaxy (Fig.~\ref{fig:DM-anillo}) or even a ring (Fig.~\ref{fig:DM-anillo2}), this phenomenon is known as gravitational lensing.\\

Because of these observations and others, the existence of dark matter is accepted today, and particle physics is trying to find an explanation. However, there are no clues about the nature of dark matter, whether it is fermionic or bosonic. If we assume that dark matter is stable and only has gravitational interaction, there are several candidates where we can highlight the weakly interacting massive particle or WIMP, however, they are still candidates since observational data are needed to confirm or reject the dark matter candidate and among the methods of searching for dark matter are direct detection, indirect detection and searching in colliders. A diagram of these methods is shown in Fig.~\ref{fig:DM-busqueda} however, in general terms, we can say that:
\begin{itemize}
\item \textbf{Direct detection:} Looks for a small recoil in atomic nuclei upon colliding with an incident dark matter particle.
\item \textbf{Indirect detection:} It is the detection of particles as a byproduct of the annihilation of dark matter particles. Some of the byproducts are antiparticles, neutrinos and gamma rays.
\item \textbf{Collider:} This method is done in particle accelerators, such as the LHC, where it is hoped that colliding known particles at high energies can produce dark matter particles.
\end{itemize}

\newpage
$\ $
\thispagestyle{empty} 
\chapter{Neutrino Physics}\label{cap.neutrino}
\markboth{NEUTRINO PHYSICS}{NEUTRINO PHYSICS}

Neutrinos are fundamental uncharged particles theorized in 1930 by Wolfgang Pauli~\cite{pauli1930letter} to explain the apparent energy loss and momentum of beta decay. However, this particle had to be massless, without strong or electromagnetic interaction, so its discovery took more than 20 years until 1953 by Reines and Cowan~\cite{Reines:1953pu}. However, in the 1960s\cite{Davis:1968cp}, a discrepancy in the flux of solar neutrinos between what was predicted by theory and those detected by experiments were discovered, leading to the idea that the flavors of the Neutrinos should oscillate and, therefore, must have masses.\\

Bruno Pontecorvo was the first to propose neutrino oscillation in 1957~\cite{Pontecorvo:1957cp,Pontecorvo:1957qd}, studying neutrino-antineutrino mixing, similar to the kaon oscillation $\left(K^0\rightarrow \overline{K}^0\right)$, later, in 1962 Maki, Nakagawa and Sakata studied the oscillation of the interaction eigenstates~\cite{Maki:1962mu} and then Pontecorvo expanded this idea of neutrino oscillation due to the presence of massive neutrinos. However, it was not until the late 1990s that experimental evidence of the oscillation of atmospheric neutrino was obtained by the Super-Kamiokande~\cite{Super-Kamiokande:1998kpq} and later, SNO and KamLAND measured the solar neutrino oscillation~\cite{SNO:2002tuh,KamLAND:2002uet}.

\section{Neutrino oscillation formalism}
\lhead[\thepage]{\thesection. Neutrino oscillation formalism} 

Because neutrinos are massive, they oscillate, where we can relate the mass and flavor eigenstates by a unitary mixing matrix. If we define $\ket{\nu_{\alpha}}$ (with $\alpha=e,\mu,\tau$) as the flavor quantum state and $\ket{\nu_k}$ (with $k=1,2,3$) as the mass quantum state, we can write the state $\ket{\nu_{\alpha}}$ in terms of $\ket{\nu_i}$ using the following unitary transformation:
\begin{equation}
\ket{\nu_{\alpha}}= U_{\alpha k}^*\ket{\nu_k}
\label{eq:estado-nu1}
\end{equation}

\noindent
where $U_{\alpha k}$ is the Pontecorvo, Maki, Nakagawa, and Sakata (PMNS) matrix or lepton mixing matrix. For three neutrino case, the PMNS can be commonly parameterized as the product of three different rotations,
\begin{align}
U_{\text{PMNS}}&= \begin{pmatrix}
1 & 0 & 0 \\
 0 & c_{23} & s_{23} \\
 0 & -s_{23} & c_{23}
\end{pmatrix} \begin{pmatrix}
c_{13} & 0 & s_{13} e^{-i \delta
   _{\text{CP}}} \\
 0 & 1 & 0 \\
 s_{13} \left(-e^{i \delta
   _{\text{CP}}}\right) & 0 & c_{13}
\end{pmatrix} \begin{pmatrix}
c_{12} & s_{12} & 0 \\
 -s_{12} & c_{12} & 0 \\
 0 & 0 & 1
\end{pmatrix} \begin{pmatrix}
1 & 0 & 0 \\
0 & e^{-i\alpha_1/2} & 0 \\
 0 & 0 & e^{-i\alpha_2/2}
\end{pmatrix}, \notag\\[5pt]
U_{\text{PMNS}}&= \begin{pmatrix}
c_{12} c_{13} & c_{13} s_{12} &
   s_{13} e^{-i \delta _{\text{CP}}}
   \\
 -c_{23} s_{12}-c_{12} s_{13} s_{23}
   e^{i \delta _{\text{CP}}} &
   c_{12} c_{23}-s_{12} s_{13}
   s_{23} e^{i \delta _{\text{CP}}}
   & c_{13} s_{23} \\
 s_{12} s_{23}-c_{12} c_{23} s_{13}
   e^{i \delta _{\text{CP}}} &
   -c_{12} s_{23} -c_{23} s_{12}
   s_{13} e^{i \delta
   _{\text{CP}}} & c_{13}
   c_{23}
\end{pmatrix}\begin{pmatrix}
1 & 0 & 0 \\
0 & e^{-i\alpha_1/2} & 0 \\
 0 & 0 & e^{-i\alpha_2/2}
\end{pmatrix},
\end{align}

\noindent
where $c_{ij}$ and $s_{ij}$ are the simplified ways of writing $\cos\theta_{ij}$ and $\sin\theta_{ij}$, respectively, $\delta_{\text{CP}}$ is the CP violation phase. In turn, $\alpha_1$ and $\alpha_2$ are additional Majorana phases, which, for Dirac neutrinos, are equal to zero.\\

Let us consider the propagation of massive neutrinos in a space-time interval ($x,t$), where their propagator can be represented as~\cite{giunti2007fundamentals}:
\begin{equation}
\ket{\nu_{\alpha}(x^{\mu})}= e^{iP^{\mu}\cdot x^{\mu}}\ket{\nu_{\alpha}}\label{eq:estado-nu2}
\end{equation}

\noindent
where $P^{\mu}$ is the energy-momentum operator. If we want to know the probability of a state $\ket{\nu_{\alpha}}\rightarrow \ket{\nu_{\beta}}$, we must obtain the transition amplitude,
\begin{align}
\mathcal{A}_{\nu_{\alpha}\rightarrow \nu_{\beta}}\left(x^{\mu}\right)&= \braket{\nu_{\beta}|\nu_{\alpha}(x^{\mu})} \notag\\
\mathcal{A}_{\nu_{\alpha}\rightarrow \nu_{\beta}}\left(x^{\mu}\right)&= \bra{\nu_{\beta}}e^{iP^{\mu}\cdot x^{\mu}}\ket{\nu_{\alpha}}
\end{align}

Massive neutrinos have defined energy and momentum, so in the plane wave approximation, $\nu_k$ is an eigenstate of the operator $P^{\mu}$, i.e., $P^{\mu}\ket{\nu_k}=p_k^{\mu}\ket{\nu_k}$, with:
\begin{equation}
p_k^0=E_k=\sqrt{\abs{\vec{p}_k}^2+m_k^2}
\label{eq:auto-p}
\end{equation}
\noindent
Therefore, replacing Eqs.~\eqref{eq:estado-nu1} and \eqref{eq:auto-p} in \eqref{eq:estado-nu2}, the transition amplitude is expressed as:
\begin{equation}
\mathcal{A}_{\nu_{\alpha}\rightarrow \nu_{\beta}}\left(x^{\mu}\right)= U_{\alpha k}^*e^{-iE_k t+i\vec{p}_k\cdot\vec{x}}U_{\beta k}\label{eq:amplitud1}
\end{equation}

If we assume neutrinos propagate along the ''$x$'' axis and are ultra-relativistic particles, i.e. $p_k\gg m_k$ and $x\simeq t$, from Eq.~\eqref{eq:auto-p}, we obtain the following approximations:
\begin{align}
E_k^2&=p_k^2+ m_k^2\notag\\
E_k^2-p_k^2&=m_k^2\notag\\
E_k-p_k&= \frac{m_k^2}{E_k+p_k}\simeq \frac{m_k^2}{2E}
\end{align}
\noindent
Therefore, from Eq.~\eqref{eq:amplitud1},
\begin{align}
-E_k t+p_kL&= -(E_k-p_k)L\notag\\
-E_k t+p_kL&\simeq -\left(\frac{m_k^2}{2E}\right)L
\end{align}
\noindent
So, our transition amplitude is expressed as:
\begin{equation}
\mathcal{A}_{\nu_{\alpha}\rightarrow \nu_{\beta}}\left(L\right)= U_{\alpha k}^*U_{\beta k}e^{-i\left(\frac{m_k^2}{2E}\right)L}\label{eq:amplitud2}
\end{equation}
\noindent
The transition probability will be:
\begin{align}
P_{\nu_{\alpha}\rightarrow \nu_{\beta}}\left(L\right)&=\abs{\mathcal{A}_{\nu_{\alpha}\rightarrow \nu_{\beta}}\left(L\right)}^2 \notag\\
P_{\nu_{\alpha}\rightarrow \nu_{\beta}}\left(L\right)&= U_{\alpha k}^*U_{\beta k}U_{\alpha j}U_{\beta j}^*e^{-i\left(\frac{\Delta m_{kj}^2}{2E}\right)L}
\label{eq:probabilidad-osci}
\end{align}

\noindent
where $\Delta m_{kj}=m_k^2-m_j^2$ is the squared mass differences.

\subsection{Neutrino mass hierarchy}

\begin{figure}
\centering
\includegraphics[scale=0.6]{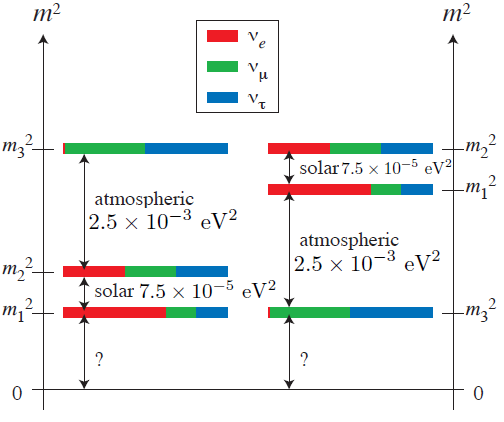}
\caption{The left side represents the Normal Hierarchy (NH), while the right side represents the Inverted Hierarchy (IH). The probability that a neutrino mass state contains each of the flavors $(\nu_e, \nu_{\mu}, \nu_{\tau})$ is proportional to the length of the respective color band.}
\label{fig:ordenJ}
\end{figure}

As we can see from the Eq.~\eqref{eq:probabilidad-osci}, the neutrino oscillation is sensitive to the squared mass differences and not to the absolute value of the neutrino mass scale, which gives us two possible orders of hierarchy, known as normal hierarchy (NH) and inverted hierarchy (IH). In Fig.~\ref{fig:ordenJ} we can see the relationship between the absolute neutrino masses $(m_1<m_2<m_3)$, which we can represent as:
\begin{equation}
m_1= m_0, \qquad m_2=\sqrt{m_0+\Delta m_{21}}, \qquad m_3= \sqrt{m_0+\Delta m_{31}},
\end{equation}
\noindent
where $m_0$ is the absolute neutrino mass scale. This scale can be restricted according to current experiments, such as cosmological observations, which put a limit on the sum of neutrino masses, whose value must be~\cite{Planck:2018vyg}
\begin{equation}
\sum_i m_i < 0.12\ \text{eV}.
\end{equation}

For Majorana neutrinos, we can obtain an effective mass for the neutrinoless double beta decay $\left( 0\nu\beta\beta\right)$, which is proportional to:
\begin{equation}
m_{ee}= \sum_k \abs{m_k U_{ek}}^2,
\label{eq:2beta}
\end{equation}

where Eq.~\eqref{eq:2beta} depends on the masses, mixtures and the CP violating phases. However, since this process has not yet been observed, the absolute neutrino mass scale can be constrained, the current limit being~\cite{KamLAND-Zen:2024eml},
\begin{equation}
m_{ee}\leq 50\ \text{meV}.
\end{equation}
\begin{table}[h!]
\centering
\begin{tabular}{c|ccc}
\hline\hline
\textbf{parameter} & \textbf{best fit $\pm 1\sigma$} & \textbf{2$\sigma$ range} & \textbf{3$\sigma$ range} \\ \hline\hline\\[-5pt]
$\Delta m^2_{21} [10^{-5}\mathrm{eV}^2]$ & $7.50^{+0.22}_{-0.20}$ & $7.12–7.93$ & $6.94–8.14$ \\[5pt]
$|\Delta m^2_{31}| [10^{-3}\mathrm{eV}^2]$ (NO) & $2.55^{+0.02}_{-0.03}$ & $2.49–2.60$ & $2.47–2.63$ \\ 
$|\Delta m^2_{31}| [10^{-3}\mathrm{eV}^2]$ (IO) & $2.45^{+0.02}_{-0.03}$ & $2.39–2.50$ & $2.37–2.53$ \\[5pt]
$\sin^2 \theta_{12} / 10^{-1}$ & $3.18 \pm 0.16$ & $2.86–3.52$ & $2.71–3.69$ \\ 
$\theta_{12} / ^\circ$ & $34.3 \pm 1.0$ & $32.3–36.4$ & $31.4–37.4$ \\[5pt]
$\sin^2 \theta_{23} / 10^{-1}$ (NO) & $5.74 \pm 0.14$ & $5.41–5.99$ & $4.34–6.10$ \\ 
$\theta_{23} / ^\circ$ (NO) & $49.26 \pm 0.79$ & $47.37–50.71$ & $41.20–51.33$ \\ 
$\sin^2 \theta_{23} / 10^{-1}$ (IO) & $5.78^{+0.10}_{-0.17}$ & $5.41–5.98$ & $4.33–6.08$ \\ 
$\theta_{23} / ^\circ$ (IO) & $49.46^{+0.60}_{-0.97}$ & $47.35–50.67$ & $41.16–51.25$ \\[5pt] 
$\sin^2 \theta_{13} / 10^{-2}$ (NO) & $2.200^{+0.069}_{-0.062}$ & $2.069–2.337$ & $2.000–2.405$ \\ 
$\theta_{13} / ^\circ$ (NO) & $8.53^{+0.13}_{-0.12}$ & $8.27–8.79$ & $8.13–8.92$ \\ 
$\sin^2 \theta_{13} / 10^{-2}$ (IO) & $2.225^{+0.064}_{-0.070}$ & $2.086–2.356$ & $2.018–2.424$ \\ 
$\theta_{13} / ^\circ$ (IO) & $8.58^{+0.12}_{-0.14}$ & $8.30–8.83$ & $8.17–8.96$ \\[5pt] 
$\delta / \pi$ (NO) & $1.08^{+0.13}_{-0.12}$ & $0.84–1.42$ & $0.71–1.99$ \\ 
$\delta / ^\circ$ (NO) & $194^{+24}_{-22}$ & $152–255$ & $128–359$ \\
$\delta / \pi$ (IO) & $1.58^{+0.15}_{-0.16}$ & $1.26–1.85$ & $1.11–1.96$ \\ 
$\delta / ^\circ$ (IO) & $284^{+26}_{-28}$ & $226–332$ & $200–353$ \\ \hline\hline
\end{tabular}
\caption{Global fit for neutrino oscillation parameters. It is taken from Ref.~\cite{deSalas:2020pgw}.}
\label{tab:neutrino-para}
\end{table}

\section{Neutrino mass}
\lhead[\thepage]{\thesection. Neutrino mass}

In the SM, neutrinos are massless since we only have Left-Handed (LH) neutrinos. However, as discussed above, they do have mass due to the discovery of neutrino oscillation. Therefore, below, we will briefly review the main methods to explain neutrino mass.

\subsection{Dirac Neutrino}

One of the options to explain the neutrino mass is to include massless Right-Handed (RH) neutrinos in the SM and singlet neutrinos under the group $SU(2)_L$. Thus, the neutrino mass can be generated in the same way as for charged fermions by the Yukawa term:
\begin{equation}
-\mathcal{L}_{D}= Y_{{ij}}^{(\nu)}\overline{l}_{L_i}\tilde{\phi}\nu_{R_j}+h.c.\ ,
\end{equation}

where $\tilde{\phi}=i\sigma_2\phi^*$. If neutrinos satisfy this solution, they are known as Dirac neutrinos, where the mass will be proportional to:
\begin{equation}
m_{\nu}=Y_{\nu}\frac{v}{\sqrt{2}}.
\label{eq:mnu1}
\end{equation}

However, this method does not seem natural, since for the neutrino mass to be within the experimental limits, the coupling $Y_{\nu}$ must be very small compared to the charged fermion couplings, with about 17 orders of magnitude between the neutrino coupling and the top quark coupling.

\subsection{Majorana Neutrino}

Ettore Majorana constructed a mass term with only the LH component since the RH component can be written in terms of the LH~\cite{illana2017model}:
\begin{align}
\psi_R&= C\overline{\psi}_L^T=\psi_L^C,
\end{align}

where $C=i\gamma_0\gamma_2$, is the charge conjugation operator, then $\psi_R=\psi_R^C$. This means that Majorana particles are their antiparticle. Therefore, we can write our field only in terms of $\psi_L$:
\begin{align}
\psi&= \psi_L+\psi_R, \notag\\
\psi&= \psi_L+\psi_L^C.
\end{align}

\noindent
Therefore, the Majorana mass term is:
\begin{equation}
-\mathcal{L}_M= \frac{1}{2}m\left(\psi_L\psi_L^C+ h.c\right)
\end{equation}

\noindent
However, the mass term is obtained through the Weinberg operator~\cite{Weinberg:1979sa},
\begin{equation}
\mathcal{O}^W=\frac{1}{\Lambda}LLHH,
\end{equation}

\noindent
So, the Majorana Lagrangian is:
\begin{equation}
-\mathcal{L}_W= \frac{1}{\Lambda}\left(\overline{l}_{L_i}\tilde{\phi}\right)Y_{ij}^{(M)}\left(\tilde{\phi}^T l_{L_j}^C\right)+ h.c.
\end{equation}

$\Lambda$ is the new physics energy scale and $Y_{ij}^{(M)}$ is the Yukawa matrix. Therefore, the mass term after spontaneous symmetry breaking will be:
\begin{equation}
m_{\nu}=  Y_M \frac{v}{\Lambda}
\end{equation}

The Weinberg operator respects the SM symmetry but generates a non-renormalizable theory. Thus, we have an effective theory at low energy, and hence, ultraviolet completeness is needed.

\section{Seesaw Mechanism}
\lhead[\thepage]{\thesection. Seesaw Mechanism}

The Seesaw mechanism was proposed in the 1970s, which attempts to explain the neutrino masses observed in neutrino oscillation experiments~\cite{Mohapatra:1979ia}, which, as discussed above, are several orders of magnitude smaller than the masses of charged quarks and leptons.\\

The literature describes several seesaw mechanisms, however, this section will discuss the mechanisms known as Type I Seesaw and Inverse Seesaw.

\subsection{Type I Seesaw}

\begin{figure}
\centering
\subfigure[]{\includegraphics[scale=0.4]{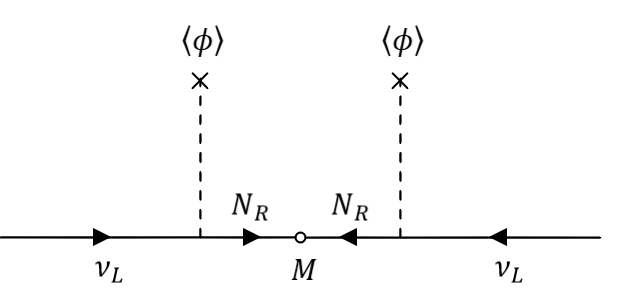}} \quad
\subfigure[]{\includegraphics[scale=0.4]{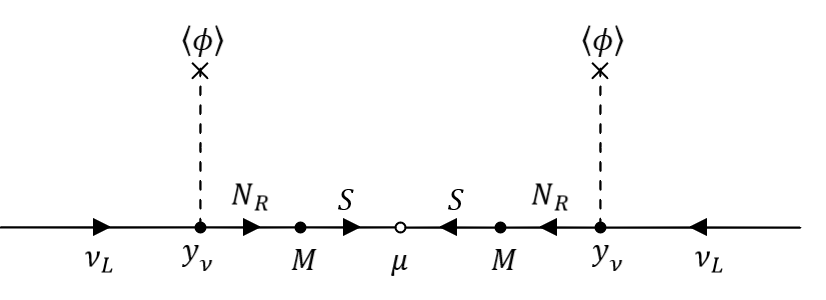}}
\caption{Feynman diagrams for a type I seesaw (a) and for an inverse seesaw (b).}
\label{fig:seesaw}
\end{figure}

The type I seesaw mechanism includes right-handed singlet neutrinos under the SM symmetry $(N_R)$. With this extra field, one can create a Majorana mass term in the Yukawa Lagrangian, whose mass generation is given according to the Feynman diagram in Fig.~\ref{fig:seesaw}. Then, for the single generation case, our Yukawa Lagrangian is:
\begin{equation}
-\mathcal{L}_Y= Y_{\nu}\overline{l}_L \tilde{\phi}N_R+ \frac{1}{2} \overline{N}_R^C M_R N_R+ h.c.\ ,
\end{equation}
\noindent
where $M_R$ is the Majorana mass term for $N_R$. After the symmetry spontaneous break, we find that:
\begin{equation}
-\mathcal{L}_Y= \overline{\nu}_L m_D N_R+ \frac{1}{2} \overline{N}_R^C M_R N_R+ h.c.\ .
\label{eq:lang-ss}
\end{equation}
\noindent
If we introduce $\nu= \left(\nu_L\quad N_R^C\right)$ the Eq.~\eqref{eq:lang-ss} can be written in terms of matrices as,
\begin{equation}
-\mathcal{L}_Y= \frac{1}{2}\overline{\nu} M_{\nu}\nu_C+ h.c.\ ,
\end{equation}
\noindent
where,
\begin{equation}
M_{\nu}= \begin{pmatrix}
0 & m_D \\
m_D & M_R
\end{pmatrix}.
\label{eq:matrix-neu1}
\end{equation}
\noindent
The matrix \eqref{eq:matrix-neu1} can be diagonalized by the following orthogonal matrix:
\begin{equation}
\mathcal{O}= \left(
\begin{array}{cc}
\cos\theta & \sin\theta \\
-\sin\theta & \cos\theta
\end{array}
\right), \quad \tan2\theta=\frac{2m_{D}}{M_{R}}, \quad \cos2\theta=\frac{M_{R}}{\sqrt{M_{R}^{2}+4m_{D}^{2}}}\ .
\label{eq:matriz-diago}
\end{equation}

\noindent
After diagonalizing the matrix \eqref{eq:matrix-neu1} we obtain the following eigenvalues:
\begin{equation}
m_{1,2}= \frac{1}{2}\left(M_R \mp \sqrt{M_R^2+4m_D^2}\right).
\end{equation}

In the limit $M_{R}\gg m_{D}$, we obtain a light neutrino ($\nu$) and a very heavy one $(N)$ with opposite CP-parities and a very small mixing angle.
\begin{align}
m_{\nu}&\simeq m_{1}\simeq \frac{m_{D}^{2}}{M_{R}},\\
m_{N}&\simeq m_{2}\simeq M_{R}\gg m_{\nu}, \\
\theta&\simeq \sqrt{m_{\nu}/m_{N}}\ll 1 \;.
\label{eq:masas-seesaw}
\end{align}

We know that there are $n=3$ generations of LH neutrinos $\left[\nu_{iL} \ (i= 1,2,3)\right]$ and there can be an arbitrary number $n_{R}$ 
of RH fields $\left[\nu_{jR} \; (j= 1,\ldots, n_{R}\right]$. The mass matrix is then the complex, symmetric and square matrix $(3 + n_{R})\times(3+ n_{R})$:
\begin{align}
M_{\nu}=\begin{pmatrix}
0 & m_{D}^{T} \\
m_{D} & M_{R}
\end{pmatrix}
\label{eq:matriz2-seesaw}
\end{align}

\noindent
where $m_{D}$ and $M_{R}$ are matrices of $n_{R}\times n$ and $n_{R}\times n_{R}$, respectively.\\

Now, if we assume that $m_{D}$ is of the order of the electroweak scale and the scale of $M_{R}$ is very high (of the order of the grand unification scale, $M_{R} \sim 10^{16}$ GeV) we obtain $n=3$ light neutrinos ($\nu_{i}$) with masses $m_{\nu} \sim (10^{-2} - 10^{-1})$ eV, which is just the correct order of magnitude to explain the tiny neutrino masses compatible with oscillation experiments, and $n_{R}$ extremely heavy neutrinos ($N_{j}$) that would play a crucial role in generating the baryonic asymmetry of the universe from their non-equilibrium decays (leptogenesis). However, masses for heavy neutrinos can be obtained on a lower energy scale, on the order of TeV, and for this we must consider smaller Yukawa couplings $(Y_{\nu}<1)$.

\subsection{Inverse Seesaw}

In the simplified case of the type I seesaw, it can explain the neutrino masses. However, this would be very difficult to test since it would occur almost at the grand unification scale. A possible solution to this problem, different from the one discussed at the end of the previous section, is the application of what is known as inverse seesaw model~\cite{Mohapatra:1986aw}. This model introduces two Majorana fermion pairs to the SM, $N_{R_i}$ and $S_{L_j}$ ($i,j=1,2,3$). These fields transform as singlets under the SM gauge group and carry a lepton number of $+1$. Then, after electroweak symmetry breaking the Lagrangian in the neutrino sector is given by
\begin{equation}
\mathcal{L_\nu}= m_D \overline{\nu}_L N_R + M \overline{N}_R S_L + \mu S^T_L C^{-1} S_L + h.c.,
\label{eq:lag1}  
\end{equation}
where $m_D$ and $M$ are Dirac $3\times3$ matrices, while $\mu$ is a Majorana $3\times3$ matrix breaking lepton number explicitly. The latter can have a dynamical origin~\cite{Mohapatra:1979ia}. As usual, $C$ denotes the charge conjugation matrix. The neutrino mass matrix in the $(\nu_L, N_R, S_L)$ basis   turns out to be
\begin{equation}
M_{\nu}=\left( 
\begin{array}{ccc}
0 & m_{D} & 0 \\ 
m_{D}^{T} & 0 & M \\ 
0 & M^{T} & \mu 
\end{array}%
\right) .
\label{MnuIS}
\end{equation}%
Taking the limit $\mu_{ij}<<\left(m_D\right)_{ij}<< M_{ij}$ ($i,j=1,2,3$) leads to a $3\times3$ matrix for light neutrinos given by
\begin{equation}
m_\nu\simeq m_D \frac{1}{M}\mu\frac{1}{M^T}m_D^T.
\label{mnuIS0}
\end{equation}

Note that the lightness of neutrinos could be associated only with the smallness of $\mu$, which is naturally protected from large radiative corrections by the $U(1)_L$-symmetry of lepton number conservation, restored in the limit $\mu\to 0$ in Eq.~(\ref{eq:lag1}). Obviously, neutrinos become massless in this limit.\\

Since active neutrinos have masses in the sub-eV range, $m_D$ is found to be in the electroweak range, $M$ in the TeV range, and 
$\mu$ in the keV range~\cite{Mohapatra:1986aw}. 
In this case, RH neutrinos can develop masses around the TeV range, and their mixing with standard neutrinos is modulated by the relation 
$m_D M^{-1}$.\\

The complete neutrino mass matrix given in the equation~\eqref{MnuIS} can be diagonalized by the following rotation matrix \cite{Catano:2012kw}:
\begin{equation}
\mathbb{{R}}=%
\begin{pmatrix}
{\bf R}_{\nu } & {\bf R}_{1}{\bf R}_{M}^{\left( 1\right) } & {\bf R}_{2}{\bf R}_{M}^{\left( 2\right) } \\ 
-\frac{({\bf R}_{1}^{\dagger }+{\bf R}_{2}^{\dagger })}{\sqrt{2}}{\bf R}_{\nu } & \frac{(1-{\bf S})}{%
\sqrt{2}}{\bf R}_{M}^{\left( 1\right) } & \frac{(1+{\bf S})}{\sqrt{2}}{\bf R}_{M}^{\left(
2\right) } \\ 
-\frac{({\bf R}_{1}^{\dagger }-{\bf R}_{2}^{\dagger })}{\sqrt{2}}{\bf R}_{\nu } & \frac{(-1-{\bf S})%
}{\sqrt{2}}{\bf R}_{M}^{\left( 1\right) } & \frac{(1-{\bf S})}{\sqrt{2}}{\bf R}_{M}^{\left(
2\right) }%
\end{pmatrix}%
,  \label{U}
\end{equation}

\noindent
where:
\begin{equation}
{\bf S}=-\frac{1}{4}{\bf M}^{-1}\bm{\mu} ,\hspace{1cm}\hspace{1cm}{\bf R}_{1}\simeq {\bf R}_{2}\simeq 
\frac{1}{\sqrt{2}}{\bf m}_{D}^{\ast }{\bf M}^{-1},
\end{equation}
\noindent
and ${\bf R}_{\nu }$, ${\bf R}_{M}^{\left( 1\right) }$ and ${\bf R}_{M}^{\left( 2\right)}$ are the rotation matrices, which diagonalize the active neutrino matrix and the sterile neutrino matrices, respectively.\\

\newpage
$\ $
\thispagestyle{empty} 
\chapter{Dark Matter from a Radiative Inverse Seesaw Majoron Model}\label{cap.modelo1}
\markboth{INVERSE SEESAW MAJORON MODEL}{RADIATIVE INVERSE SEESAW MAJORON MODEL}

The following chapters is based on the first paper of this thesis~\ref{paper:DM}, where we will analyze models in which we extend the symmetry of the SM and use low-scale seesaw mechanisms to generate light-active neutrino masses. In each model, we will extend the symmetry and the particle content of the SM, starting with a model with a minimal extension of the SM (this chapter) by adding a global symmetry, then moving to an extension based on discrete and cyclic symmetry, and finally to a model with a more complex SM extension, including several additional symmetries and fields. The neutrino masses are generated in all models by implementing a seesaw mechanism, in two cases radiatively, and then we will analyze the phenomenological implications in each model.

\section{Intruduction}
\lhead[\thepage]{\thesection. Intruduction}

As discussed above, the SM does not explain the neutrino mass nor gives any clues about the dark matter content in the universe. For these reasons, there are several proposals to explain these questions, where we can highlight the extension of the SM symmetry and the implementation of the seesaw mechanism for the neutrino mass. In this chapter, we propose a minimal extension to the SM symmetry by assuming a new symmetry $U(1)_X$ and the particle content is extended by adding new scalar and fermionic fields, where we will study the scalar sector and analyze the existence of a scalar and fermionic dark matter candidate.\\

In addition, the implication of model in processes that violate the leptonic flavor will be studied, where the neutrino masses will be generated through a radiative inverse seesaw mechanism since the parameter $\mu$ will be generated in a loop. We will see if the model obtains values for the branching ratio for $\mu \rightarrow e\gamma$ below the experimental limits.

\section{The model}\label{model1} 
\lhead[\thepage]{\thesection. The model}

We consider a model that adds to the SM, two complex scalars $\sigma $ and $\eta $ and six Majorana fermions $\nu _{R_k}$, $N_{R_k}$ and $\Omega _{R_k}$ ($k=1,2$). All these new fields are $SU(2)$ gauge singlets and carry neutral electric charge. In addition, the existence of a global $U(1) _{X}$ symmetry is assumed. This symmetry breaks down to a $\mathcal{Z}_2$ symmetry when the singlet scalar gets a vacuum expectation value (vev) $\langle\sigma\rangle=v_{\sigma}$. Table~\ref{tab:scalar-fermions} shows the charge assignments of scalars and leptons under the $SU\left( 2\right) _{L}\otimes U\left( 1\right) _{Y}\otimes U\left( 1\right) _{X}$ symmetry. 

\begin{table}[h!]
\centering
\begin{tabular}{|c|ccc|ccccc|} 
\hline
                         & $\Phi$ & $\sigma$ & $\eta$ & $L_{L_i}$ & $l_{R_i}$ & $\nu _{R_k}$ & $N_{R_k}$ & $\Omega _{R_k}$  \\ 
\hline\hline
$SU\left( 2\right) _{L}$ & $2$    & $1$      & $1$    & $2$       & $1$       & $1$          & $1$       & $1$              \\
$U\left( 1\right)_{Y}$   & $1/2$  & $0$      & $0$    & $-1/2$    & $-1$      & $0$          & $0$       & $0$              \\
$U\left( 1\right) _{X}$  & $0$    & $-1$     & $1/2$  & $-1$      & $-1$      & $-1$         & $1$       & $-1/2$           \\
\hline
\end{tabular}
\caption{Charge assignments of scalar and lepton fields. Here $i=1,2,3$ and $k=1,2$.}
\label{tab:scalar-fermions}
\end{table}

In fact, after electroweak symmetry breaking the unbroken symmetry is $SU(3)_C\otimes U(1)_{EM}\otimes \mathcal{Z}_2$ where $\mathcal{Z}_2$ turns out to be the symmetry that stabilizes the dark matter candidate of the theory. Schematically, the symmetry breaking chain goes as follows,
\begin{eqnarray}
&&\mathcal{G}=SU(3)_{C}\otimes SU\left( 2\right) _{L}\otimes U\left( 1\right) _{Y}\otimes U\left( 1\right) _{X}  \notag \\
&&\hspace{35mm}\Downarrow v_{\sigma }  \notag \\[0.12in]
&&\hspace{15mm}SU(3)_{C}\otimes SU\left( 2\right) _{L}\otimes U\left( 1\right)
_{Y}\times \mathcal{Z}_2  \notag \\[0.12in]
&&\hspace{35mm}\Downarrow v_{\Phi}  \notag \\[0.12in]
&&\hspace{15mm}SU(3)_{C}\otimes U\left( 1\right) _{Q}\otimes \mathcal{Z}_2
\end{eqnarray}
\noindent
where the Higgs vev is represented by $\langle\Phi^0\rangle=v_\Phi$.\\

Given the charge assignments shown in Table~\ref{tab:scalar-fermions} we have that the singlet's vev is invariant under the following transformation, $e^{2\pi i \hat{X}}\langle\sigma\rangle=\langle\sigma\rangle$, where $\hat{X}$ is the $U(1)_X$ charge operator. 
This implies the existence of a residual discrete symmetry 
$(-1)^{2\hat{X}}\in\mathcal{Z}_2$
surviving spontaneous breaking of the global $U(1)_X$ group. 
Therefore, to all fields  
are assigned the corresponding $\mathcal{Z}_2$-parities 
$(-1)^{2\hat{Q}_X}$ according to their $U(1)_X$ charges $Q_X$ in Table~\ref{tab:scalar-fermions}.
The particles $\eta$ and $\Omega_{R_k}$ $(k=1,2)$  have odd $\mathcal{Z}_2$-parities and form the dark sector of the model.

\section{Scalar sector}\label{scalarsector}
\lhead[\thepage]{\thesection. Scalar sector}

The scalar potential invariant under the symmetry group $\mathcal{G}$ is given by

\begin{eqnarray}
\label{eq:pot}
V(\Phi ,\sigma ,\eta )&=& -\frac{\mu _{\Phi }^{2}}{2}|\Phi |^{2}+\frac{\lambda _{\Phi }}{2}%
|\Phi |^{4}-\frac{\mu _{\sigma }^{2}}{2}|\sigma |^{2}+\frac{\lambda _{\sigma
}}{2}|\sigma |^{4}+\frac{\mu _{\eta }^{2}}{2}|\eta |^{2}+\frac{\lambda
_{\eta }}{2}|\eta |^{4} \notag\\
&+&\lambda _{1}|\Phi |^{2}|\sigma |^{2}+\lambda _{2}|\Phi |^{2}|\eta
|^{2}+\lambda _{3}|\sigma |^{2}|\eta |^{2}+ \frac{\mu _{4}}{\sqrt{2}}\sigma \eta ^{2}+h.c., 
\end{eqnarray}%
where the quartic couplings $\lambda_a$ are dimensionless parameters whereas the $\mu_a$ are dimensionful. In our analysis we will impose perturbativity ($\lambda_a<\sqrt{4\pi}$) and the boundedness conditions given in section~\ref{apx:stabilityconds}. \\

The singlet $\sigma$ and the neutral component of the doublet $\Phi=(\phi^+,\phi^0)^T$ acquire vacuum expectation
values (vevs). Here the singlet's vev $v_\sigma$ is responsible for the breaking of the global $U(1)_X$ symmetry while the double's vev $v_\Phi$ triggers electroweak symmetry breaking. Therefore we shift the fields as
\begin{equation}
\phi^0 =\frac{1}{\sqrt{2}}(v_{\Phi }+\phi_R+i\phi_I),\quad \sigma =\left( \frac{%
v_{\sigma }+\sigma_R+i\sigma_I }{\sqrt{2}}\right) .
\end{equation}%

Evaluating the second derivatives of the scalar potential at the minimum one finds the CP-even, $M_R^2$, and CP-odd, $M_I^2$, mass matrices. The CP-even mass matrix $M_R^2$ mixes $\phi_R$ and $\sigma_R$ and its eigenvalues correspond to the squared masses of the physical scalar states. They are
given by
\begin{equation}
\label{higgsmasses}
m_{h_{1},h_{2}}^{2}=\frac{1}{2}\left( \lambda _{\sigma }v_{\sigma
}^{2}+\lambda _{\Phi }v_{\Phi }^{2}\mp \frac{\lambda _{\sigma }v_{\sigma
}^{2}-\lambda _{\Phi }v_{\Phi }^{2}}{\cos 2\theta }\right) ,
\end{equation}%
where we identify $h_{1}$ with the 125 GeV Higgs boson, $v_{\Phi }=246$ GeV,
and the mixing angle $\theta $ fulfilling 
\begin{equation}
\label{higgsmixing}
\tan 2\theta =\frac{2\lambda _{1}v_{\Phi }v_{\sigma }}{\lambda _{\sigma
}v_{\sigma }^{2}-\lambda _{\Phi }v_{\Phi }^{2}}.
\end{equation}%
Moreover, the flavor and physical bases are connected through out the 
following relations,
\begin{eqnarray}
 \sigma_R&=&-h_{1}\sin \theta +h_{2}\cos \theta, \notag \\
 \phi_R&=& h_{1}\cos \theta +h_{2}\sin \theta. \label{eq:srot} 
\end{eqnarray}
The CP-odd mass matrix $M_I^2$ has two null eigenvalues. One of them corresponds to the would-be Goldstone boson which becomes the longitudinal component of the $Z$-boson by virtue of the Higgs mechanism. The other one is the physical Goldstone boson resulting from spontaneous breaking of the global $U(1)_X$ symmetry, similar to the singlet Majoron model in Ref.~\cite{Mohapatra:1979ia}. Notice that given that this Goldstone is an SM singlet, it is invisible, and its phenomenological impact in the model is not dangerous. It includes any cosmological manifestations. Let us also stress that the mass of this field is non-vanishing, at least due to the well-known quantum gravity effects. It can also gain mass due to non-perturbative QCD effects in an extended version of our model in which the fields in the quark sector have non-trivial $U(1)_X$ assignments entailing a mixed $(SU(3)_C)^2U(1)_X$-anomaly. This is what defines our Majoron model variant. 
The masses of $Z_2$-odd scalar components $\eta =\eta _{R}+i\eta_{I}$ turn out to be 
\begin{eqnarray}
m_{\eta _{R}}^{2} &=&\mu _{\eta }^{2}+\frac{1}{2}\left( \lambda _{2}v_{\Phi
}^{2}+\lambda _{3}v_{\sigma }^{2}\right) +\mu _{4}v_{\sigma },  \label{rel1}
\\
m_{\eta _{I}}^{2} &=&\mu _{\eta }^{2}+\frac{1}{2}\left( \lambda _{2}v_{\Phi
}^{2}+\lambda _{3}v_{\sigma }^{2}\right) -\mu _{4}v_{\sigma },
\end{eqnarray}%
where the mass splitting of these two components can be recasted as $\mu _{4}=(m_{\eta _{R}}^{2}-m_{\eta_{I}}^{2})/(2v_{\sigma })$. Note that $\eta _{R}$ and $\eta _{I}$ are degenerate when $\mu _{4}\rightarrow 0$. Here, the lightest of these two components, $\eta_R$
and $\eta_I$, can be the stable dark matter.\\

\subsection{Boundedness Conditions}
\label{apx:stabilityconds}

The boundedness conditions of the scalar potential in Eq.~(\ref{eq:pot}) are
derived assuming that the quartic terms dominate over at high energies. In order to do so, we define the following bilinears,
\begin{equation}
a=|\Phi |^{2}\quad ;\quad b=|\sigma |^{2}\quad ;\quad c=|\eta |^{2},
\label{eq:bcpar}
\end{equation}
and rewrite the quartic terms of the scalar potential. Then, using the expressions in Eq.~(\ref{eq:bcpar}) one gets
\begin{eqnarray}
V_{q} &=&\frac{1}{2}(\sqrt{\lambda _{\Phi }}a-\sqrt{\lambda _{\sigma }}%
b)^{2}+\frac{1}{2}(\sqrt{\lambda _{\Phi }}a-\sqrt{\lambda _{\eta }}c)^{2}+%
\frac{1}{2}(\sqrt{\lambda _{\sigma }}b-\sqrt{\lambda _{\eta }}%
c)^{2}+(\lambda _{1}+\sqrt{\lambda _{\Phi }\lambda _{\sigma }})ab  \notag \\
&&+(\lambda _{2}+\sqrt{\lambda _{\Phi }\lambda _{\eta }})ac+(\lambda _{3}+%
\sqrt{\lambda _{\sigma }\lambda _{\eta }})bc-\frac{1}{2}(\lambda _{\Phi
}a^{2}+\lambda _{\eta }c^{2}).  \label{ec:potentialcuartic}
\end{eqnarray}

Following  Refs.~\cite{Maniatis:2006fs,Bhattacharyya:2015nca} the boundedness conditions
of the model turn out to be,
\begin{eqnarray}
\lambda _{\Phi } &\geq &0\quad ;\quad \lambda _{\sigma }\geq 0\quad ;\quad
\lambda _{\eta }\geq 0  \label{ec:condition1} \\
\lambda _{1}+\sqrt{\lambda _{\Phi }\lambda _{\sigma }} &\geq &0\quad ;\quad
\lambda _{2}+\sqrt{\lambda _{\Phi }\lambda _{\eta }}\geq 0\quad ;\quad
\lambda _{3}+\sqrt{\lambda _{\sigma }\lambda _{\eta }}\geq 0
\label{ec:condition2}
\end{eqnarray}

\section{Neutrino sector}
\lhead[\thepage]{\thesection. Neutrino sector}

\begin{figure}[h!]
\begin{center}
\includegraphics[width=0.5\textwidth]{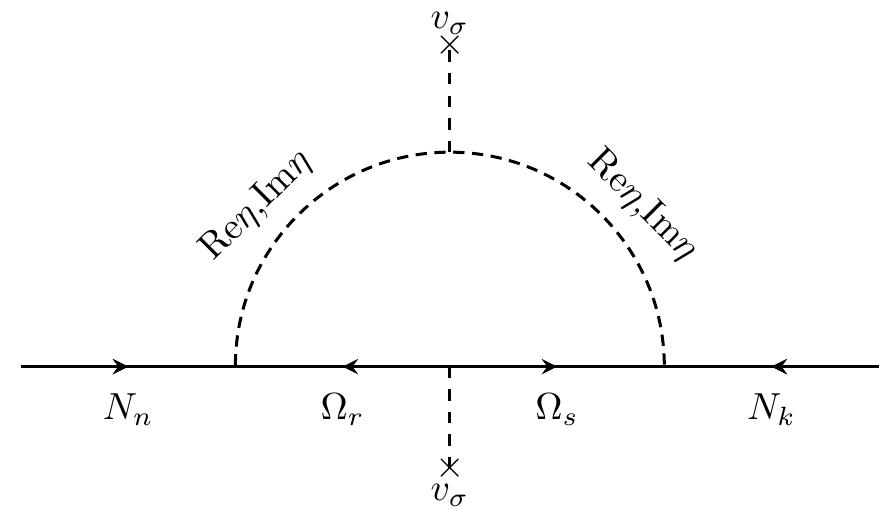}
\end{center}
\caption{One-loop Feynman diagram contributing to the Majorana neutrino mass
in Eq.~(\ref{Mnu}).}
\label{Neutrinoloopdiagram}
\end{figure}

Using Table~\ref{tab:scalar-fermions}, the invariant lepton Yukawa Lagrangian is given by
\begin{eqnarray}
\label{eq:yukL}
-\mathcal{L}_{Y}^{\left( l\right) } &=&\sum_{i=1}^{3}\sum_{j=1}^{3}\left(
y_{l}\right) _{ij}\overline{L}_{L_i}l_{R_j}\Phi+\sum_{i=1}^{3}\sum_{k=1}^{2}\left( y_{\nu }\right) _{ik}\overline{L}%
_{L_i}\nu _{R_k}\widetilde{\Phi }+\sum_{n=1}^{2}\sum_{k=1}^{2}M_{nk}\overline{%
\nu }_{R_n}N_{R_k}^{c}  \notag \\
&&+\sum_{n=1}^{2}\sum_{k=1}^{2}\left( y_{N}\right) _{nk}\overline{N}%
_{R_n} \Omega _{R_k}^{c}\eta+\sum_{n=1}^{2}\sum_{k=1}^{2}\left( y_{\Omega
}\right) _{nk}\overline{\Omega }_{R_n} \Omega _{R_k}^{c}\sigma+h.c.,  \notag
\end{eqnarray}
\noindent
where $\psi^c=C \overline{\psi}^T$ and $\widetilde{\Phi }=i\sigma_2 \Phi^\ast$.
After spontaneous symmetry breaking (SSB), the neutrino mass matrix has the form,
 \begin{equation}
 M_{\nu }=\left( 
 \begin{array}{ccc}
 0_{3\times 3} & m_{D} & 0_{3\times 2} \\ 
 m_{D}^{T} & 0_{2\times 2} & M \\ 
 0_{2\times 3} & M^{T} & \mu 
 \end{array}%
 \right) ,  \label{Mnu}
 \end{equation}%
 where $m_D$ is the tree-level Dirac mass term 
 \begin{equation}
 \left( m_{D}\right) _{ik} =\left( y_{\nu }\right) _{ik}\frac{v_{\Phi }%
 }{\sqrt{2}},
 \label{ec:mdirac}
 \end{equation}
 with $i=1,2,3$ and $k=1,2$. The submatrix $\mu$ in Eq.~(\ref{Mnu}) is generated at
 one-loop level, 
 \begin{eqnarray} 
 \label{muloop}
 \mu _{sp} &=&\sum_{k=1}^{2}\frac{\left( y_{N}\right) _{sk}\left(
 y_{N}^{T}\right) _{kp}m_{\Omega _{k}}}{16\pi ^{2}}\left[ \frac{m_{\eta
 _{R}}^{2}}{m_{\eta _{R}}^{2}-m_{\Omega _{k}}^{2}}\ln \left( \frac{m_{\eta
 _{R}}^{2}}{m_{\Omega _{k}}^{2}}\right) -\frac{m_{\eta _{I}}^{2}}{m_{\eta
 _{I}}^{2}-m_{\Omega _{k}}^{2}}\ln \left( \frac{m_{\eta _{I}}^{2}}{m_{\Omega
 _{k}}^{2}}\right) \right] , 
 \label{eq:matrix-mu}
 \end{eqnarray}
with $s,p=1,2$. The Feynman diagram of $\mu$ is depicted in Figure~\ref{Neutrinoloopdiagram}.
One can see from Eq.~\eqref{muloop} that the $\mu$ term vanishes when the 
scalars $\eta _{R}$ and $\eta _{I}$ are degenerate. This implies that neutrino masses
go to zero in the limit $\mu\rightarrow 0$. 

Then one has that active light neutrino masses are generated via an inverse seesaw
mechanism at the one-loop level. Physical neutrino mass matrices are
given by\footnote{The diagonalization of the neutrino mass matrix in Eq.~(\ref{Mnu}) can be followed from Ref.~\cite{Catano:2012kw}}: 
\begin{eqnarray}
\label{Mnutilde}
\widetilde{M}_{\nu } &=&m_{D}\left( M^{T}\right) ^{-1}\mu M^{-1}m_{D}^{T},\hspace{0.7cm}   \\
M_{\nu }^{\left( -\right) } &=&-\frac{1}{2}\left( M+M^{T}\right) +\frac{1}{2}%
\mu ,\hspace{0.7cm} \label{Mnuminus} \\
M_{\nu }^{\left( +\right) } &=&\frac{1}{2}\left( M+M^{T}\right) +\frac{1}{2}%
\mu \label{Mnuplus}.  
\end{eqnarray}%
Now $\widetilde{M}_{\nu}$ is the mass matrix for active light
neutrinos ($\nu _{a}$), whereas $M_{\nu }^{(-)}$ and $M_{\nu }^{(+)}$ are
the mass matrices for sterile neutrinos. From Eq.~(\ref{Mnutilde}) one can see that 
active light neutrinos are massless in the limit $\mu\to0$ which implies that lepton number is a conserved quantity. Eqs.~(\ref{Mnuminus}) and (\ref{Mnuplus}) tell us that the smallness of
the parameter $\mu$ (small mass splitting) induces pseudo-Dirac pairs of sterile neutrinos.\\ 

\begin{figure}[h!]
\centering
\includegraphics[width=0.9\textwidth]{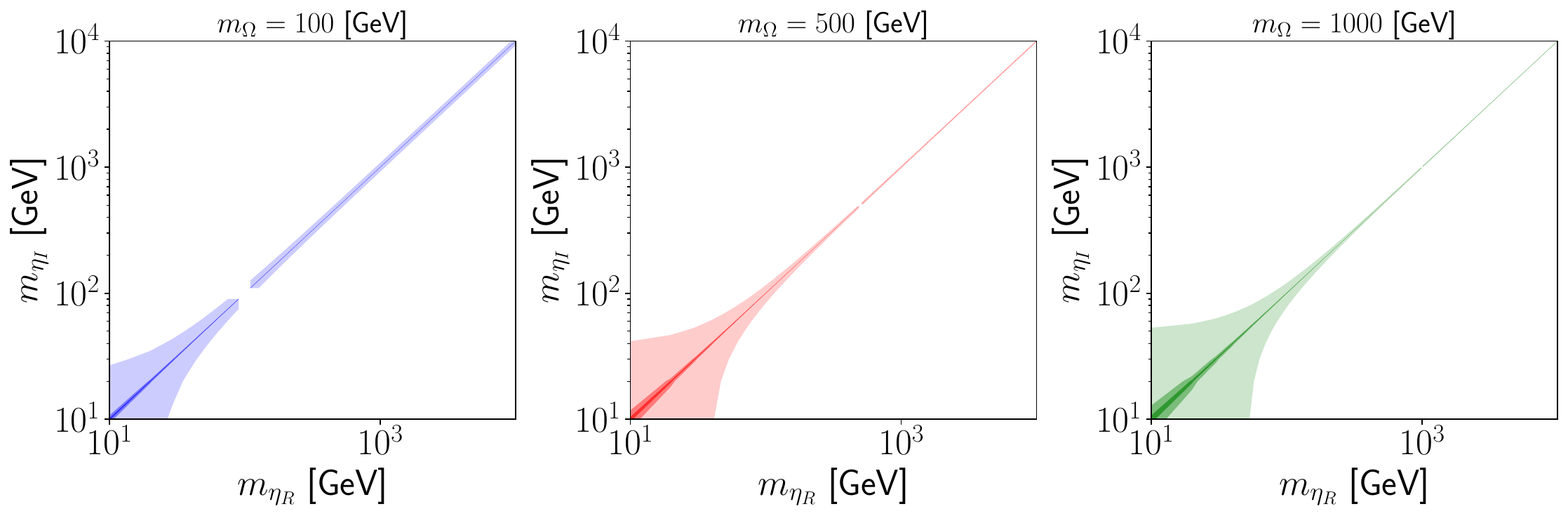}
\caption{Parameter space fulfilling $-10$ keV $\leq \mu \leq 10$ keV, for the DM masses indicated above each plot. The color in each plot, from light to dark, represents $y_N = 0.01, 0.05$ and 0.1, respectively. Here we
assume that $m_{\Omega_2} \gg m_\Omega$. The discontinuity appears when a degenerate mass spectrum is reached in Eq.~(\ref{muloop}).}
\label{fig:etaI-etaR}
\end{figure}

From Eq.~(\ref{Mnutilde}), one can see that a sub-eV neutrino mass scale can be linked to a small lepton number breaking parameter $\mu$ which depends on the Yukawas $y_N^2$, $y_\Omega$  and the masses of the particles running in the loop $(m_{\eta_R},m_{\eta_I},m_\Omega)$. This parameter is further suppressed by the loop factor, see Eq.~(\ref{muloop}). Figure~\ref{fig:etaI-etaR} shows the allowed parameter space regions for fixed Yukawa couplings $y_N$ and masses of the $Z_2$-odd Majorana fermions $\Omega _{k}$. Each plot is generated using Eq.~(\ref{muloop}), varying the masses $(m_{\eta_R},m_{\eta_I})$, fixing $m_\Omega$ and $y_N$. Then, from left to right, Figure~\ref{fig:etaI-etaR} shows the parameter space that fulfills $-10$ keV $\leq \mu \leq 10$ keV in the $(m_{\eta_R},m_{\eta_I})$-plane, considering $m_\Omega=100, 500,$ and  $1000$~GeV, respectively. In all panels, $y_N=0.01,0.05,$ and $0.1$, the smaller the Yukawa value, the lighter the region. The discontinuity appears when the mass spectrum in Eq.~(\ref{muloop}) is degenerate.

 As we have mentioned, the dark sector is formed by the $Z_2$-odd particles, see Table~\ref{tab:scalar-fermions}. The dark matter candidate of the model is the lightest component of either the singlet scalar $\eta$ or the Majorana fermion $\Omega$. 
The phenomenological consequences of having the lightest component of the scalar singlet $\eta$ as dark matter candidate is similar to what have been discussed in Refs.~\cite{Ma:2006km,Ahriche:2016acx,Bernal:2017xat,Mandal:2019oth,Abada:2021yot}. For this reason, in what follows we discuss only the constraints and projections of the model for the case in which the DM candidate is the Majorana fermion $\Omega$. 

\section{Fermion Dark Matter}\label{DM}
\lhead[\thepage]{\thesection. Fermion Dark Matter}
\begin{figure}[t!]
\centering
\begin{tikzpicture}[line width=1.0 pt, scale=0.6]

\begin{scope}[shift={(0,0)}]
	\draw[fermion](-2.5,1) -- (-1,0);
	\draw[fermionbar](-2.5,-1) -- (-1,0);
	\draw[scalarnoarrow](-1,0) -- (1,0);
	\draw[line](1,0) -- (2.5,1);
	\draw[line](1,0) -- (2.5,-1);
    \node at (-3,1.0) {$\Omega$};
	\node at (-3,-1.0) {$\Omega$};
    \node at (-0.1,0.46) {$(h_1,h_2)$};
	\node at (1.4,1.3) {SM};
    \node at (1.4,-1.3) {SM}; 
    \node at (-0,-3) {$\textit{(a)}$};
\end{scope}

\begin{scope}[shift={(7,0)}]
	\draw[fermion](-2.5,1) -- (-1,0);
	\draw[fermionbar](-2.5,-1) -- (-1,0);
	\draw[scalarnoarrow](-1,0) -- (1,0);
	\draw[scalarnoarrow](1,0) -- (2.5,1);
	\draw[scalarnoarrow](1,0) -- (2.5,-1);
    \node at (-3,1.0) {$\Omega$};
	\node at (-3,-1.0) {$\Omega$};
    \node at (-0.1,0.46) {$(h_1,h_2)$};
	\node at (1.4,1.1) {$\chi$};
    \node at (1.4,-1.1) {$\chi$}; 
    \node at (-0,-3) {$\textit{(b)}$};
\end{scope}

\begin{scope}[shift={(14,0)}]
	\draw[fermion](-2.5,1) -- (-1,0);
	\draw[fermionbar](-2.5,-1) -- (-1,0);
	\draw[scalarnoarrow](-1,0) -- (1,0);
	\draw[scalarnoarrow](1,0) -- (2.5,1);
	\draw[scalarnoarrow](1,0) -- (2.5,-1);
    \node at (-3,1.0) {$\Omega$};
	\node at (-3,-1.0) {$\Omega$};
    \node at (-0.1,0.46) {$\chi$};
	\node at (1.4,1.1) {$\chi$};
    \node at (1.4,-1.1) {$(h_1,h_2)$}; 
    \node at (-0,-3) {$\textit{(c)}$};
\end{scope}

\begin{scope}[shift={(21,0)}]
 \draw[fermion](-3,1) -- (-1,1);
	\draw[fermionbar](-3,-1) -- (-1,-1);
	\draw[fermion](-1,1) -- (-1,-1);
	\draw[scalarnoarrow](-1,1) -- (1,1);
	\draw[scalarnoarrow](-1,-1) -- (1,-1);
    \node at (-3.6,1.0) {$\Omega$};
	\node at (-3.6,-1.0) {$\Omega$};
    \node at (-0.4,0) {$\Omega$};
	\node at (1.5,1) {$\sigma$};
    \node at (1.5,-1) {$\sigma$};   
    \node at (-0.8,-3) {$\textit{(d)}$};
\end{scope}

\begin{scope}[shift={(0,-6)}]
 \draw[fermion](-3,1) -- (-1,1);
	\draw[fermionbar](-3,-1) -- (-1,-1);
	\draw[scalarnoarrow](-1,1) -- (-1,-1);
	\draw[fermion](-1,1) -- (1,1);
	\draw[fermionbar](-1,-1) -- (1,-1);
    \node at (-3.6,1.0) {$\Omega$};
	\node at (-3.6,-1.0) {$\Omega$};
    \node at (-0,0) {$\eta_{R, I}$};
	\node at (1.8,1) {$N_R$};
    \node at (1.8,-1) {$N_R$};   
    \node at (-0.8,-3) {$\textit{(e)}$};
\end{scope}

\begin{scope}[shift={(7,-6)}]
 \draw[fermion](-3,1) -- (-1,1);
	\draw[fermionbar](-3,-1) -- (-1,-1);
	\draw[fermion](-1,1) -- (-1,-1);
	\draw[scalarnoarrow](-1,1) -- (1,1);
	\draw[scalarnoarrow](-1,-1) -- (1,-1);
    \node at (-3.6,1.0) {$\Omega$};
	\node at (-3.6,-1.0) {$\Omega$};
    \node at (-0.4,0) {$N_R$};
	\node at (1.8,1) {$\eta_{R,I}$};
    \node at (1.8,-1) {$\eta_{R,I}$};   
    \node at (-0.8,-3) {$\textit{(f)}$};
\end{scope}

\begin{scope}[shift={(14,-6)}]
 \draw[fermion](-3,1) -- (-1,1);
	\draw[scalarnoarrow](-3,-1) -- (-1,-1);
	\draw[scalarnoarrow](-1,1) -- (-1,-1);
	\draw[fermion](-1,1) -- (1,1);
	\draw[scalarnoarrow](-1,-1) -- (1,-1);
    \node at (-3.6,1.0) {$\Omega$};
	\node at (-3.6,-1.0) {$\eta_{R,I}$};
    \node at (-0.2,0) {$\eta_{R,I}$};
	\node at (1.8,1) {$N_R$};
    \node at (1.8,-1) {$\sigma$};   
    \node at (-0.8,-3) {$\textit{(g)}$};
\end{scope}

\begin{scope}[shift={(21,-6)}]
 \draw[fermion](-3,1) -- (-1,1);
	\draw[fermionbar](-3,-1) -- (-1,-1);
	\draw[scalarnoarrow](-1,1) -- (-1,-1);
	\draw[fermion](-1,1) -- (1,1);
	\draw[fermion](-1,-1) -- (1,-1);
    \node at (-3.6,1.0) {$\Omega$};
	\node at (-3.6,-1.0) {$N$};
    \node at (0.2,0) {$(h_1,h_2)$};
	\node at (1.6,1) {$\Omega$};
    \node at (1.6,-1) {$N$};   
    \node at (-0.8,-3) {$\textit{(h)}$};
\end{scope}

\end{tikzpicture}
\caption{\textit{Diagrams (a)-(g) are relevant for the freeze-out of $\Omega$. The diagram (h) is relevant for direct detection, with $N$ representing the nucleons. Here to simplify notation we have used $\sigma$ to denote either of $(h_1, h_2, \chi)$.}}\label{diagrams2}
\end{figure}
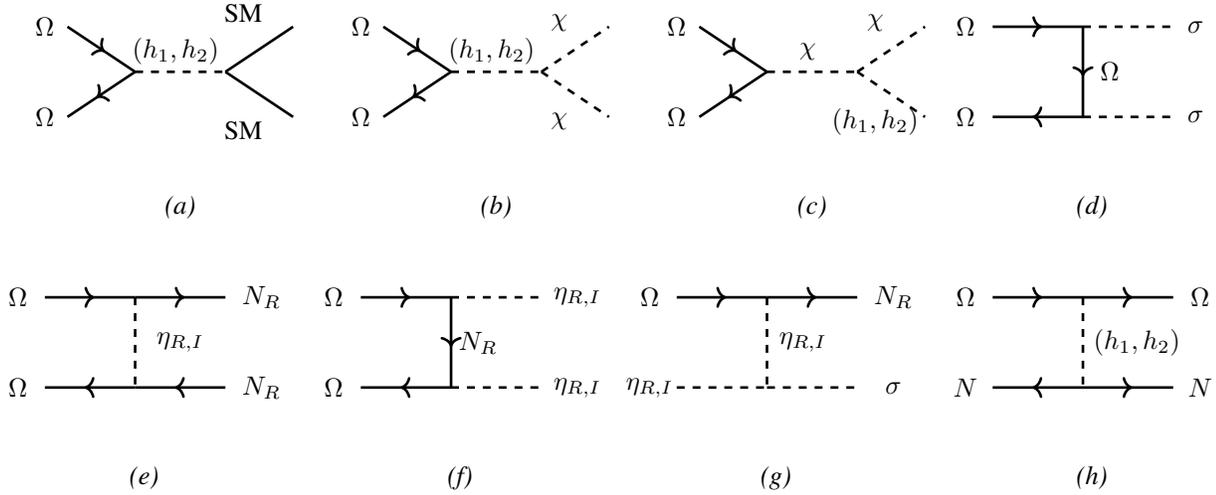


For simplicity, we consider the case in which $y_\Omega$ in Eq.~(\ref{eq:yukL}) is a diagonal matrix  and assume that $\Omega_{R_1}$ is the lightest $\mathcal{Z}_2$-odd state. That is, $\Omega_{R_1}$ is the fermion DM candidate accounting for the 80\% of the matter content of the universe. According to the Planck collaboration the DM relic abundance is~\cite{planck2018}\begin{equation}
\label{DMrelic}
\Omega_c h^2=0.1200\pm 0.0012\ \ \text{at} \ \ 68\% \text{C.L.}    
\end{equation}
In our setup  the Lagrangian providing the relevant interactions of $\Omega_{R_1}$ is given by
\begin{eqnarray}\label{omegalag}
 \mathcal{L} &\supset & y_{N_1}\bar{N}_R \Omega_{R_1}^c \eta+ y_{\Omega_1}\overline{\Omega}_{R_1} \Omega_{R_1}^c \sigma+ h.c.. 
 \end{eqnarray}
After SSB Eq.~(\ref{omegalag}) becomes
\begin{eqnarray}
 \mathcal{L} \supset (y_{N1}\bar{N}_{R} \Omega^c_{1R} \eta  + h.c.) + m_{\Omega 1}\overline{\Omega}\Omega + y_{\Omega 1}\overline{\Omega}\Omega(-h_1\sin\theta + h_2\cos\theta ) + y_{\Omega 1}i\overline{\Omega}\gamma^5\Omega \chi ,
\end{eqnarray}
where $m_{\Omega_1} = y_{\Omega_1} v_\sigma/\sqrt{2}$. We have defined $\Omega \equiv (\Omega_{1R})^c + \Omega_{1R}$, $N_R \equiv (-N_R^{+} + N_R^{-})/\sqrt{2}$ and $\chi\equiv\sigma_I$.\\

Taking into account the assumptions above, the relic abundance of $\Omega$ is determined by the annihilation channels shown in Figure~\ref{diagrams2}. Given that $\Omega$ is the DM candidate of the theory then $m_\Omega < (m_{N_R}, m_{\eta_R}, m_{\eta_I})$. In this case, the main annihilation channels are those s-channels depicted by diagrams (a), (b) and (c) in Figure~\ref{diagrams2}. We further simplify the analysis by considering that the annihilation channels mediated by the Higgs via the dimensionless parameters $\lambda_2$ and $\lambda_3$ are subleading. That is, we set $\lambda_2 = \lambda_3 = 0$. 
Therefore, the independent parameters to be used in the numerical analysis turn out to be $(m_\Omega, m_{\eta_R}, m_{\eta_I}, m_{h_2}, m_{N_R}, y_{N_1}, y_{\Omega_1}, \theta)$.\\

Let us note that in this inverse seesaw model, the Majorana dark matter candidate can interact with the atomic nucleons at tree-level. For this reason, our model gets restricted by direct detection constraints. As a matter of fact, these constraints come from the $t$-channel exchange of $h_1$ and $h_2$ shown by Figure~\ref{diagrams2}-(h). Then, here the spin-independent (SI) tree-level DM-nucleon scattering cross section is, approximately,~\cite{Garcia-Cely:2013wda, Garcia-Cely:2013nin} 
\begin{eqnarray}\label{dd}
 \sigma_\Omega \approx \frac{f_p^2 m_N^4m_\Omega^2}{4\pi v_\Phi^2(m_\Omega + m_N)^2}\left(\frac{1}{m_{h_1}^2} - \frac{1}{m_{h_2}^2}\right)^2(y_{\Omega_1} \sin 2\theta)^2,
\end{eqnarray}
where $m_N$ denotes the nucleon mass and the nuclear elements $f_p \approx 0.27$. The approximation given in Eq.~(\ref{dd}) does not take into account the finite width of both Higgs scalars, although the outputs of this expression match the numerical results from Micromegas code v5.3.35~\cite{Belanger:2013ywg} which do include these finite widths.

\subsection{Analysis and results}

In what follows, we compute the relic abundance of the Majorana fermion $\Omega$ assuming freeze-out mechanism, the direct detection via non-relativistic scattering, and indirect detection prospects today. For our calculation, we make use of Micromegas code v5.3.35~\cite{Belanger:2013ywg}.

\begin{figure}[t!]
\centering
\includegraphics[width=0.42\textwidth]{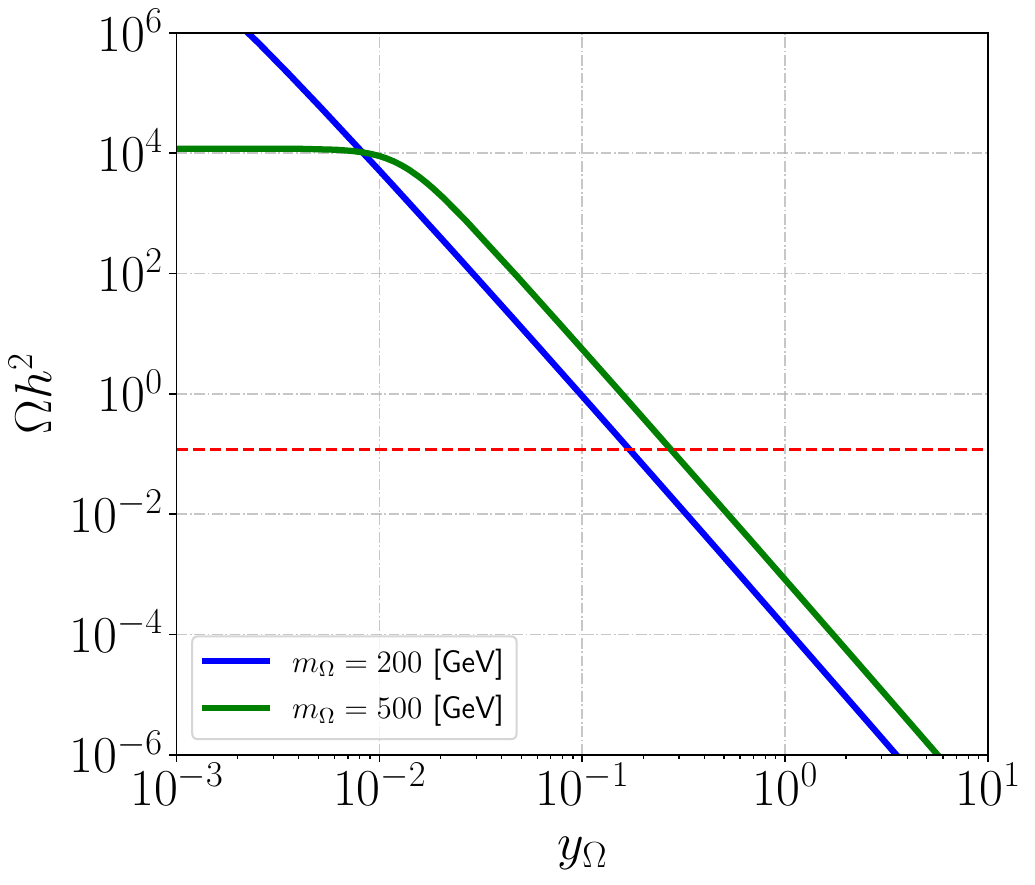}\quad
\includegraphics[width=0.42\textwidth]{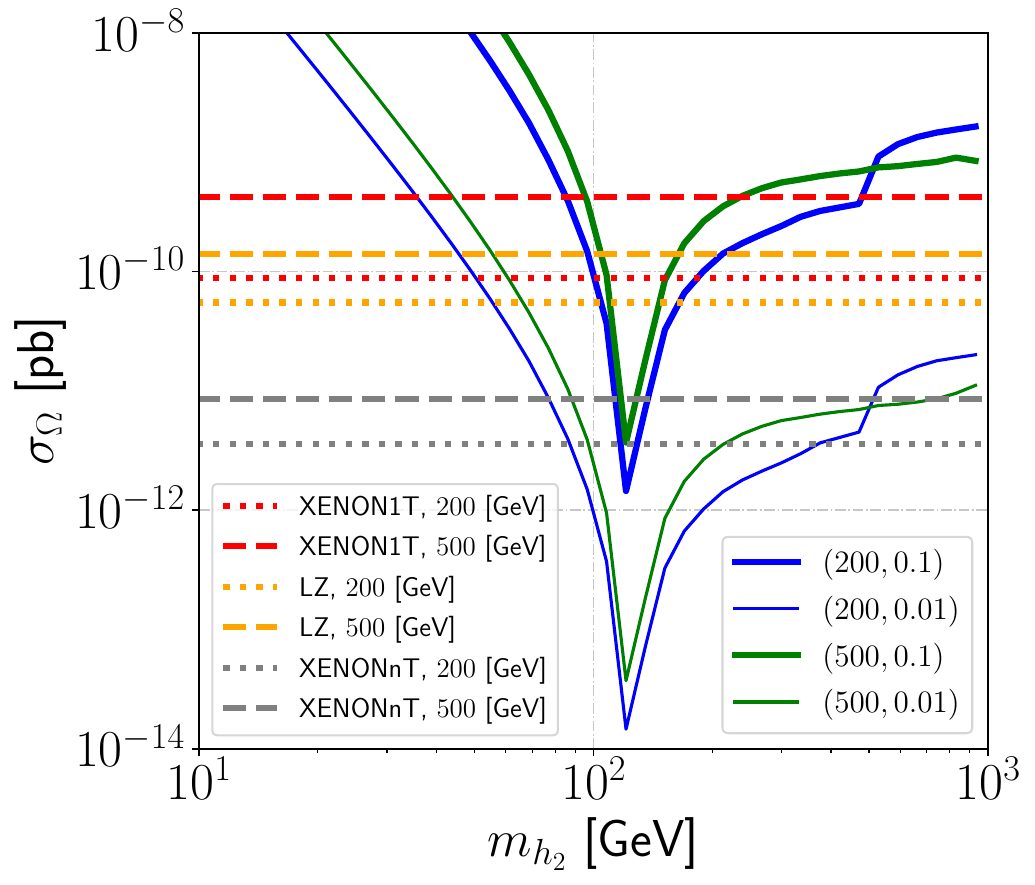}
\caption{Left-panel: Relic abundance as a function of $y_{\Omega}$ for $m_\Omega = 200$ and 500 GeV. Here we have set $m_{h_2} = 120$ GeV and $\theta = 0.1$, with the rest of the parameters specified in the text. Right-panel: Direct detection cross section as a function of $m_{h_2}$ and different combinations of $(m_\Omega \text{[GeV]}, \tan\theta)$. The horizontal red and orange lines are the current DD upper limits for each value of $m_\Omega$, whereas the grey curves are the XENONnT projections. The solid curves depicted here fulfill the correct relic abundance.}
\label{figb}
\end{figure}

As mentioned, here we explore the case where $\lambda_{1}\neq0$, $\lambda_2 = \lambda_3 = 0$. Therefore, the contribution to the relic abundance coming from the annihilation of $\Omega$ into SM particles happens only via the Higgs portal associated to $\lambda_1$. 
The left panel in Figure~\ref{figb} shows the relic abundance of $\Omega$ as a function of $y_\Omega$, for $m_\Omega = 200$ (solid blue) and $500$ GeV (solid greed), assuming $m_{h_2}= 120$ GeV and $\theta = 0.1$\footnote{Collider searches of additional scalars restrict the doublet-singlet mixing angle to be $\theta \lesssim 0.2$~\cite{Falkowski:2015iwa}}. This benchmark considers $y_N = 0.1$, $m_{\eta_R} = 2000$ GeV, $m_{\eta_I} = 2001$ GeV and $m_N = 300$ GeV (these parameters will be fixed to such values from now on). The limit of the DM relic abundance given in Eq.~(\ref{DMrelic}) is represented by the red dashed line. One can see from Figure~\ref{figb} that the relic abundance has a strong dependence on the DM mass and the parameter $y_\Omega$. For a small Yukawa coupling $y_\Omega \lesssim 10^{-2}$ and $m_\Omega > m_N$, the process $\Omega\Omega \rightarrow N N$ is kinematically allowed and dominates over the other annihilation channels. In the case of $m_\Omega < m_N$, the leading contributions to the relic abundance are those processes which involve the fields $(\chi, h_1, h_2)$ (diagrams $b$ and $c$ in Figure~\ref{diagrams2}). In such a case, the relic abundance turns out to be inversely proportional to $y_\Omega^2$. For this reason, as shown on the left-panel in Figure~\ref{figb}, the solid blue (green) curve decreases when the value of Yukawa increases. It is evident that, for a given $y_\Omega$, the relic abundance grows as the DM mass decreases. This behaviour is expected since $\Omega_\Omega h^2$ is inversely proportional to the annihilation cross section which depends on the center-of-mass energy of the colliding non-relativistic DM particles, i.e. $s \approx 4m_\Omega^2$. Here, we are focusing on the case of small doublet-singlet mixing $\theta$, i.e. $\lesssim 0.1$. For this reason, the DM relic abundance turns out to be "blind" to this parameter and is completely determined by the interaction of fields belonging to the dark sector (namely, $\chi$ and/or $h_2$). The DM annihilation channel into a pair of $\chi$ is always present unless further assumptions are made to suppress it.

In contrast to the situation previously described, the DM direct detection given by the SI cross section $\sigma_\Omega$, Eq.~(\ref{dd}), is sensitive to the values of the doublet-singlet mixing $\theta$. This is shown by the plot on the right-hand side of Figure~\ref{figb} where $\sigma_\Omega$ is depicted as a function of the mass of second $\text{CP-even}$ scalar $m_{h_2}$ (considering  $m_\Omega=200$ and 500~GeV, and each case with $\theta = 0.1, 0.01$). The blue and green curves represent the points in the parameter space fulfilling the correct relic abundance while the red, yellow and gray dashed horizontal lines correspond to the experimental limits provided by XENON1T~\cite{XENON:2018voc} and LUX-ZEPLIN (LZ)~\cite{LZ:2022lsv}, and the projections by XENONnT~\cite{XENON:2020kmp}, respectively. 
From Eq.~(\ref{dd}) one can notice that the cross section rests on the existence of a mixing between the scalar doublet $\Phi$ and the singlet $\sigma$. This dependence can be observed from Figure~\ref{figb} which shows the sensitivity of the cross section to $\theta$ variations. Furthermore, we can see that when the CP-even scalars $h_1$ and $h_2$ are (semi)-degenerate, i.e. $m_{h_2} \approx m_{h_1}$, there is a numerical cancellation that generates the inverted peak in the SI cross section. This allows to elude the experimental bounds when $h_1$ and $h_2$ are close in mass. 
Another possibility to relax the experimental constraints, including the one coming from XENONnT, happens by shrinking the value of doublet-singlet mixing as  depicted in Figure~\ref{figb}. This parameter space limit is reached, for instance, when lepton number gets broken at energies much higher than the electroweak scale. \\

Let us note that this model is characterized by the presence of the process $\Omega\bar{\Omega}\rightarrow \chi h_{1,2}$, with $h_{1,2}$ decaying into SM particles~\cite{Garcia-Cely:2013wda}. These s-wave processes are velocity independent channels not present in other models~\cite{Ahriche:2016acx,Mandal:2019oth} and give good prospects for indirect DM detection especially when $\chi$ is a pseudo-Goldstone boson, i.e. $m_\chi\neq0$. For this reason, we look into regions of the parameter space where the processes $\Omega\bar{\Omega}\rightarrow \chi h_{1,2}$ dominate the DM annihilation and provide the limits coming from the Alpha Magnetic Spectrometer
(AMS) experiment as well as the future sensitivities of the Cherenkov Telescope Array (CTA) experiment. For convenience, we assume $m_\chi>m_{h_2}/2$ so that the decay $h_2 \to 2\chi$ is kinematically disallowed. In this way, the DM candidate does not annihilate primarily into invisible channels. \\

Using Eq.~(\ref{eq:srot}), one can express the $h_2$ branching fraction into SM particles as~\cite{Robens:2015gla}, 
\begin{eqnarray}\label{breq}
 \text{BR} (h_2\rightarrow \text{SM}) = \sin^2\theta \left[\frac{\Gamma (h_2\rightarrow \text{SM})}{\Gamma_{\text{tot}}}\right]
\end{eqnarray}
where $\Gamma (h_2\rightarrow \text{SM})$ corresponds to the partial decay width of the scalar boson $h_2$ (with mass $m_{h_2}$) into SM states, and the total decay width is given by 
\begin{eqnarray}
 \Gamma_\text{tot} = \sin^2\theta \times \Gamma^{\text{SM}}_{\text{tot}} + \Gamma (h_2\rightarrow 2h_1) + \Gamma (h_2\rightarrow 2\Omega) + \Gamma (h_2\rightarrow 2\chi).
 \label{Gtot}
\end{eqnarray}
Here $\Gamma^\text{SM}_{\text{tot}}$ corresponds to the total decay width of $h_2$ into SM states~\cite{LHCHiggsCrossSectionWorkingGroup:2011wcg}. 
For simplicity, we focus on the case in which $m_{h_2} < 2m_\Omega$. Furthermore, in order to assure observability, via $h_2$ decaying into SM particles, we assume $m_\chi\neq0$ as well as $m_{h_2} < 2m_\chi$. Therefore, the third and fourth terms in Eq.~(\ref{Gtot}) are not present in our study.\\

\begin{figure}[t!]
\centering
\includegraphics[width=0.8\textwidth]{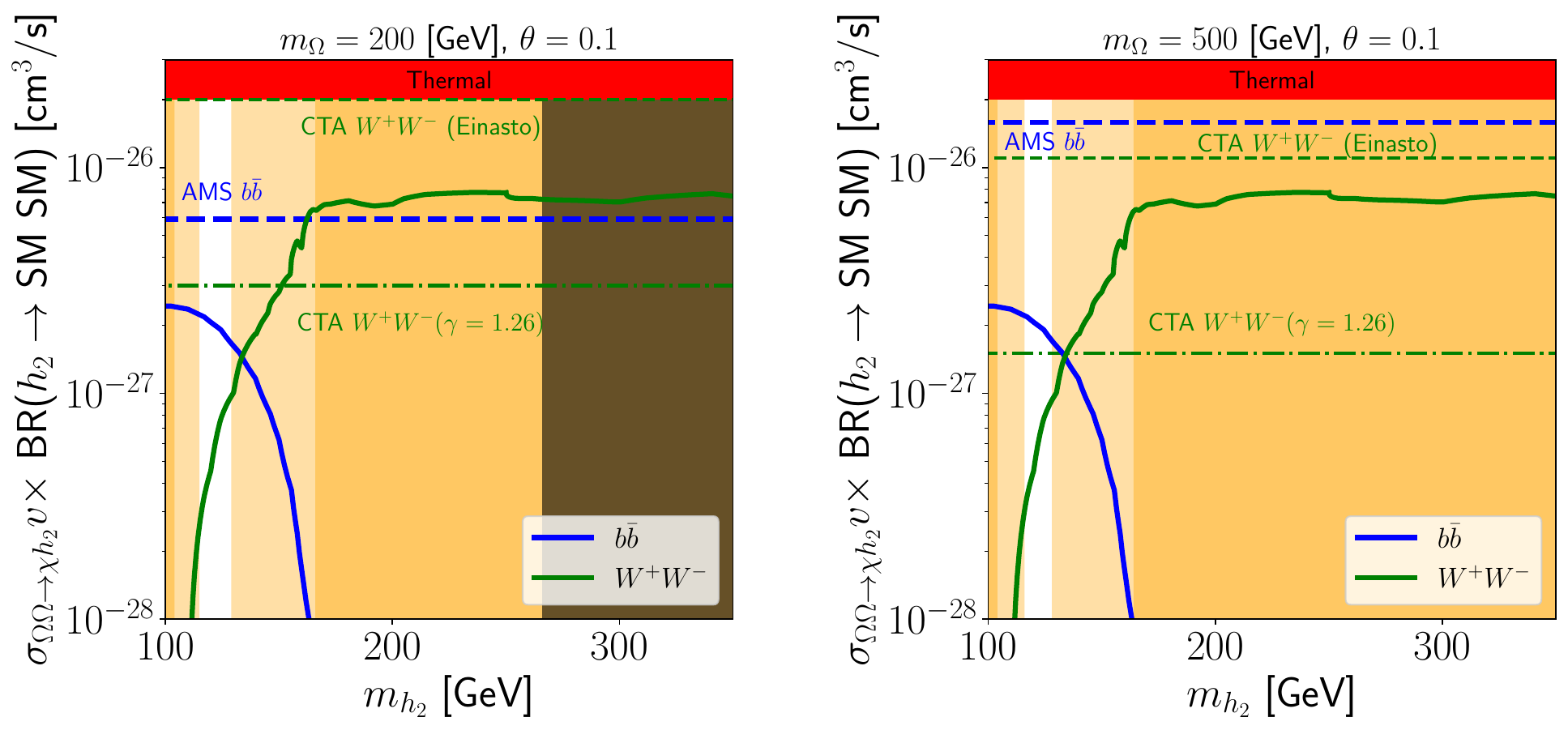} \\
\includegraphics[width=0.8\textwidth]{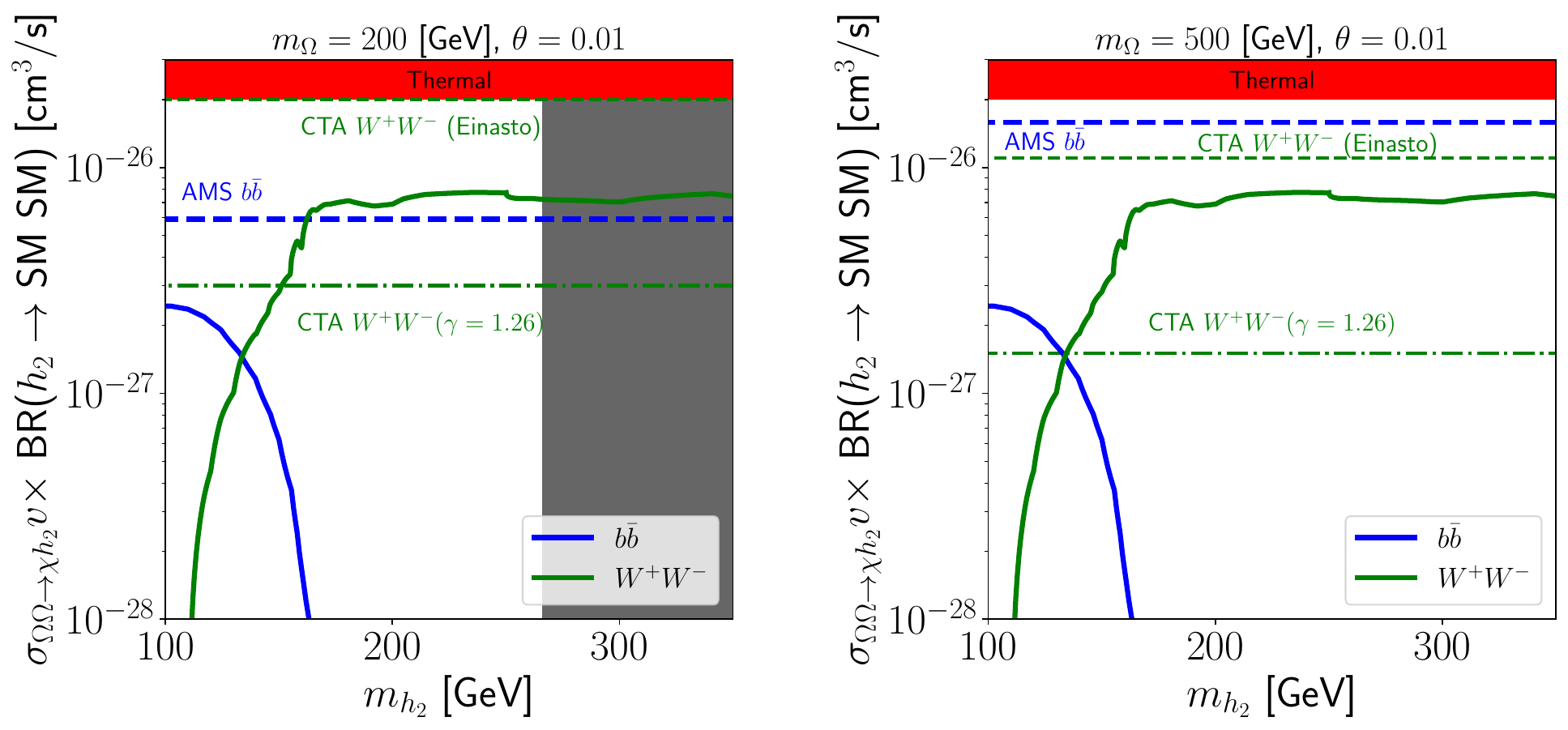}
\caption{Parameter space restrictions from direct and indirect dark matter. The thermal value of $\braket{\sigma v}_{\Omega\Omega\rightarrow h_2\chi}$ is given by the red band on top of the plots. It is assumed that $h_2$ decays into either $b\bar{b}$ (blue solid line) or $W^+W^-$ (green solid line). The dashed blue horizontal line represents the latest bound on the DM annihilation cross section into $b\bar{b}$ from AMS-02. The green dashed and dotted-dashed are the projected sensitivities of CTA depending on the DM profile. The dark and light orange areas correspond to the LZ exclusion region and the future sensitivity of XENONnT, respectively. The dark gray area in the left panels satisfies $m_{h_2} > 2 m_\chi$ and is then forbidden.}
\label{id}
\end{figure}

Taking into account the considerations previously stated, we perform a numerical analysis. Figure~\ref{id} shows the predictions for indirect detection signals coming from the DM annihilation into a pseudo-Goldstone $\chi$ and $h_2$. The CP-even scalar subsequently decays into SM particles, see Eq.~(\ref{breq}). Then, the DM annihilation would be $\Omega\bar{\Omega}\rightarrow \chi h_{2}\rightarrow \chi \text{SM}$. The assumed thermal value $2\times 10^{-26}$ cm$^3/$s$\leq\braket{\sigma v}_{\Omega\bar{\Omega}\rightarrow \chi h_2}\leq 3 \times 10^{-26}$ cm$^3/$s is depicted by the red region at the top of each panel in Figure.~\ref{id}. The left-panels consider $m_\Omega = 200$ GeV for $\theta = 0.1$ and $0.01$, while the panels on the right take $m_\Omega = 500$ GeV for the same values of $\theta$. In all panels, the Yukawa coupling is $y_N = 0.1$, and the dark scalar and pseudoscalar masses are fixed as $m_{\eta_R} = 2000$ GeV and $m_{\eta_I} = 2001$ GeV, respectively. The solid blue (green) line corresponds to the DM matter annihilation into $b\bar{b}$ ($W^+ W^-$). The dashed blue horizontal line is the upper bound of AMS-02~\cite{Reinert:2017aga} for DM annihilation into $b\bar{b}$, whereas the dashed green horizontal one represents the future sensitivity of CTA~\cite{CTA:2020qlo} in the $W^+W^-$ channel. We also consider the bound projected by CTA for DM annihilation searches with $W^+W^-$ in the final state assuming a gNFW profile with a slope parameter $\gamma=1.26$~\cite{CTA:2020qlo}. Figure~\ref{id} includes direct detection bounds at fixed $\theta$ and DM mass. The dark orange is the exclusion region coming from the LZ results on DD searches. The light orange area represents the future sensitivity of XENONnT. Therefore, only the light orange and white parts in each panel are allowed by current DM direct detection constraints. In addition, the dark gray area in the left panels of Figure~\ref{id} satisfies $m_{h_2} > 2 m_\chi$ and is then forbidden. This is because we are working under the assumption that $m_{h_2} < 2 m_\chi$ or $m_\chi=2 m_\Omega - m_{h_2}> 2/3 m_\Omega$, i.e. $h_2$ does not decay into $2\chi$. This guarantees that the fermion DM candidate $\Omega$ annihilates into observable modes.\\

 Figure~\ref{id} shows the parameter space fraction that CTA would be able to test in the future by looking for $W^+W^-$ products. One can see that the CTA projections do not reach the model predictions if the DM density distribution follows an Einasto profile. On the other hand, if the DM posses a gNFW profile, CTA searches could prove masses $m_{h_2}\gtrsim 150$ GeV. In addition, we have that AMS-02 data impose parameter space restrictions over the parameter space from DM annihilation into $b\bar{b}$ searches. Notice that, as expected from Eq.~(\ref{dd}), the direct detection constraints get weaker when the singlet-doublet mixing takes smaller values. 

\section{Charged Lepton Flavor Violation}\label{clfvandlepto} 
\lhead[\thepage]{\thesection. CLFV}

In this section we analyze charged lepton flavor violation (cLFV) processes present due to the mixing between active and heavy sterile neutrinos. Here we focus in the one-loop decays $l_{i}\rightarrow
l_{j}\gamma$ whose branching ratios are given by~\cite{Langacker:1988up,Lavoura:2003xp,Hue:2017lak}
\begin{eqnarray}
\text{BR}\left( l_{i}\rightarrow l_{j}\gamma \right)  
&=&\frac{\alpha
_{W}^{3}s_{W}^{2}m_{l_{i}}^{5}}{256\pi ^{2}m_{W}^{4}\Gamma _{i}}\left\vert
G_{ij}\right\vert ^{2}\label{Brmutoegamma1} \\
G_{ij} &\simeq &\sum_{k=1}^{3}\left( \left[
\left( 1-RR^{\dagger }\right) U_{\nu }\right] ^{\ast }\right) _{ik}\left(
\left( 1-RR^{\dagger }\right) U_{\nu }\right) _{jk}G_{\gamma }\left( \frac{%
m_{\nu _{k}}^{2}}{m_{W}^{2}}\right) +2\sum_{l=1}^{2}\left(R^{\ast }\right)
_{il}\left( R\right) _{jl}G_{\gamma }\left( \frac{m_{N_{R_l}}^{2}}{m_{W}^{2}}%
\right), \label{Brmutoegamma2}\\
G_{\gamma } (x) &=&\frac{10-43x+78x^{2}-49x^{3}+18x^{3}\ln x+4x^{4}}{%
12\left( 1-x\right) ^{4}},  \notag
\end{eqnarray}
where $\Gamma _{\mu }=3\times 10^{-19}$ GeV is the total muon decay width, $U_{\nu}$ is the matrix that diagonalizes the light neutrinos mass matrix which, in our case, is equal to the Pontecorvo–Maki–Nakagawa–Sakata (PMNS) matrix since the charged lepton mixing matrix is equal to the identity $U_{\ell}=\mathbb{I}$. In addition, the matrix $R$ is given by
\begin{equation}
R=\frac{1}{\sqrt{2}}m_D^{*}M^{-1},
\label{eq:Rneutrino}
\end{equation}
where $M$ and $m_D$ are the heavy Majorana mass matrix and the Dirac neutrino mass matrix, respectively. We provide in section~\ref{apx:mutoegamma} all assumptions made for computing $R$ in Eq.~(\ref{eq:Rneutrino}). Table~\ref{tab:parameter} contains the benchmarks points used to compute $ \mu \rightarrow e\gamma$.\\

Then, feeding Eq.~(\ref{Brmutoegamma1}) with the values of the dimensionless parameters and masses given in Table~\ref{tab:parameter}, one gets the following branching fractions,
\begin{eqnarray}
\text{BR}^{(a)}(\mu \rightarrow e\gamma) \simeq 2.02\times 10^{-13} \ \ \text{and} \ \
\text{BR}^{(b)}(\mu \rightarrow e\gamma) \simeq 1.13\times 10^{-13},
\label{eq:brsbenchmarks}
\end{eqnarray}
where $(a)$ is for $m_{\Omega}=200\text{ GeV}$ and $(b)$ is for $m_{\Omega}=500\text{ GeV}$.\\

\begin{figure}[h!]
\centering
\subfigure[]{
\includegraphics[scale=0.28]{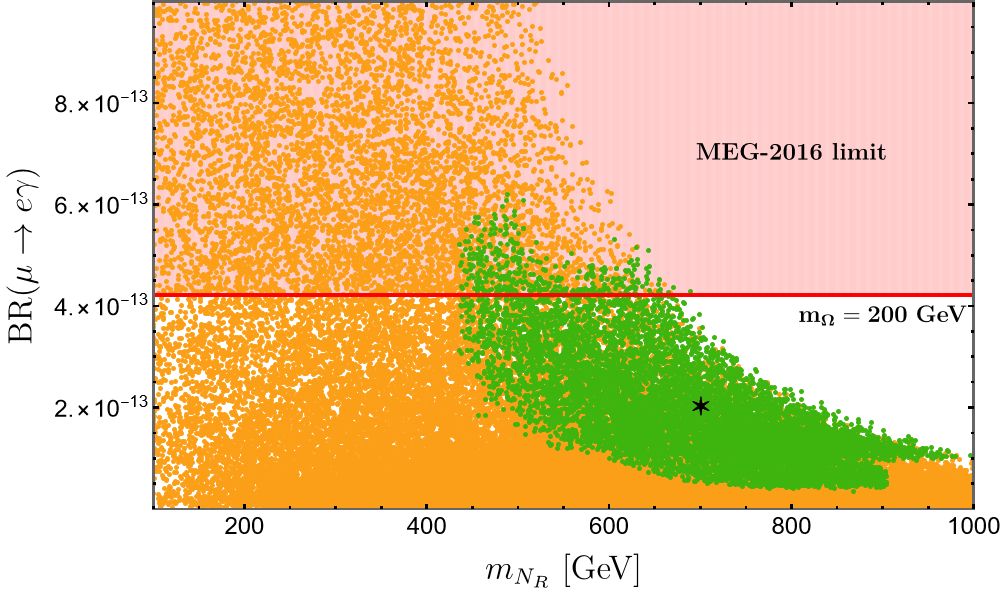}
}
\subfigure[]{
\includegraphics[scale=0.28]{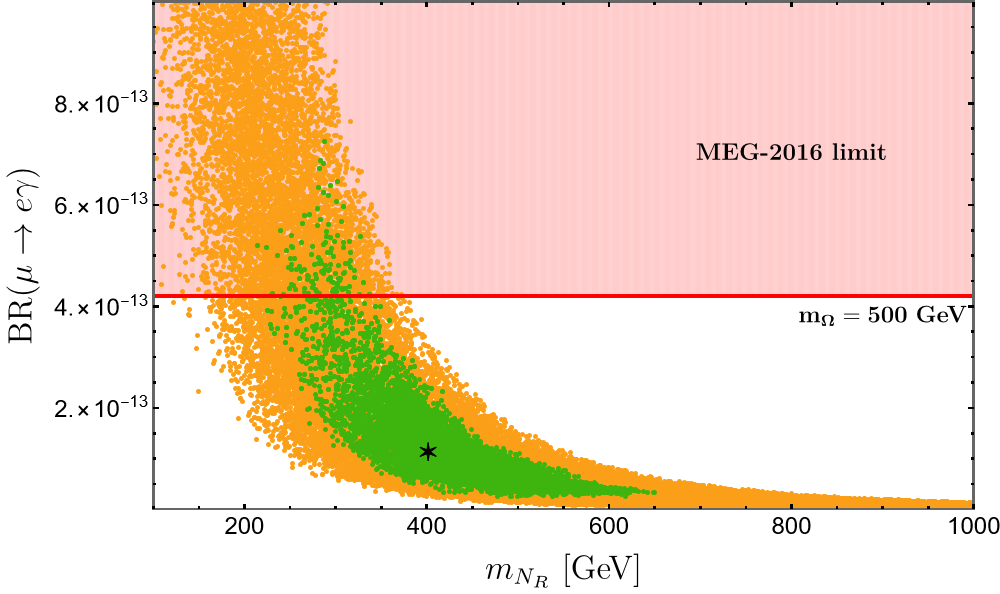}
}
\caption{Branching ratio $\text{BR}\left(\mu \rightarrow e\gamma \right)$ as a function of the mass of the lightest RH Majorana neutrino $N_R$. The shadowed region is excluded excluded by MEG~\cite{MEG:2016leq}. The black star corresponds to the prediction of the best-fit point of the model, for $m_{\Omega}=200\; \text{GeV}$ (left-panel) and $m_{\Omega}=500\; \text{GeV}$ (right-panel). The green points are compatible with current neutrino oscillation experimental limits at $3\sigma$. The orange point are out of the $3\sigma$ range and, hence, excluded by neutrino oscillation data.}
\label{mutoegammas2}
\end{figure}

Figure~\ref{mutoegammas2} shows the correlation between the branching ratio $\text{BR}\left( \mu \rightarrow e\gamma \right)$ and the mass of the lightest RH Majorana neutrino $N_R$. One observes that the branching ratio decreases as the mass of $N_R$ increases. In both plots, the red horizontal line and the shaded region represent the latest experimental constraint provided by the MEG~\cite{MEG:2016leq} collaboration,
\begin{equation}
\text{BR}\left( \mu \rightarrow e\gamma \right)^{\text{exp}}<4.2\times 10^{-13}.
\end{equation}

The black stars in Figure~\ref{mutoegammas2} correspond to the branching ratio predicted, Eq.~(\ref{eq:brsbenchmarks}), by the best-fit points of the model for $m_{\Omega}=200\; \text{GeV}$ (left-panel) and $m_{\Omega}=500\; \text{GeV}$ (right-panel). The scatter plots come from a random variation of the dimensionless parameters up to 30\% around the best-fit value. The green points are compatible with current neutrino oscillation experimental limits at $3\sigma$. One can see that neutrino oscillation data restrict the lightest RH neutrino mass to be in the range $436.9\;\text{GeV}\leq m_{ N_R}\leq 996.5 \; \text{GeV}$ for $m_{\Omega}=200\; \text{GeV}$, and $204.5\;\text{GeV}\leq m_{ N_R}\leq 649.9 \; \text{GeV}$ for $m_{\Omega}=500\; \text{GeV}$. All orange points are out of the $3\sigma$ range and, hence, excluded by neutrino oscillation data. 

\subsection{Benchmarks for $\mu\to e \gamma$ }
\label{apx:mutoegamma}

The cLFV process $\mu \rightarrow e\gamma$ is computed using Eq.~(\ref{Brmutoegamma2}) and taking as inputs the model outputs that minimize the $\chi^2$ function given by,
\begin{equation}
\chi ^{2}=\frac{\left[ \Delta m_{21}^{2\,(\exp)}-\Delta m_{21}^{2\,(\text{th})}\right] ^{2}}{\sigma
_{\Delta m_{21}^2}^{2}}+\frac{\left[ \Delta m_{31}^{2\,(\exp)}-\Delta m_{31}^{2\,(\text{th})}\right] ^{2}}{\sigma
_{\Delta m_{31}^2}^{2}}+\sum_{i<j}\frac{\left[ s_{{ij}}^{(\exp)}-s_{ij}^{(\text{th})}\right] ^{2}}{\sigma _{s_{ij}}^{2}}+\frac{\left[ \delta _{CP}^{(\exp) }-\delta
_{CP}^{(\text{th})}\right] ^{2}}{\sigma _{\delta_{CP} }^{2}}\;,  \label{eq:chisq}
\end{equation}
where $s_{ij}\equiv \sin\theta_{ij}$ (with $i,j=1,2,3$), $\delta_{CP}$ is the leptonic CP violating phase, the label (th) are used to identify the model outputs, while the ones with label (exp) correspond to the experimental values, and $\sigma_a $ represent the experimental errors. Table~\ref{table:neutrinos_data} shows best fit values and $1\sigma-3\sigma$ intervals reported by neutrino oscillation global fits~\cite{deSalas:2020pgw} \footnote{Table~\ref{table:neutrinos_data} correspond to normal neutrino mass ordering. The inverted mass ordering can be consulted in ~\cite{deSalas:2020pgw}. For other fits of neutrino oscillation parameters we refer the reader to Refs.~\cite{Capozzi:2017ipn, Esteban:2020cvm}.}.

\begin{table}[h!]
\begin{tabular}{c|cccccc} 
\hline\hline
\text{Observable}      & \text{$\Delta m_{21}^2[10^{-5}\;\text{eV}]$} & \text{$\Delta m_{31}^2[10^{-3}\;\text{eV}]$} & \text{$\sin^2\theta_{12}/10^{-1}$} & \text{$\sin^2\theta_{23}/10^{-1}$} & \text{$\sin^2\theta_{13}/10^{-2}$} & \text{$\delta_{\text{CP}}/^{\circ}$}  \\ 
\hline\hline
\text{ Best fit $\pm1\sigma$} & $7.50_{-0.20}^{+0.22}$                         & $2.55_{-0.03}^{+0.02}$                         & $3.18\pm 0.16$                       & $5.74\pm 0.14$                       & $2.200_{-0.062}^{+0.069}$            & $194_{-22}^{+24}$                   \\
\text{$3\sigma$ range} & $6.94-8.14$                                    & $2.47-2.63$                                    & $2.71-3.69$                          & $4.34-6.10$                          & $2.00-2.405$                         & $128-359$                           \\
\hline\hline
\end{tabular}
\caption{Neutrino oscillation parameters from global fits~\cite{deSalas:2020pgw}.}
\label{table:neutrinos_data}
\end{table}

In order to compute all neutrino oscillation parameters we perform a random scan of the free parameters in the lepton sector and make assumptions about the flavor structure of the Dirac mass matrix $m_{D}$, and the Majorana matrices $M$ and $\mu$ in Eq.~(\ref{Mnu}). Using the Casas–Ibarra parametrization~\cite{Casas:2001sr}, the matrix $m_{D}$ in Eq.~\eqref{ec:mdirac}, reads as follows~\cite{Casas:2001sr,Ibarra:2003up,Restrepo:2019ilz,Cordero-Carrion:2019qtu,Dolan:2018qpy,Hernandez:2021xet},
\begin{equation}
m_{D}=  \frac{y_\nu v_\Phi}{\sqrt{2}}=U_{\text{PMNS}}\left(\hat{m}_{\nu}\right)^{1/2}\hat{R}\mu^{-1/2}M\;,  \label{eq:mD-casa}
\end{equation}%
where $U_{\text{PMNS}}\equiv U_\ell^{\dagger}U_\nu$, $\hat{m}_{\nu}=\text{diag}(m_1,m_2,m_3)$ is the diagonal neutrino mass matrix and $\hat{R}$ is a rotation matrix given by,
\begin{eqnarray}
\hat{R}=\left(
\begin{tabular}{cc}
0 & 0 \\
$\cos\hat{\theta}$ & $\sin\hat{\theta}$ \\
$-\sin\hat{\theta}$ & $\cos\hat{\theta}$
\end{tabular}
\right)\ \ \text{with} \ \ \hat{\theta}\in[0,2\pi]. \label{ec:Rot}
\end{eqnarray}

The matrices $M$ and $\mu $ are assumed to be diagonal,
\begin{eqnarray}
M=\left(
\begin{tabular}{cc}
$M_1$ & 0 \\
0 & $M_2$
\end{tabular}
\right)\ \ \text{and} \ \
\mu=\left(
\begin{tabular}{cc}
$\mu_1$ & 0 \\
0 & $\mu_2$
\end{tabular}
\right)
.\label{ec:benchmark}
\end{eqnarray}

For the matrix $M$ in Eq.~(\ref{ec:benchmark}) we varied $M_1$ within the range $100\;\text{GeV} \leq M_{1} \leq 1\; \text{TeV}$ and considered $M_2= 10 M_1$, where $M_1$ is the mass of the lightest RH neutrino, $M_1=m_{N_R}$. In the case of the $\mu$ matrix we used Eq.~\eqref{eq:matrix-mu} to compute $\mu_1$ and assumed $\mu_2=10\mu_1$. This matrix depends on $y_N$, $m_{\Omega}$, $m_{\eta_{R}}$ and $m_{\eta_{I}}$. Then, following the study made in section~\ref{DM}, we analyze two situations: one with DM mass of $m_{\Omega}=200$ and the other with $m_{\Omega}=500\; \text{GeV}$. In both cases, the Yukawa $y_N$ was varied in the range $10^{-2}\leq y_N \leq 1$ and the masses of the $\mathcal{Z}_2$-odd scalars were fixed to
\begin{eqnarray}
 m_{\eta_R}=2000\; \text{GeV}, \ \ m_{\eta_I}=2001\; \text{GeV}.  
 \end{eqnarray}

We also take the charged lepton mass matrix as a diagonal matrix, i.e. $M_{l}=\text{diag}(m_e,m_\mu,m_\tau)$, which implies $U_{\text{PMNS}}=U_\nu$ in Eq.~(\ref{eq:mD-casa}).\\

 Table~\ref{tab:parameter} shows the dimensionless parameters that minimize the $\chi^2$ function in Eq.~(\ref{eq:chisq}) and that are used to compute $\mu\to e\gamma$, Eq.~(\ref{Brmutoegamma2}). The best fits of the model, for $m_{\Omega}=200\text{ GeV}$ and $m_{\Omega}=500\text{ GeV}$ are presented in Table~\ref{table:modelfit}.\\

\begin{table}[h!]
\centering
\begin{tabular}{l|l} 
\hline\hline
\multicolumn{2}{c}{\textbf{Dimensionless parameters $(a)$}}                \\ 
\hline\hline
$y_{\nu_{11}}=0.0109e^{1.87i}$  & $y_{\nu_{31}}=0.0124e^{1.49i}$  \\
$y_{\nu_{12}}=0.0105e^{-1.31i}$ & $y_{\nu_{32}}=0.0808e^{1.77i}$  \\
$y_{\nu_{21}}=0.0270e^{1.70i}$  & $y_N=2.04\times 10^{-2}$        \\
$y_{\nu_{22}}=0.0463e^{1.56i}$  & $\hat{\theta}=1.303\; \text{rad}$     \\ 
\hline\hline
\end{tabular}
\qquad
\begin{tabular}{l|l} 
\hline\hline
\multicolumn{2}{c}{\textbf{Dimensionless parameters $(b)$}}                  \\ 
\hline\hline
$y_{\nu_{11}}=0.00525e^{-1.51i}$ & $y_{\nu_{31}}=0.0177e^{1.572i}$  \\
$y_{\nu_{12}}=0.0151e^{-1.59i}$  & $y_{\nu_{32}}=0.00424e^{1.61i}$  \\
$y_{\nu_{21}}=0.0166e^{1.57i}$   & $y_N=1.22\times 10^{-2}$         \\
$y_{\nu_{22}}=0.0354e^{-1.58i}$  & $\hat{\theta}=-4.98\; \text{rad}$      \\ 
\hline\hline
\end{tabular}
\caption{Dimensionless parameters used to compute $\text{BR}(\mu\rightarrow e\gamma)$ compatible with neutrino oscillation data. Case (a) considers $m_{\Omega}=200$  GeV and case (b) is for $m_{\Omega}=500$  GeV.}
\label{tab:parameter}
\end{table}

\begin{table}[h!]
\resizebox{16.8cm}{!}{
\begin{tabular}{c|cccccccc}
\hline\hline
Observable & $\mu_1[\text{keV}]$ & $m_{N_R}[\text{GeV}]$ & $\Delta m_{21}^{2}$[$10^{-5}$eV$^{2}$]
& $\Delta m_{31}^{2}$[$10^{-3}$eV$^{2}$] & $\sin\theta^{(l)}_{12}/10^{-1}$
& $\sin\theta^{(l)}_{13}/10^{-3}$ & $\sin\theta^{(l)}_{23}/10^{-1}$ & $%
\delta^{(l)}_{CP} (^{\circ })$ \\ \hline\hline
Best fit case (a)  &  $-0.562$ & $700.5$ & $7.53$ & $2.51$ & $3.53$ & $2.11$ & $5.66$ & $195.3$\\ \hline
Best fit case (b) & $-0.4099$ & $401.3$ & $7.50$ & $2.55$ & $3.25$ & $2.22$ & $5.64$ & $174.8$\\ \hline\hline
\end{tabular}
}
\caption{Best fit values of the model. Case (a) considers $m_{\Omega}=200\text{ GeV}$ and case (b) is for $m_{\Omega}=500\text{ GeV}$.}
\label{table:modelfit}
\end{table}

\newpage
$\ $
\thispagestyle{empty} 
\chapter{Phenomenological of extended multiHiggs doublet models with $S_4$ family symmetry}\label{cap.modelo3HDM}
\markboth{EXTENDED MULTIHIGGS DOUBLET MODEL}{EXTENDED MULTIHIGGS DOUBLET MODEL}

The next chapter is based on the second paper of this thesis~\ref{paper:S4}. In this chapter we propose extended 3HDM and 4HDM models where the SM gauge symmetry is
enlarged by the inclusion of the $S_4\times Z_2\times Z_4$ discrete group
and the scalar and SM fermion sectors are augmented by several scalar singlets and
right-handed Majorana neutrinos, respectively. We employ the $S_{4}$ family symmetry because it is the smallest non abelian group having a doublet, triplet and singlet irreducible representations, thus it naturally accommodates the number of fermion generations of the SM. This non abelian discrete $S_4$ group yields viable leptonic and quark mass matrices that allow to successfully fit the experimental values of the charged lepton masses, neutrino mass squared splittings, quark and leptonic mixing angles and CP phases. This is due to the fact that three families of left-handed leptonic doublets can be grouped into a $S_{4}$ triplet irreducible representation, whereas two generations of quark doublets are unified in a $S_4$ doublet, and the remaining one is assigned as a $S_4$ singlet. Given that the quark sector is more restrictive than the lepton sector as the physical observables associated with the former are measured with much more experimental precision than the ones corresponding to the latter, more degree of flexibility is needed in the quark sector. Because of this reason, in this work, the two generations of quark doublets are unified in a $S_4$ doublet and the remaining one is assigned as a $S_4$ singlet, whereas the three generations of left-handed lepton doublets are grouped in a $S_4$ triplet. This makes the choice of the $S_4$ group in this work more convenient than $S_3$ or $A_4$. Other non abelian discrete groups will either be larger than $S_4$ or will only have doublets and singlets in their irreducible representations. The $S_4$ discrete group \cite{Lam:2008rs,Altarelli:2009gn,Bazzocchi:2009da,Bazzocchi:2009pv,deAdelhartToorop:2010vtu,Patel:2010hr,Morisi:2011pm,Altarelli:2012bn,Mohapatra:2012tb,BhupalDev:2012nm,deMedeirosVarzielas:2012apl,Ding:2013hpa,Ishimori:2010fs,Ding:2013eca,Hagedorn:2011un,Campos:2014zaa,Dong:2010zu,Vien:2015fhk,deAnda:2017yeb,deAnda:2018oik,CarcamoHernandez:2019eme,Chen:2019oey,deMedeirosVarzielas:2019cyj,DeMedeirosVarzielas:2019xcs,CarcamoHernandez:2019iwh, Garcia-Aguilar:2022gfw} has been shown to provide a nice description for the observed pattern of SM fermion masses and mixing angles.\\

In here, the masses of the light active neutrinos are produced by a radiative seesaw mechanism at one loop level mediated by right-handed Majorana neutrinos and electrically neutral scalars. All gauge singlet scalars will be part of $S_4$ triplets, excepting one scalar field, which in the extended 3HDM is assigned as trivial $S_4$ singlet. The gauge singlet $S_4$ triplet scalars are needed to yield a viable texture for the neutrino sector consistent with the experimental data on neutrino oscillations, whereas the right-handed Majorana neutrinos are the fermionic mediators participating in the radiative seesaw mechanism that produces the tiny masses of the light active neutrinos. In addition, the $S_4$ and $Z_4$ discrete groups are spontaneously broken, whereas the $Z_2$ symmetry is preserved, thus preventing tree level masses for light active neutrinos and allow them to appear at one loop level. Furthermore, the preserved $Z_2$ symmetry ensures the stability of fermionic and scalar dark matter candidates as well as the radiative nature of the seesaw mechanism that produces the tiny active neutrino masses. The $Z_4$ discrete symmetry distinguishes the different generations of right-handed charged leptonic fields and is needed to yield a nearly diagonal charged lepton mass matrix, so that the leptonic mixing will mainly arise from the neutrino sector. This reduces the number of lepton sector model parameters and suppresses the flavor changing neutral scalar interactions in the charged lepton sector.

\section{The models}
\lhead[\thepage]{\thesection. The models}

\label{model3} Our proposed models are extensions of the 3HDM and 4HDM
theories based on the $S_{4}$ family symmetry. In the first model, which
corresponds to an extended 3HDM theory, an electrically neutral gauge
singlet scalar field odd under a preserved $Z_{2}$ discrete symmetry is
introduced to generate light active neutrino masses via a radiative seesaw
mechanism at one loop level mediated by two right handed Majorana neutrinos.
In the second model, which corresponds to an extended 4HDM theory, there is
no such gauge singlet inert scalar in the particle spectrum, however one of
the scalar doublets is inert and allows a successful implementation of a
radiative seesaw mechanism that produces the tiny active neutrino masses.
These two models have the same common feature in both quark and lepton
sectors. Furthermore, we have included the gauge singlet scalars in non
trivial representations of the $S_{4}$ discrete group in order to build the
neutrino Yukawa terms invariant under the $S_{4}$ symmetry, necessary to
give rise to a
light active neutrino mass matrix consistent with
the neutrino oscillation experimental data. 
\subsection{Model 1}

Model 1 is an extended 3HDM theory where the tiny active neutrino masses
are generated  by a radiative seesaw mechanism mediated by right
handed Majorana neutrinos $N_{kR}$ ($k=1,2$) and a gauge  scalar singlet $%
\varphi $, charged under a preserved $Z_{2}$ symmetry. The preserved $Z_{2}$ symmetry ensures the stability of the dark matter candidate and prevents the appearance of tree-level active neutrino masses. Furthermore, this setup allows to have an one loop level scotogenic realization of active neutrino masses where the lightest of the seesaw mediators corresponds to a dark matter candidate, whose stability is guaranteed by the preserved $Z_2$ symmetry. Moreover, a $Z_4$ symmetry is needed to yield a nearly diagonal charged lepton mass matrix thus allowing to have a predictive pattern of lepton mixing, which will be mainly governed by the neutrino sector and to suppress flavor changing neutral scalar interactions associated with the charged scalar sector. In this setup the rates for flavor changing leptonic Higgs decays, as well as the ones corresponding to charged lepton flavor violating decays, can acquire values smaller than their experimental upper limits.
The particle assignments with respect to the symmetry group are
summarized in Table \ref{table:fermionasig}, but for clarity of the
notation let us write explicitly the field content. 
Our proposed models are consistent with the $S_{4}\times Z_2\times Z_4$ discrete symmetry, as indicated in Tables I and II.
The scalar fields in
model 1 have the following $S_{4}\times Z_{2}\times Z_{4}$ assignments: 
\begin{eqnarray}
\Xi _{I} &=&\left( \Xi _{1},\Xi _{2}\right) \sim \left( \mathbf{2}%
,0,0\right) ,\hspace{1.5cm}\Xi _{3}\sim \left( \mathbf{1}_{1},0,0\right) ,%
\hspace{1.5cm}\varphi \sim \left( \mathbf{1}_{1},1,0\right),  \notag \\
\chi &=&\left( \chi _{1},\chi _{2},\chi _{3}\right) \sim \left( \mathbf{3}%
_{2},0,0\right) ,\hspace{0.8cm}\eta =\left( \eta _{1},\eta _{2},\eta
_{3}\right) \sim \left( \mathbf{3}_{2},0,0\right) , \hspace{0.8cm}\rho =\left( \rho _{1},\rho _{2},\rho
_{3}\right) \sim \left( \mathbf{3}_{2},0,0\right) \mathbf{,} \notag\\
\Phi _{e} &=&\left( \Phi _{e}^{\left( 1\right) },\Phi _{e}^{\left( 2\right)
},\Phi _{e}^{\left( 3\right) }\right) \sim \left( \mathbf{3}_{2},0,-1\right)
,\hspace{1.5cm}
\Phi _{\mu } =\left( \Phi _{\mu }^{\left( 1\right) },\Phi _{\mu }^{\left(
2\right) },\Phi _{\mu }^{\left( 3\right) }\right) \sim \left( \mathbf{3}%
_{2},0,-2\right),\notag\\
\Phi _{\tau }&=&\left( \Phi _{\tau }^{\left(
1\right) },\Phi _{\tau }^{\left( 2\right) },\Phi _{\tau }^{\left( 3\right)
}\right) \sim \left( \mathbf{3}_{2},0,0\right) .
\end{eqnarray}
where $\Xi _{i}$ ($i=1,2,3$) are $SU(2)$ scalar doublets, whereas the remaining scalar fields are SM gauge singlets. The low energy scalar potential for the active $\Xi _{i}$ ($i=1,2,3$) $SU(2)$ scalar doublets is shown in Appendix \ref{appPot}. As it will be shown below, the charged lepton Yukawa terms in models 1 and 2 are the same and give rise to the same mass matrix for charged leptons. In this work, motivated by the alignment limit, we consider the scenario where $v_3 >> v_1 = v_2$, provided that the quartic scalar coupling values are very similar. Here $v_i$ corresponds to the vacuum expectation value of the neutral component of the $\Xi_i$ scalar doublet. In the scenario $v_3 >> v_1 = v_2$, the charged lepton mass matrix is nearly diagonal and has a negligible impact in the leptonic mixing parameters, thus implying that the PMNS leptonic mixing matrix mainly arises from the neutrino sector. It is worth mentioning that three $S_4$ scalar triplets, i.e. $\chi$, $\eta$ and $\rho$, which do acquire different VEV patterns (as indicated below) are introduced in the neutrino sector in order to generate a viable light active neutrino mass matrix that will allow to successfully reproduce the current the measured neutrino mass squared splittings, leptonic mixing parameters and leptonic Dirac CP phase. Having only one $S_4$ scalar triplet in the neutrino sector, would imply a VEV pattern, which will not be a natural solution of the minimization conditions of the scalar potential for a large region of parameter space. On the other hand, the three families of right handed leptonic fields as well as the three $S_4$ scalar triplets $\Phi _{e}$, $\Phi _{\mu }$ and $\Phi _{\tau }$ will be distinguished by their $Z_{4}$ assignments, thus resulting in a nearly diagonal mass matrix for charged leptons.

The fermionic fields in model 1 have the following assignments under the $%
S_{4}\times Z_{2}\times Z_{4}$ discrete group: 
\begin{eqnarray}
q_{L} &=&\left( q_{1L},q_{2L}\right) \sim \left( \mathbf{2},0,0\right) ,%
\hspace{1.5cm}q_{3L}\sim \left( \mathbf{1}_{1},0,0\right) ,\hspace{1.5cm}%
d_{R}=\left( d_{1R},d_{2R}\right) \sim \left( \mathbf{2},0,0\right) ,  \notag
\\
u_{R} &=&\left( u_{1R},u_{2R}\right) \sim \left( \mathbf{2},0,0\right) ,%
\hspace{1.5cm}u_{3R}\sim \left( \mathbf{1}_{1},0,0\right) ,\hspace{1.5cm}%
d_{3R}\sim \left( \mathbf{1}_{1},0,0\right) ,\hspace{1.5cm} \notag\\
l_{L} &\sim &\left( \mathbf{3}_{1},0,0\right) ,\hspace{1.5cm}l_{1R}\sim
\left( \mathbf{1}_{2},0,1\right) ,\hspace{1.5cm}l_{2R}\sim \left( \mathbf{1}%
_{2},0,2\right) \hspace{1.5cm}l_{3R}\sim \left( \mathbf{1}_{2},0,0\right),\notag \\
N_{1R} &\sim &\left( \mathbf{1}_{2},1,0\right) ,\hspace{1.5cm}N_{2R}\sim
\left( \mathbf{1}_{2},1,0\right) .
\end{eqnarray}
As shown in Appendix \ref{apptrip}, the following vacuum expectation value (VEV) configurations for the $S_4$ triplets scalars%
\begin{eqnarray}
\left\langle \chi \right\rangle &=&v_{\chi }\left( 0,0,1\right) ,\hspace{%
1.5cm}\left\langle \eta \right\rangle =\frac{v_{\eta }}{\sqrt{2}}\left(
1,1,0\right) ,\hspace{1.5cm}\left\langle \rho \right\rangle =\frac{v_{\rho }%
}{\sqrt{2}}\left( 1,-1,0\right),\label{eq:trip1} \\
\left\langle \Phi _{e}\right\rangle &=&v_{\Phi _{e}}\left( 1,0,0\right) ,%
\hspace{1.5cm}\left\langle \Phi _{\mu }\right\rangle =v_{\Phi _{\mu }}\left(
0,1,0\right) ,\hspace{1.5cm}\left\langle \Phi _{\tau }\right\rangle =v_{\Phi
_{\tau }}\left( 0,0,1\right) .\label{eq:trip2}
\end{eqnarray}
are consistent with the scalar potential minimization conditions for a large region of the parameter space. These VEV patterns given above, similar to the ones considered in \cite{Ivanov:2014doa,CarcamoHernandez:2019iwh}, allows us to obtain a viable pattern of lepton masses and mixings as it will be shown in the following sections of this article.

With the above specified particle content, the following $S_{4}\times Z_{2}\times Z_{4}$ invariant
Yukawa terms arise:
\begin{eqnarray}
\mathcal{L} _{Y} &=&y_{1}^{d}\left[ \bar{q}_{1L}\left( \Xi _{1}d_{2R}+\Xi
_{2}d_{1R}\right) +\bar{q}_{2L}\left( \Xi _{1}d_{1R}-\Xi _{2}d_{2R}\right) %
\right] +y_{2}^{d}\left[ \bar{q}_{1L}\Xi _{3}d_{1R}+\bar{q}_{2L}\Xi
_{3}d_{2R}\right] +y_{3}^{d}\left[ \bar{q}_{1L}\Xi _{1}+\bar{q}_{2L}\Xi _{2}%
\right] d_{3R}  \notag \\
&&+y_{4}^{d}\bar{q}_{3L}\left[ \Xi _{1}d_{1R}+\Xi _{2}d_{2R}\right]
+y_{5}^{d}\bar{q}_{3L}\Xi _{3}d_{3R}+y_{1}^{u}\left[ \bar{q}_{1L}\left(\widetilde{\Xi }_{1}u_{2R}+\widetilde{\Xi}_{2}u_{1R}\right)+\bar{q}_{2L}\left(\widetilde{\Xi}_{1}u_{1R}-\widetilde{\Xi}_{2}u_{2R}\right) %
\right]
\notag \\
&&+y_{2}^{u}\left[ \bar{q}_{1L}\widetilde{\Xi }_{3}u_{1R}+\bar{q}_{2L}\widetilde{\Xi}_{3}u_{2R}\right]+y_{3}^{u}\left[ \bar{q}_{1L}\widetilde{\Xi } _{1}+\bar{q}_{2L}\widetilde{\Xi } _{2}%
\right] u_{3R}+y_{4}^{u}\bar{q}_{3L}\left[ \widetilde{\Xi }_{1}u_{1R}+\widetilde{\Xi }_{2}u_{2R}\right]
+y_{5}^{u}\bar{q}_{3L}\widetilde{\Xi }_{3}u_{3R}\notag \\
&&+\frac{y_{1}^{l}}{\Lambda }\bar{l}_{L}\Xi _{3}l_{1R}\Phi _{e}+\frac{%
y_{2}^{l}}{\Lambda }\bar{l}_{L}\Xi _{3}l_{2R}\Phi _{\mu }+\frac{y_{3}^{l}}{%
\Lambda }\bar{l}_{L}\Xi _{3}l_{3R}\Phi _{\tau }+\frac{x_{1}^{l}}{\Lambda }%
\bar{l}_{L}\Xi _{I}l_{1R}\Phi _{e}+\frac{x_{2}^{l}}{\Lambda }\bar{l}_{L}\Xi
_{I}l_{2R}\Phi _{\mu }+\frac{x_{3}^{l}}{\Lambda }\bar{l}_{L}\Xi
_{I}l_{3R}\Phi _{\tau }\label{yuk1} \\
&&+\sum_{k=1}^{2}y_{1k}^{\left( N\right) }\bar{l}_{L}\widetilde{\Xi }%
_{3}N_{kR}\frac{\eta \varphi }{\Lambda ^{2}}+\sum_{k=1}^{2}y_{2k}^{\left(
N\right) }\bar{l}_{L}\widetilde{\Xi }_{3}N_{kR}\frac{\rho \varphi }{\Lambda
^{2}}+\sum_{k=1}^{2}y_{3k}^{\left( N\right) }\bar{l}_{L}\widetilde{\Xi }%
_{3}N_{kR}\frac{\chi \varphi }{\Lambda ^{2}}  \notag \\
&&+\sum_{k=1}^{2}x_{1k}^{\left( N\right) }\bar{l}_{L}\widetilde{\Xi }%
_{I}N_{kR}\frac{\eta \varphi }{\Lambda ^{2}}+\sum_{k=1}^{2}x_{2k}^{\left(
N\right) }\bar{l}_{L}\widetilde{\Xi }_{I}N_{kR}\frac{\chi \varphi }{\Lambda
^{2}}+\sum_{k=1}^{2}x_{3k}^{\left( N\right) }\bar{l}_{L}\widetilde{\Xi }%
_{I}N_{kR}\frac{\rho \varphi }{\Lambda ^{2}}+m_{N_{1}}N_{1R}\overline{%
N_{1R}^{C}}+m_{N_{2}}N_{2R}\overline{N_{2R}^{C}}+h.c.  \notag
\end{eqnarray}

\begin{table}[tbp]
\centering
\resizebox{16.8cm}{!}{
\begin{tabular}{c|cccccc|ccccc|ccccccccc}
\hline\hline
& $q_{L}$ & $q_{3L}$ & $u_{R}$ & $u_{3R}$ & $d_{R}$ & $d_{3R}$ & $%
l_{L}$ & $l_{1R}$ & $l_{2R}$ & $l_{3R}$ & $N_{iR}$ & $\Xi_I$ & $\Xi_3$ & 
$\varphi$ & $\chi$ & $\eta$ & $\rho$ & $\Phi_e$ & $\Phi_{\mu}$ & $\Phi_{\tau}$ \\ 
\hline\hline
$SU(3)_C$ & $\mathbf{3}$ & $\mathbf{3}$ & $\mathbf{3}$ & $\mathbf{3}$ & $\mathbf{3}$ & $\mathbf{3}$ & $\mathbf{1}$ & $\mathbf{1}$ & $\mathbf{1}$ & $\mathbf{1}$ & $\mathbf{1}$ & 
$\mathbf{1}$ & $\mathbf{1}$ & $\mathbf{1}$ & $\mathbf{1}$ & $\mathbf{1}$ & $\mathbf{1}$ & $\mathbf{1}$ & $\mathbf{1}$ & $\mathbf{1}$ \\
$SU(2)_L$ & $\mathbf{2}$ & $\mathbf{2}$ & $\mathbf{1}$ & $\mathbf{1}$ & $\mathbf{1}$ & $\mathbf{1}$ & $\mathbf{2}$ & $\mathbf{1}$ & $\mathbf{1}$ & $\mathbf{1}$ & $\mathbf{1}$ & 
$\mathbf{2}$ & $\mathbf{2}$ & $\mathbf{1}$ & $\mathbf{1}$ & $\mathbf{1}$ & $\mathbf{1}$ & $\mathbf{1}$ & $\mathbf{1}$ & $\mathbf{1}$ \\
$U(1)_Y$ & $\mathbf{\frac{1}{6}}$ & $\mathbf{\frac{1}{6}}$ & $\mathbf{\frac{2}{3}}$ & $\mathbf{\frac{2}{3}}$ & $\mathbf{-\frac{1}{3}}$ & $\mathbf{-\frac{1}{3}}$ & $\mathbf{-\frac{1}{2}}$ & $\mathbf{-1}$ & $\mathbf{-1}$ & $\mathbf{-1}$ & $\mathbf{0}$ & 
$\mathbf{\frac{1}{2}}$ & $\mathbf{\frac{1}{2}}$ & $\mathbf{0}$ & $\mathbf{0}$ & $\mathbf{0}$ & $\mathbf{0}$ & $\mathbf{0}$ & $\mathbf{0}$ & $\mathbf{0}$ \\ \hline\hline
$S_4$ & $2$ & $1_1$ & $2$ & $1_1$ & $2$ & $1_1$ & $3_1$ & $1_1$ & $1_1$
& $1_1$ & $1_1$ & 2 & $1_1$ & $1_1$ & $3_2$ & $3_2$ & $3_2$ & $3_1$ & $3_1$ & $3_1$
\\ 
$Z_2$ & 0 & 0 & 0 & 0 & 0 & 0 & 0 & 0 & 0 & 0 & 1 & 0 & 0 & 1 & 0 & 0 & 0 & 0 & 0
& 0 \\ 
$Z_4$ & 0 & 0 & 0 & 0 & 0 & 0 & 0 & 1 & 2 & 0 & 0 & 0 & 0 & 0 & 0 & 0 & 0 & -1 & 
-2 & 0 \\ \hline\hline
\end{tabular}%
}
\caption{Fermion and scalar assignments under the group $S_4\times Z_2\times Z_4$ for model 1. Here the numbers in boldface are dimensions of $SU(3)_C\times SU(2)_L\times U(1)_Y$ representations.}
\label{table:fermionasig}
\end{table}

\begin{table}[tbp]
\centering
\resizebox{16.8cm}{!}{
\begin{tabular}{c|cccccc|ccccc|ccccccccccc}
\hline\hline
& $q_{L}$ & $q_{3L}$ & $u_{R}$ & $u_{3R}$ & $d_{R}$ & $d_{3R}$ & $%
l_{L}$ & $l_{1R}$ & $l_{2R}$ & $l_{3R}$ & $N_{iR}$ & $\Xi_I$ & $\Xi_3$ & $\Xi_4$ & $\chi$ & $%
\eta$ & $\rho$ & $\Phi_e$ & $\Phi_{\mu}$ & $\Phi_{\tau}$  \\ \hline\hline
$SU(3)_C$ & $\mathbf{3}$ & $\mathbf{3}$ & $\mathbf{3}$ & $\mathbf{3}$ & $\mathbf{3}$ & $\mathbf{3}$ & $\mathbf{1}$ & $\mathbf{1}$ & $\mathbf{1}$ & $\mathbf{1}$ & $\mathbf{1}$ & 
$\mathbf{1}$ & $\mathbf{1}$ & $\mathbf{1}$ & $\mathbf{1}$ & $\mathbf{1}$ & $\mathbf{1}$ & $\mathbf{1}$ & $\mathbf{1}$ & $\mathbf{1}$ \\
$SU(2)_L$ & $\mathbf{2}$ & $\mathbf{2}$ & $\mathbf{1}$ & $\mathbf{1}$ & $\mathbf{1}$ & $\mathbf{1}$ & $\mathbf{2}$ & $\mathbf{1}$ & $\mathbf{1}$ & $\mathbf{1}$ & $\mathbf{1}$ & 
$\mathbf{2}$ & $\mathbf{2}$ & $\mathbf{2}$ & $\mathbf{1}$ & $\mathbf{1}$ & $\mathbf{1}$ & $\mathbf{1}$ & $\mathbf{1}$ & $\mathbf{1}$ \\
$U(1)_Y$ & $\mathbf{\frac{1}{6}}$ & $\mathbf{\frac{1}{6}}$ & $\mathbf{\frac{2}{3}}$ & $\mathbf{\frac{2}{3}}$ & $\mathbf{-\frac{1}{3}}$ & $\mathbf{-\frac{1}{3}}$ & $\mathbf{-\frac{1}{2}}$ & $\mathbf{-1}$ & $\mathbf{-1}$ & $\mathbf{-1}$ & $\mathbf{0}$ & 
$\mathbf{\frac{1}{2}}$ & $\mathbf{\frac{1}{2}}$ & $\mathbf{\frac{1}{2}}$ & $\mathbf{0}$ & $\mathbf{0}$ & $\mathbf{0}$ & $\mathbf{0}$ & $\mathbf{0}$ & $\mathbf{0}$ \\ \hline\hline
$S_4$ & $2$ & $1_1$ & $2$ & $1_1$ & $2$ & $1_1$ & $3_1$ & $1_1$ & $1_1$
& $1_1$ & $1_1$ & $2$ & $1_1$ & $1_2$ & $3_2$ & $3_2$ & $3_2$ & $3_1$ & $3_1$ & $3_1$  \\ 
$Z_2$ & 0 & 0 & 0 & 0 & 0 & 0 & 0 & 0 & 0 & 0 & 1 & $0$ & $0$ & 1 & 0 & 0 & 0 & 0 & 0 & 0   \\ 
$Z_4$ & 0 & 0 & 0 & 0 & 0 & 0 & 0 & 1 & 2 & 0 & 0 & $0$ & $0$ & 0 & 0 & 0 & 0 & -1 & -2 & 0  \\ \hline\hline
\end{tabular}%
}
\caption{Fermion and scalar assignments under the group $S_4\times Z_2\times Z_4$ for model 2. Here the numbers in boldface are dimensions of $SU(3)_C\times SU(2)_L\times U(1)_Y$ representations}
\label{table:fermionasig2}
\end{table}

\subsection{Model 2}

Model 2 is very similar to model 1
(though they differ in the form of the scalar potential as discussed later), the particle assignments can be seen in Table \ref{table:fermionasig2}.
The only difference is the lack of the
inert gauge singlet scalar field, which is replaced by an inert scalar
doublet $\Xi _{4}$ which triggers a radiative seesaw mechanism at one loop
level to generate the tiny masses of the light active neutrinos. The inclusion of the inert scalar doublet in model 2 allows to have four dark scalar fields which will open the possibility of having co-annihilations during the freezout, that makes annihilations of the dark sector more effective when the scalar masses are very similar, thus yielding regions of parameter space consistent with the Planck limit of the dark matter relic abundance as well as with the experimental constraints on direct detection. This will be shown in detail in section \ref{dark-sector}, where the numerical analysis indicates that the case of the inert doublet, corresponding to the model 2 is more favoured than the one of the inert singlet of model 1 since the allowed region of parameter space consistent with the dark matter constraints is larger in the former than in the latter. This, together with having a radiative mechanism of active neutrino masses where the lightest of the seesaw mediators is identified with a dark matter candidate, provides a motivation for considering model 2.  
The neutrino Yukawa terms in model 2 have the form:\
\begin{equation}
-\mathcal{L}_{Y}^{\left( \nu \right) }=\sum_{k=1}^{2}y_{1k}^{\left( N\right)
}\bar{l}_{L}\widetilde{\Xi }_{4}N_{kR}\frac{\eta }{\Lambda }%
+\sum_{k=1}^{2}y_{2k}^{\left( N\right) }\bar{l}_{L}\widetilde{\Xi }_{4}N_{kR}%
\frac{\rho }{\Lambda ^{2}}+\sum_{k=1}^{2}y_{3k}^{\left( N\right) }\bar{l}_{L}%
\widetilde{\Xi }_{4}N_{kR}\frac{\chi }{\Lambda }+m_{N_{1}}N_{1R}\overline{%
N_{1R}^{C}}+m_{N_{2}}N_{2R}\overline{N_{2R}^{C}}+h.c.
\end{equation}

\section{Lepton masses and mixings}\label{lepton}
\lhead[\thepage]{\thesection. Lepton masses and mixings}
\subsection{ Neutrino sector}

The neutrino Yukawa interactions of model 1 are:
\begin{eqnarray}
-\mathcal{L}_{Y\left( 1\right) }^{\left( \nu \right) }
&=&\sum_{k=1}^{2}\left( Y_{1k}^{\left( N\right) }+Y_{2k}^{\left( N\right)
}\right) \overline{\nu }_{1L}\varphi N_{kR}+\sum_{k=1}^{2}\left(
Y_{1k}^{\left( N\right) }-Y_{2k}^{\left( N\right) }\right) \overline{\nu }%
_{2L}\varphi N_{kR}  \notag \\
&&+\sum_{k=1}^{2}Y_{3k}^{\left( N\right) }\overline{\nu }_{3L}\varphi
N_{kR}+m_{N_{1}}N_{1R}\overline{N_{1R}^{C}}+m_{N_{2}}N_{2R}\overline{%
N_{2R}^{C}}+h.c  \label{ec:lagangian-neutrino}
\end{eqnarray}

where:
\begin{equation*}
Y_{1k}^{\left( N\right) }=\frac{y_{1k}^{(N)}v_{\eta }v_{\Xi _{3}}}{2\Lambda
^{2}},\hspace{0.7cm}\hspace{0.7cm}Y_{2k}^{(N)}=\frac{y_{2k}^{(N)}v_{\rho
}v_{\Xi _{3}}}{2\Lambda ^{2}},\hspace{0.7cm}\hspace{0.7cm}Y_{3k}^{(N)}=\frac{%
y_{3k}^{(N)}v_{\chi }v_{\Xi _{3}}}{\sqrt{2}\Lambda ^{2}},\hspace{0.7cm}%
\hspace{0.7cm}k=1,2
\end{equation*}

Then, due the preserved $Z_{2}$ symmetry the light active neutrino masses
are generated from at one loop level radiative seesaw mechanism. The light
active neutrino mass matrix in model 1 has the form:
\begin{eqnarray}
\mathbf{m}_{\nu } &\simeq &\sum_{k=1}^{2} f_k%
\begin{pmatrix}
\left( Y_{1k}^{\left( N\right) }+Y_{2k}^{\left( N\right) }\right) ^{2} & 
\left( Y_{1k}^{\left( N\right) }+Y_{2k}^{\left( N\right) }\right) \left(
Y_{1k}^{\left( N\right) }-Y_{2k}^{\left( N\right) }\right) & Y_{3k}^{\left(
N\right) }\left( Y_{1k}^{\left( N\right) }+Y_{2k}^{\left( N\right) }\right),
\notag\\ 
\left( Y_{1k}^{\left( N\right) }+Y_{2k}^{\left( N\right) }\right) \left(
Y_{1k}^{\left( N\right) }-Y_{2k}^{\left( N\right) }\right) & \left(
Y_{1k}^{\left( N\right) }-Y_{2k}^{\left( N\right) }\right) ^{2} & 
Y_{3k}^{\left( N\right) }\left( Y_{1k}^{\left( N\right) }-Y_{2k}^{\left(
N\right) }\right) \\ 
Y_{3k}^{\left( N\right) }\left( Y_{1k}^{\left( N\right) }+Y_{2k}^{\left(
N\right) }\right) & Y_{3k}^{\left( N\right) }\left( Y_{1k}^{\left( N\right)
}-Y_{2k}^{\left( N\right) }\right) & \left( Y_{3k}^{\left( N\right) }\right)
^{2}%
\end{pmatrix}%
 \\
&=&\left( 
\begin{array}{ccc}
W^{2} & WX\cos \varphi & WY\cos \left( \varphi -\varrho \right) \\ 
WX\cos \varphi & X^{2} & XY\cos \varrho \\ 
WY\cos \left( \varphi -\varrho \right) & XY\cos \varrho & Y^{2}%
\end{array}%
\right)\label{ec:model1-neutrino}
\end{eqnarray}%
where the loop functions $f_{k}$ ($k=1,2$) are given by: 
\begin{equation}
f_{k}=\frac{m_{N_{k}}}{16\pi ^{2}}\left[ \frac{m_{\varphi _{R}}^{2}}{%
m_{\varphi _{R}}^{2}-m_{N_{k}}^{2}}\ln \left( \frac{m_{\varphi _{R}}^{2}}{%
m_{N_{k}}^{2}}\right) -\frac{m_{\varphi _{I}}^{2}}{m_{\varphi
_{I}}^{2}-m_{N_{k}}^{2}}\ln \left( \frac{m_{\varphi _{I}}^{2}}{m_{N_{k}}^{2}}%
\right) \right] ,\hspace{0.7cm}\hspace{0.7cm}k=1,2,
\end{equation}
\noindent
and the effective parameters $W$, $X$ and $Y$, $\varphi $ and $\varrho $
fulfill the following relations:
\begin{align}
W =\left\vert \overrightarrow{W}\right\vert &=\allowbreak \sqrt{%
\sum_{k=1}^{2}\left( Y_{1k}^{\left( N\right) }+Y_{2k}^{\left( N\right)
}\right) ^{2}f_{k}}\;,  & \overrightarrow{W} &=\left( \left( Y_{11}^{\left( N\right) }+Y_{21}^{\left(
N\right) }\right) \sqrt{f_{1}},\left( Y_{12}^{\left( N\right)
}+Y_{22}^{\left( N\right) }\right) \sqrt{f_{2}}\right), \quad \cos \varphi =\frac{\overrightarrow{W}\cdot \overrightarrow{X}}{\left\vert \overrightarrow{%
W}\right\vert \left\vert \overrightarrow{X}\right\vert },\notag \\[5pt]
X=\left\vert \overrightarrow{X}\right\vert&=\sqrt{\sum_{k=1}^{2}\left( Y_{1k}^{\left( N\right) }-Y_{2k}^{\left(
N\right) }\right) ^{2}f_{k}}\;, & \overrightarrow{X}&=\left( \left( Y_{11}^{\left( N\right) }-Y_{21}^{\left(
N\right) }\right) \sqrt{f_{1}},\left( Y_{12}^{\left( N\right)
}-Y_{22}^{\left( N\right) }\right) \sqrt{f_{2}}\right), \quad \cos\left( \varphi -\varrho \right) =\frac{\overrightarrow{W}\cdot 
\overrightarrow{Y}}{\left\vert \overrightarrow{W}\right\vert \left\vert 
\overrightarrow{Y}\right\vert },\notag\\[5pt]
Y=\left\vert \overrightarrow{Y}\right\vert &=\sqrt{\sum_{k=1}^{2}\left( Y_{3k}^{\left( N\right) }\right)
^{2}f_{k}}\;, & \overrightarrow{Y} &=\left( Y_{31}^{\left( N\right) }\sqrt{f_{1}}%
,Y_{32}^{\left( N\right) }\sqrt{f_{2}}\right), \qquad \cos \varrho =\frac{%
\overrightarrow{X}\cdot \overrightarrow{Y}}{\left\vert \overrightarrow{X}%
\right\vert \left\vert \overrightarrow{Y}\right\vert } ~.
\end{align}
On the other hand, the neutrino Yukawa terms of model 2 can be rewritten as
follows: 
\begin{eqnarray}
-\mathcal{L}_{Y}^{\left( \nu \right) } &=&\sum_{k=1}^{2}z_{1k}^{\left(
N\right) }\left( \bar{l}_{1L}+\bar{l}_{2L}\right) \widetilde{\Xi }%
_{4}N_{kR}+\sum_{k=1}^{2}z_{2k}^{\left( N\right) }\left( \bar{l}_{1L}-\bar{l}%
_{2L}\right) \widetilde{\Xi}_{4}N_{kR}+\sum_{k=1}^{2}z_{3k}^{\left( N\right) }%
\bar{l}_{3L}\widetilde{\Xi }_{4}N_{kR}\notag \\
&&+m_{N_{1}}N_{1R}\overline{N_{1R}^{C}}+m_{N_{2}}N_{2R}\overline{N_{2R}^{C}}%
+h.c.
\end{eqnarray}%
where the effective neutrino couplings have the form: 
\begin{equation}
z_{1k}^{\left( N\right) }=\frac{y_{1k}^{\left( N\right) }v_{\eta }}{\sqrt{2}%
\Lambda },\hspace{0.7cm}\hspace{0.7cm}z_{2k}^{\left( N\right) }=\frac{%
y_{2k}^{\left( N\right) }v_{\rho }}{\sqrt{2}\Lambda },\hspace{0.7cm}\hspace{%
0.7cm}z_{3k}^{\left( N\right) }=\frac{y_{3k}^{\left( N\right) }v_{\chi }}{%
\Lambda },\hspace{0.7cm}\hspace{0.7cm}k=1,2.
\end{equation}

Due to the preserved $Z_{2}$ symmetry, the mass matrix for the light active
neutrinos is radiatively generated and is given by
\begin{eqnarray}
\mathbf{m}_{\nu } &=&\sum_{k=1}^{2} \widetilde{f}_{k}%
\begin{pmatrix}
\left( z_{1k}^{\left( N\right) }+z_{2k}^{\left( N\right) }\right) ^{2} & 
\left( z_{1k}^{\left( N\right) }+z_{2k}^{\left( N\right) }\right) \left(
z_{1k}^{\left( N\right) }-z_{2k}^{\left( N\right) }\right) & z_{3k}^{\left(
N\right) }\left( z_{1k}^{\left( N\right) }+z_{2k}^{\left( N\right) }\right),
\notag\\ 
\left( z_{1k}^{\left( N\right) }+z_{2k}^{\left( N\right) }\right) \left(
z_{1k}^{\left( N\right) }-z_{2k}^{\left( N\right) }\right) & \left(
z_{1k}^{\left( N\right) }-z_{2k}^{\left( N\right) }\right) ^{2} & 
z_{3k}^{\left( N\right) }\left( z_{1k}^{\left( N\right) }-z_{2k}^{\left(
N\right) }\right) \\ 
z_{3k}^{\left( N\right) }\left( z_{1k}^{\left( N\right) }+z_{2k}^{\left(
N\right) }\right) & z_{3k}^{\left( N\right) }\left( z_{1k}^{\left( N\right)
}-z_{2k}^{\left( N\right) }\right) & \left( z_{3k}^{\left( N\right) }\right)
^{2}%
\end{pmatrix}%
. \\
&=&\left( 
\begin{array}{ccc}
\widetilde{W}^{2} & \widetilde{W}\widetilde{X}\cos \varphi & \widetilde{W}%
\widetilde{Y}\cos \left( \varphi -\varrho \right) \\ 
\widetilde{W}\widetilde{X}\cos \varphi & \widetilde{X}^{2} & \widetilde{X}%
\widetilde{Y}\cos \varrho \\ 
\widetilde{W}\widetilde{Y}\cos \left( \varphi -\varrho \right) & X\widetilde{%
Y}\cos \varrho & \widetilde{Y}^{2}%
\end{array}%
\right)\label{ec:model2-neutrino}
\end{eqnarray}
\noindent
where the loop functions $\widetilde{f}_{k}$ ($k=1,2$) take the form:
\begin{equation}
\widetilde{f}_{k}=\frac{m_{N_{k}}}{16\pi ^{2}}\left[ \frac{m_{H_{4R}^{0}}^{2}%
}{m_{H_{4R}^{0}}^{2}-m_{N_{k}}^{2}}\ln \left( \frac{m_{H_{4R}^{0}}^{2}}{%
m_{N_{k}}^{2}}\right) -\frac{m_{H_{4I}^{0}}^{2}}{%
m_{H_{4I}^{0}}^{2}-m_{N_{k}}^{2}}\ln \left( \frac{m_{H_{4I}^{0}}^{2}}{%
m_{N_{k}}^{2}}\right) \right] ,\hspace{0.7cm}\hspace{0.7cm}k=1,2,
\end{equation}
\noindent
and the effective parameters $\widetilde{W}$, $\widetilde{X}$ and $%
\widetilde{Y}$, $\varphi $ and $\varrho $ fulfill the following relations:
\begin{align}
\widetilde{W} =\left\vert \overrightarrow{\widetilde{W}}\right\vert &=\allowbreak \sqrt{\sum_{k=1}^{2}\left( z_{1k}^{\left( N\right)
}+z_{2k}^{\left( N\right) }\right) ^{2}\widetilde{f}_{k}}, & \overrightarrow{\widetilde{W}} &=\left( \left( Y_{11}^{\left( N\right)
}+Y_{21}^{\left( N\right) }\right) \sqrt{\widetilde{f}_{1}},\left(
Y_{12}^{\left( N\right) }+Y_{22}^{\left( N\right) }\right) \sqrt{\widetilde{f}_{2}}\right),\quad \cos \varphi =\frac{\overrightarrow{\widetilde{W}}\cdot 
\overrightarrow{\widetilde{X}}}{\left\vert \overrightarrow{\widetilde{W}}%
\right\vert \left\vert \overrightarrow{\widetilde{X}}\right\vert },\notag \\[5pt]
\widetilde{X}=\left\vert \overrightarrow{\widetilde{X}}\right\vert &=\sqrt{%
\sum_{k=1}^{2}\left( z_{1k}^{\left( N\right) }-z_{2k}^{\left( N\right)
}\right) ^{2}\widetilde{f}_{k}}, & \overrightarrow{X}&=\left( \left( Y_{11}^{\left(
N\right) }-Y_{21}^{\left( N\right) }\right) \sqrt{\widetilde{f}_{1}},\left(
Y_{12}^{\left( N\right) }-Y_{22}^{\left( N\right) }\right) \sqrt{\widetilde{f%
}_{2}}\right) ,\quad \cos \left( \varphi -\varrho \right) =\frac{\overrightarrow{\widetilde{%
W}}\cdot \overrightarrow{\widetilde{Y}}}{\left\vert \overrightarrow{%
\widetilde{W}}\right\vert \left\vert \overrightarrow{\widetilde{Y}}%
\right\vert },\notag \\[5pt]
\widetilde{Y}=\left\vert \overrightarrow{\widetilde{Y}}\right\vert &=\sqrt{\sum_{k=1}^{2}\left(
Y_{3k}^{\left( N\right) }\right) ^{2}\widetilde{f}_{k}}, & \overrightarrow{\widetilde{Y}} &=\left( Y_{31}^{\left( N\right) }\sqrt{%
\widetilde{f}_{1}},Y_{32}^{\left( N\right) }\sqrt{\widetilde{f}_{2}}\right) , \qquad \cos \varrho =\frac{\overrightarrow{\widetilde{X}%
}\cdot \overrightarrow{\widetilde{Y}}}{\left\vert \overrightarrow{\widetilde{%
X}}\right\vert \left\vert \overrightarrow{\widetilde{Y}}\right\vert }.
\end{align}
\begin{table}[tp]
\begin{tabular}{c|c|cccccc}
\toprule[0.13em] Observable & range & $\Delta m_{21}^{2}$ [$10^{-5}$eV$^{2}$]
& $\Delta m_{31}^{2}$ [$10^{-3}$eV$^{2}$] & $\sin\theta^{(l)}_{12}/10^{-1}$
& $\sin\theta^{(l)}_{13}/10^{-3}$ & $\sin\theta^{(l)}_{23}/10^{-1}$ & $%
\delta^{(l)}_{CP} (^{\circ })$ \\ \hline
Experimental & $1\sigma$ & $7.50_{-0.20}^{+0.22}$ & $2.55_{-0.03}^{+0.02}$ & 
$3.18\pm 0.16$ & $2.200_{-0.062}^{+0.069}$ & $5.74\pm 0.14$ & $%
194_{-22}^{+24}$ \\ 
Value & $3\sigma$ & $6.94-8.14$ & $2.47-2.63 $ & $2.71-3.69$ & $2.000-2.405$
& $4.34-6.10$ & $128-359$ \\ \hline
Fit & $1\sigma-2\sigma$ & $7.53$ & $2.55$ & $3.21$ & $2.19$ & $5.75$ & $180$
\\ 
\bottomrule[0.13em] &  &  &  &  &  &  & 
\end{tabular}%
%
%
\caption{Model predictions for the scenario of normal order (NO) neutrino
mass. The experimental values are taken from Ref. \protect\cite{deSalas:2020pgw}}
\label{table:neutrinos_valueS4}
\end{table}

\begin{figure}[tbp]
\centering
\includegraphics[scale=0.41]{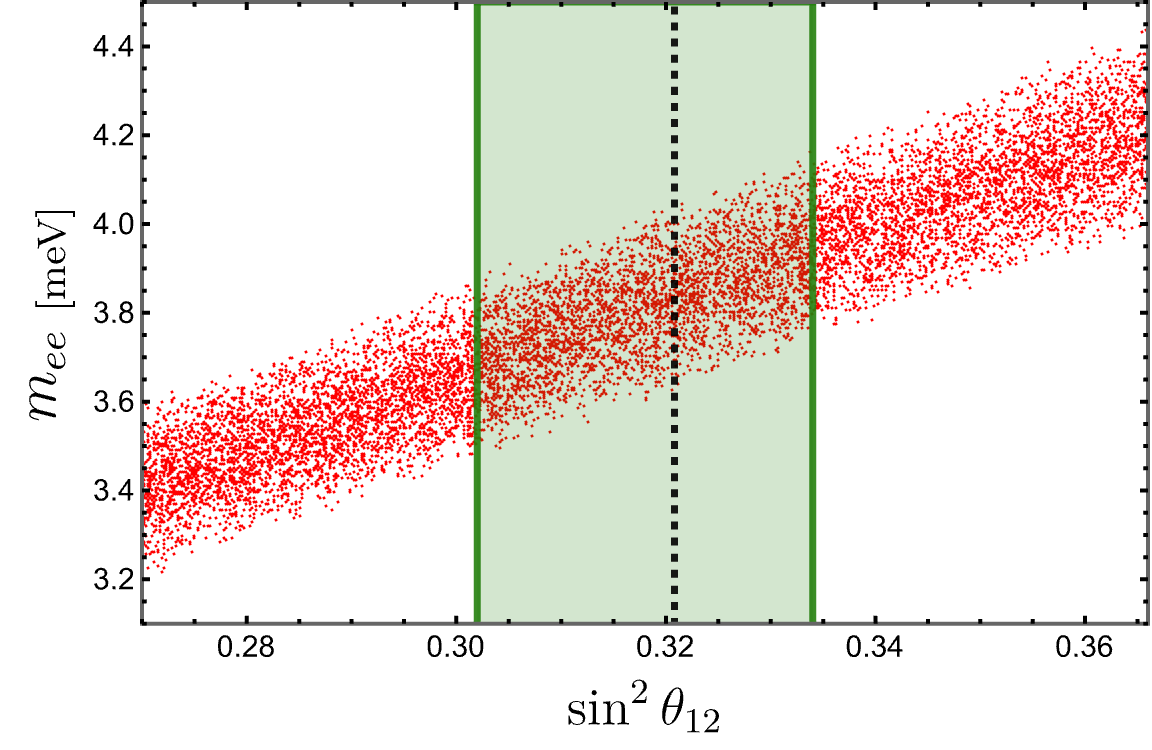}  
\caption{Correlation between the solar mixing parameter $\sin^2\theta_{12}$ and the effective Majorana neutrino mass parameter $m_{ee}$ (red). The green bands represent the range $1\sigma$ in the experimental values and the dotted line (black) represents the best
experimental value.}
\label{fig:neutrinocorrelation}
\end{figure}

Models 1 and 2 yield the same light active neutrino mass matrix as shown by Eqs. \eqref{ec:model1-neutrino} and \eqref{ec:model2-neutrino}, where it follows that our model yields a neutrino mass matrix texture different than the one corresponding to the cobimaximal pattern. The obtained neutrino mass matrix successfully reproduces the neutrino mass squared differences i.e, $\Delta m_{21}^2$ and $\Delta m_{31}^2$, the mixing angles $\sin^2\theta_{12}$, $\sin^2\theta_{23}$, $\sin^2\theta_{13}$ and the leptonic CP violating phase, whose obtained values are consistent with the neutrino oscillation experimental data, as indicated in Table \ref{table:neutrinos_valueS4}. We successfully reproduced the experimental values for these observables through a fit of the free parameters of our model, finding the \textquotedblleft best-fit point\textquotedblright\ by minimizing the following $\chi^2$ function:
\begin{equation}
\chi_{\nu}^{2}=\frac{\left( m_{21}^{\exp }-m_{21}^{th}\right) ^{2}}{\sigma
_{m_{21}}^{2}}+\frac{\left( m_{31}^{\exp }-m_{31}^{th}\right) ^{2}}{\sigma
_{m_{31}}^{2}}+\frac{\left( s_{\theta _{12}}^{\exp }-s_{\theta
_{12}}^{th}\right) ^{2}}{\sigma _{s_{12}}^{2}}+\frac{\left( s_{\theta
_{23}}^{\exp }-s_{\theta _{23}}^{th}\right) ^{2}}{\sigma _{s_{23}}^{2}}+%
\frac{\left( s_{\theta _{13}}^{\exp }-s_{\theta _{13}}^{th}\right) ^{2}}{%
\sigma _{s_{13}}^{2}}+\frac{\left( \delta _{CP}^{\exp }-\delta
_{CP}^{th}\right) ^{2}}{\sigma _{\delta }^{2}}\;,  \label{ec:funtion_errorS4}
\end{equation}
where $m_{i1}$ is the difference of the square of the neutrino masses (with $i=2,3$), $s_{\theta _{jk}}$ is the sine function of the mixing angles (with $ j,k=1,2,3$) and $\delta _{CP}$ is the CP violation phase. The superscripts represent the experimental (\textquotedblleft exp\textquotedblright) and theoretical (\textquotedblleft th\textquotedblright) values and the $\sigma $ are the experimental errors. Therefore, the minimization of $\chi_{\nu}^2$ gives us the following value,
\begin{equation}
\chi_{\nu}^{2}=0.0531\label{eq:chi-nu}.
\end{equation}

Where the numerical values of our effective parameters of the mass matrices of Eqs. \eqref{ec:model1-neutrino} and \eqref{ec:model2-neutrino} that minimize Eq. \eqref{eq:chi-nu} are:
\begin{eqnarray}
W=\widetilde{W}&=&0.0616\; \text{eV} \quad;\quad X=\widetilde{X}=0.174\; \text{eV} \quad;\quad Y=\widetilde{Y}=0.158\; \text{eV} \notag\\[5pt]
\varphi -\varrho &=&1.73\; \text{rad} \quad ; \quad \varrho =0.670\; \text{rad} ~,
\end{eqnarray}

Since the structure of the matrices \eqref{ec:model1-neutrino} and \eqref{ec:model2-neutrino} are the same and we are working with the effective values and not with the input parameters, we obtain the same values for both parameters as well as for the $\chi_{\nu}^2$ function. Let us note that $W$, $X$ and $Y$ are effective parameters in the neutrino sector for model 1, whereas  $\widetilde{W}$, $\widetilde{X}$ and $\widetilde{Y}$ correspond to model 2. The analytical dependence of the effective neutrino sector parameters on the input parameters is different in both models since the dark scalar sector of models 1 and 2 has an inert scalar singlet and an inert scalar doublet, respectively. Notice that there are five effective free parameters in the neutrino sector of both models that allow to successfully accommodate the experimental values of the physical observables of the neutrino sector: the two neutrino mass squared splittings, the three leptonic mixing angles and the leptonic Dirac CP violating phase.

Furthermore, in our model, another observable can be obtained. This is the effective Majorana neutrino mass parameter of the neutrinoless double beta decay, which gives information on the Majorana nature of the neutrinos. This mass parameter has the form:
\begin{equation}
m_{ee}=\left| \sum_i \mathbf{U}_{ei}^2m_{\nu i}\right|\;,
\label{ec:mee}
\end{equation}

where $\mathbf{U}_{ei}$ and $m_{\nu i}$ are the matrix elements of the PMNS leptonic mixing matrix and the light active neutrino masses, respectively. The neutrinoless double beta ($0\nu\beta\beta$) decay amplitude is proportional to $m_{ee}$. Fig. \ref{fig:neutrinocorrelation} shows the correlation between the effective Majorana neutrino mass parameter $m_{ee}$ and the solar mixing parameter $\sin\theta_{12}$, where the neutrino sector model parameters were randomly generated in a range of values where the neutrino mass squared splittings and the mixing parameters are inside the $3\sigma$ experimentally allowed range. As
seen from Fig.~\ref{fig:neutrinocorrelation}, our models predict a solar mixing parameter $\sin\theta_{12}$ in the range $0.27\lesssim \sin^2\theta_{12} \lesssim 0.37$ and an effective Majorana neutrino mass parameter in the range $3.2\; meV\lesssim m_{ee}\lesssim 4.4\; meV$ for the scenario of normal neutrino mass hierarchy. The current most stringent experimental upper bound on the effective Majorana neutrino mass parameter, i.e., $m_{ee}\leq 50\; meV$ arises from the KamLAND-Zen limit on the $^{136}X_e\; 0\nu\beta\beta$ decay half-life $T_{1/2}^{0\nu\beta\beta}(^{136}X_e) >2.0\times 10^{26}$ yr~\cite{KamLAND-Zen:2022tow}.

\section{Quark masses and mixings.}\label{quarks}
\lhead[\thepage]{\thesection. Quark masses and mixings}

From the Yukawa interactions of Eq.(\ref{yuk1}), the quark  mass term is given by
\begin{equation}
-\mathcal{L}_{q}=\bar{q}_{iL}\left( \mathbf{M}_{q}\right) _{ij}q_{jR}+h.c.
\label{eq7}
\end{equation}
where the quark mass matrix is explicitly written as
\begin{equation}
\mathbf{M}_{q}= 
\begin{pmatrix}
a_{q}+b_{q}^{\prime } & b_{q} & c_{q} \\ 
b_{q} & a_{q}-b_{q}^{\prime } & c_{q}^{\prime } \\ 
f_{q} & f_{q}^{\prime } & g_{q}%
\end{pmatrix} ~,%
\end{equation}

with the $q=u,d$. The quark matrix elements are given by 
\begin{eqnarray}
a_{q} =y_{2}^{q}v_{3},\quad b_{q}^{\prime }=y_{1}^{q}v_{2},\quad
b_{q}=y_{1}^{q}v_{1},\quad c_{q}=y_{3}^{q}v_{1},\quad c_{q}^{\prime
}=y_{3}^{q}v_{2},\quad f_{q}=y_{4}^{q}v_{1},\quad f_{q}^{\prime }
=y_{4}^{q}v_{2},\quad g_{q}=y_{5}^{q}v_{3} .
\end{eqnarray}
These free parameters are reduced substantially by imposing an alignment of the vacuum expectation values, in particular, $v_{1}=v_{2}$ which is a solution of the scalar potential with three Higgs doublets with assignment $\left(\Xi_{1},\Xi_{2}\right)\sim \textbf{2}$ and $\Xi_{3}\sim \mathbf{1}_1$~\cite{Kubo:2003iw,Beltran:2009zz}. In consequence, $c_{q}=c^{\prime}_{q}$ and $f_{q}=f^{\prime}_{q}$. On the other hand, these mass matrices are complex and  can be diagonalized by
the unitary matrices ${\mathbf{U}_{u(L,R)}}$ and ${\mathbf{U}_{d(L,R)}}$
such that 
\begin{equation}
\hat{\mathbf{M}}_{d}=\text{diag.}\left( m_{d},m_{s},m_{b}\right) =\mathbf{U}%
_{dL}^{\dagger }\mathbf{M}_{d}\mathbf{U}_{dR},\hspace{1.5cm}\hat{\mathbf{M}}%
_{u}=\text{diag.}\left( m_{u},m_{c},m_{t}\right) =\mathbf{U}_{uL}^{\dagger }%
\mathbf{M}_{u}\mathbf{U}_{uR}.
\end{equation}

To simplify our analysis we consider a particular benchmark
scenario where the matrices $\mathbf{M}_{u}$ and $\mathbf{M}_{d}$ are
symmetric. 
Therefore, we have
\begin{equation}
\mathbf{M}_{q}=%
\begin{pmatrix}
a_{q}+b_{q} & b_{q} & c_{q} \\ 
b_{q} & a_{q}-b_{q} & c_{q} \\ 
c_{q} & c_{q} & g_{q}%
\end{pmatrix}%
.
\end{equation}
In addition, we make the
following rotations $\mathbf{U}_{q}=\mathbf{U}_{\pi/4}\mathbf{u}_{q(L, R)}$ so
that $\hat{ \mathbf{M}}_{q}=\text{diag.}\left(m_{q_{1}}, m_{q_{2}},
m_{q_{3}}\right)=\mathbf{u}^{\dagger}_{q L}\mathbf{m}_{q}\mathbf{u}_{q R}$.
Then, we obtain 
\begin{align}  \label{NNI2}
\mathbf{m}_{q}=\mathbf{U}^{T}_{\pi/4}\mathbf{M}_{q} \mathbf{U}_{\pi/4}= 
\begin{pmatrix}
A_{q} & b_{q} & 0 \\ 
b_{q} & B_{q} & C_{q} \\ 
0 & C_{q} & g_{q}%
\end{pmatrix}
,\,\, \mathbf{U}_{\pi/4}= 
\begin{pmatrix}
\frac{1}{\sqrt{2}} & \frac{1}{\sqrt{2}} & 0 \\ 
-\frac{1}{\sqrt{2}} & \frac{1}{\sqrt{2}} & 0 \\ 
0 & 0 & 1%
\end{pmatrix}%
,
\end{align}
where $A_{q}=a_{q}-b_{q}$, $B_{q}=a_{q}+b_{q}$ and $C_{q}=\sqrt{2}c_{q}$. As
one can notice, the phases can be factorized as $\mathbf{m}_{q}=\mathbf{P}_{q} \mathbf{\bar{%
m}}_{q}\mathbf{P}_{q}$
with 
\begin{equation}
\mathbf{\bar{m}}_{q}= 
\begin{pmatrix}
\vert A_{q}\vert  & \vert b_{q}\vert  & 0 \\ 
\vert b_{q}\vert  & \vert B_{q}\vert  & \vert C_{q}\vert \\ 
0 & \vert C_{q}\vert & \vert g_{q}\vert %
\end{pmatrix}%
, \qquad \mathbf{P}_{q}=%
\begin{pmatrix}
e^{i\eta_{q_{1}}} & 0 & 0 \\ 
0 & e^{i\eta_{q_{2}}} & 0 \\ 
0 & 0 & e^{i\eta_{q_{3}}}%
\end{pmatrix}%
,
\end{equation}
where
\begin{equation}
    \eta_{q_{1}}=\frac{\textrm{arg}.(A_{q})}{2},\quad  \eta_{q_{2}}=\frac{\textrm{arg}.(B_{q})}{2},\quad \eta_{q_{3}}=\frac{\textrm{arg}.(g_{q})}{2},\quad \eta_{q_{1}}+\eta_{q_{2}}=\textrm{arg}. (b_{q}),\quad \eta_{q_{2}}+\eta_{q_{3}}=\textrm{arg}. (C_{q})
\end{equation}

Then, $\mathbf{u}_{q L }=\mathbf{P}_{q}%
\mathbf{O}_{q}$ and $\mathbf{u}_{q R }=\mathbf{P}^{\dagger}_{q}%
\mathbf{O}_{q}$, here $\mathbf{O}_{q}$ is an orthogonal matrix that
diagonalizes the real symmetric mass matrix, $\mathbf{\bar{m}}_{q}$. Thus,
we get $\hat{ \mathbf{M}}_{q}=\mathbf{O}^{T}_{q}\mathbf{\bar{m}}_{q}\mathbf{O%
}_{q}$. The real orthogonal matrix is given by \footnote{%
See Appendix \ref{quarks-app} for more details about the diagonalization process of the quark mass matrices.} 
\begin{equation}
\mathbf{O}_{q}=%
\begin{pmatrix}
\sqrt{\frac{\left(\vert g_{q}\vert -m_{q_{1}}\right)\left(m_{q_{2}}-\vert A_{q}\vert\right)
\left(m_{q_{3}}-\vert A_{q}\vert \right)}{\mathcal{M}_{q_{1}}}} & \sqrt{\frac{
\left(\vert g_{q}\vert -m_{q_{2}}\right)\left(m_{q_{3}}-\vert A_{q}\vert \right)\left(
\vert A_{q}\vert -m_{q_{1}}\right)}{\mathcal{M}_{q_{2}}}} & \sqrt{\frac{
\left(m_{q_{3}}-\vert g_{q}\vert\right)\left(m_{q_{2}}-\vert A_{q}\vert\right)\left(
\vert A_{q}\vert -m_{q_{1}}\right)}{\mathcal{M}_{q_{3}}}} \\ 
-\sqrt{\frac{\left(\vert g_{q}\vert-\vert A_{q}\vert \right)\left(\vert g_{q}\vert-m_{q_{1}}\right)
\left(\vert A_{q}\vert-m_{q_{1}}\right)}{\mathcal{M}_{q_{1}}}} & \sqrt{\frac{
\left(\vert g_{q}\vert-\vert A_{q}\vert \right)\left(\vert g_{q}\vert-m_{q_{2}}\right)\left(m_{q_{2}}-\vert A_{q}\vert 
\right)}{\mathcal{M}_{q_{2}}}} & \sqrt{\frac{\left(\vert g_{q}\vert-\vert A_{q}\vert\right)
\left(m_{q_{3}}-\vert g_{q}\vert \right)\left(m_{q_{3}}-\vert A_{q}\vert \right)}{\mathcal{M}%
_{q_{3}} }} \\ 
\sqrt{\frac{\left(\vert g_{q}\vert-m_{q_{2}}\right)\left(m_{q_{3}}-\vert g_{q}\vert\right)\left(
\vert A_{q}\vert-m_{q_{1}}\right)}{\mathcal{M}_{q_{1}}}} & -\sqrt{\frac{
\left(\vert g_{q}\vert-m_{q_{1}}\right)\left(m_{q_{2}}-\vert A_{q}\vert\right)
\left(m_{q_{3}}-\vert g_{q}\vert\right)}{\mathcal{M}_{q_{2}}}} & \sqrt{\frac{
\left(\vert g_{q}\vert-m_{q_{1}}\right)\left(\vert g_{q}\vert-m_{q_{2}}\right)
\left(m_{q_{3}}-\vert A_{q}\vert \right)}{\mathcal{M}_{q_{3}}}}%
\end{pmatrix}%
\label{eq:matrixO}
\end{equation}
with 
\begin{eqnarray}
\mathcal{M}_{q_{1}}&=&\left(\vert g_{q}\vert -\vert A_{q}\vert \right)\left(m_{q_{2}}-m_{q_{1}}%
\right)\left(m_{q_{3}}-m_{q_{1}}\right)  \notag \\
\mathcal{M}_{q_{2}}&=&\left(\vert g_{q}\vert -\vert A_{q}\vert \right)\left(m_{q_{2}}-m_{q_{1}}%
\right)\left(m_{q_{3}}-m_{q_{2}}\right)  \notag \\
\mathcal{M}_{q_{3}}&=&\left(\vert g_{q}\vert -\vert A_{q}\vert \right)\left(m_{q_{3}}-m_{q_{1}}%
\right)\left(m_{q_{3}}-m_{q_{2}}\right).
\end{eqnarray}

Actually, there is a hierarchy which has to be satisfied, this is, $%
m_{q_{3}}>\vert g_{q}\vert >m_{q_{2}}>\vert A_{q}\vert >m_{q_{1}}$. 
Having obtained the relevant matrices that take place 
in the CKM matrix, we have that $\mathbf{V}_{CKM}=\mathbf{U}^{\dagger}_{u L}%
\mathbf{U}_{d L}=\mathbf{O}^{T}_{u}\mathbf{\bar{P}_{q}} \mathbf{O}_{d}$ where $%
\mathbf{\bar{P}_{q}}=\mathbf{P}^{\dagger}_{u}\mathbf{P}_{d}=\text{diag}%
.\left( e^{i\bar{\eta}_{q_{1}}}, e^{i\bar{\eta}_{q_{2}}}, e^{i\bar{\eta}_{q_{3}}}\right)$ with $\bar{\eta}_{q_{i}}%
=\eta_{d_{i}}-\eta_{u_{i}}$.

In short, we have considered a benchmark where the quark mass matrix is complex and symmetric
so that the free parameters are reduced a little bit. At the end of the
day, the CKM mixing matrix has seven free parameter namely $\vert g_{q}\vert $, $\vert A_{q}\vert$
and three CP-violating phases. Nonetheless, two effective phases are relevant in the CKM matrix as we can see in the Appendix~\ref{quarks-app}. These free parameters must be fixed by a 
statistical method.

Finally, we compare the theoretical CKM mixing matrix with the standard
parametrization one to get the following formulae 
\begin{eqnarray}
\sin{\theta_{13}}&=&\vert \mathbf{V}^{ub}_{CKM} \vert,  \notag \\
\sin{\theta_{23}}&=&\frac{\vert \mathbf{V}^{cb}_{CKM} \vert}{\sqrt{1-\vert 
\mathbf{V}^{ub}_{CKM} \vert^{2}}},  \notag \\
\sin{\theta_{12}}&=&\frac{\vert \mathbf{V}^{us}_{CKM} \vert}{\sqrt{1-\vert 
\mathbf{V}^{ub}_{CKM} \vert^{2}}}.
\end{eqnarray}

\begin{table}[th]
\begin{center}
\begin{tabular}{c|l|l}
\hline\hline
Observable & Model value & Experimental value \\ \hline
$\sin \theta _{12}$ & \quad $0.224$ & \quad $0.22452\pm 0.00044$ \\ \hline
$\sin \theta _{23}$ & \quad $0.0433$ & \quad $0.04182_{-0.00074}^{+0.00085}$ \\ \hline
$\sin \theta _{13}$ & $\quad 0.00360$ & \quad $0.00369\pm 0.00011$ \\ \hline
$J_{q}$ & $\quad 3.32\times 10^{-5}$ & $\left( 3.05_{-0.13}^{+0.15}\right) \times
10^{-5}$ \\ \hline
\end{tabular}%
\end{center}
\caption{Model and experimental values of the CKM parameters. The experimental values are taken from
Ref. \cite{Workman:2022ynf}}
\label{Tab:quarks}
\end{table}

\begin{figure}
\centering
\subfigure[]  {\ \includegraphics[scale=0.3]{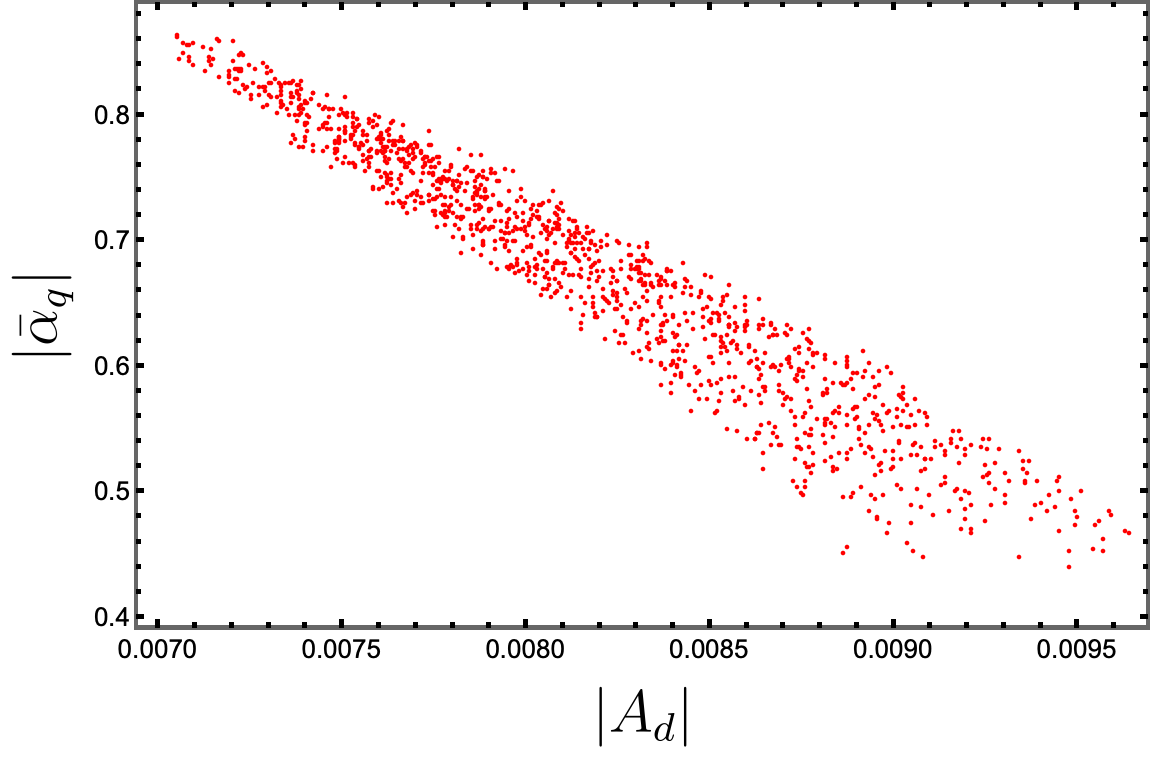} }
\subfigure[]  {\  \includegraphics[scale=0.3]{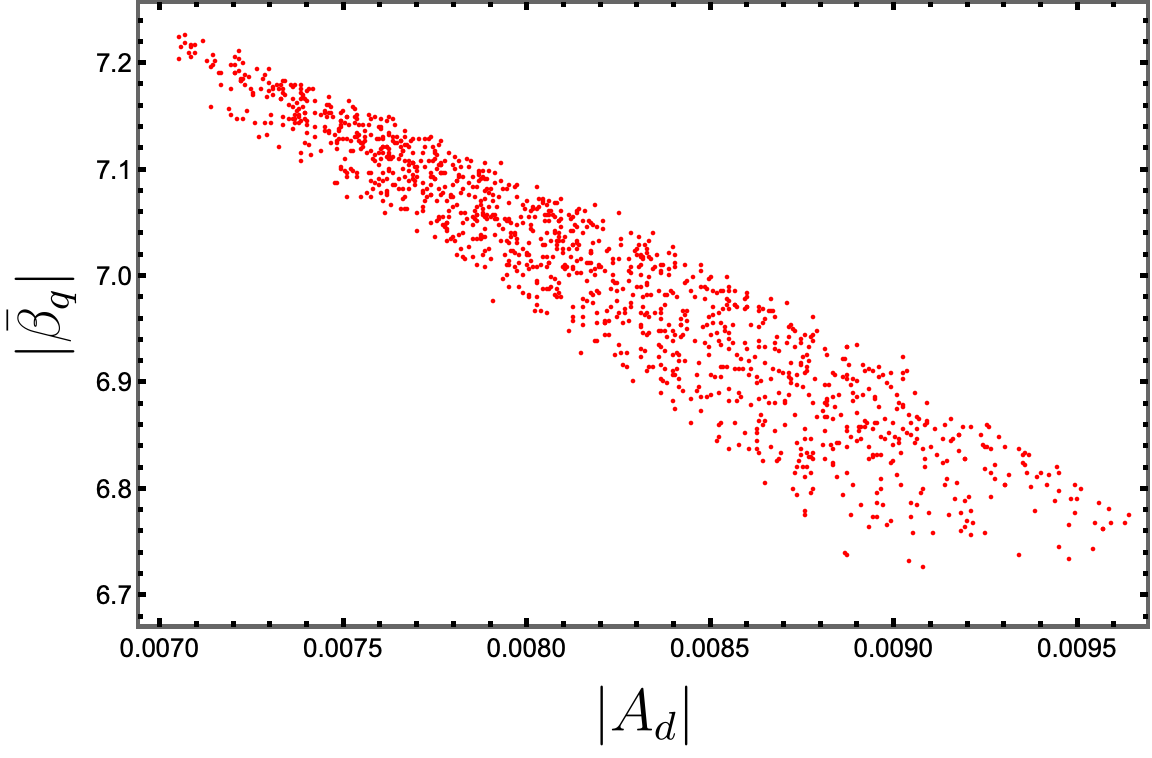}  }\\
\subfigure[]  {\  \includegraphics[scale=0.3]{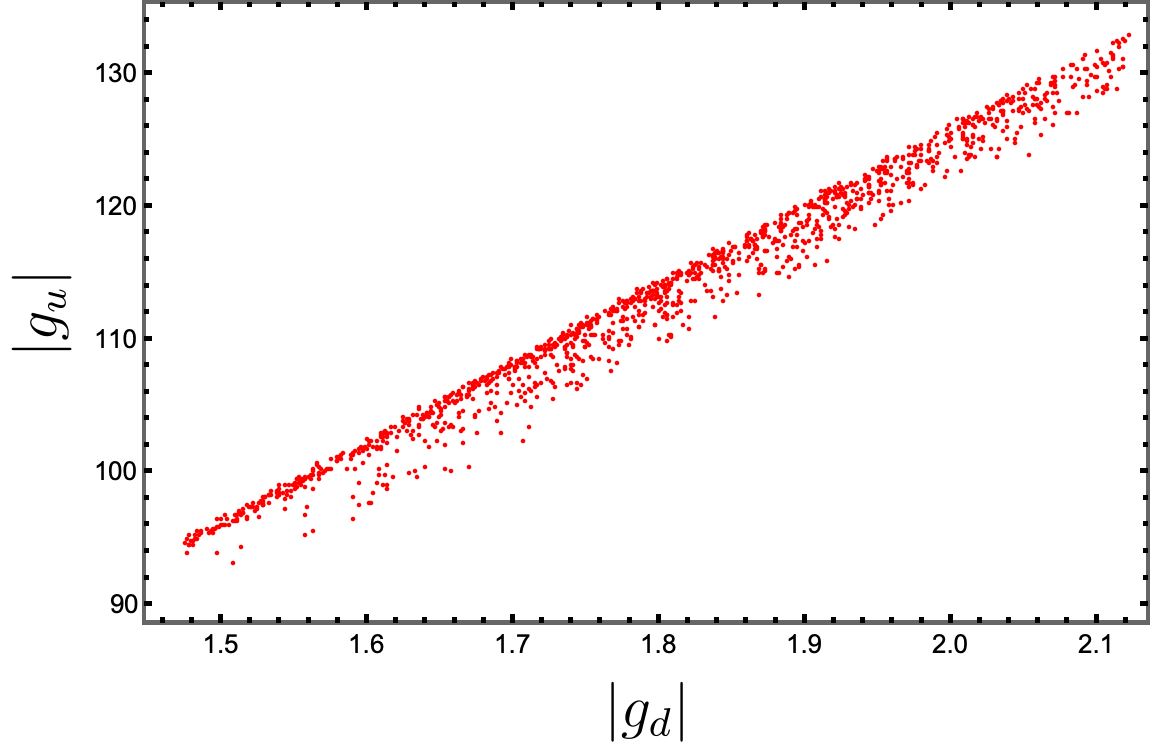}  }
\subfigure[]  {\  \includegraphics[scale=0.3]{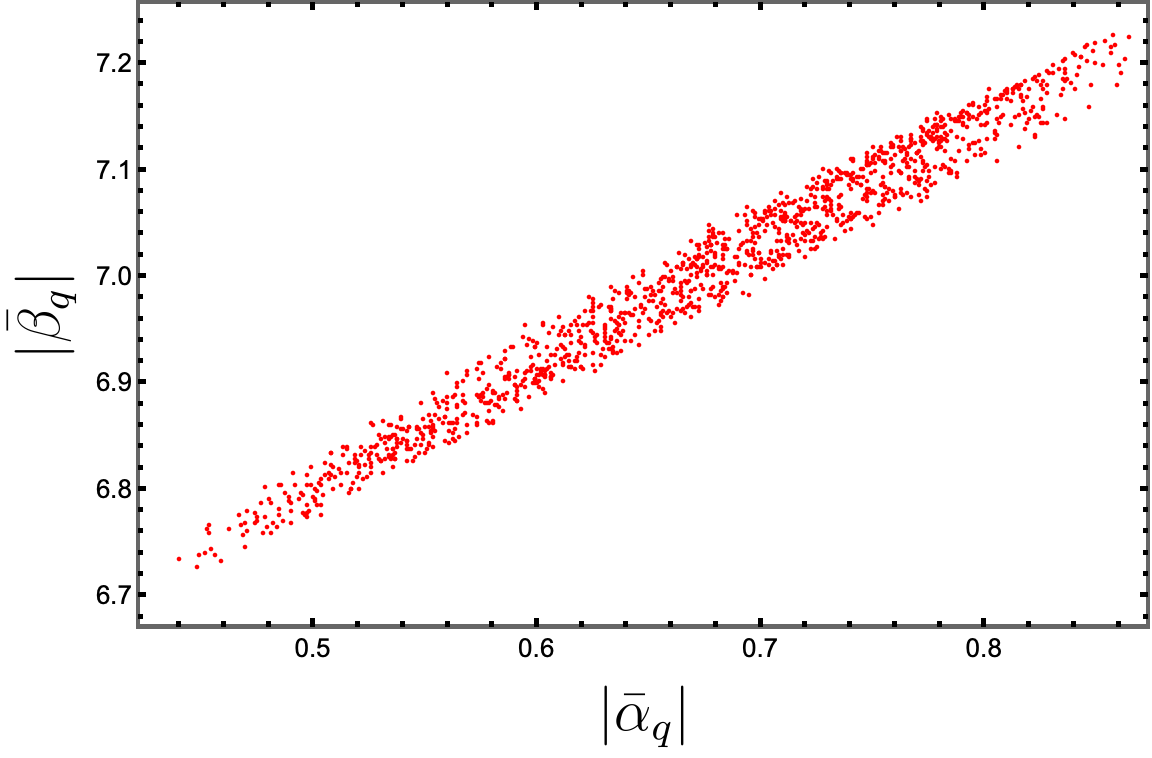}  }
\caption{Scatter plot between the effective model parameters for the quark sector.}
\label{fig:scat-phase}
\end{figure}

Therefore, to fit our quark sector parameters, we again minimize the $\chi^2$ function (defined in similar way as Eq.~\eqref{table:neutrinos_valueS4}). However, this function has now been defined with only the quark mixing angles and the Jarlskog invariant, as follows,
\begin{equation}
\chi_q^{2}=\frac{\left( s_{\theta _{12}}^{\exp }-s_{\theta
_{12}}^{th}\right) ^{2}}{\sigma _{s_{12}}^{2}}+\frac{\left( s_{\theta
_{23}}^{\exp }-s_{\theta _{23}}^{th}\right) ^{2}}{\sigma _{s_{23}}^{2}}+%
\frac{\left( s_{\theta _{13}}^{\exp }-s_{\theta _{13}}^{th}\right) ^{2}}{%
\sigma _{s_{13}}^{2}}+\frac{\left( J_q^{\exp }-J_q^{th}\right) ^{2}}{\sigma _{J_q }^{2}}\;,  \label{ec:funtion_error-q}
\end{equation}
where $s_{\theta _{jk}}$ is the sine function of the mixing angles (with $ j,k=1,2,3$) and $J_q$ is the Jarlskog invariant. The superscripts represent the experimental (\textquotedblleft exp\textquotedblright) and theoretical (\textquotedblleft th\textquotedblright) values and the $\sigma $ are the experimental errors. So, after minimizing $\chi_q^2$, we obtain the following result: 
\begin{equation}
\chi_q^{2}= 0.470,\label{eq:chi-q}
\end{equation}

while the values of the free parameters of \eqref{eq:matrixO} that yield the result of the $\chi_q^2$ of Eq. \eqref{eq:chi-q} and correspond to the best-fit point of our benchmark scenario are
\begin{eqnarray}
\vert A_u \vert &=&0.0190\; \text{GeV} \quad;\quad \vert A_d\vert =0.00832\; \text{GeV} \quad;\quad \vert g_u\vert =113.6\; \text{GeV} \notag\\[5pt]
\vert g_d\vert  &=&1.80\; \text{GeV} \quad ; \quad \vert\bar{\alpha}_{q}\vert =0.651\; \text{rad} \quad ; \quad \vert\bar{\beta}_{q}\vert =6.98\; \text{rad} ~,
\end{eqnarray}
where the best-fit values of the quark mixing angles and the Jarlskog invariant together with their corresponding experimental values (within the $1\sigma$ range) are shown in table \ref{Tab:quarks}.

Fig.~\ref{fig:scat-phase} shows a scatter plot between the effective parameters of our model for the quark sector, whose dependence can be seen in Appendix~\ref{quarks-app}. For all parameter values shown in Fig.~\ref{fig:scat-phase}, the mixing angles in the quark sector can be reproduced within the experimental range, where we obtain the following ranges of values for each observable: $2.23\times 10^{-1}\lesssim \sin\theta_{12}\lesssim 2.26\times10^{-1}$, $3.99\times 10^{-2}\lesssim \sin\theta_{23}\lesssim 4.44\times10^{-2}$, $3.30\times 10^{-3}\lesssim \sin\theta_{13}\lesssim 4.01\times10^{-3}$ and $2.73\times 10^{-5}\lesssim J\lesssim 3.63\times10^{-1}$.

\section{Meson mixings}\label{KKbar}
\lhead[\thepage]{\thesection. Meson mixings}

In this section we discuss the implications of our model in
the Flavour Changing Neutral Current (FCNC) interactions in the down type
quark sector. These FCNC down type quark Yukawa interactions produce $K^{0}-%
\bar{K}^{0}$, $B_{d}^{0}-\bar{B}_{d}^{0}$ and $B_{s}^{0}-\bar{B}_{s}^{0}$
meson oscillations, whose corresponding effective Hamiltonians are: 
\begin{equation}
\mathcal{H}_{eff}^{\left( K\right) }\mathcal{=}\sum_{j=1}^{3}\kappa
_{j}^{\left( K\right) }\left( \mu \right) \mathcal{O}_{j}^{\left( K\right)
}\left( \mu \right) ,
\end{equation}%
\begin{equation}
\mathcal{H}_{eff}^{\left( B_{d}\right) }\mathcal{=}\sum_{j=1}^{3}\kappa
_{j}^{\left( B_{d}\right) }\left( \mu \right) \mathcal{O}_{j}^{\left(
B_{d}\right) }\left( \mu \right) ,
\end{equation}%
\begin{equation}
\mathcal{H}_{eff}^{\left( B_{s}\right) }\mathcal{=}\sum_{j=1}^{3}\kappa
_{j}^{\left( B_{s}\right) }\left( \mu \right) \mathcal{O}_{j}^{\left(
B_{s}\right) }\left( \mu \right) ,
\end{equation}
In our analysis of meson oscillations we follow the approach of \cite{Dedes:2002er,Aranda:2012bv}. The $K^{0}-\bar{K}^{0}$, $B_{d}^{0}-\bar{B}_{d}^{0}$ and $B_{s}^{0}-\bar{B}_{s}^{0}$
meson oscillations in our model are induced by the tree level exchange of neutral CP even and CP odd scalars, then yielding the operators:
\begin{eqnarray}
\mathcal{O}_{1}^{\left( K\right) } &=&\left( \overline{s}_{R}d_{L}\right)
\left( \overline{s}_{R}d_{L}\right) ,\hspace{0.7cm}\hspace{0.7cm}\mathcal{O}%
_{2}^{\left( K\right) }=\left( \overline{s}_{L}d_{R}\right) \left( \overline{%
s}_{L}d_{R}\right) ,\hspace{0.7cm}\hspace{0.7cm}\mathcal{O}_{3}^{\left(
K\right) }=\left( \overline{s}_{R}d_{L}\right) \left( \overline{s}%
_{L}d_{R}\right) ,  \label{op3fS4} \\
\mathcal{O}_{1}^{\left( B_{d}\right) } &=&\left( \overline{d}%
_{R}b_{L}\right) \left( \overline{d}_{R}b_{L}\right) ,\hspace{0.7cm}\hspace{%
0.7cm}\mathcal{O}_{2}^{\left( B_{d}\right) }=\left( \overline{d}%
_{L}b_{R}\right) \left( \overline{d}_{L}b_{R}\right) ,\hspace{0.7cm}\hspace{%
0.7cm}\mathcal{O}_{3}^{\left( B_{d}\right) }=\left( \overline{d}%
_{R}b_{L}\right) \left( \overline{d}_{L}b_{R}\right) , \\
\mathcal{O}_{1}^{\left( B_{s}\right) } &=&\left( \overline{s}%
_{R}b_{L}\right) \left( \overline{s}_{R}b_{L}\right) ,\hspace{0.7cm}\hspace{%
0.7cm}\mathcal{O}_{2}^{\left( B_{s}\right) }=\left( \overline{s}%
_{L}b_{R}\right) \left( \overline{s}_{L}b_{R}\right) ,\hspace{0.7cm}\hspace{%
0.7cm}\mathcal{O}_{3}^{\left( B_{s}\right) }=\left( \overline{s}%
_{R}b_{L}\right) \left( \overline{s}_{L}b_{R}\right) ,
\end{eqnarray}
and the Wilson coefficients take the form: 
\begin{eqnarray}
\kappa _{1}^{\left( K\right) } &=&\frac{x_{H_{3}^{0}\overline{s}_{R}d_{L}}^{2}}{%
m_{H_{3}^{0}}^{2}}+\sum_{n=1}^{2}\left( \frac{x_{H_{n}^{0}\overline{s}_{R}d_{L}}^{2}}{%
m_{H_{n}^{0}}^{2}}-\frac{x_{A_{n}^{0}\overline{s}_{R}d_{L}}^{2}}{m_{A_{n}^{0}}^{2}}%
,\right) \\
\kappa _{2}^{\left( K\right) } &=&\frac{x_{H_{3}^{0}\overline{s}_{L}d_{R}}^{2}}{%
m_{H_{3}^{0}}^{2}}+\sum_{n=1}^{2}\left( \frac{x_{H_{n}^{0}\overline{s}_{L}d_{R}}^{2}}{%
m_{H_{n}^{0}}^{2}}-\frac{x_{A_{n}^{0}\overline{s}_{L}d_{R}}^{2}}{m_{A_{n}^{0}}^{2}}%
\right) ,\hspace{0.7cm}\hspace{0.7cm} \\
\kappa _{3}^{\left( K\right) } &=&\frac{x_{H_{3}^{0}\overline{s}_{R}d_{L}}x_{H_{3}^{0}%
\overline{s}_{L}d_{R}}}{m_{H_{3}^{0}}^{2}}+\sum_{n=1}^{2}\left( \frac{x_{H_{n}^{0}%
\overline{s}_{R}d_{L}}x_{H_{n}^{0}\overline{s}_{L}d_{R}}}{m_{H_{n}^{0}}^{2}}-\frac{%
x_{A_{n}^{0}\overline{s}_{R}d_{L}}x_{A_{n}^{0}\overline{s}_{L}d_{R}}}{m_{A_{n}^{0}}^{2}}%
\right) ,
\end{eqnarray}%
\begin{eqnarray}
\kappa _{1}^{\left( B_{d}\right) } &=&\frac{x_{H_{3}^{0}\overline{d}_{R}b_{L}}^{2}}{%
m_{H_{3}^{0}}^{2}}+\sum_{n=1}^{2}\left( \frac{x_{H_{n}^{0}\overline{d}_{R}b_{L}}^{2}}{%
m_{H_{n}^{0}}^{2}}-\frac{x_{A_{n}^{0}\overline{d}_{R}b_{L}}^{2}}{m_{A_{n}^{0}}^{2}}%
\right) , \\
\kappa _{2}^{\left( B_{d}\right) } &=&\frac{x_{H_{3}^{0}\overline{d}_{L}b_{R}}^{2}}{%
m_{H_{3}^{0}}^{2}}+\sum_{n=1}^{2}\left( \frac{x_{H_{n}^{0}\overline{d}_{L}b_{R}}^{2}}{%
m_{H_{n}^{0}}^{2}}-\frac{x_{A_{n}^{0}\overline{d}_{L}b_{R}}^{2}}{m_{A_{n}^{0}}^{2}}%
\right) , \\
\kappa _{3}^{\left( B_{d}\right) } &=&\frac{x_{H_{3}^{0}\overline{d}_{R}b_{L}}x_{H_{3}^{0}%
\overline{d}_{L}b_{R}}}{m_{H_{3}^{0}}^{2}}+\sum_{n=1}^{2}\left( \frac{x_{H_{n}^{0}%
\overline{d}_{R}b_{L}}x_{H_{n}^{0}\overline{d}_{L}b_{R}}}{m_{H_{n}^{0}}^{2}}-\frac{%
x_{A_{n}^{0}\overline{d}_{R}b_{L}}x_{A_{n}^{0}\overline{d}_{L}b_{R}}}{m_{A_{n}^{0}}^{2}}%
\right) ,
\end{eqnarray}%
\begin{eqnarray}
\kappa _{1}^{\left( B_{s}\right) } &=&\frac{x_{H_{3}^{0}\overline{s}_{R}b_{L}}^{2}}{%
m_{H_{3}^{0}}^{2}}+\sum_{n=1}^{2}\left( \frac{x_{H_{n}^{0}\overline{s}_{R}b_{L}}^{2}}{%
m_{H_{n}^{0}}^{2}}-\frac{x_{A_{n}^{0}\overline{s}_{R}b_{L}}^{2}}{m_{A_{n}^{0}}^{2}}%
\right) , \\
\kappa _{2}^{\left( B_{s}\right) } &=&\frac{x_{H_{3}^{0}\overline{s}_{L}b_{R}}^{2}}{%
m_{H_{3}^{0}}^{2}}+\sum_{n=1}^{2}\left( \frac{x_{H_{n}^{0}\overline{s}_{L}b_{R}}^{2}}{%
m_{H_{n}^{0}}^{2}}-\frac{x_{A_{n}^{0}\overline{s}_{L}b_{R}}^{2}}{m_{A_{n}^{0}}^{2}}%
\right) , \\
\kappa _{3}^{\left( B_{s}\right) } &=&\frac{x_{H_{3}^{0}\overline{s}_{R}b_{L}}x_{H_{3}^{0}%
\overline{s}_{L}b_{R}}}{m_{H_{3}^{0}}^{2}}+\sum_{n=1}^{2}\left( \frac{x_{H_{n}^{0}%
\overline{s}_{R}b_{L}}x_{H_{n}^{0}\overline{s}_{L}b_{R}}}{m_{H_{n}^{0}}^{2}}-\frac{%
x_{A_{n}^{0}\overline{s}_{R}b_{L}}x_{A_{n}^{0}\overline{s}_{L}b_{R}}}{m_{A_{n}^{0}}^{2}}%
\right) ,
\end{eqnarray}%
where we have used the notation of section \ref{scalarS4} for the physical
scalars, assuming $H_{3}^{0}$ is the lightest of the CP-even ones and
corresponds to the SM Higgs.
The $K-\bar{K}$, $B_{d}^{0}-\bar{B}_{d}^{0}$ and $B_{s}^{0}-\bar{B}_{s}^{0}$%
\ meson mass splittings read: 
\begin{equation}
\Delta m_{K}=\Delta m_{K}^{\left( SM\right) }+\Delta m_{K}^{\left( NP\right)
},\hspace{1cm}\Delta m_{B_{d}}=\Delta m_{B_{d}}^{\left( SM\right) }+\Delta
m_{B_{d}}^{\left( NP\right) },\hspace{1cm}\Delta m_{B_{s}}=\Delta
m_{B_{s}}^{\left( SM\right) }+\Delta m_{B_{s}}^{\left( NP\right) },
\label{DeltamS4}
\end{equation}

\begin{figure}[tbp]
\centering
\subfigure[] {\includegraphics[scale=0.3]{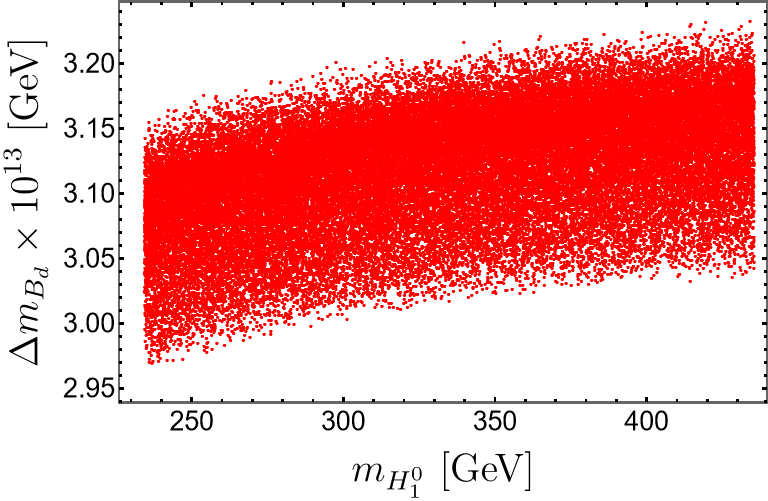}}
\quad \subfigure[] {%
\includegraphics[scale=0.3]{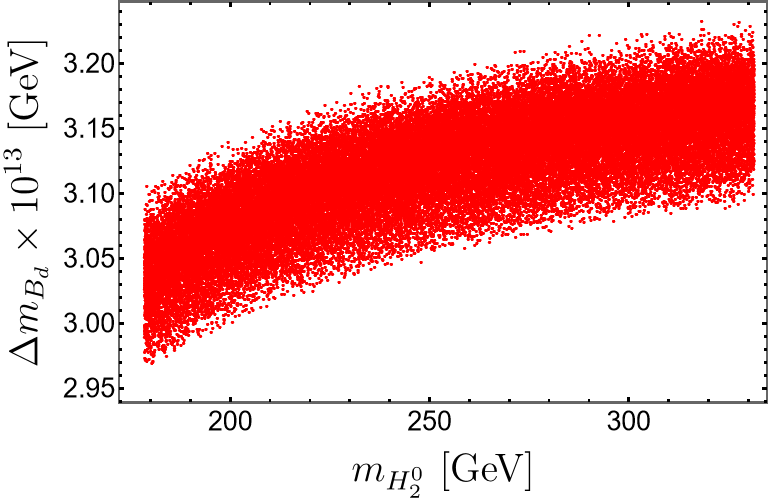}} \quad %
\subfigure[] {\includegraphics[scale=0.3]{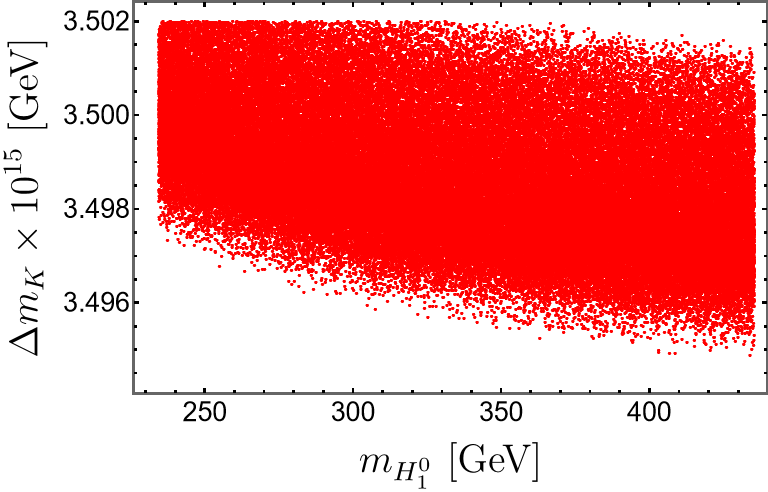}%
} \quad \subfigure[] {%
\includegraphics[scale=0.3]{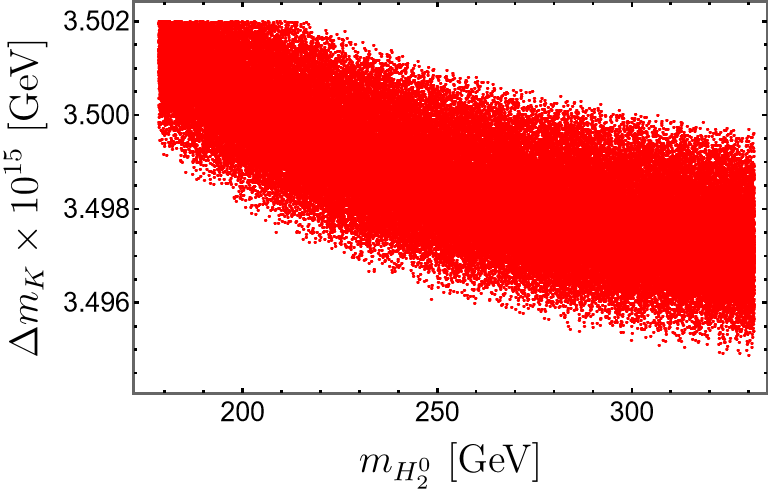}}
\caption{a) Correlation between the $\Delta m_{B_d}$ mass splitting and the
CP even scalar mass $m_{H_2^0}$, b) between the $\Delta m_{B_d}$ mass
splitting and the 
CP even scalar mass $m_{H_1^0}$, c) Correlation
between the $\Delta m_{B_k}$ mass splitting and the 
CP even scalar
mass $m_{H_2^0}$, d) between the $\Delta m_{B_k}$ mass splitting and the
CP even scalar mass $m_{H_1^0}$.}
\label{fig:mesonmixingS4}
\end{figure}
where $\Delta m_{K}^{\left( SM\right) }$, $\Delta m_{B_{d}}^{\left(
SM\right) }$ and $\Delta m_{B_{s}}^{\left( SM\right) }$ correspond to the SM
contributions, while $\Delta m_{K}^{\left( NP\right) }$, $\Delta
m_{B_{d}}^{\left( NP\right) }$ and $\Delta m_{B_{s}}^{\left( NP\right) }$
are due to new physics effects. Our model predicts the following new physics
contributions for the $K-\bar{K}$, $B_{d}^{0}-\bar{B}_{d}^{0}$ and $%
B_{s}^{0}-\bar{B}_{s}^{0}$ meson mass differences: 
\begin{equation}
\Delta m_{K}^{\left( NP\right) }=\frac{8}{3}f_{K}^{2}\eta _{K}B_{K}m_{K}%
\left[ r_{2}^{\left( K\right) }\kappa _{3}^{\left( K\right) }+r_{1}^{\left(
K\right) }\left( \kappa _{1}^{\left( K\right) }+\kappa _{2}^{\left( K\right)
}\right) \right]~,
\end{equation}%
\begin{equation}
\Delta m_{B_{d}}^{\left( NP\right) }=\frac{8}{3}f_{B_{d}}^{2}\eta
_{B_{d}}B_{B_{d}}m_{B_{d}}\left[ r_{2}^{\left( B_{d}\right) }\kappa
_{3}^{\left( B_{d}\right) }+r_{1}^{\left( B_{d}\right) }\left( \kappa
_{1}^{\left( B_{d}\right) }+\kappa _{2}^{\left( B_{d}\right) }\right) \right]~,
\end{equation}%
\begin{equation}
\Delta m_{B_{s}}^{\left( NP\right) }=\frac{8}{3}f_{B_{s}}^{2}\eta
_{B_{s}}B_{B_{s}}m_{B_{s}}\left[ r_{2}^{\left( B_{s}\right) }\kappa
_{3}^{\left( B_{s}\right) }+r_{1}^{\left( B_{s}\right) }\left( \kappa
_{1}^{\left( B_{s}\right) }+\kappa _{2}^{\left( B_{s}\right) }\right) \right]~.
\end{equation}%
Using the following numerical values of the meson parameters \cite{Jubb:2016mvq,Artuso:2015swg,HFLAV:2019otj,Wang:2018csg,CPLEAR:1998zfe,Lenz:2019lvd,FlavourLatticeAveragingGroupFLAG:2021npn,Workman:2022ynf}:
\begin{eqnarray}
\left(\Delta m_{K}\right)_{\exp }&=&\left( 3.484\pm 0.006\right) \times
10^{-12}\, \mathrm{{MeV},\hspace{1.5cm}\left( \Delta m_{K}\right)
_{SM}=3.483\times 10^{-12}\, {MeV}}  \notag \\
f_{K} &=&155.7\, \mathrm{{MeV},\hspace{1.5cm}B_{K}=0.85,\hspace{1.5cm}\eta
_{K}=0.57,}  \notag \\
r_{1}^{\left( K\right) } &=&-9.3,\hspace{1.5cm}r_{2}^{\left(K\right) }=30.6,%
\hspace{1.5cm}m_{K}=\left(497.611\pm 0.013\right)\, \mathrm{{MeV},}
\end{eqnarray}%
\begin{eqnarray}
\left( \Delta m_{B_{d}}\right) _{\exp } &=&\left(3.334\pm 0.013\right)
\times 10^{-10}\, \mathrm{{MeV},\hspace{1.5cm}\left( \Delta m_{B_{d}}\right)
_{SM}=\left(3.653\pm 0.037\pm 0.019\right)\times 10^{-10}\, {MeV},}  \notag
\\
f_{B_{d}} &=&188\, \mathrm{{MeV},\hspace{1.5cm}B_{B_{d}}=1.26,\hspace{1.5cm}%
\eta _{B_{d}}=0.55,}  \notag \\
r_{1}^{\left( B_{d}\right) } &=&-0.52,\hspace{1.5cm}r_{2}^{\left(
B_{d}\right) }=0.88,\hspace{1.5cm}m_{B_{d}}=\left(5279.65\pm 0.12\right)\,%
\mathrm{{MeV},}
\end{eqnarray}%
\begin{eqnarray}
\left( \Delta m_{B_{s}}\right) _{\exp } &=&\left(1.1683\pm 0.0013\right)
\times 10^{-8}\, \mathrm{{MeV},\hspace{1.5cm}\left( \Delta m_{B_{s}}\right)
_{SM}=\left(1.1577\pm 0.022\pm 0.051\right) \times 10^{-8}\, {MeV},}  \notag
\\
f_{B_{s}} &=&225\, \mathrm{{MeV},\hspace{1.5cm}B_{B_{s}}=1.33,\hspace{1.5cm}%
\eta _{B_{s}}=0.55,}  \notag \\
r_{1}^{\left( B_{s}\right) } &=&-0.52,\hspace{1.5cm}r_{2}^{\left(
B_{s}\right) }=0.88,\hspace{1.5cm}m_{B_{s}}=\left(5366.9\pm 0.12\right)\, 
\mathrm{{MeV}.}
\end{eqnarray}
where the experimental values of the meson masses are taken from \cite{CPLEAR:1998zfe,Artuso:2015swg,Jubb:2016mvq,Wang:2018csg,Lenz:2019lvd,HFLAV:2019otj,FlavourLatticeAveragingGroupFLAG:2021npn}, whereas those corresponding to the bag parameters from \cite{Workman:2022ynf,FlavourLatticeAveragingGroupFLAG:2021npn}. Furthermore, the values for the $r_{k}^{\left(K\right) }$, $r_{k}^{\left( B_{d}\right) }$ and $r_{k}^{\left( B_{s}\right) }$ ($k=1,2$) parameters are taken from \cite{Dedes:2002er}.

Fig. \ref{fig:mesonmixingS4}a and Fig. \ref{fig:mesonmixingS4}b display the
correlations between the $\Delta m_{B_d}$ mass splitting and the 
CP even scalar masses $m_{H_2^0}$ and $m_{H_1^0}$, respectively. On the other hand, Fig. \ref%
{fig:mesonmixing}c and Fig. \ref{fig:mesonmixingS4}d display the correlations
between the $\Delta m_{B_k}$ mass splitting and the 
CP even scalar masses $m_{H_2^0}$ and $m_{H_1^0}$, respectively. As seen from these figures, the obtained values for the meson mass splittings feature a small variation of less than  $10\%$ in the considered ranges of CP even and CP odd scalar masses. In our numerical analysis, for
the sake of simplicity, we have consider the couplings of the flavor-changing
neutral Yukawa interactions that produce the $(B_d^0-\overline{B}_d^0)$ and $%
(K^0-\overline{K}^0)$ oscillations of the same order of magnitude 
and we perform a random scan over the Yukawa couplings and scalar masses. We find that the couplings of the flavor-changing
neutral Yukawa interactions that produce the $(B_d^0-\overline{B}_d^0)$ and $%
(K^0-\overline{K}^0)$ meson mixings should be of the order of $10^{-4}$ and $10^{-6}$ respectively, in order to successfully comply with meson oscillation constraints.
Furthermore, we have varied the masses around 30\% from their central values obtained in the scalar sector analysis shown in the plots of
Fig. \ref{plotScalars}. As indicated in Fig. \ref{fig:mesonmixingS4}, the experimental
constraints arising from $(B_d^0-\overline{B}_d^0)$ and $(K^0-\overline{K}%
^0) $ meson oscillations are successfully fulfilled for the aforementioned
range of parameter space. We have numerically checked that in the above
described range of masses, the obtained values for the $\Delta m_{B_s}$ mass
splitting are consistent with the experimental data on meson oscillations
for flavor violating Yukawa couplings equal to $2.5\times 10^{-4}$. %

\section{Oblique $T$, $S$ and $U$ parameters \label{TnS}}
\lhead[\thepage]{\thesection. Oblique $T$, $S$ and $U$ parameters}

The extra scalars affect the oblique corrections of the SM, and these values
are measured in high precision experiments. Consequently, they act as a
further constraint on the validity of our model. The oblique corrections are
parametrized in terms of the three well-known quantities $T$, $S$ and $U$. In
this section we calculate one-loop contributions to the oblique parameters $%
T $, $S$\ and $U$ defined as~\cite%
{Peskin:1991sw,Altarelli:1990zd,Barbieri:2004qk} 
\begin{eqnarray}
T &=&\frac{\Pi _{33}\left( q^{2}\right) -\Pi _{11}\left( q^{2}\right) }{%
\alpha _{EM}(M_{Z})M_{W}^{2}}\biggl|_{q^{2}=0},\ \ \ \ \ \ \ \ \ \ \ S=\frac{%
2\sin 2{\theta }_{W}}{\alpha _{EM}(M_{Z})}\frac{d\Pi _{30}\left(
q^{2}\right) }{dq^{2}}\biggl|_{q^{2}=0},  \label{T-S-definition} \\
U &=&\frac{4\sin ^{2}\theta _{W}}{\alpha _{EM}(M_{Z})}\left( \frac{d\Pi
_{33}\left( q^{2}\right) }{dq^{2}}-\frac{d\Pi _{11}\left( q^{2}\right) }{%
dq^{2}}\right) \biggl|_{q^{2}=0}
\end{eqnarray}%
where $\Pi _{11}\left( 0\right) $, $\Pi _{33}\left( 0\right) $, and $\Pi
_{30}\left( q^{2}\right) $ are the vacuum polarization amplitudes with $%
\{W_{\mu }^{1},W_{\mu }^{1}\}$, $\{W_{\mu }^{3},W_{\mu }^{3}\}$ and $%
\{W_{\mu }^{3},B_{\mu }\}$ external gauge bosons, respectively, and $q$ is
their momentum. We note that in the definitions of the $T$, $S$ and $U$
parameters, the new physics is assumed to be heavy when compared to $M_{W}$ and $M_{Z}$.

In order to simplify our numerical analysis we restrict to the scenario of
the alignment limit where the neutral CP even part of the $SU\left(
2\right) $ scalar doublet $\Xi_{3}$ is identified with the 126 GeV SM like
Higgs boson. We further restrict to the region of parameter space where the
neutral CP odd and electrically charged components of $\Xi_{3}$ correspond to
the SM Goldstone bosons. In that simplified benchmark scenario, the non SM
physical scalar states relevant at low energies will arise from the $\Xi_{1}$
and $\Xi_{2}$ scalar doublets. We further assume that the gauge singlet scalars acquire very large vacuum expectation values (VEVs), which implies that the mixing angles of these fields with the $\Xi_{1}$
 and $\Xi_{2}$ scalar doublets are very small since they are suppressed by the ratios of their VEVs (assumed that the quartic scalar couplings are of the same order of magnitude), which is a consequence of the method of recursive expansion proposed in \cite{Grimus:2000vj}. Because of this reason, the singlet scalar fields do not have a relevant impact in the electroweak precision observables since they do not couple with the SM gauge bosons and their mixing angles with the neutral components of the scalar doublets are very small. Therefore, under the aforementioned considerations, in the alignment limit scenario where the $126$ GeV Higgs boson is identified with the CP even neutral part of $\Xi_{3}$, the oblique $T$, $S$ and $U$ parameters will receive new physics contributions arising from the electrically neutral and electrically charged scalar fields arising from the $SU(2)$ scalar doublets $\Xi_{1}$ and $\Xi_{2}$. The oblique $T$, $S$ and $U$ parameters have been computed in the framework of multiHiggs doublet models in \cite{Grimus:2007if,Grimus:2008nb,CarcamoHernandez:2015smi}. Then, the contributions arising from new physics to the $T$, $S$ and $U$ parameters in model 1 are:%
\begin{eqnarray}
T &\simeq &\frac{1}{16\pi ^{2}v^{2}\alpha _{EM}(M_{Z})}\left\{
\sum_{i=1}^{2}\sum_{k=1}^{2}\left( \left( R_{C}\right) _{ik}\right)
^{2}m_{H_{k}^{\pm }}^{2}+\sum_{i=1}^{2}\sum_{j=1}^{2}\sum_{k=1}^{2}\left(
\left( R_{H}\right) _{ki}\right) ^{2}\left( \left( R_{A}\right) _{kj}\right)
^{2}F\left( m_{H_{i}^{0}}^{2},m_{A_{j}^{0}}^{2}\right) \right.   \notag \\
&&-\left. \sum_{i=1}^{2}\sum_{j=1}^{2}\sum_{k=1}^{2}\left( \left(
R_{H}\right) _{ki}\right) ^{2}\left( \left( R_{C}\right) _{kj}\right)
^{2}F\left( m_{H_{i}^{0}}^{2},m_{H_{j}^{\pm }}^{2}\right)
-\sum_{i=1}^{2}\sum_{j=1}^{2}\sum_{k=1}^{2}\left( \left( R_{A}\right)
_{ki}\right) ^{2}\left( \left( R_{C}\right) _{kj}\right) ^{2}F\left(
m_{A_{i}^{0}}^{2},m_{H_{j}^{\pm }}^{2}\right) \right\} 
\end{eqnarray}%
\begin{equation}
S\simeq \sum_{i=1}^{2}\sum_{j=1}^{2}\sum_{k=1}^{2}\frac{\left( \left(
R_{H}\right) _{ki}\right) ^{2}\left( \left( R_{A}\right) _{kj}\right) ^{2}}{%
12\pi }K\left( m_{H_{i}^{0}}^{2},m_{A_{j}^{0}}^{2},m_{H_{k}^{\pm
}}^{2}\right) ,
\end{equation}%
\begin{eqnarray}
U &\simeq &-S+\sum_{i=1}^{2}\sum_{j=1}^{2}\sum_{k=1}^{2}\left( \left( R_{A}\right)
_{ki}\right) ^{2}\left( \left( R_{C}\right) _{kj}\right) ^{2}K_{2}\left(
m_{A_{i}^{0}}^{2},m_{H_{j}^{\pm }}^{2}\right)   \notag \\
&&+\sum_{i=1}^{2}\sum_{j=1}^{2}\sum_{k=1}^{2}\left( \left( R_{H}\right) _{ki}\right)
^{2}\left( \left( R_{C}\right) _{kj}\right) ^{2}K_{2}\left(
m_{H_{i}^{0}}^{2},m_{H_{j}^{\pm }}^{2}\right) ,
\end{eqnarray}
where we introduced the functions \cite{CarcamoHernandez:2015smi} 
\begin{equation}
F\left( m_{1}^{2},m_{2}^{2}\right) =\frac{m_{1}^{2}m_{2}^{2}}{%
m_{1}^{2}-m_{2}^{2}}\ln \left( \frac{m_{1}^{2}}{m_{2}^{2}}\right) ,\hspace{%
1.5cm}\hspace{1.5cm}\lim_{m_{2}\rightarrow m_{1}}F\left(
m_{1}^{2},m_{2}^{2}\right) =m_{1}^{2}.
\end{equation}%
\begin{eqnarray}
K\left( m_{1}^{2},m_{2}^{2},m_{3}^{2}\right) &=&\frac{1}{\left(
m_{2}^{2}-m_{1}^{2}\right) {}^{3}}\left\{ m_{1}^{4}\left(
3m_{2}^{2}-m_{1}^{2}\right) \ln \left( \frac{m_{1}^{2}}{m_{3}^{2}}\right)
-m_{2}^{4}\left( 3m_{1}^{2}-m_{2}^{2}\right) \ln \left( \frac{m_{2}^{2}}{%
m_{3}^{2}}\right) \right.  \notag \\
&&-\left. \frac{1}{6}\left[ 27m_{1}^{2}m_{2}^{2}\left(
m_{1}^{2}-m_{2}^{2}\right) +5\left( m_{2}^{6}-m_{1}^{6}\right) \right]
\right\} ,
\end{eqnarray}%
with the properties 
\begin{eqnarray}
\lim_{m_{1}\rightarrow m_{2}}K(m_{1}^{2},m_{2}^{2},m_{3}^{2})
&=&K_{1}(m_{2}^{2},m_{3}^{2})=\ln \left( \frac{m_{2}^{2}}{m_{3}^{2}}\right) ,
\notag \\
\lim_{m_{2}\rightarrow m_{3}}K(m_{1}^{2},m_{2}^{2},m_{3}^{2})
&=&K_{2}(m_{1}^{2},m_{3}^{2})=\frac{%
-5m_{1}^{6}+27m_{1}^{4}m_{3}^{2}-27m_{1}^{2}m_{3}^{4}+6\left(
m_{1}^{6}-3m_{1}^{4}m_{3}^{2}\right) \ln \left( \frac{m_{1}^{2}}{m_{3}^{2}}%
\right) +5m_{3}^{6}}{6\left( m_{1}^{2}-m_{3}^{2}\right) ^{3}},  \notag \\
\lim_{m_{1}\rightarrow m_{3}}K(m_{1}^{2},m_{2}^{2},m_{3}^{2})
&=&K_{2}(m_{2}^{2},m_{3}^{2}).
\end{eqnarray}
Here $R_{H}$, $R_{A}$ and $R_{C}$\ are the rotation matrices diagonalizing
the squared mass matrices for the non SM CP-even, CP-odd and electrically charged
scalars. It is worth mentioning that, from the properties of the loop functions appearing in the expressions for the oblique $S$, $T$ and $U$ parameters given in \cite{Grimus:2007if,Grimus:2008nb,CarcamoHernandez:2015smi}, it follows that in multiHiggs doublet models, the contributions to these parameters arising from new physics will vanish in the limit of degenerate heavy non SM scalars. Thus, in multiHiggs doublet models, a spectrum of non SM scalars with a moderate mass splitting will be favoured by electroweak precision tests.

On the other hand, the contributions arising from new physics to the $T$, $S$
and $U$ parameters in model 2 are:
\begin{eqnarray}
T &\simeq &\frac{1}{16\pi ^{2}v^{2}\alpha _{EM}(M_{Z})}\left\{
\sum_{i=1}^{2}\sum_{k=1}^{2}\left( \left( R_{C}\right) _{ik}\right)
^{2}m_{H_{k}^{\pm }}^{2}+\sum_{i=1}^{2}\sum_{j=1}^{2}\sum_{k=1}^{2}\left(
\left( R_{H}\right) _{ki}\right) ^{2}\left( \left( R_{A}\right) _{kj}\right)
^{2}F\left( m_{H_{i}^{0}}^{2},m_{A_{j}^{0}}^{2}\right) \right.   \notag \\
&&-\left. \sum_{i=1}^{2}\sum_{j=1}^{2}\sum_{k=1}^{2}\left( \left(
R_{H}\right) _{ki}\right) ^{2}\left( \left( R_{C}\right) _{kj}\right)
^{2}F\left( m_{H_{i}^{0}}^{2},m_{H_{j}^{\pm }}^{2}\right)
-\sum_{i=1}^{2}\sum_{j=1}^{2}\sum_{k=1}^{2}\left( \left( R_{A}\right)
_{ki}\right) ^{2}\left( \left( R_{C}\right) _{kj}\right) ^{2}F\left(
m_{A_{i}^{0}}^{2},m_{H_{j}^{\pm }}^{2}\right) \right\}   \notag \\
&&+\frac{1}{16\pi ^{2}v^{2}\alpha _{EM}(M_{Z})}\left\{ F\left(
m_{H_{4}^{0}}^{2},m_{A_{4}^{0}}^{2}\right) +m_{H_{4}^{\pm }}^{2}-F\left(
m_{H_{4}^{0}}^{2},m_{H_{4}^{\pm }}^{2}\right) -F\left(
m_{A_{4}^{0}}^{2},m_{H_{4}^{\pm }}^{2}\right) \right\} 
\end{eqnarray}%
\begin{equation}
S\simeq \sum_{i=1}^{2}\sum_{j=1}^{2}\sum_{k=1}^{2}\frac{\left( \left(
R_{H}\right) _{ki}\right) ^{2}\left( \left( R_{A}\right) _{kj}\right) ^{2}}{%
12\pi }K\left( m_{H_{i}^{0}}^{2},m_{A_{j}^{0}}^{2},m_{H_{k}^{\pm
}}^{2}\right) +\frac{1}{12\pi }K\left(
m_{H_{4}^{0}}^{2},m_{A_{4}^{0}}^{2},m_{H_{4}^{\pm }}^{2}\right) ,
\end{equation}%
\begin{eqnarray}
U &\simeq &-S+\sum_{i=1}^{2}\sum_{j=1}^{2}\sum_{k=1}^{2}\left( \left(
R_{A}\right) _{ki}\right) ^{2}\left( \left( R_{C}\right) _{kj}\right)
^{2}K_{2}\left( m_{A_{i}^{0}}^{2},m_{H_{j}^{\pm }}^{2}\right)   \notag \\
&&+\sum_{i=1}^{2}\sum_{j=1}^{2}\sum_{k=1}^{2}\left( \left( R_{H}\right)
_{ki}\right) ^{2}\left( \left( R_{C}\right) _{kj}\right) ^{2}K_{2}\left(
m_{H_{i}^{0}}^{2},m_{H_{j}^{\pm }}^{2}\right)   \notag \\
&&+K_{2}\left( m_{A_{4}^{0}}^{2},m_{H_{4}^{\pm }}^{2}\right) +K_{2}\left(
m_{H_{4}^{0}}^{2},m_{H_{4}^{\pm }}^{2}\right) ,
\end{eqnarray}%
where $H^{0}_{4}$, $A^{0}_{4}$ and $H^{\pm }_{4}$ are the physical scalar fields arising
from the inert doublet $\Xi_{4}$.

Besides that, the experimental values of $T$, $S$ and $U$ are constrained to
be in the ranges \cite{Lu:2022bgw}:
\begin{equation}
T=-0.01\pm 0.10,\ \ \ \ \ \ \ \ \ \ \ S=0.03\pm 0.12,\ \ \ \ \ \ \ \ \ \ \
U=0.02\pm 0.11
\end{equation}%
We have numerically checked that both models can successfully reproduce the allowed experimental values for the oblique $T$, $S$ and $U$ parameters. Furthermore, we checked the existence of a parameter space consistent with both scenarios where the $W$ mass anomaly is absent or present. This is consistent with models with Higgs multiplets, as it has been analyzed in \cite{Tran:2022yrh}.

\section{Scalar sector}\label{scalarS4}
\lhead[\thepage]{\thesection. Scalar sector}

In the present section we address the discussion of the phenomenology of the scalar sectors in the
low energy regime. Both models share the same effective low energy scalar potential with respect to
the active $Z_2$ even scalar doublets $\Xi_i$, $i=1,2,3$, but in each model the inert scalar couples
differently to these fields. 
In addition, at the loop level, the difference in DM particle content between the models has significant impact on the
observables in the scalar sector, such as the possibility of a large contribution to the $h\rightarrow \gamma\gamma$
decay width from the inert doublet charged Higgs boson, see e.g.~\cite{Aiko:2023nqj,Degrassi:2023eii}.
On the other hand, since the quark and charged
lepton sectors of both models are identical while the corresponding neutrino sectors bear no tangible
influence on the kind of phenomenological analysis considered here, 
we mainly focus on collider limits for the new scalars predicted by the inclusion of the
extra Higgs doublets. For the numerical calculations we neglect the masses of the first and second family 
of fermions and also off-diagonal entries in the Yukawa matrices. We expect deviations of the matter sector
relative to the SM to be of negligible influence in the phenomenology of the
scalar sector at present accelerator searches.
This argument is fully justified, for example, SM Higgs boson branching ratios are dominated by the $b\bar{b}$,
$WW$, $gg$ and $\tau\bar{\tau}$ channels, with other channels contributing less than $\sim 3\times 10^{-2}$,
see e.g. figure 9 of reference \cite%
{LHCHiggsCrossSectionWorkingGroup:2016ypw}. In our analysis, we ensure that the lightest of the CP-even scalars
satisfies the alignment limit and hence its couplings and decay rates 
coincide with those of the SM Higss.
For the heavier scalars, the situation is similar, with the dominating channels being $t\bar{t}, b\bar{b}$
and $\tau\bar{\tau}$, see e.g. figures 17 and 18 of reference \cite%
{Spira:2016ztx} in the context of the MSSM.
In other words, contributions from the first two fermion families to the decay branching ratios of the 
physical scalars are completely negligible.
Similarly, for the scalar boson production at the LHC, the gluon fusion mechanism \cite%
{Georgi:1977gs}
dominates for the scalar mass ranges considered here, with
the gluon coupling to the Higgs bosons mediated most importantly by triangular top- and bottom-quark loops. Notice that we do not consider the effect of the several active singlet scalar fields as they are assumed to acquire very large vacuum expectation values, much larger than the electroweak symmetry breaking scale, thus allowing to decouple them in the low energy effective field theory. As previously mentioned the mixing angles of the singlet scalar fields with the scalar doublets are very small as they are suppressed by the ratio between the electroweak symmetry breaking scale and the scale of spontaneous breaking of the $S_4\times Z_4$ discrete symmetry.
Under these assumptions we see from Eq. (\ref{yuk1}) that the third generation of quarks couples only to
$\Xi_{3}$. For the charged lepton case we further assume that $x_3^l << y_3^l$ so that the same is true
for the third generation of charged leptons (note that the cubic and quartic vertices involving $\Phi_\tau$
are highly suppressed by the high energy scale $\Lambda$ so that this $S_4$ triplet scalar is decoupled
at the energies considered here).

\section{Low energy scalar potential and scalar mass spectrum}\label{appPot}
\lhead[\thepage]{\thesection. Low energy scalar potential}

In order to simplify our analysis, we restrict to the simplified benchmark
scenario where the scalar singlets of the model do not feature mixings with
the three $SU\left( 2\right) _{L}$ scalar doublets. The low energy scalar
potential of the model then corresponds to the $S_4$ symmetric scalar potential of the three 
$SU\left( 2\right) _{L}$ scalar doublets plus a soft-breaking mass term and has the form:
\begin{eqnarray}
V &=&-\mu _{1}^{2}\left( \Xi _{1}^{\dagger }\Xi _{1}\right) -\mu
_{2}^{2}\left( \Xi _{2}^{\dagger }\Xi _{2}\right) -\mu _{3}^{2}\Xi
_{3}^{\dagger }\Xi _{3}+\lambda _{1}\left( \Xi _{I}^{\dagger }\Xi
_{I}\right) _{\mathbf{1}_{1}}\left( \Xi _{I}^{\dagger }\Xi _{I}\right) _{%
\mathbf{1}_{1}}+\lambda _{2}\left( \Xi _{I}^{\dagger }\Xi _{I}\right) _{%
\mathbf{1}_{2}}\left( \Xi _{I}^{\dagger }\Xi _{I}\right) _{\mathbf{1}_{2}}  \notag \\
&&+\lambda _{3}\left( \Xi _{I}^{\dagger }\Xi _{I}\right) _{\mathbf{2}}\left(
\Xi _{I}^{\dagger }\Xi _{I}\right) _{\mathbf{2}}+\lambda _{4}\left\{ \left[
\left( \Xi _{I}^{\dagger }\Xi _{I}\right) _{\mathbf{2}}\Xi _{I}^{\dagger }%
\right] _{\mathbf{1}_{1}}\Xi _{3}+h.c\right\} +\lambda _{5}\left( \Xi
_{3}^{\dagger }\Xi _{3}\right) \left( \Xi _{I}^{\dagger }\Xi _{I}\right) _{%
\mathbf{1}_{1}} \\
&&+\lambda _{6}\left[ \left( \Xi _{3}^{\dagger }\Xi _{I}\right) \left( \Xi
_{I}^{\dagger }\Xi _{3}\right) \right] _{\mathbf{1}_{1}}+\lambda _{7}\left[
\left( \Xi _{3}^{\dagger }\Xi _{I}\right) \left( \Xi _{3}^{\dagger }\Xi
_{I}\right) +h.c\right] +\lambda _{8}\left( \Xi _{3}^{\dagger }\Xi
_{3}\right) ^{2} \notag 
\end{eqnarray}
where we have included soft breaking mass terms in the scalar potential. They can arise at high energy scale from the quartic scalar interaction $\kappa \left( \chi \chi \right) _{2}\left( \Xi _{I}^{\dagger }\Xi
_{I}\right)_{\mathbf{2}}$. Note that only bilinear soft-breaking mass terms and not trilinear terms are allowed in the scalar potential, as follows from gauge invariance.\footnote{Notice that the low energy scalar potential without the soft-breaking mass terms has an underlying $S_3$ discrete symmetry. This is due to the fact that $S_3$ is a subgroup of $S_4$ as well as to our choice of irreducible representations to accommodate the active $SU(2)$ scalar doublets.}\\

After the spontaneous breaking of the $S_{4}$ discrete symmetry, the low
energy scalar potential of the model under consideration takes the form: 
\begin{eqnarray}
V &=&-\mu _{1}^{2}\left( \Xi _{1}^{\dagger }\Xi _{1}\right) -\mu
_{2}^{2}\left( \Xi _{2}^{\dagger }\Xi _{2}\right) -\mu _{3}^{2}\left( \Xi
_{3}^{\dagger }\Xi _{3}\right) +\lambda _{1}\left( \Xi _{1}^{\dagger }\Xi
_{1}+\Xi _{2}^{\dagger }\Xi _{2}\right) ^{2}+\lambda _{2}\left( \Xi
_{2}^{\dagger }\Xi _{1}-\Xi _{1}^{\dagger }\Xi _{2}\right) ^{2}  \notag \\
&&+\lambda _{3}\left[ \left( \Xi _{1}^{\dagger }\Xi _{2}+\Xi _{2}^{\dagger
}\Xi _{1}\right) ^{2}+\left( \Xi _{1}^{\dagger }\Xi _{1}-\Xi _{2}^{\dagger
}\Xi _{2}\right) ^{2}\right] \notag  \\
&&+\lambda _{4}\left[ \left( \Xi _{1}^{\dagger }\Xi _{2}+\Xi _{2}^{\dagger
}\Xi _{1}\right) \left( \Xi _{1}^{\dagger }\Xi _{3}+\Xi _{3}^{\dagger }\Xi
_{1}\right) +\left( \Xi _{1}^{\dagger }\Xi _{1}-\Xi _{2}^{\dagger }\Xi
_{2}\right) \left( \Xi _{2}^{\dagger }\Xi _{3}+\Xi _{2}\Xi _{3}^{\dagger
}\right) \right] \\  \label{APP_scalar_pot1}
&&+\lambda _{5}\left( \Xi _{3}^{\dagger }\Xi _{3}\right) \left( \Xi
_{1}^{\dagger }\Xi _{1}+\Xi _{2}^{\dagger }\Xi _{2}\right) +\lambda _{6} 
\left[ \left( \Xi _{3}^{\dagger }\Xi _{1}\right) \left( \Xi _{1}^{\dagger
}\Xi _{3}\right) +\left( \Xi _{3}^{\dagger }\Xi _{2}\right) \left( \Xi
_{2}^{\dagger }\Xi _{3}\right) \right]  \notag \\
&&+\lambda _{7}\left[ \left( \Xi _{3}^{\dagger }\Xi _{1}\right) ^{2}+\left(
\Xi _{3}^{\dagger }\Xi _{2}\right) ^{2}+\left( \Xi _{3}\Xi _{1}^{\dagger
}\right) ^{2}+\left( \Xi _{3}\Xi _{2}^{\dagger }\right) ^{2}\right] +\lambda
_{8}\left( \Xi _{3}^{\dagger }\Xi _{3}\right) ^{2} \notag ~.
\end{eqnarray}

The minimization conditions of the scalar potential are given by:
\begin{eqnarray}
\mu _{1}^{2} &=&\frac{1}{2}\left( 4\lambda _{1}v_{1}^{2}+4\lambda
_{3}v_{1}^{2}+6\lambda _{4}v_{3}v_{1}+\lambda _{5}v_{3}^{2}+\lambda
_{6}v_{3}^{2}+2\lambda _{7}v_{3}^{2}\right) , \\
\mu _{2}^{2} &=&\frac{4\lambda _{1}v_{1}^{3}+4\lambda _{3}v_{1}^{3}+\lambda
_{5}v_{3}^{2}v_{1}+\lambda _{6}v_{3}^{2}v_{1}+2\lambda _{7}v_{3}^{2}v_{1}}{%
2v_{1}} \\
\mu _{3}^{2} &=&\frac{-\lambda _{4}v_{2}^{3}+\lambda
_{5}v_{3}v_{2}^{2}+\lambda _{6}v_{3}v_{2}^{2}+2\lambda
_{7}v_{3}v_{2}^{2}+3\lambda _{4}v_{1}^{2}v_{2}+\lambda
_{5}v_{1}^{2}v_{3}+\lambda _{6}v_{1}^{2}v_{3}+2\lambda
_{7}v_{1}^{2}v_{3}+2\lambda _{8}v_{3}^{3}}{2v_{3}}~,
\end{eqnarray}

and the resulting squared mass matrices for the CP even neutral, CP odd
neutral and electrically charged scalar fields are given by: 
\begin{equation}
\mathbf{M}_{CP-\text{even}}^{2}=\left( 
\begin{array}{ccc}
2\left( \lambda _{1}+\lambda _{3}\right) v_{1}^{2} & 2\left( \lambda
_{1}+\lambda _{3}\right) v_{1}^{2}+3\lambda _{4}v_{3}v_{1} & 
v_{1}\left( 3\lambda _{4}v_{1}+\left( \lambda _{5}+\lambda
_{6}+2\lambda _{7}\right) v_{3}\right) \\ 
2\left( \lambda _{1}+\lambda _{3}\right) v_{1}^{2}+3\lambda
_{4}v_{3}v_{1} & 2\left( \lambda _{1}+\lambda _{3}\right) v_{1}^{2}-3%
\lambda _{4}v_{1}v_{3} & \left( \lambda _{5}+\lambda
_{6}+2\lambda _{7}\right) v_{1}v_{3} \\ 
v_{1}\left( 3\lambda _{4}v_{1}+\left( \lambda _{5}+\lambda
_{6}+2\lambda _{7}\right) v_{3}\right) & \left( \lambda
_{5}+\lambda _{6}+2\lambda _{7}\right) v_{1}v_{3} & 2\lambda _{8}v_{3}^{2}-%
\frac{\lambda _{4}v_{1}^{3}}{v_{3}} \\ 
&  & \label{CPEven}
\end{array}%
\right)
\end{equation}

\begin{equation}
\mathbf{M}_{CP-\text{odd}}^{2}=\left( 
\begin{array}{ccc}
-2\left(\lambda _2+\lambda _3\right) v_1^2-2\lambda _4 v_3 v_1-2\lambda _7 v_3^2
&  v_1 \left(2 \left(\lambda _2+\lambda _3\right) v_1+\lambda _4
v_3\right) &  v_1 \left(\lambda _4 v_1+2 \lambda _7 v_3\right) \\ 
 v_1 \left(2 \left(\lambda _2+\lambda _3\right) v_1+\lambda _4
v_3\right) & -2\left(\lambda _2+\lambda _3\right) v_1^2- \lambda
_4 v_3 v_1-2\lambda _7 v_3^2 & 2\lambda _7 v_1 v_3 \\ 
 v_1 \left(\lambda _4 v_1+2 \lambda _7 v_3\right) & 2\lambda _7
v_1 v_3 & -\frac{\lambda _4 v_1^3}{v_3}-4 \lambda _7 v_1^2 \\ 
&  & 
\end{array}
\right)
\end{equation}
{\small
\begin{equation}
\mathbf{M}_{\text{charged}}^{2}=\left( 
\begin{array}{ccc}
-2 \lambda _3 v_1^2-2 \lambda _4 v_3 v_1-\frac{1}{2} \left(\lambda _6+2
\lambda _7\right) v_3^2 & v_1 \left(2 \lambda _3 v_1+\lambda _4 v_3\right) & 
\frac{1}{2} v_1 \left(2 \lambda _4 v_1+\left(\lambda _6+2 \lambda _7\right)
v_3\right) \\ 
v_1 \left(2 \lambda _3 v_1+\lambda _4 v_3\right) & -2 \lambda _3
v_1^2-\lambda _4 v_3 v_1-\frac{1}{2} \left(\lambda _6+2 \lambda _7\right)
v_3^2 & \frac{1}{2} \left(\lambda _6+2 \lambda _7\right) v_1 v_3 \\ 
\frac{1}{2} v_1 \left(2 \lambda _4 v_1+\left(\lambda _6+2 \lambda _7\right)
v_3\right) & \frac{1}{2} \left(\lambda _6+2 \lambda _7\right) v_1 v_3 & -%
\frac{v_1^2 \left(\lambda _4 v_1+\left(\lambda _6+2 \lambda _7\right)
v_3\right)}{v_3} \\ 
&  & 
\end{array}
\right)
\end{equation}
}
The last two mass matrices can be diagonilized analytically, we find for the pseudo-scalars and the charged scalars the mass spectra:

\begin{equation}
    m^2_{A_1^0} = a_1 -  \frac{v_1}{2v_3}\sqrt{a_2},\ \ \ \ \ \ \ \ \ \ \ m^2_{A_2^0} = a_1 +  \frac{v_1}{2v_3}\sqrt{a_2},\ \ \ \ \ \ \ \ \ \ \ m^2_{G_Z} = 0,
\end{equation}

\begin{equation}
    m^2_{H_1^\pm} = c_1 - \frac{v_1}{2v_3}\sqrt{c_2},\ \ \ \ \ \ \ \ \ \ \ m^2_{H_2^\pm} = c_1 +  \frac{v_1}{2v_3}\sqrt{c_2},\ \ \ \ \ \ \ \ \ \ \ m^2_{G_W^\pm} = 0,
\end{equation}
where:

\begin{equation}
    a_1 = -2v_1^2(\lambda_2 + \lambda_3 + \lambda_7) -2v_3^2 \lambda_7 -\frac{1}{2} \lambda_4 v_1 \left(  3v_3+\frac{v_1^2}{v_3} \right),
\end{equation}

\begin{equation}
    a_2 = 8v_1 v_3 \lambda_4 (\lambda_2 + \lambda_3 - \lambda_7) (2v_3^2 - v_1^2) + (v_1^4 + 5 v_3^4) \lambda_4^2 + 2 v_1^2 v_3^2 (8(\lambda_2 + \lambda_3 - \lambda_7)^2 - \lambda_4^2),
\end{equation}

\begin{equation}
    c_1 = - \frac{1}{2} v_1^2 (4\lambda_3 + \lambda_6 + 2\lambda_7) - \frac{1}{2} v_3^2 (\lambda_6 + 2 \lambda_7) - \frac{1}{2} \lambda_4 v_1 \left(  3v_3+\frac{v_1^2}{v_3} \right),
\end{equation}

\begin{equation}
    c_2 = (v_1^4 + 5 v_3^4) \lambda_4^2 + 2 v_1 v_3 \lambda_4 (4\lambda_3 - \lambda_6 - 2\lambda_7) (2v_3^2 - v_1^2) + v_1^2 v_3^2 [ (4\lambda_3 - \lambda_6 - 2\lambda_7)^2 - 2 \lambda_4^2 )].
\end{equation}
From the squared scalar mass matrices given above and considering that the quartic scalar couplings can take values up to $4\pi$, which is the upper bound of these couplings allowed by perturbativity, one can succesfully accomodate masses of non SM scalars in the subTeV and few TeV range.\\

Additionally, analytical stability conditions are
calculated in section \ref{appStab} but nevertheless during the numerical calculations we
employ the public tool \texttt{EVADE} \cite%
{Hollik:2018wrr,Ferreira:2019iqb},
which features the minimization of the scalar potential through 
polynomial homotopy continuation \cite{Maniatis:2012ex}, and an estimation of the decay rate of
a false vacuum \cite{Coleman:1977py,Callan:1977pt}.
From the expression for the potential (\ref{APP_scalar_pot1}) we obtain the square mass matrices for the CP-even scalars
$H_1^0$, $H_2^0$, $H_3^0$, the pseudo-scalars $A_1^0$, $A_2^0$ and the charged scalars $H_1^\pm$ and $H_2^\pm$,
where we define $H_3^0$ as the SM-like Higgs ($A_3^0$ and $H_3^\pm$ denote the EWSB nonphysical Goldstone bosons).
We will continue to assume the vev alignment $v_1=v_2$ and
here we mainly discuss analytical approximations for the CP-even scalars masses, let us denote the mass matrix
by the expression:
\begin{equation}
	\mathbf{M}_{CP-\text{even}}^{2}=\left( 
	\begin{array}{ccc}
		a & d & f \\ 
		d & b & e \\ 
		f & e & c%
	\end{array}%
	\right) ,  \label{mhh}
\end{equation}%
the specific entries in terms of the parameters of the potential can be read off from Eq. (\ref{CPEven}).
With the exception of cases where one or several entries of this matrix are zero or cases where there are
degenerate eigenvalues, we can approximate the masses of these physical scalars by the expressions \cite%
{deledalle:hal-01501221}:
\begin{eqnarray}\label{masasH}
	m_{H_3^0}^2 &=& \frac{1}{3} 
	\left(
	a + b + c - 2 \sqrt{x_1} \cos{[\Xi_s/3]}
	\right)~, \notag \\
	m_{H_1^0}^2 &=& \frac{1}{3} 
	\left(
	a + b + c + 2 \sqrt{x_1} \cos{[(\Xi_s-\pi)/3]}
	\right)~,  \\
	m_{H_2^0}^2 &=& \frac{1}{3} 
	\left(
	a + b + c + 2 \sqrt{x_1} \cos{[(\Xi_s+\pi)/3]}
	\right)~, \notag 
\end{eqnarray}
where
\begin{equation}
	x_1 = a^2 + b^2 + c^2 - a b - ac - bc + 3(d^2 + f^2 + e^2)
\end{equation}
and 
\begin{equation}
	\Xi_s = \left\{
	\begin{array}{lcc}
		\arctan \left(\frac{\sqrt{4x_1^3 - x_2^2}}{x_2} \right)
		& , & x_2 >0  \\
		\pi/2
		& , & x_2 = 0 \\
		\arctan \left(\frac{\sqrt{4x_1^3 - x_2^2}}{x_2} \right) + \pi
		& , & x_2 <0 
	\end{array}
	\right.
\end{equation}
with
\begin{eqnarray}\label{x2}
	x_2 &=& -(2a - b - c)(2b - a - c)(2c - a -  b) \notag \\
	& & + 9[(2c - a -b) d^2 + (2b -a - c) f^2 + (2a -b - c) e^2] - 54 d e f ~.
\end{eqnarray}
We will explore in detail a region of parameter space where
$H_3^0$ is the lightest of the three CP-even scalars, and we will assume it is the SM-like Higgs.
Due to the nontrivial dependence of $\Xi_s$ on the parameters, it is not possible to invert the above equations
and trade couplings for squared masses in the general case.
This represents a disadvantage at the numerical level since we have to enforce the constraint that
the mass of $H_3^0$ has to be very close to $125.5$ GeV. If we do this, it would result in very inefficient scans of the parameter space because
a large proportion of the test points in parameter space do not yield such value for the mass of the SM Higgs-like scalar.
However, there is one particular slice of parameter space where we can eliminate some of the couplings in
favor of the squared masses, as we describe below.\\

In an effort to trade generality for the possibility to perform a thorough exploration of a region of parameter space
compatible with the value of the Higgs mass, we enforce the equation
\begin{equation}\label{x20}
	x_2=0
\end{equation}
by suitable choosing one of the quartic couplings ($\lambda_{5}$) so that Eq. (\ref{x20}) is satisfied. This can
always be done since this equation is  quadratic in $\lambda_{5}$, and we choose this coupling since it does not appear
in the expressions of the masses of the pseudo-scalars nor the charged scalars.
Henceforth we will be presenting a numerical analysis of the parameter slice $\Xi_s = \pi/2$. In this hyper-region of
parameter space the equations for the masses (see Eq.(\ref{masasH})) not only take a simple form, but also allow to eliminate
two more quartic couplings ($\lambda_1$ and $\lambda_8$) in favor of $m_{H_2^0}$ and $m_{H_3^0}$. In this way we gain
control over the values of these masses, and from the relation:
\begin{equation}\label{Deltaaa}
	\Delta \equiv \sqrt{x_1/3} = m_{H_1^0}^2 - m_{H_2^0}^2 = m_{H_2^0}^2 - m_{H_3^0}^2~,
\end{equation}
which follows from the simplified equations of the masses, we see that in the explored slice of parameter space we have
the hierarchy $m_{H_1^0}^2 > m_{H_2^0}^2 > m_{H_3^0}^2$ and that these squared masses are separated by the same mass gap
$\Delta$. We shall refer to this slice as the symmetric gap region.
Having control over the value of these masses allows us to perform a scan of parameter space in which we choose the mass of
$H_3^0$ to be in a small interval (given by the current experimental error bars) around $125.5$ GeV. We then vary the mass of
$H_2^0$ in the interval $m_{H_3^0}<m_{H_2^0}<1$ TeV, while that of $H_1^0$ is determined from the value of $\Delta$ and
$m_{H_2^0}^2$.\\

For the numerical computations we implement the model in 
\texttt{SARAH}~\cite{Staub:2013tta,Staub:2009bi,Staub:2010jh,Staub:2012pb}
from which we generate corresponding model files for some of the other tools using the
\texttt{SARAH-\allowbreak SPheno} framework~\cite{Staub:2015kfa,Porod:2003um,Porod:2011nf}.
When testing a given point of parameter space, for positivity
and stability of the scalar potential we use \texttt{EVADE}, while exclusion limits from scalar searches at Tevatron, LEP
and the LHC are implemented with the aid of \texttt{HiggsBounds} \cite%
{Bechtle:2020pkv}. We impose hard cuts discarding points not complying with these constraints.
For points not filtered by the previous hard cuts we calculate numerically the model predicted observables
that are used to construct a composite likelihood function. We calculate the couplings and decay branching ratios of the $125$ GeV SM Higgs-like
and the rest of the scalars with the help of the \texttt{SARAH} generated \texttt{SPheno} code. 
In particular, we use the decay probabilities of the heavy scalars and pseudo-scalars into pairs of $\tau^+ \tau^-$
leptons in order to compare these predictions with the recent search of the ATLAS collaboration involving these
type of resonances decaying into $\tau$-lepton pairs~\cite{ATLAS:2020zms}. This specific ATLAS search was 
motivated because such decay modes can be enhanced in multi-Higgs models relative to the SM predictions.
A higher cross section for Higgs boson production in association with $b$ quarks ($bbH$) can also occur in such
scenarios, making this production channel competitive with the main gluon fusion production ($gg$F).
We calculate $bbH$ and $gg$F cross section productions for all neutral scalars using \texttt{SusHi}~\cite%
{Harlander:2012pb,Harlander:2016hcx}. While \texttt{SusHi} features these calculations for the Two Higgs Doublet Model (2HDM)
and the Minimal Supersymmetric Standard Model (MSSM), it uses a strategy of calculation based on the observation that
for example, assuming that the SM Higgs-like is the lightest scalar, it is possible to efficiently estimate the
next to leading order (NLO) production cross section for a given CP-even scalar in the model from the known
NLO production cross section for a SM Higgs of the same mass by rescaling with the LO coupling ratios
(see e.g.~\cite{Craig:2012vn}):
\begin{equation}
	\sigma_{\text{NLO}}(X \rightarrow Y) \simeq \sigma^{\text{SM}}_{\text{NLO}}(X \rightarrow Y) \times \frac{\sigma_{\text{LO}}(X \rightarrow Y)}{\sigma^{\text{SM}}_{\text{LO}}(X \rightarrow Y)} ~,
\end{equation}
the leading order ratio in this equation involves only the tree-level couplings of the scalars in the model relative to the SM Higgs boson couplings.
This is only an estimation because at the loop level it is assumed that only SM particles run in the loops. It is straightforward to use the
capabilities of \texttt{SusHi} for other multi-Higgs models by simply changing the rescaling factors appropriately.
This is also true for pseudo-scalars which only require the rescaling of
the calculation of the production rate of a pseudo-scalar Higgs boson motivated by BSM models such as the MSSM
\cite{Kauffman:1993nv,Spira:1993bb,Spira:1995rr,Harlander:2002vv,Anastasiou:2002wq}.
We use the above predictions of the model to construct the composite likelihood function:
\begin{equation}
	\label{likeScalar}
	\log {\mathcal L}_{\text{scalar}} = \log {\mathcal L}_{\text{Higgs}} + \log {\mathcal L}_{\text{ATLAS}} + \log {\mathcal L}_{h\rightarrow\gamma\gamma}
\end{equation}
using public numerical tools. We obtain the likelihood $\log {\mathcal L}_{\text{Higgs}}$ that measures how well the couplings of the SM Higgs-like 
$H^0_3$ resemble that of the already discovered SM Higgs using \texttt{HiggsSignals}~\cite{Bechtle:2020uwn}.
Specifically\footnote{
We thank an anonymous referee for this and other useful suggestions to improve the analysis.
}
, we define:
\begin{equation}
    -2\log \left( {\mathcal L}_{\text{Higgs}} /  {\mathcal L}^{\text{max}}_{\text{Higgs}}\right) = 
            \chi^2_{\text{Higgs}}
\end{equation}
where $\chi^2_{\text{Higgs}}$ is constructed to minimize the quantity:
\begin{equation}
    \left| \chi^2_{\text{SM}} - \chi^2_{\text{S4}} \right|
\end{equation}
here $\chi^2_{\text{SM}}$ refers to the total chi-square of the LHC rate measurements of the observed Higgs boson
while $\chi^2_{\text{S4}}$ is the prediction of the model under study here, both of these quantities are calculated with
\texttt{HiggsSignals}.
In this manner, the scan of the parameter space yields model predictions that are ensured to
be contained mostly on an interval close to the SM prediction which is well in agreement with 
the LHC measurements.
Note that \texttt{HiggsSignals} uses information from the public LHC repository database, specifically concerning
the discovered Higgs mass and signal strength from the LHC Run-1 observables, and therefore the reported likelihood
is basically a measurement of the so called alignment limit. 
In other words, parameter space points maximizing this likelihood 
(or in the neighboring regions of the maximum)
satisfy the conditions of the alignment limit.\\

As discussed earlier, most analysis in the matter sector have negligible impact in the
phenomenology of the scalar and dark matter sectors considered here. For this reason, we do not
consider constraints arising from both FCNC and CKM fits 
simultaneously with the ones from the log-likelihood 
observables defined in Eq. \eqref{likeScalar}.
Moreover, the consistency of this analysis of the scalar sector with that of the previous sections is ensured 
since in the latter case the alignment limit is explicitly assumed while in the former it is enforced by the 
likelihood $\log {\mathcal L}_{\text{Higgs}}$ present in the above same equation.
In a similar manner, the observables studied in the previous sections can be regarded 
as weakly correlated with
the considered observables in the dark sector (see next section), 
such as the relic abundance which 
depends on annihilation cross sections that are dominated by channels involving the heaviest 
fermion family or the gauge bosons, or DM-nucleon dispersion cross sections which are not sensitive
to details of the quark masses or mixings, but rather to the values of the measured nucleon
masses and nuclear form factors. 

For the likelihood $\log {\mathcal L}_{\text{ATLAS}}$ which implements the public data from the ATLAS search mentioned before 
 we make use of the capabilities of \texttt{HiggsBounds}
\cite{Bechtle:2015pma,Bechtle:2020pkv},
whilst for the likelihood regarding the branching ratio of the SM-like Higgs into two photons
we use the experimental value~\cite{ParticleDataGroup:2024cfk}:
\begin{equation}
    \text{BR}^\text{exp}_{h\rightarrow\gamma\gamma} = (2.5\pm 0.20)\times 10^{-3}
\end{equation}
to construct a simple chi-square function defining
$-2 \log \left( {\mathcal L}_{h\rightarrow\gamma\gamma} / {\mathcal L}^{\text{max}}_{h\rightarrow\gamma\gamma}  \right)$.
To the likelihood (\ref{likeScalar}) we add the corresponding ones from the dark sector discussed in the 
next section and maximize this composite likelihood:
\begin{equation}
    \log {\mathcal L} = \log {\mathcal L}_{\text{scalar}} + \log {\mathcal L}_{\text{DM}}  
\end{equation}
Finally, we perform the scan of the
parameter space and construct the likelihood profiles using \texttt{Diver} \cite{Martinez:2017lzg,DarkMachinesHighDimensionalSamplingGroup:2021wkt,Scott:2012qh} (in standalone
mode).
\begin{figure}[tbp]
	\centering
	\includegraphics[width=0.7\textwidth]{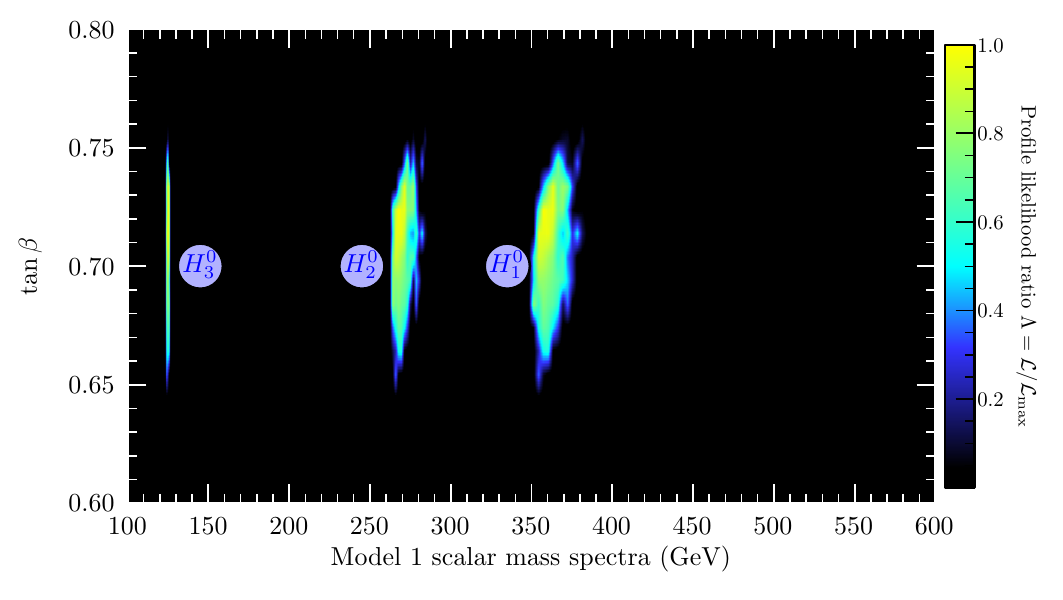}\newline%
	\includegraphics[width=0.7\textwidth]{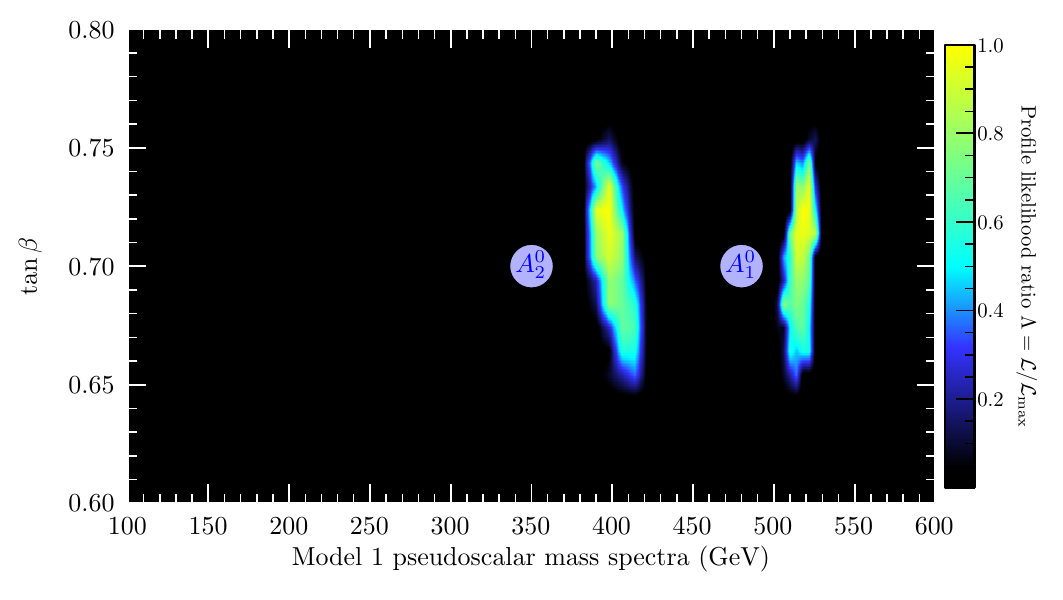}\newline%
	\includegraphics[width=0.7\textwidth]{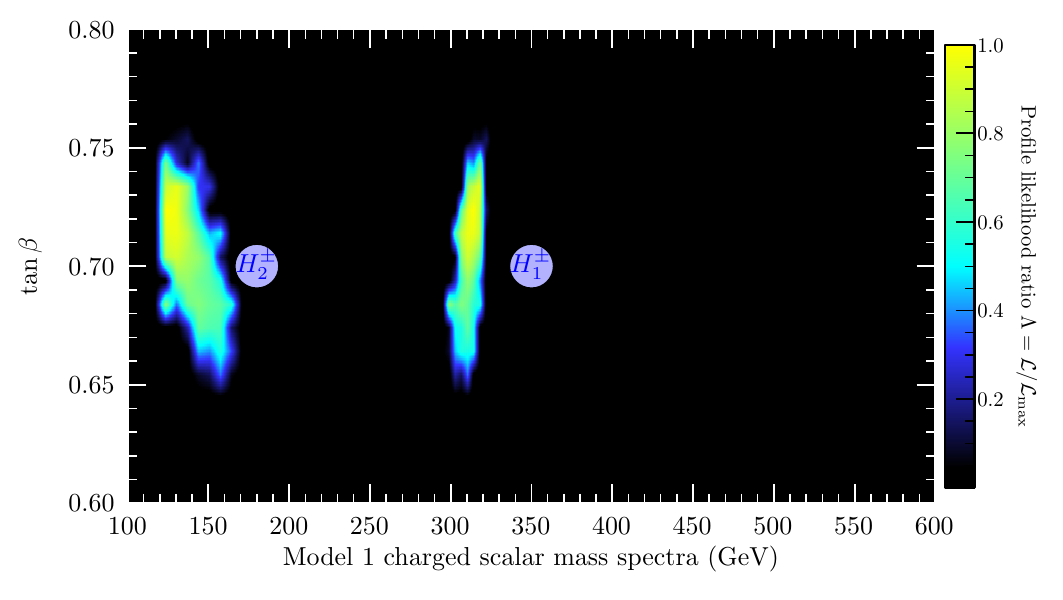} \newline%
	\caption{
		Model 1 composite likelihood as a function of
		the CP-even scalar masses (top panel), pseudo-scalar masses (middle panel), charge scalar masses (bottom panel) and $\tan\beta$. Bright regions are most compatible with observations, while dark regions are completely excluded.
  }
	\label{plotScalarsSinglet}
\end{figure}
\begin{figure}[tbp]
	\centering
	\includegraphics[width=0.7\textwidth]{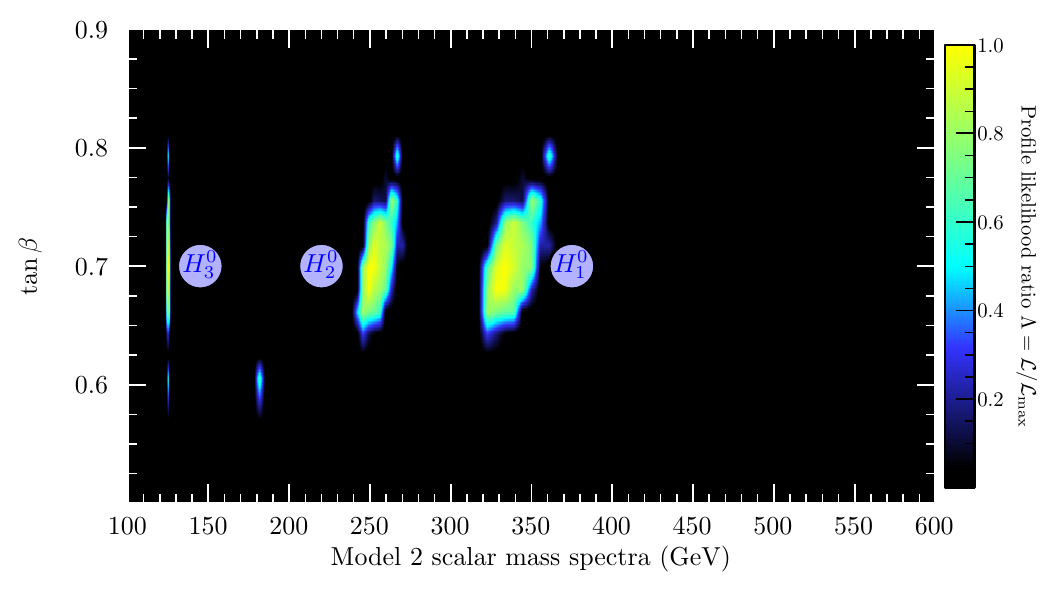}\newline%
	\includegraphics[width=0.7\textwidth]{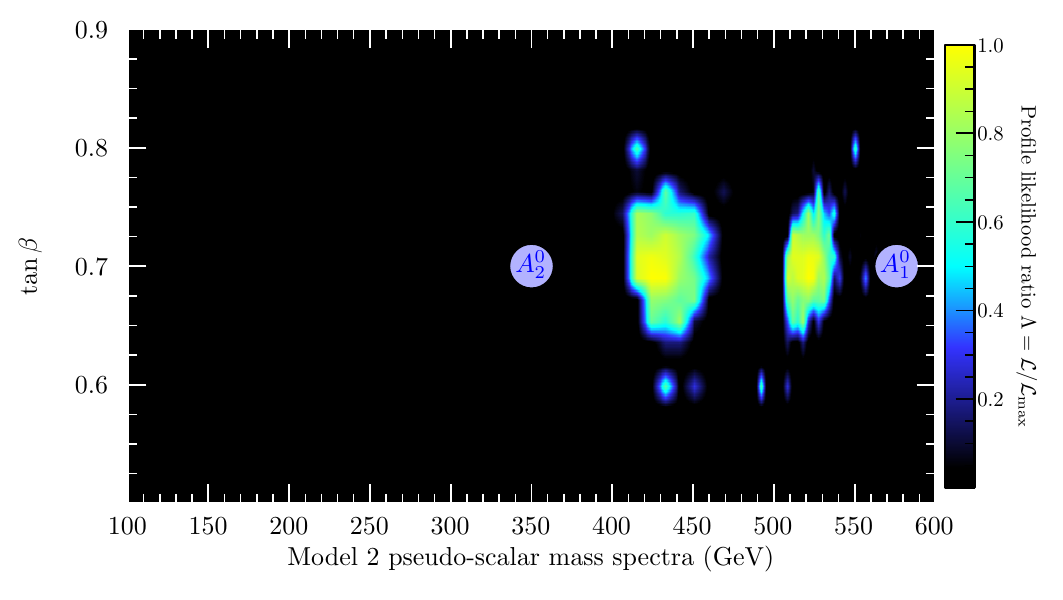}\newline%
	\includegraphics[width=0.7\textwidth]{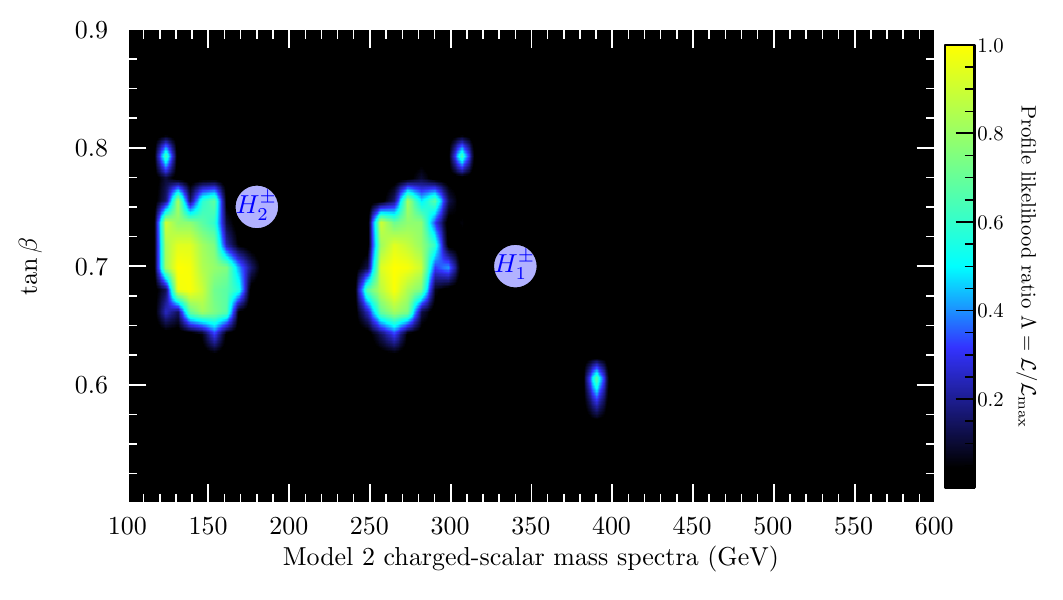} \newline%
	\caption{
		Model 2 composite likelihood as a function of
		the CP-even scalar masses (top panel), pseudo-scalar masses (middle panel), charge scalar masses (bottom panel) and $\tan\beta$. Bright regions are most compatible with observations, while dark regions are completely excluded.
  }
	\label{plotScalarsDoublet}
\end{figure}
Fig.~\ref{plotScalarsSinglet} and \ref{plotScalarsDoublet} show the obtained profiles with 
respect to the full composite likelihood ${\mathcal L}$ for each model respectively,
showing the spectra of masses of the scalars 
and its correlation with the value of $\tan\beta \equiv \sqrt{2} v1/v3$. 
We note that the phenomenological analysis results in the model's consistency with observations
only for small values of $\tan\beta$, concretely, this observable appears to be constrained to 
take values in between $\sim 0.65$ and $\sim 0.75$ in both models at the 
preferred values of the masses.\\

For model 1 the masses of $H_2^0$ and $H_1^0$ most favored lie in $\sim 270$ GeV and $\sim 360$ GeV respectively,
those of $A_2^0$ and $A_1^0$ in $\sim 400$ GeV and $\sim 520$ GeV and those 
of $H_2^pm$ and $H_1^pm$ in $\sim 130$ GeV and $\sim 315$ GeV respectively. 
For model 2 the corresponding masses of $H_2^0$ and $H_1^0$ lie in $\sim 255$ GeV and $\sim 335$ GeV,
those of $A_2^0$ and $A_1^0$ in $\sim 430$ GeV and $\sim 520$ GeV and those 
of $H_2^pm$ and $H_1^pm$ in $\sim 135$ GeV and $\sim 270$ GeV respectively.

\section{The scalar potential for a $S_4$ triplet}\label{apptrip}
\lhead[\thepage]{\thesection. scalar potential $S_4$ triplet}

The relevant terms determining the VEV directions of any $S_4$ scalar triplet are:
\begin{eqnarray}
V_T &= & -g_{\Psi}^2 \left( \Psi \Psi^*\right)_\textbf{1}+k_1 \left( \Psi \Psi^*\right)_\textbf{1}\left( \Psi \Psi^*\right)_\textbf{1} +k_2 \left( \Psi \Psi^*\right)_{3_1}\left( \Psi \Psi^*\right)_{3_1}+ k_3 \left( \Psi \Psi^*\right)_{3_2}\left( \Psi \Psi^*\right)_{3_2} \\
\nonumber & &+ k_4 \left( \Psi \Psi^*\right)_\textbf{2} \left( \Psi \Psi^*\right)_\textbf{2}
+H.c.
\label{VTS4}
\end{eqnarray} 
where $\Psi=\xi, \eta, \rho, \Phi_e,\Phi_{\mu},\Phi_{\tau}$. Note that we restrict to a particular simplified benchmark scenario where the mixings between $S_4$ scalar triplets is neglected. The part of the scalar potential for each $S_4$ scalar triplet has five free parameters: one bilinear and
four quartic couplings. The minimization conditions of the scalar potential for a $S_4$ triplet yield the
following relations:
\begin{eqnarray} \label{vev-S4}
\nonumber \frac{\partial \langle V_T \rangle}{\partial v_{\Psi_1}}   & = & -2g_{\Psi}^2 v_{\Psi_1} +8 k_2 v_{\Psi_1}  \left(v_{\Psi_2}^2 +v_{\Psi_3}^2 \right) + 4k_1 v_{\Psi_1} \left( v_{\Psi_1}^2 + v_{\Psi_2}^2 + v_{\Psi_3}^2 \right) - \frac{4}{3} k_4 v_{\Psi_1} \left( -2v_{\Psi_1}^2 + v_{\Psi_2}^2 + v_{\Psi_3}^2 \right)  \\
 \nonumber & = & 0 \\
 \nonumber & & \\
\frac{\partial \langle V_T \rangle}{\partial v_{\Psi_2}}   & = & -2g_{\Psi}^2 v_{\Psi_2} + 8 k_2 v_{\Psi_2} \left( v_{\Psi_1}^2 + v_{\Psi_3}^2  \right)+ 4 k_1 v_{\Psi_2} \left( v_{\Psi_1}^2 + v_{\Psi_2}^2 + v_{\Psi_3}^2 \right) +2k_4  v_{\Psi_2}  \left( v_{\Psi_2}^2 - v_{\Psi_3}^2\right) \\
\nonumber & & + \frac{2}{3} k_4 v_{\Psi_2} \left(-2v_{\Psi_1}^2 +v_{\Psi_2}^2 + v_{\Psi_3}^2 \right) \\
 \nonumber & = & 0 \\
\nonumber  & & \\
\nonumber \frac{\partial \langle V_T \rangle}{\partial v_{\Psi_3}}   & = & -2g_{\Psi}^2 v_{\Psi_3}
+8 k_2 v_{\Psi_3}   \left( v_{\Psi_1}^2 + v_{\Psi_2}^2 \right) + 4k_1 v_{\Psi_3} \left( v_{\Psi_1}^2 + v_{\Psi_2}^2 + v_{\Psi_3}^2 \right)- 2k_4  v_{\Psi_3}  \left( v_{\Psi_2}^2 - v_{\Psi_3}^2\right) \\
\nonumber & & + \frac{2}{3} k_4v_{\Psi_3} \left(-2v_{\Psi_1}^2 +v_{\Psi_2}^2 + v_{\Psi_3}^2 \right) \\
 \nonumber & = & 0.
\end{eqnarray}
From the scalar potential minimization equations, for the $S_4$ scalar triplet $\Phi_e$, we obtain the following relation:
\begin{eqnarray}
g_{\Phi_e}^2 & = & \frac{2}{3\left(3k_1 + 2k_4 \right)  }v_{\Phi_e}^2
\end{eqnarray}
This shows that the VEV configuration of the $S_4$ triplet $\Phi_e$
given in Eqs. \eqref{eq:trip1} and \eqref{eq:trip2}, is in accordance with the scalar potential
minimization condition of Eq. \eqref{vev-S4}. The VEV configurations of the other $S_4$ triplets in our model are also consistent with the scalar potential minimization conditions, which can be demonstrated by using the same procedure described in this appendix. These results show that the VEV directions for the $S_4$ triplets $\xi$, $\eta$, $\rho$, $\Phi_e$, $\Phi_{\mu}$, $\Phi_{\tau}$ are consistent with a global minimum of the scalar potential for a large region of parameter space.

\section{Stability conditions}\label{appStab}
\lhead[\thepage]{\thesection. Stability conditions}

The stability conditions of the low energy scalar potential will be given by its quartic terms since these will be the dominant ones for large values of the field components. Therefore, we define the following bilinear
conventions of the scalar fields:
\begin{eqnarray}
a &=&\Xi _{1}^{\dag }\Xi _{1}\ \ ;\ \ b=\Xi _{2}^{\dag }\Xi _{2}\ \ ;\ \
c=\Xi _{3}^{\dag }\Xi _{3} \\
d &=&\Xi _{1}^{\dag }\Xi _{2}+\Xi _{2}^{\dag }\Xi _{1}\ \ ;\ \ e=i(\Xi
_{1}^{\dag }\Xi _{2}-\Xi _{2}^{\dag }\Xi _{1}) \\
\Xi _{1}^{\dag }\Xi _{3} &=&f+ig\ \ ;\ \ \Xi _{2}^{\dag }\Xi _{3}=h+ik
\end{eqnarray}%
by using this new definition, we can rewrite the quartic terms of the scalar
potential as follows:
\begin{eqnarray}
V_4&=&
(\lambda_1+\lambda_3)(a^2+b^2)+(\lambda_6+\lambda_7)(f^2+g^2+h^2)+2\left(%
\lambda_1-\lambda_3+\frac{\lambda_5}{2}\right)ab  \notag \\
&&+\left(\sqrt{\lambda_3}d-\sqrt{\lambda_7}f\right)^2+\left(\sqrt{%
\lambda_3\lambda_7}+\lambda_4\right)df +(\sqrt{\lambda_5}b-\sqrt{\lambda_8}%
c)^2 +\lambda_4(ah-bh)  \notag \\
&&+\lambda_7(h^2-3g^2-2k^2)+\sqrt{\lambda_5\lambda_8}bc+\lambda_5(c-b)b-e^2%
\lambda_2 +g^2\lambda_6 ~.
\end{eqnarray}
By using the method employed in~\cite{Bhattacharyya:2015nca,Abada:2021yot}, we find the following stability conditions for the low energy scalar potential: 
\begin{align}
\lambda_3&\geq 0 & \lambda_7&\geq 0 & \lambda_5&\geq 0 & \lambda_8\geq 0&,\hspace{1.5cm}\lambda_2\leq 0 \\
\lambda_1+\lambda_3&\geq 0 & \lambda_6+\lambda_7&\geq 0 & 
\lambda_1-\lambda_3+\frac{\lambda_5}{2}&\geq 0 & \sqrt{\lambda_3\lambda_7}%
+\lambda_4&\geq 0 & 
\end{align}

\section{Dark sectors}\label{dark-sector}
\lhead[\thepage]{\thesection. Dark sectors}

We now describe the dark sectors of both models introduced in previous sections.
We couple the $Z_2$ charged scalar fields to the active scalars in a minimalistic way
and consistent with their $S_4$ assignments. The scalar potential for each model is taken
as the sum of the active scalars' potential of the previous section with the respective
one containing the dark scalars. Denoting by $V_{\text{m1}}$ and $V_{\text{m2}}$
the model 1 and model 2 total scalar potentials, we take:
\begin{equation}
	V_{\text{m1}} = V - \mu_\phi^2\phi^2 + \lambda_\phi \, \phi^4 + \lambda_9 \, \phi^2 \left( \Xi _{I}^{\dagger }\Xi
	_{I}\right) _{\mathbf{1}_{1}} + \lambda_{10} \, \phi^2 \left( \Xi_{3}^{\dagger }\Xi _{3}\right)
\end{equation}

\begin{eqnarray}
	V_{\text{m2}} &=&V - \mu^2_4 \left( \Xi_{4}^{\dagger }\Xi _{4}\right)
	+\lambda _{11}\left( \Xi_{4}^{\dagger }\Xi _{4}\right) \left( \Xi _{I}^{\dagger }\Xi _{I}\right) _{%
		\mathbf{1}_{1}}+\lambda _{12}\left[ \left( \Xi _{4}^{\dagger }\Xi _{I}\right) \left( \Xi
	_{I}^{\dagger }\Xi _{4}\right) \right] _{\mathbf{1}_{1}}  \notag  \\
	&&+\lambda _{13}\left[
	\left( \Xi _{4}^{\dagger }\Xi _{I}\right) \left( \Xi _{4}^{\dagger }\Xi
	_{I}\right) +h.c.\right] +\lambda _{14}\left( \Xi _{4}^{\dagger }\Xi
	_{4}\right) ^{2} +\lambda _{15}\left( \Xi _{3}^{\dagger }\Xi
	_{3}\right)  \left( \Xi _{4}^{\dagger }\Xi
	_{4}\right)  ~, 
\end{eqnarray}
where for simplicity we have assumed $\phi$ to be real, and $V$ is given by Eq. (\ref{APP_scalar_pot1}).
We keep checking the stability of each potential numerically and maintain the hard cuts described 
in the previous section. Both models offer the possibility of a scalar or fermion dark matter candidate,
however, the right handed neutrinos do not couple to quarks at tree level and therefore only indirect 
DM detection observables would be of phenomenological interest when one of the fermions is the lightest
($Z_2$) odd particle (LOP), a typical example of this would be the scotogenic model, see e.g. \cite%
{Tao:1996vb,Ma:2006km,deBoer:2021pon}.
Nevertheless, in our proposals it is easy to check that for a fermion LOP, annihilation 
is only possible to pairs of neutrinos in model 1 and 2, or pairs of charged leptons in model 2. 
Therefore, in both models we deem much more interesting the case of a scalar LOP, where thanks to
the couplings of the dark scalars to the active ones it is possible to have tree level scattering
amplitudes between a scalar LOP and quarks, allowing the phenomenological analysis of direct detection (DD)
of such candidates, these type of models are commonly referred to as Higgs portals, see e.g. \cite%
{Arcadi:2019lka}.
Presumably, the DD constraints are currently the most stringent ones compared to
anti-matter signals (fermion LOP) or gamma ray fluxes (scalar LOP) from annihilation 
of DM in these models,
for a thorough review of these topics see e.g.~\cite%
{Bertone:2004pz}.
In model 1 we take $\phi$ as the DM candidate and in model 2 we assume that from the components of the
inert doublet $\Xi_4$, which we denote $H_4^0$, $A_4^0$ and $H_4^\pm$, $H_4^0$ is the lightest.\\

With this rationale, for both models we construct a log-likelihood function involving the observables in the (visible) scalar sector
of the previous section, and the DD and relic abundance observables:
\begin{figure}[tbp]
	\centering
	\includegraphics[scale=0.6]{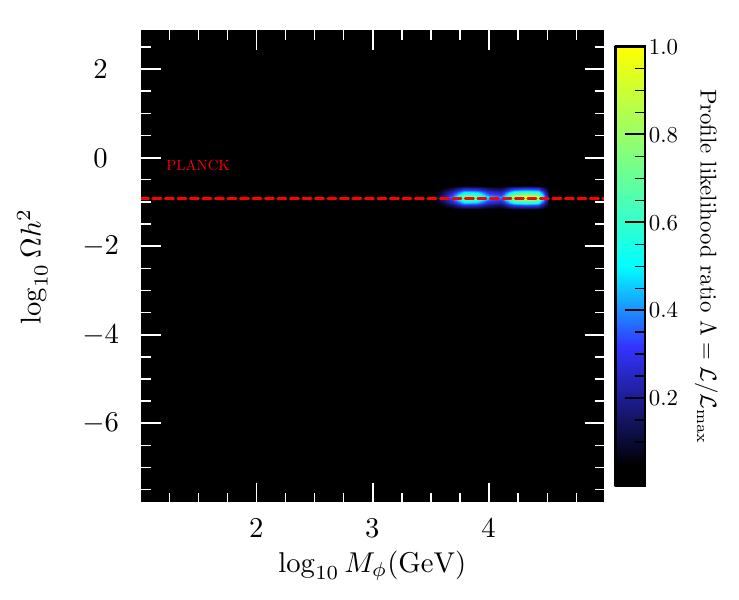}%
	\includegraphics[scale=0.6]{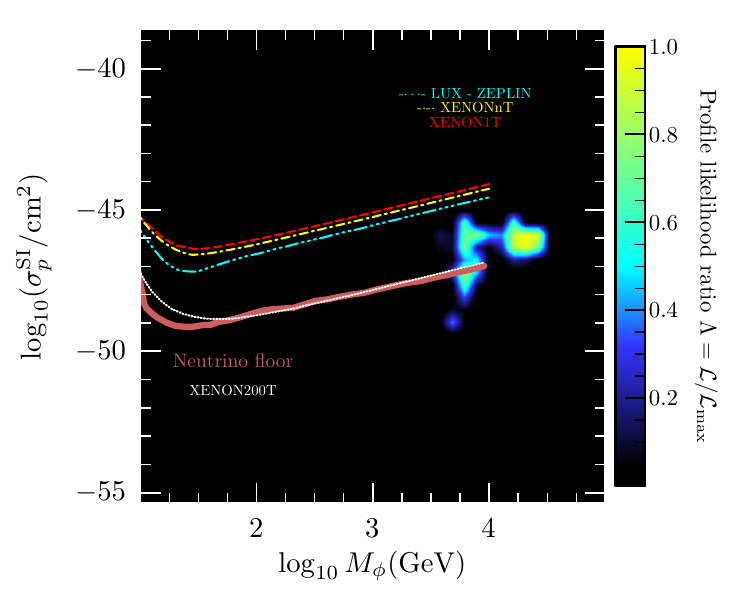}\newline%
	\includegraphics[scale=0.6]{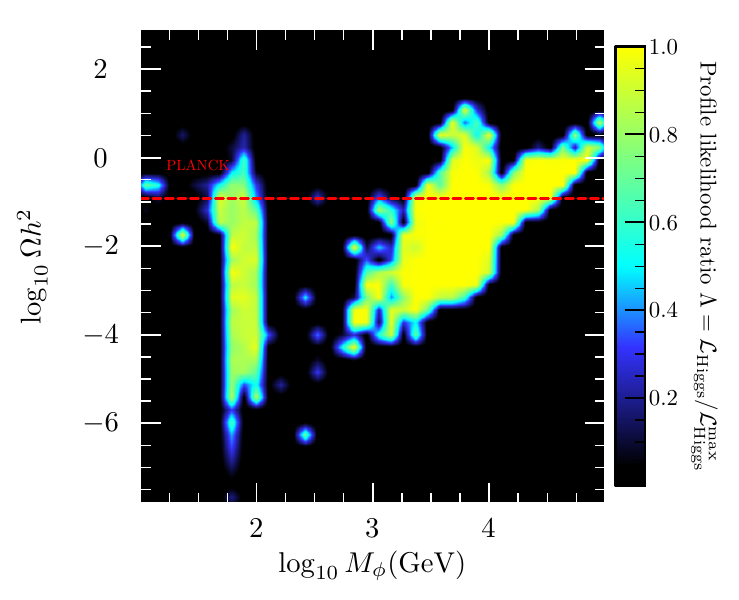}%
	\includegraphics[scale=0.6]{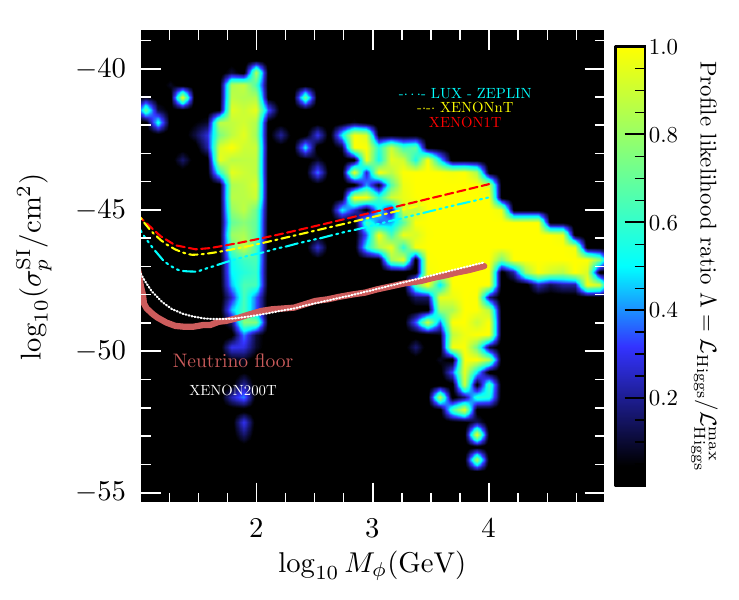}\newline%
	\caption{
		Composite likelihood as a function of model 1
		DM candidate mass, DM abundance (left panels) and SI DM-proton cross section (right panels).
        Top panels show the profiles with respect to the 
        full likelihood function ${\mathcal{L}}$ while
        the bottom panels show the corresponding profiles 
        with the same sampled points but with respect to the
        partial likelihood ${\mathcal{L}}_{\text{Higgs}}$.
        For more details see the main text.
	}
	\label{model1DM}
\end{figure}

\begin{equation}
	\log {\mathcal L} = \log {\mathcal L}_{\text{scalar}} + \log {\mathcal L}_{\text{DD}} + \log {\mathcal L}_{\Omega h^2}
	\label{likeTot} ~.
\end{equation}
For the numerical calculation of the relic density, as well as the DM-nucleon scattering cross sections, we use the capabilities of \texttt{Micromegas} \cite%
{Belanger:2013oya,Belanger:2014vza,Barducci:2016pcb,Belanger:2018ccd}. We construct ${\mathcal L}_{\Omega h^2}$ as a basic Gaussian likelihood with respect
to the PLANCK \cite%
{Planck:2018vyg} measured value, while the likelihood ${\mathcal L}_{\text{DD}}$ involves publicly available data from
the direct detection XENON1T experiment \cite{XENON:2018voc}. We use
the numerical tool \texttt{DDCalc}
to compute the Poisson likelihood given by
\begin{equation}
	\mathcal{L}_\text{DD} = \frac{(b+s)^o \exp{\{-(b+s)\}}}{o!}
\end{equation}
where $o$ is the number of observed events in the detector and $b$ is the
expected background count. From the model's predicted DM-nucleon scattering cross
sections as input, \texttt{DDCalc} computes the number
of expected signal events $s$ for given DM local halo and velocity
distribution models (we take the tool's default ones, for specific details
on the implementation such as simulation of the detector efficiencies and
acceptance rates, possible binning etc. see \cite%
{GAMBITDarkMatterWorkgroup:2017fax,GAMBIT:2018eea}).\\

In Fig. \ref{model1DM} top panel we show the main results for the DM sector of model 1 with respect to the DM abundance 
(left) and the DM-proton scattering cross section (right).
It is instructive to analyze the same set of points shown here
taking into account the partial likelihood 
${\mathcal L}_{\text{Higgs}}$,
with respect to this partial likelihood, we show the likelihood profiles of the same observables in the bottom panel of the same
figure.
As expected, the top-left plot with the full log-likelihood
is just a slim horizontal bright band around the Planck measured value in the regions where the DM candidate can accommodate
for $100\%$ of the observed DM abundance, while in the
bottom-left plot we can observe vast regions of the parameter space in which the candidate can only be a fraction of 
such abundance.
Notice that there is a region with masses below $100$ GeV
that predict the correct DM abundance, but these regions
as well as most of the regions above $100$ GeV turn out to 
have very low likelihood with respect to the Direct Detection
experiment.\\

The plots in the right panels show the dependence
of the likelihood on the DM mass and the DM-proton spin independent (SI)
cross section,
we also depict the 90\% CL upper limit on the SI cross section from the
XENON1T (1t $\times$ yr)  \cite{XENON:2018voc}, 
the XENONnT \cite{XENON:2023cxc} and the 
LUX-ZEPLIN (LZ) \cite{LZ:2022lsv}
experiments, alongside with the XENON experiment
multi ton-scale time projection to 200 t $\times$ yr of reference
\cite%
{Schumann:2015cpa} 
	(for better comparison with the other curves we extrapolated linearly the
	data available from this reference from 1 TeV up to 10 TeV)
and an estimation of the neutrino floor \cite%
{Billard:2013qya}.
We can see from the top-right panel of this figure 
that the DM candidate of model 1
is strongly constrained by the analysis. There is only a very small region of parameter
space  with a likelihood ratio above $\sim 0.8$ in the neighborhood of slightly above the value 
$M_\phi \sim 10$ TeV. 
Due to the constraints from the XENON1T observations, the
allowed region lies below the respective exclusion curve, but we can see from the 
comparison with the projected exclusion limits of the 200 ton upgrade that it will
be possible to exclude this region and its surroundings in the near future.
Notice from the bottom right panel that there are large regions
of parameter space that are consistent with a SM-like Higgs
and are not excluded by the DD exclusion limits, but turn out
to be not compatible with the other observables, this shows the
importance of using a composite likelihood for the analysis.\\
\begin{figure}[tbp]
	\centering
	\includegraphics[scale=0.6]{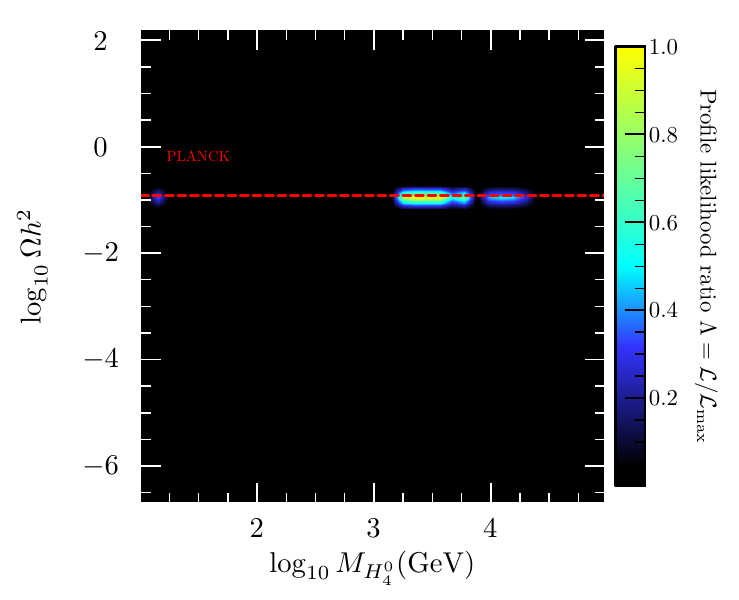}%
	\includegraphics[scale=0.6]{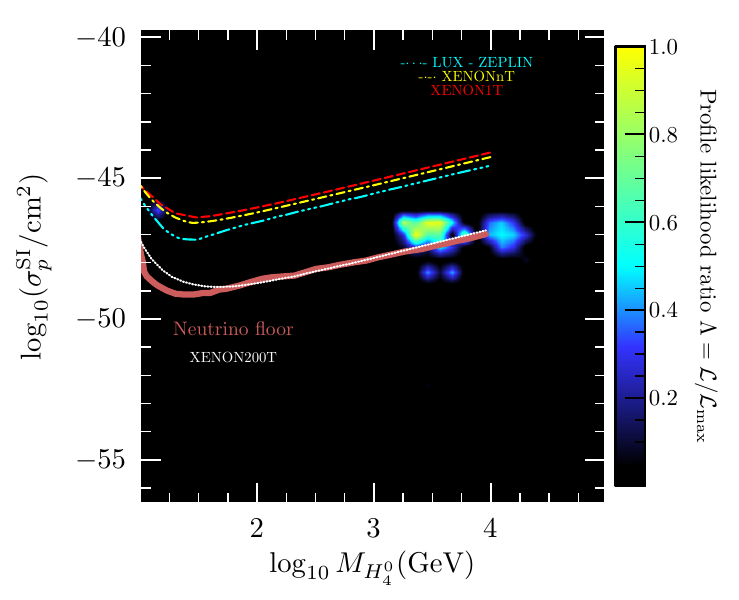}\newline%
	\includegraphics[scale=0.6]{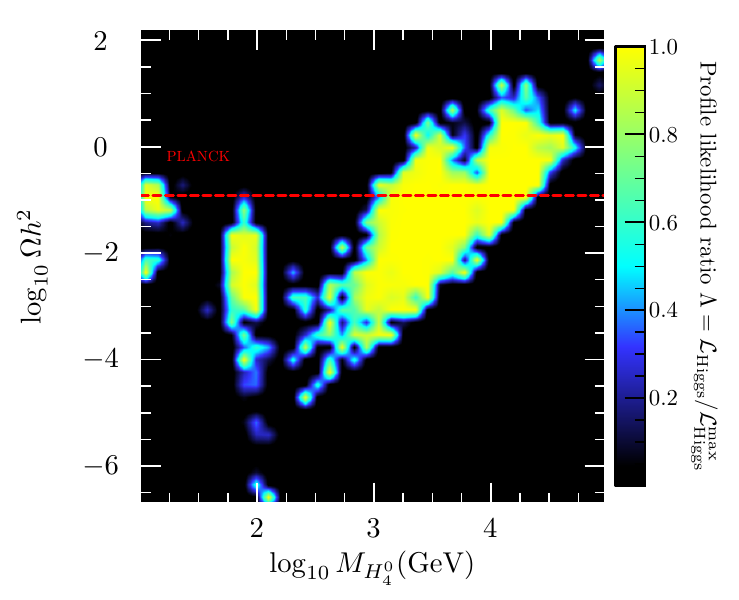}%
	\includegraphics[scale=0.6]{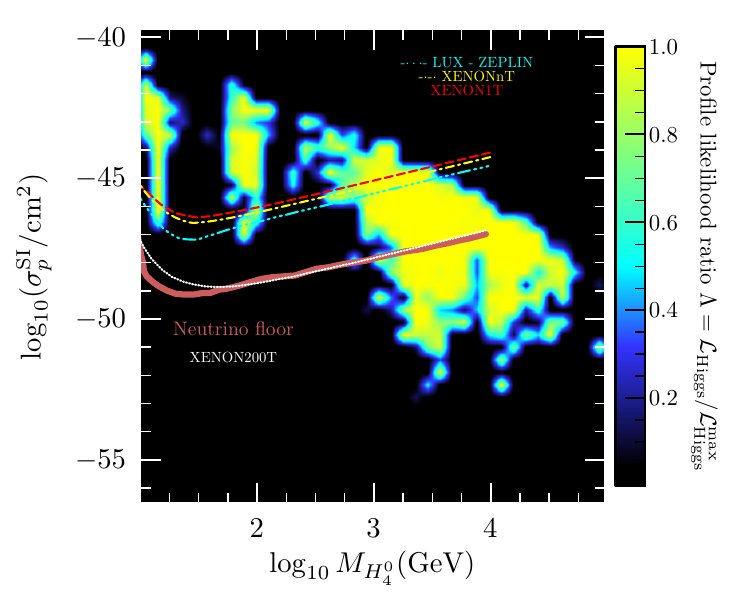}\newline%
	\caption{
		Composite likelihood as a function of model 2
		DM candidate mass, DM abundance (left panels) and SI DM-proton cross section (right panels).
        Top panels show the profiles with respect to the 
        full likelihood function ${\mathcal{L}}$ while
        the bottom panels show the corresponding profiles 
        with the same sampled points but with respect to the
        partial likelihood ${\mathcal{L}}_{\text{Higgs}}$.
        For more details see the main text. }
	\label{model2DM}
\end{figure}

In Fig. \ref{model2DM} we present the corresponding plots
for model 2. The situation is very similar to the case of 
the singlet, the full analysis strongly constrains the
parameter space of the model with only a small region
surviving current experimental observations.
The biggest difference occurs in the location of the
maximum of the composite likelihood function, which
is located below 10 TeV.
This characteristic presumably stems from the fact that in model 2
there are other DM particles ($A_4^0$ and $H_4^\pm$) that have influence on the 
value of the DM abundance by means of the co-annihilation process.
When the mass of these particles is close to the LOP mass their contribution 
to the annihilation cross section at the freeze out epoch is non-negligible and
enhances it, making the annihilation of DM particles more efficient and thus
diminishing the final DM abundance.
Comparing both models, this seem to be particularly
important for masses just above 1 TeV, where the doublet
has regions which predict the correct abundance and the singlet does not and these regions are below the current exclusion
limits from DD.\\

Finally, we mention that in both models there are regions below
100 GeV that predict a correct (or close to correct) DM 
abundance but are excluded by current DD limits.
This is in sharp contrast with models featuring a scalar
DM candidate but with no extra Higgses, e.g.~\cite{Ilnicka:2015jba, Kanemura:2016sos,Khojali:2022squ},
which can accommodate a viable DM candidate with mass below 
100 GeV.
In the present model this characteristic stems from the fact
that we have analyzed a region of parameter space where active scalars with masses below 600 GeV are present
and their couplings to the DM scalar induce a contribution
to the DM-nucleon scattering amplitude which, according to our
results, enhances the cross section in such a way that makes
it nonviable with respect to current exclusion limits.
Of course we could relax the condition (\ref{x20})
and analyze a region where the extra active scalars are
very massive or are decoupled from the low mass particles,
in such a case their contribution to the scattering amplitudes
should be negligible and we would recreate the situation
of those models which do not have extra Higgses.

\newpage
$\ $
\thispagestyle{empty} 
\chapter{Phenomenological aspects of the fermion and scalar sectors of a $S_4$ flavored 3-3-1 model}\label{cap.modelo331}
\markboth{331 MODEL}{331 MODEL}

The next chapter is based on the third paper of this thesis~\ref{paper:331}, where we will see an extension see an extension of the Standard Model, in what is known in the literature as the 331 models, where the SM symmetry is extended to a $SU(3)_C\times SU(2)_L\times U(1)_X$. These models have been widely studied since they provide an explanation for the number of families in the fermionic sector and there is cancellation of chiral anomalies, which occurs when there are non-universal gauge assignments under the group $U(1)_X$ for the left quark fields and the number of the fermionic triplet and antitriplet of $SU(3)_L$ are equal, however, for this to be fulfilled, the fermion family number must be a multiple of three. Along with the above, the 331 models also have the characteristics that the Peccei-Quinn (PQ) symmetry can be obtained naturally, which provides a possible solution to the strong CP problem. In addition, these models have several sources of CP violation and give an explanation for the quantization of the electric charge.\\

In this chapter, we propose an extension of the SM through a 331 model with (RH) neutrinos, also adding a global lepton symmetry $U(1)_{Lg}$ to ensure the conservation of the lepton number, a discrete non-abelian symmetry $S_4$ to reproduce the masses and mixing of the fermionic sector and three other auxiliary cyclic symmetries $Z_4\times Z_4^{\prime}\times Z_2$, where the $Z_4$ symmetry is introduced to obtain a texture with zeros in the entries of the up type quark mass matrix, in addition, the symmetry $Z_4^{\prime}$ together with the VEV pattern of the scalar triplets associated with the sector of charged leptons is necessary to get a diagonal charged lepton mass matrix. Finally, the $Z_2$ symmetry is necessary to obtain all the complete entries of the third column of the mass matrix of the down quarks and thus with the values of the VEVs of the scalars that participate in the Yukawa terms of this sector. Given that the charged lepton mass matrix is predicted to be diagonal, the lepton mixing entirely arises from the neutrino sector, where we will use an inverse seesaw mechanism \cite{Mohapatra:1986bd} mediated by right-handed heavy neutrinos to generate the tiny active neutrinos masses. The $S_4$ symmetry group is a compelling choice due to its unique properties and its ability to efficiently describe the observed pattern of fermion masses and mixing angles in the standard model. As the smallest non-abelian group with irreducible representations of doublet, triplet, and singlet, $S_4$ allows for an elegant accommodation of the three fermion families. Furthermore, its cyclic structure and spontaneous breaking provide a suitable framework for generating fermion masses through mechanisms such as the Froggatt-Nielsen mechanism \cite{Froggatt:1978nt} for charged fermions and the inverse seesaw mechanism \cite{Mohapatra:1986bd} for light active neutrinos. The application of $S_4$ has demonstrated success in describing the observed patterns of SM fermion masses and mixings 
\cite{Altarelli:2009gn,Bazzocchi:2009da,Bazzocchi:2009pv,deAdelhartToorop:2010vtu,Patel:2010hr,Morisi:2011pm,Altarelli:2012bn,Mohapatra:2012tb,BhupalDev:2012nm,deMedeirosVarzielas:2012apl,Ding:2013hpa,Ishimori:2010fs,Ding:2013eca,Hagedorn:2011un,Campos:2014zaa,Dong:2010zu,Vien:2015fhk,deAnda:2017yeb,deAnda:2018oik,CarcamoHernandez:2019eme,CarcamoHernandez:2019iwh,deMedeirosVarzielas:2019cyj,DeMedeirosVarzielas:2019xcs,Chen:2019oey,Garcia-Aguilar:2022gfw,CarcamoHernandez:2022vjk}.\\

The non-abelian symmetry $S_4$ provides a solid theoretical framework for constructing scalar fields and understanding various phenomenologies. Scalar fields, transforming according to the representations of $S_4$, allow us to determine the phenomenon of meson oscillations. These oscillations are of great interest as they provide valuable information about flavor symmetry violations in the Standard Model. In addition, the scalar mass spectrum is analyzed and discussed in detail. Its analysis allows to study the implications of our model in the decay of the Standard Model like Higgs boson into a photon pair as well as in meson oscillations. Finally, besides providing new physics contribution to meson mixingz and Higgs decay into two photons, the considered model succesfully satisfies the constraints imposed by the oblique parameters, where the corresponding analysis is performed consideing the low energy effective field theory below the scale of spontaneous breaking of the $SU(3)_L\times U(1)_X\times U(1)_{L_g}$ symmetry. These parameters, derived from precision measurements in electroweak physics, are essential for evaluating the consistency between the theoretical model and experimental data. The consistency of our model with the oblique parameters demonstrates its ability to reproduce experimental observations in the context of electroweak physics for energy scales below $1$ TeV.

\section{\label{model2}The model}
\lhead[\thepage]{\thesection. The model}

 The model is based on the $SU(3)_C \times SU(3)_L \times U(1)_X $ gauge symmetry, supplemented by the $S_4$ family symmetry and the $Z_4 \times Z'_{4} \times Z_2$ auxiliary cyclic symmetries, whose spontaneous breaking generates the observed SM fermion mass and mixing pattern. We also introduce a global $U(1)_{L_g}$ of the generalized leptonic number $L_g$ \cite{Chang:2006aa,CarcamoHernandez:2017cwi,CarcamoHernandez:2019lhv}. That global $U(1)_{L_g}$ lepton number symmetry will be spontaneously broken down to a residual discrete lepton number symmetry $Z_2^{(L_g)}$ by a VEV of the gauge-singlet scalars $\varphi$ and $\xi$ to be introduced below. The correspoding massless Goldstone bosons, the Majoron, are phenomenologically harmless since they are gauge singlets. It is worth mentioning that under the discrete lepton number symmetry $Z_2^{(L_g)}$, the leptons are charged and the other particles are neutral, thus implying that in any interaction leptons can appear only in pair, thus, forbidding proton decay. 
  The $S_4$ symmetry is the smallest non-Abelian discrete symmetry group having five irreducible representations (irreps), explicitly, two singlets (trivial and no-trivial), one doublet and two triplets (3 and 3') \cite{Ishimori:2010au}. While the auxiliary cyclic symmetries $Z_4$, $Z'_4$ and $Z_2$ select the allowed entries of the SM fermion mass matrices that yield a viable pattern of SM fermion masses and mixings, and at the same time, the cyclic symmetries also allow a successful implementation of the inverse seesaw mechanism. The $\mathcal{G}$ chosen symmetry exhibits the following three-step spontaneous breaking:
\begin{gather}
\mathcal{G}=SU(3)_{C}\times SU(3)_{L}\times U(1)_{X}\times
U(1)_{L_{g}}\times S_{4}\times Z_{4}\times Z_{4}^{\prime }\times Z_{2} \\
\Downarrow \Lambda _{\text{int }}  \notag \\
SU(3)_{C}\times SU(3)_{L}\times U(1)_{X}\times U(1)_{L_{g}}\notag \\
\Downarrow v_{\chi }  \notag \\
SU(3)_{C}\times SU(2)_{L}\times U(1)_{Y}\times Z_2^{(L_g)}  \notag \\
\ \ \ \ \ \Downarrow v_{\eta_2},v_{\rho }  \notag \\
SU(3)_{C}\times U(1)_{Q}\times Z_2^{(L_g)}  \notag
\end{gather}%
where the different symmetry breaking scales satisfy the following hierarchy:
\begin{eqnarray}
 v=246 \text{GeV} = \sqrt{v_\rho^2+v_{\eta_2}^2} \ll v_\chi \sim \mathcal{O}\left(9.9 \right)\ \text{TeV}  \ll \Lambda_{\textbf{int}},
\end{eqnarray}
which corresponds in our model to the VEVs of the scalar fields.\\

The electric charge operator is defined \cite{Valle:1983dk,Dong:2010zu} in terms of the $SU(3)$ generators $T_3$ and $T_8$ and the identity $I_{3\times 3}$ as follows:
\begin{eqnarray}
Q=T_3+\beta T_8+ I_{3\times 3}X.
\end{eqnarray}
where for our model we choose $\beta = -1/\sqrt{3}$, and $X$ are the charge associated with gauge group $U(1)_{X}$.

The fermionic content in this model under $SU(3)_{C}\times SU(3)_{L}\times U(1)_{X}$ \cite{Valle:1983dk,Long:1995ctv,CarcamoHernandez:2017cwi}. The leptons are in triplets of flavor, in which the third component is an RH neutrino. The three generations of leptons for anomaly cancellation are:
\begin{equation}
L_{iL}=\begin{pmatrix}
\nu _{i} \\ 
l_{i} \\ 
\nu _{i}^{c} \\ 
\end{pmatrix}_{\hspace{-0.15cm}L}\sim \left(\textbf{1},\textbf{3},-1/3\right), \quad l_{iR} \sim \left(\textbf{1},\textbf{1},-1\right), \quad  N_{i R} \sim \left(\textbf{1},\textbf{1},0 \right),
\end{equation}
where $i=1,2,3$ is the family index. Here $\nu^c\equiv \nu_R^c$ is the RH neutrino and $\nu_{iL}$ are the family of neutral leptons, while $l_{iL}\ (e_L,\mu_L, \tau_L)$ is the family of charged leptons, and $N_{i R}$ are three right handed Majorana neutrinos, singlets under the $3$-$3$-$1$ group. Regarding the quark content, the first two generations are antitriplet of flavor, while the third family is a triplet of flavor; note that the third generation has different gauge content compared with the first two generations, which is required by the anomaly cancellation.
\begin{gather}
Q_{nL}=\begin{pmatrix}
d_{n} \\ 
-u_{n} \\ 
J_{n} \\ 
\end{pmatrix}_{\hspace{-0.15cm}L} \sim \left( \textbf{3},\overline{\textbf{3}},0\right),
\quad Q_{3L}=\begin{pmatrix}
u_{3} \\ 
d_{3} \\ 
T \\ 
\end{pmatrix}_{\hspace{-0.15cm}L}\sim \left( \textbf{3},\textbf{3},1/3\right), \quad n=1,2.\\
\nonumber u_{iR} \sim \left(\textbf{3},\textbf{1},2/3\right), \quad d_{iR} \sim \left(\textbf{3},\textbf{1},-1/3\right), \quad J_{nR} \sim \left(\textbf{3},\textbf{1},-1/3\right), \quad T_R \sim \left(\textbf{3},\textbf{1},2/3\right).
\end{gather}
We can observe that the $d_{iR}$ and $J_{iR}$ quarks have the same $X$ quantum number, and so are the $u_{iR}$ and $T_R$ quarks. Here $u_{i L}$ and $d_{i L}$ are the LH up and down type quarks fields in the flavor basis, respectively. Furthermore, $u_{i R}$ and $d_{i R}$ are the RH SM quarks, and $J_{n R}$ and $T_R$ are the RH exotic quarks. And the scalar sector contains four scalar triplets of flavor,
\begin{eqnarray}\label{triplet}
\nonumber \rho &=&\begin{pmatrix}
\rho _{1}^{+} \\ 
\frac{1}{\sqrt{2}}(v_{\rho }+\xi _{\rho }\pm i\zeta _{\rho }) \\ 
\rho _{3}^{+}%
\end{pmatrix}\sim \left( \textbf{1},\textbf{3},\frac{2}{3}\right) ,\quad  \chi\ =\ \begin{pmatrix}
\chi _{1}^{0} \\ 
\chi _{2}^{-} \\ 
\frac{1}{\sqrt{2}}(v_{\chi }+\xi _{\chi }\pm i\zeta _{\chi })%
\end{pmatrix}%
\sim \left( \textbf{1},\textbf{3},-\frac{1}{3}\right) ,
\\
 &&\\
\nonumber  \eta _{2} &=& \begin{pmatrix}
\frac{1}{\sqrt{2}}(v_{\eta_2}+\xi _{\eta_2}\pm i\zeta _{\eta_2}) \\ 
\eta _{22}^{-} \\ 
\eta _{32}^{0}
\end{pmatrix}\sim \left( \textbf{1},\textbf{3},-\frac{1}{3}\right), \quad  \eta_{1} \ =\ \begin{pmatrix}
\frac{1}{\sqrt{2}}(\xi _{\eta_1}\pm i\zeta _{\eta_1}) \\ 
\eta _{21}^{-} \\ 
\eta _{31}^{0}
\end{pmatrix}\sim \left( \textbf{1},\textbf{3},-\frac{1}{3}\right).
 \end{eqnarray}
where $\eta_1$ is an inert triplet scalar, on the other hand, the $SU(3)_L$ scalars $\rho$, $\chi$, and $\eta_2$ acquire the following vacuum expectation value (VEV) patterns:
\begin{equation}
\left\langle \chi \right\rangle^T= \left(0,0, v_{\chi}/\sqrt{2}\right) , \quad \left\langle \rho \right\rangle^T= \left(0, v_{\rho}/\sqrt{2}, 0\right), \quad 
\left\langle \eta_2 \right\rangle^T= \left( v_{\eta_2}/\sqrt{2},0, 0\right).
\end{equation}
In addition, some singleton scalars are introduced: $\left\lbrace   \sigma, \Theta_n, \zeta_n, \varphi , S_{k} , \Phi , \xi \right\rbrace $, $\left( k=e,\mu,\tau\right)$ where all fields transform $\left( \textbf{1},\textbf{1},0\right)$ under $SU(3)_{C}\times SU(3)_{L}\times U(1)_{X}$. 

With respect to the global leptonic symmetry defined as \cite{Chang:2006aa}:
\begin{eqnarray}
L=\frac{4}{\sqrt{3}} T_8+I_{3\times 3} L_g, 
\end{eqnarray}
where $L_g$ is a conserved charge corresponding to the $U(1)_{L_{g}}$ global symmetry, which commutes with the gauge symmetry. The difference between the $SU(3)_L$ Higgs triplet  can be explained using different charges $L_g$ the generalized lepton number. The lepton and anti-lepton are in the triplet, the leptonic number operator $L$ does not commute with gauge symmetry. \\
The choice of the $S_4$ symmetry group containing irreducible triplet, doublet, trivial singlet and non-trivial singlet representations presented in the Appendix \ref{S4}, allows us to naturally group the three charged left lepton families, and the three right-handed majonara neutrinos into $S_4$ triplets; $\left( L_{1L},L_{2L},L_{2L}\right) \sim \textbf{3}$, $N_R = \left( N_{1R},N_{2N},N_{2N}\right)\sim \textbf{3}$, while the first two families of left SM quarks, and the second and third families of right SM quarks in $S_4$ doublets;  $Q_{L} = \left( Q_{1L},Q_{2L}\right) \sim \mathbf{2}$, $U_{R}=\left( u_{2R},u_{3R}\right) \sim \mathbf{2}$ respectively, as well as the following exotic quarks; $J_{R}=\left( J_{1R},J_{2R}\right) \sim  \textbf{2} $, the remaining fermionic fields $Q_{3L}$, $T_R$, and  $l_{iR}$ as trivial singlet, $u_{1R}$ and $d_{iR}$ non-trivial singlet of $S_4$. The assignments under $S_4$, the scalar fields $S_k$, $\Phi$, and $\xi$ are grouped into triplets. In our model, the $S_k$ fields play a fundamental role in the vacuum configurations for the $S_4$ triplets leading to diagonal mass matrices for the charged leptons of the standard model. The inert field $\eta_1$ and $\eta_2$ in doublet $\eta=\left( \eta_1,\eta_2\right) \sim \textbf{2}$, there are three nontrivial singlets; $\{\sigma, \Theta_2,\ \zeta_2\} \sim \textbf{1'}$ and five trivial singlets $\{\rho, \chi, \Theta_1, \varphi, \zeta_1 \}$.

Using the particle spectrum and symmetries given in Tables \ref{Table1} and \ref{Table2}, we can write the Yukawa interactions for the quark and lepton sectors:
\begin{eqnarray}
 -\mathcal{L}_{Y}^{\left( q\right) } &=&y_{1}^{\left( T\right) }\overline{Q}%
_{3L}\chi T_{R}+y^{\left( J\right) }\left( \overline{Q}%
_{L}\chi ^{\ast }J_{R}\right) _{\mathbf{1}}+y_{3}^{\left( u\right) }%
\overline{Q}_{3L}\left( \eta U_{R}\right) _{\mathbf{1}} +y_{2}^{\left( u\right) }\varepsilon_{abc}\left( \overline{Q}_{L}^{a}\eta
^{b}\right) _{\mathbf{2}}\chi ^{c}U_{R}\frac{\sigma }{\Lambda ^{2}}\label{ec:lag-quarks}\\
\nonumber &&+y_{1}^{\left( u\right) }\varepsilon _{abc}\left( \overline{Q}_{L}^{a}\eta
^{b}\right) _{\mathbf{1^{\prime }}}\chi ^{c}u_{1R}\frac{\sigma ^{2}}{\Lambda
^{3}} +y_{13}^{\left( d\right) }\left( \overline{Q}_{L}\eta ^{\ast }\right) _{%
\mathbf{1^{\prime }}}d_{3R}\frac{\zeta _{1}}{\Lambda }+y_{12}^{\left(
d\right) }\left( \overline{Q}_{L}\eta ^{\ast }\right) _{\mathbf{1^{\prime }}%
}d_{2R}\frac{\Theta _{1}}{\Lambda }\\
\nonumber&&+y_{11}^{\left( d\right) }\left( 
\overline{Q}_{L}\eta ^{\ast }\right) _{\mathbf{1^{\prime }}}d_{1R}\frac{%
\sigma ^{2}}{\Lambda ^{2}} +y_{23}^{\left( d\right) }\left( \overline{Q}_{L}\eta ^{\ast }\right) _{%
\mathbf{1}}d_{3R}\frac{\zeta _{2}}{\Lambda }+y_{22}^{\left( d\right) }\left( 
\overline{Q}_{L}\eta ^{\ast }\right) _{\mathbf{1}}d_{2R}\frac{\Theta _{2}}{%
\Lambda } +y_{33}^{\left( d\right) }\overline{Q}_{3L}\rho d_{3R},
\end{eqnarray}
\begin{eqnarray}
-\mathcal{L}_{Y}^{\left( l\right) } &=&y_{1}^{\left( L\right) }\overline{L}_{L}\rho l_{1R}\frac{S_{e}}{\Lambda }+y_{2}^{\left( L\right) }\overline{L}_{L}\rho l_{2R}\frac{S_{\mu }}{\Lambda }+y_{3}^{\left( L\right) }\overline{L}_{L}\rho l_{3R}\frac{S_{\tau }}{\Lambda }+y_{\chi }^{\left( L\right) }\left( 
\overline{L}_{L}\chi N_{R}\right) _{\mathbf{\mathbf{1}}} \label{ec:lag-lep}  \\
\nonumber &&+y_{\nu }\varepsilon _{abc}\left( \overline{L}_{L}^{a}\rho ^{*c}\left(
L_{L}^{C}\right) ^{b}\right) _{\mathbf{3}}\frac{\Phi }{\Lambda }
+h_{1N}\left( N_{R}\overline{N_{R}^{C}}\right) _{\mathbf{\mathbf{1}}}\varphi 
\frac{\sigma ^{2}}{\Lambda ^{2}}+h_{2N}\left( N_{R}\overline{N_{R}^{C}}\right)_{\mathbf{\mathbf{3}}}\xi \frac{\sigma ^{2}}{\Lambda ^{2}}+H.c.,
\end{eqnarray}
where the $y^{(I)}$ (with $I= T, J, u, d, L$) represent the Yukawa couplings, which are dimensionless. The superscripts of these couplings indicate the specific contribution of the particles involved; $(T)$ and $(J)$, refer to the Yukawa couplings of the exotic $T$ and $J$ quarks respectively, $(u)$ and $(d)$ refer to the couplings of the quarks that give rise to the mass matrices of the up and down sectors respectively, and the index $(L)$ indicates the couplings of the charged leptonic Yukawa interactions. In the quark sector, under the $S_4$ group the couplings determine the interactions between the quark doublets $Q_L$, the quark singlets $u_{iR}$ and $d_{iR}$, the doublet $U_R$, and the scalar fields. The $h_{nN}$ terms are associated with operators involving heavy neutrinos, $h_{nN}$ refer to the interactions between the right-handed neutrinos and the scalar fields, crucial for the generation of neutrino masses. And the subscripts $\textbf{1}$, $\textbf{1}^{\prime}$, $\textbf{2}$ and $\textbf{3}$ denote the irreducible representations under the discrete symmetry group $S_4$, whose tensor products are shown in Appendix \ref{S4}.

We use the following VEV configurations (see section~\ref{ScalarS4doubletandtriplet}) for the $S_4$ triplets
\begin{eqnarray}
\left\langle S_{e}\right\rangle &=&v_{S_{e}}\left( 1,0,0\right) ,\hspace{1cm}%
\left\langle S_{\mu }\right\rangle =v_{S_{\mu }}\left( 0,1,0\right) ,\hspace{%
1cm}\left\langle S_{\tau }\right\rangle =v_{S_{\tau }}\left( 0,0,1\right) ,%
\hspace{1cm} \label{eq:vev-lep}\\
\left\langle \Phi \right\rangle &=&v_{\Phi }\left(1,r_1e^{i\theta},r_1e^{i\theta}\right),\hspace{1cm}\left\langle \xi \right\rangle =v_{\xi }\left(1,1,r_2\right) \label{eq:vev-lep2}
\end{eqnarray}
where $r_1$, $r_2$ and $\theta$ are free parameters.
Regarding the $S_4$ scalar doublet, we consider the following VEV pattern:
\begin{equation} \label{veveta}
\left\langle \eta \right\rangle =\frac{v_{\eta_2}}{\sqrt{2}}\left( 0,1\right),
\end{equation}
The above given VEV patterns are consistent with the scalar potential minimization conditions for a large region of parameter space. They allow to get a predictive and viable pattern of SM fermion masses and mixings as it will be shown in the next sections.

Furthermore, the $S_4$ singlet gauge singlet scalars have VEVs given by:
\begin{equation}
\left\langle \sigma \right\rangle  =  v_\sigma, \quad \left\langle \Theta_n \right\rangle =  v_{\Theta_n}, \quad \left\langle \varphi \right\rangle  = v_\varphi, \quad \left\langle \zeta_n \right\rangle= v_{\zeta_n}, \quad n=1,2.
\end{equation}

\begin{table}[]
\centering
\begin{tabular}{|c|c|c|c|c|c|c|c|c|c|c|c|c|c|c|}
\hline
& $\rho $ & $\eta$ & $\chi $ & $\sigma $ & $\Theta _{1}$ & $\Theta _{2}$ & $%
\varphi $ & $\zeta _{1}$ & $\zeta _{2}$ & $S_{1}$ & $S_2$ & $S_3$ & $\Phi $ & $\xi $ \\ 
\hline $U(1)_{L_g}$ & $-2/3$ & $-2/3$  & $4/3$ & $0$ & $0$ & $0$ & $2$ & $0$ & $0$ & $0$ &  $0$ &  $0$& $0$& $2$ \\ \hline
$S_{4}$ & $\mathbf{1}$ & $\mathbf{2}$ & $\mathbf{1}$ & $\mathbf{1^{\prime}}$
& $\mathbf{1}$ & $\mathbf{1}^{\prime }$ & $\mathbf{1}$ & $\mathbf{1}$ & $\mathbf{1}^{\prime }$ & $\mathbf{3}$ & $\textbf{3}$ &  $\textbf{3}$& $\mathbf{3}$ & $\mathbf{3}$ \\ \hline
$Z_{4}$ & $0$ & $1$ & $0$ & $-1$ & $0$ & $0$ & $0$ & $0$ & $0$ & $-1$ & $-1$ & $-1$ & $-2$ & $0$ \\ \hline
$Z_{4}^{\prime }$ & $0$ & $0$ & $0$ & $0$ & $0$ & $0$ & $0$ & $0$ & $0$ & $1$ & $1$ &  $1$ & $0$ & $0$\\ \hline
$Z_{2}$ & $-1$ & $0$ & $0$ & $0$ & $0$ & $0$ & $0$ & $-1$ & $-1$ & $0$ &$0$ &$0$ &  $1$ & $0$\\ \hline
\end{tabular}
\caption{Scalar assignments under $
U(1)_{L_{g}}\times S_{4}\times Z_{4}\times Z_{4}^{\prime }\times Z_{2}$.}
\label{Table1}
\end{table}

\begin{table}[]
\centering
\begin{tabular}{|c|c|c|c|c|c|c|c|c|c|c|c|c|c|c|}
\hline & $Q_{L}$ &  $Q_{3 L}$ & $u_{1 R}$ & $U_{R}$ &$d_{1R}$& $d_{2R}$ &$d_{3R}$  &  $T_R$ & $J_{ R}$ &    $L_L$ & $l_{1 R}$ & $l_{2 R}$ & $l_{3 R}$ & $N_R$ \\
\hline$U(1)_{L_g}$ & $2/3$& $-2/3$& $0$ & $0$ &$0$ &$0$ & $0$ & $-2$  & $2$ & $1/3$ & $1$ & $1$ & $1$& $-1$ \\
\hline$S_4$ & $\textbf{2}$ &   $\textbf{1}$ & $\textbf{1}^\prime$ & $\textbf{2}$ & $\textbf{1}^\prime$ & $\textbf{1}^\prime$ & $\textbf{1}^\prime$  &    $\textbf{1}$  &$\textbf{2}$ &   $\textbf{3}$ & $\textbf{1}$ & $\textbf{1}$ & $\textbf{1}$ &  $\textbf{3}$ \\
\hline$Z_4$ & $-1$ &$0$  & $0$ & $-1$ & $2$ & $0$ & $0$ &   $0$ & $-1$ & $-1$ & $0$ & $0$ & $0$ &  $1$   \\
\hline$Z_4'$ & $0$ & $0$ & $0$ & $0$ & $0$ & $0$ & $0$ & $0$ & $0$ & $0$ &  $-1$ & $-1$ & $-1$ & $0$   \\
\hline$Z_2$ & $0$ & $0$ & $0$ &  $0$ & $0$ & $0$ & $1$ &   $0$ &$ 0$ &  $0$ & $1$ & $1$ &  $1$ &  $0$  \\
\hline
\end{tabular}
\caption{Fermion assignments under $
U(1)_{L_{g}}\times S_{4}\times Z_{4}\times Z_{4}^{\prime }\times Z_{2}$.}
\label{Table2}
\end{table}


\section{Quark masses and mixings}\label{quark-sector}
\lhead[\thepage]{\thesection. Quark masses and mixings}

After the spontaneous symmetry breaking of the Lagrangian of Eq.\eqref{ec:lag-quarks}, we obtain the following $3\times 3$ low-scale quark mass matrices:
\begin{eqnarray}
M_{\text{U}}&=&\left( 
\begin{array}{ccc}
\frac{v_{\eta_2}v_{\chi}v_{\sigma}^2}{\Lambda^3}y_1^{(u)} & 0 & -\frac{v_{\eta_2}v_{\chi}v_{\sigma}}{\Lambda^2}y_2^{(u)} \\ 
0 & -\frac{v_{\eta_2}v_{\chi}v_{\sigma}}{\Lambda^2}y_2^{(u)} & 0 \\ 
0 & 0 & v_{\eta_2}y_3^{(u)}%
\end{array}%
\right) =\left( 
\begin{array}{ccc}
C & 0 & A \\ 
0 & A & 0 \\ 
0 & 0 & B%
\end{array}%
\right), \notag\\%
M_{\text{D}}&=&\left(
\begin{array}{ccc}
-\frac{v_{\eta_2}v_{\sigma}^2}{\Lambda^2}y_{12}^{(d)} & -\frac{v_{\eta_2}v_{\Theta_1}}{\Lambda}y_{12}^{(d)} & -\frac{v_{\eta_2}v_{\zeta_1}}{\Lambda}y_{13}^{(d)} \\ 
0 & \frac{v_{\eta_2}v_{\Theta_2}}{\Lambda}y_{22}^{(d)} & \frac{v_{\eta_2}v_{\zeta_2}}{\Lambda}y_{23}^{(d)} \\ 
0 & 0 & v_{\rho}y_{33}^{(d)}%
\end{array}%
\right) = \left( 
\begin{array}{ccc}
C_{1} & A_{1} & B_{1} \\ 
0 & A_{2} & B_{2} \\ 
0 & 0 & B_{3}%
\end{array}%
\right)%
, \label{MUMD}
\end{eqnarray}

\begin{table}
\centering
\begin{tabular}{c|c|c} 
\hline\hline
\textbf{Observable} & \textbf{Model Value} & \textbf{Experimental Value}                        \\ 
\hline\hline
$m_u\; (\text{MeV})$       &         $2.04$      & $2.16_{-0.26}^{+0.49}$                             \\
$m_c\; (\text{GeV})$       &         $1.26$      & $1.27\pm 0.02$                                     \\
$m_t\; (\text{GeV})$       &         $172.50$      & $172.69\pm 0.30$                                   \\
$m_d\; (\text{MeV})$       &         $4.40$      & $4.67_{-0.17}^{+0.48}$                             \\
$m_s\; (\text{MeV})$       &         $93.7$      & $93.4_{-3.4}^{+8.6}$                               \\
$m_b\; (\text{GeV})$       &         $4.18$      & $4.18_{-0.02}^{+0.03}$                             \\
$\sin\theta^{(q)}_{12}$   &         $0.22530$     & $0.22500\pm 0.00067$                               \\
$\sin\theta^{(q)}_{23}$   &         $0.04332$    & $0.04182_{-0.00074}^{+0.00085}$                    \\
$\sin\theta^{(q)}_{13}$   &        $0.00390$    & $0.00369\pm 0.00011$                               \\
$J_q$               &  $2.93\times 10^{-5}$    & $\left(3.08_{-0.13}^{+0.15}\right)\times 10^{-5}$  \\
\hline\hline
\end{tabular}
\caption{The model and experimental values for observables in the quarks sector. The model value correspond to the best fit values of the quark sector observables and we use the experimental values of the quark masses at the On-Shell scale from Ref. \cite{Workman:2022ynf}.
}
\label{tab:quarks}
\end{table}

The matrices of Eq.~\eqref{MUMD} are non hermitian and we are considering the complex coupling $B_2$. However, 
the following matrices are Hermitian,
\begin{align}
M_{\text{U}}M_{\text{U}}^T= \begin{pmatrix}
A^2+C^2 & 0 & A B \\
 0 & A^2 & 0 \\
 A B & 0 & B^2
\end{pmatrix} \quad ;\quad 
M_{\text{D}}M_{\text{D}}^{\dagger}= \begin{pmatrix}
 A_1^2+B_1^2+C_1^2 & A_1 A_2+B_1 B_2 & B_1 B_3 \\
 A_1 A_2+B_1 B_2 & A_2^2+B_2^2 & B_2 B_3 \\
 B_1 B_3 & B_2 B_3 & B_3^2
\end{pmatrix}\label{eq:msqrt}
\end{align}

Considering the unitary matrices $M_{\text{U}}M_{\text{U}}^T$ (we consider the case where all entries of $M_{\text{U}}$ are real) and $M_{\text{D}}M_{\text{D}}^{\dagger}$
, we can fit the quark sector parameters in order to successfully reproduce the experimental values of the physical observables of the quark sector. To this end, we proceed to 
minimize the following $\chi^2$ function defined as:
\begin{equation} \label{ec:funtion_error}   
\begin{aligned}
 \chi ^{2} = & \frac{\left( m_{u}^{\exp }-m_{u}^{\text{th}}\right) ^{2}}{\sigma
_{m_{u}}^{2}}+\frac{\left( m_{c}^{\exp }-m_{c}^{\text{th}}\right) ^{2}}{\sigma
_{m_{c}}^{2}}+\frac{\left( m_{t}^{\exp }-m_{t}^{\text{th}}\right) ^{2}}{\sigma
_{m_{t}}^{2}}+\frac{\left( m_{d}^{\exp }-m_{d}^{\text{th}}\right) ^{2}}{\sigma
_{m_{d}}^{2}}+\frac{\left( m_{s}^{\exp }-m_{s}^{\text{th}}\right) ^{2}}{\sigma
_{m_{s}}^{2}}\\
& +\frac{\left( m_{b}^{\exp }-m_{b}^{\text{th}}\right) ^{2}}{\sigma_{m_{b}}^{2}}
+\frac{\left( \sin{\theta}_{12}^{(q)\exp }-\sin\theta_{12}^{(q)\text{th}}\right)^{2}}{\sigma _{\sin\theta^{(q)}_{12}}^{2}}
+\frac{\left( \sin\theta_{23}^{(q)\exp }-\sin\theta _{23}^{(q)\text{th}}\right) ^{2}}{\sigma _{\sin\theta_{23}^{(q)}}^{2}}\\
& +\frac{\left( \sin\theta_{13}^{(q)\exp }-\sin\theta_{13}^{(q)\text{th}}\right) ^{2}}{%
\sigma _{\sin\theta^{(q)}_{13}}^{2}}+\frac{\left( J_q^{\exp }-J_q^{\text{th}}\right) ^{2}}{\sigma _{J_q}^{2}}\;, 
\end{aligned}
\end{equation}

where $m_i$ are the masses of the quarks ($i=u,c,t,d,s,b$), $\sin\theta^{(q)}_{ij}$ is the sine function of the quark mixing angles (with $ j , k = 1, 2, 3$) and $J_q$ is the quark Jarlskog invariant. The supra indices represent the experimental (\enquote{exp}) and theoretical (\enquote{$\text{th}$}) values, and the $\sigma$ are the experimental errors. The best-fit point of our model is shown in Table \ref{tab:quarks} together with the current experimental values, while Eq.~\eqref{eq:para} shows the benchmark point of the low energy quark sector effective parameters that allow to successfully reproduce the measured SM quark masses and CKM parameters, obtaining the following values for our best-fit point:
\begin{align}
C&=2.03\pm 4.27\times 10^{-2}\; \text{MeV} & A&= 1.26\pm 1.10\times 10^{-2}\; \text{GeV} & B&= 172.47\pm 0.19\; \text{GeV}. \notag\\
C_1&= -4.52\pm 0.10\; \text{MeV} & A_1&= 21.0\pm 0.5\; \text{MeV} & A_2&= -91.3\pm 2.1\; \text{MeV}.\label{eq:para}\\
B_1&= 14.3\pm 0.2 \; \text{MeV} & B_2&=(0.181\pm 2\times 10^{-3})e^{2.23i} \; \text{GeV} & B_3&= 4.18\pm 0.01\; \text{GeV}.\notag
\end{align}
From Eq~\eqref{MUMD} for the Up quarks, we can see that the eigenvalues are $(C,A,B)$ and from Eq~\eqref{eq:para} we have that $C\simeq m_u$, $ A= m_c$ and $B\simeq m_t$, where only the parameter $A$ is equal to the mass of the \textit{charm} quark, while the other parameters are similar, but not equal to the values of the masses of the \textit{up} quark and the \textit{top} quark, this is due to the non diagonal structure of the up type quark mass matrix, which features a mixing in the 1-3 plane, as indicated by Eq~\eqref{eq:msqrt}.
The same is true for the down quark sector, however, the structure of the matrix $M_{\text{D}}M_{\text{D}}^{\dagger}$ is more complicated than the one of $M_{\text{U}}M_{\text{U}}^{T}$ making it difficult to work with analytical expressions. For this reason, the entire analysis and diagonalization of the matrices $M_{\text{U}}M_{\text{U}}^{T}$ and $M_{\text{D}}M_{\text{D}}^{\dagger}$ have been performed numerically, obtaining the following unitary rotation matrices for our best-fit point,
\begin{align}
R_u&=
\begin{pmatrix}
 0.999 & 0 & -0.007 \\
 0 & -1. & 0 \\
 -0.007 & 0 & -0.999
\end{pmatrix}, \\
R_d&=
\begin{pmatrix}
0.974 & 0.225 & -0.003 \\
0.225 & 0.973e^{-3.14i} & 0.043e^{-0.910i} \\
0.008e^{1.24i} & 0.043^{-2.24i} & 0.999e^{-3.14i}
\end{pmatrix}.
\end{align}

\begin{figure}[H]
\centering
\subfigure[]{\includegraphics[scale=0.25]{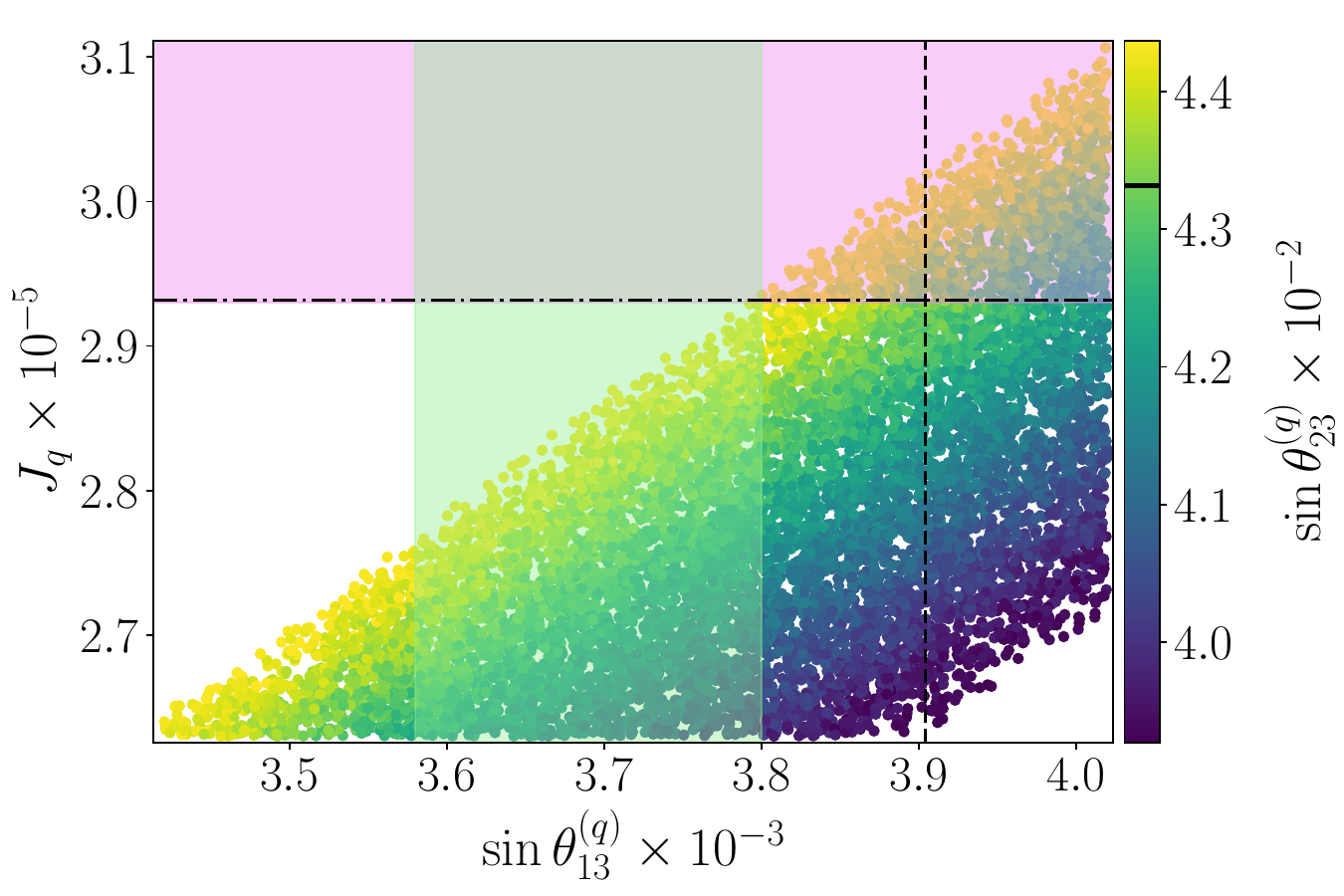}} \quad
\subfigure[]{\includegraphics[scale=0.25]{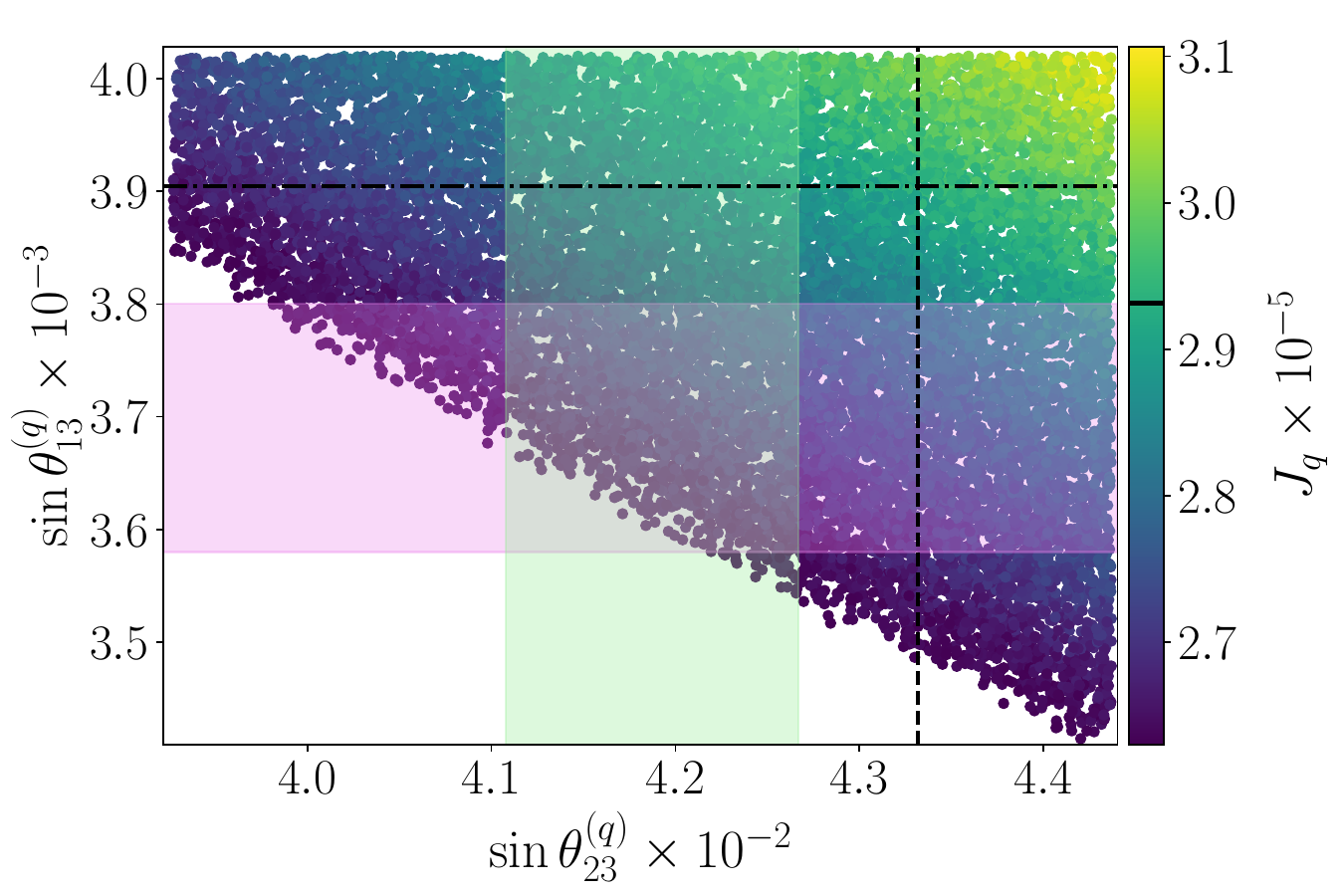}}
\caption{Correlation plot between the mixing angles of the quarks and the Jarlskog invariant obtained with our model. The green and purple bands represent the $1\sigma$ range in the experimental values, while the dotted line (black) represents the best-fit point by our model. The black line in the vertical color bar represents the best-fit point of the model for $\sin\theta_{23}^{(q)}$ and $J_q$.}
\label{fig:quarkscorr}
\end{figure}

Fig.~\ref{fig:quarkscorr} we can see the correlation plot between the quark mixing angles $\sin\theta^{(q)}_{13}$, $\sin\theta^{(q)}_{23}$ and the Jarlskog invariant. These correlation plots were obtained by varying the best-fit point of the quark sector parameters around $20\%$, whose values are shown in Eq.~\eqref{eq:para}. The dots in Fig.~\eqref{fig:quarkscorr} represent the correlation between each observable whereas the color background represents different values for  $\sin\theta_{23}^{(q)}$ (Fig.~\ref{fig:quarkscorr}a) and $J_q$ (Fig.~\ref{fig:quarkscorr}b), the vertical (green) and horizontal (purple) bars represent the $1\sigma$ range in the experimental values, while the dotted line (black) represents the value for the best-fit point of the model. The Fig.~\ref{fig:quarkscorr} a shows the correlation between $\sin\theta_{13}^{(q)}$ versus $J_q$, for different values of the $\sin\theta_{23}^{(q)}$, where the model predicts that $\sin\theta^{(q)}_{13}$ is found in the range $3.4\times 10^{-3} \lesssim \sin\theta^{(q)}_{13} \lesssim 4.0\times 10^{-3}$ in the allowed parameter space and, moreover, it increases when the Jarlskog invariant takes larger values, whose values are in the range $2.63\times 10^{-5}\lesssim J_q \lesssim 3.11\times 10^{-5}$. The plot, Fig.~(\ref{fig:quarkscorr}b), shows a correlation between $\sin\theta^{(q)}_{13}$ versus $\sin\theta^{(q)}_{23}$ for different values of $J_q$, in which the first variable takes on a wider range of values, with a lower limit decreasing while the upper limit remains constant, when the second one acquires larger values, where we can see that $\sin\theta^{(q)}_{23}$ is in the range $3.9\times 10^{-2} \lesssim \sin\theta^{(q)}_{23} \lesssim 4.4\times 10^{-2}$.

\section{Lepton masses and mixings}\label{lepton-sector}
\lhead[\thepage]{\thesection. Lepton masses and mixings}
\subsection{Charged lepton sector}
The $Z_4^{\prime}$ charge assignments of the model fields shown in the Table \ref{Table1}, as well the VEV pattern of the $S_4$ scalar triplets $S_e$, $S_{\mu }$ and $S_{\tau}$ shown in Eq. \eqref{eq:vev-lep} imply that the charged lepton Yukawa terms of Eq.~(\ref{ec:lag-lep}), yield a diagonal charged lepton mass matrix:
\begin{equation}
M_l=\left(
\begin{array}{ccc}
m_e & 0 & 0 \\
0 & m_{\mu} & 0 \\
0 & 0 & m_{\tau}
\end{array}
\right),
\end{equation}
where the masses of the SM charged leptons are given by:
\begin{equation}
m_{e}=y_{1}^{\left( L\right) }\frac{v_{S_{e}}v_{\rho }}{\Lambda }=a_{1}\frac{%
v_{\rho }}{\Lambda },\hspace{1cm}m_{\mu }=y_{2}^{\left( L\right) }\frac{%
v_{S_{\mu }}v_{\rho }}{\Lambda }=a_{2}\frac{v_{\rho }}{\Lambda },\hspace{1cm}%
m_{\tau}=y_{3}^{\left( L\right) }\frac{v_{S_{\tau }}v_{\rho }}{\Lambda }=a_{3}%
\frac{v_{\rho }}{\Lambda }.
\end{equation}

\subsection{Neutrino sector}
The neutrino Yukawa interactions of Eq. (\ref{ec:lag-lep}) give rise to the following neutrino mass terms:
\begin{equation}
-\mathcal{L}_{mass}^{\left( \nu \right) }=\frac{1}{2}\left( 
\begin{array}{ccc}
\overline{\nu _{L}^{C}} & \overline{\nu _{R}} & \overline{N_{R}}%
\end{array}%
\right) M_{\nu }\left( 
\begin{array}{c}
\nu _{L} \\ 
\nu _{R}^{C} \\ 
N_{R}^{C}%
\end{array}%
\right) +H.c,  \label{Lnu}
\end{equation}

where the neutrino mass matrix reads:
\begin{equation}
M_{\nu }=%
\begin{pmatrix}
0_{3\times 3} & m_{\nu_D} & 0_{3\times 3} \\ 
m_{\nu_D}^{T} & 0_{3\times 3} & M \\ 
0_{3\times 3} & M^{T} & \mu%
\end{pmatrix}%
,
\end{equation}

and the submatrices are given by:
\begin{eqnarray}
m_{\nu_D} &=&\frac{y_{\nu }v_{\rho }v_{\Phi }}{\sqrt{2}\Lambda }\left( 
\begin{array}{ccc}
0 & r_1e^{i\theta} & r_1e^{i\theta} \\ 
-r_1e^{i\theta} & 0 & 1 \\ 
-r_1e^{i\theta} & -1 & 0%
\end{array}%
\right) ,\hspace{1cm}\hspace{1cm}M=y_{\chi }^{\left( L\right) }\frac{%
v_{\chi }}{\sqrt{2}}\left( 
\begin{array}{ccc}
1 & 0 & 0 \\ 
0 & 1 & 0 \\ 
0 & 0 & 1%
\end{array}%
\right) ,  \notag \\
\mu &=&\left( 
\begin{array}{ccc}
h_{1N}v_{\varphi } & h_{2N}v_{\xi }r_2 & h_{2N}v_{\xi} \\ 
h_{2N}v_{\xi }r_2 & h_{1N}v_{\varphi } & h_{2N}v_{\xi } \\ 
h_{2N}v_{\xi } & h_{2N}v_{\xi } & h_{1N}v_{\varphi }%
\end{array}%
\right) \frac{v_{\sigma }^{2}}{\Lambda ^{2}}.
\label{MR}
\end{eqnarray}

The light active masses arise from an inverse seesaw mechanism and the resulting physical neutrino mass matrices take the form~\cite{Mohapatra:1986bd,Malinsky:2005bi}:
\begin{eqnarray}
\widetilde{M}_{\nu} &=& m_{\nu_D}\left(M^T\right)^{-1}\mu M^{-1}m_{\nu_D}^T, \\
\widetilde{M}_{\nu}^{(1)} &=& \frac{1}{2}\mu-\frac{1}{2}\left(M+M^T\right), \\
\widetilde{M}_{\nu}^{(2)} &=& \frac{1}{2}\mu+\frac{1}{2}\left(M+M^T\right).
\end{eqnarray}

Here, $\widetilde{M}_{\nu}$ is the mass matrix for the active neutrino ($\nu_{\alpha}$), whereas $\widetilde{M}_{\nu}^{(1)}$ and $\widetilde{M}_{\nu}^{(2)}$ are the sterile neutrinos mass matrices.\\
Thus, the light active neutrino mass matrix is given by:
\begin{equation}
\widetilde{M}_{\nu} = \left(
\begin{array}{ccc}
 2 a^2 \left(y_2+z\right) & a \left(b \left(y_2+z\right)-a \left(y_1+y_2\right)\right) & -a \left(a \left(y_1+y_2\right)+b
   \left(y_2+z\right)\right) \\
 a \left(b \left(y_2+z\right)-a \left(y_1+y_2\right)\right) & z \left(a^2+b^2\right)-2 a b y_2 & a^2 z+a b \left(y_1-y_2\right)-b^2 y_2 \\
 -a \left(a \left(y_1+y_2\right)+b \left(y_2+z\right)\right) & a^2 z+a b \left(y_1-y_2\right)-b^2 y_2 & z \left(a^2+b^2\right)+2 a b y_1 \\
\end{array}
\right),
\label{ec:matrix-neutrino}
\end{equation}

where the effective parameters are defined as:
\begin{align}
a&=\frac{y_{\nu}v_{\rho}v_{\Phi}r_1}{y_{\chi}^{(L)}v_{\chi}\Lambda}e^{i\theta}, & b&= \frac{y_{\nu}v_{\rho}v_{\Phi}}{y_{\chi}^{(L)}v_{\chi}\Lambda}, & y_1&= \frac{v_{\sigma}^2h_{2N}v_{\xi}r_2}{\Lambda^2},\notag\\
y_2&= \frac{v_{\sigma}^2h_{2N}v_{\xi}}{\Lambda^2}, & z&= \frac{v_{\sigma}^2h_{1N}v_{\varphi}}{\Lambda^2}.\label{eq:efecpara}
\end{align}

From Eq.~\eqref{eq:efecpara}, we can observe that the effective parameters are not independent, obtaining the following relation,
\begin{equation}
a=b r_1e^{i\theta} \quad;\quad y_1=y_2r_2.
\end{equation}

In order to fit the effective neutrino sector parameters to successfully reproduce the experimental values of the neutrino mass squared splittings, the leptonic mixing angles and the leptonic Dirac CP phase, we proceed to minimize the following $\chi ^{2}$ function:
\begin{eqnarray}
\chi ^{2} &= & \frac{\left( \Delta m_{21}^{2\ \exp }-\Delta m_{21}^{2\ \text{th}}\right) ^{2}}{\sigma
_{\Delta m^2_{21}}^{2}}+\frac{\left( \Delta m_{31}^{2\ \exp }-\Delta m_{31}^{2\ \text{th}}\right) ^{2}}{\sigma
_{\Delta m^2_{31}}^{2}}+
\frac{\left( \sin^2\theta_{12}^{(l)\exp }-\sin^2\theta
_{12}^{(l)\text{th}}\right) ^{2}}{\sigma _{\sin^2 \theta^{(l)}_{12}}^{2}}\label{ec:error-neu}\\
\nn & & + \frac{\left( \sin^2\theta_{23}^{(l)\exp}-\sin^2\theta_{23}^{(l)\text{th}}\right) ^{2}}{\sigma_{\sin^2\theta^{(l)}_{23}}^{2}}
 + \frac{\left( \sin^2\theta_{13}^{(l)\exp }-\sin^2\theta_{13}^{(l)\text{th}}\right) ^{2}}{\sigma_{\sin^2\theta^{(l)}_{13}}^{2}}
+\frac{\left( \delta_{\text{CP}}^{\exp }-\delta_{\text{CP}}^{\text{th}}\right) ^{2}}{\sigma _{\delta_{\text{CP}}}^{2}}\;,  \notag
\end{eqnarray}

where $\Delta m_{i1}^2$ (with $i= 2, 3)$ are the neutrino mass squared differences, $\sin\theta^{(l)}_{jk}$ is the sine function of the mixing angles (with $j,k=1,2,3$) and $\delta_{\text{CP}}$ is the CP violation phase. The supra indices represent the experimental (\enquote{exp}) and theoretical (\enquote{$\text{th}$}) values, and the $1\sigma$ are the experimental errors. However, as was done in the quark sector, the matrix \eqref{ec:matrix-neutrino} is not Hermitian.
Even so, due to its structure, the $\widetilde{M}_{\nu}\widetilde{M}_{\nu}^\dagger$ matrix is not simple to work with analytically, so numerical analysis is also carried out in the neutrino sector. Therefore, after minimizing Eq.~\eqref{ec:error-neu} using the matrix $\widetilde{M}_{\nu}\widetilde{M}_{\nu}^\dagger$, we get the following values for the model parameters:
\begin{align}
a&= (0.521\pm 0.022)e^{-3.027i}, & b&= (1.080\pm 0.041)i, & y_1&=-0.086\pm 0.007\ \text{eV},\notag\\
y_2&= 2.05\pm 0.12\ \text{eV}, & z&= -2.30\pm 0.14\ \text{eV}.\label{ec:bestpara}
\end{align}

The diagonalization of the matrix \eqref{ec:matrix-neutrino} gives us a determinant equal to zero, so we obtain the following eigenvalues:
\begin{eqnarray}
 m_1^2 &=& 0 ,\\
m_2^2 &=& \bigg| \left(a^2\left(y_2+2 z\right)+a b \left(y_1-y_2\right)+b^2 z\right. \notag\\
& & -\left.\left.\sqrt{2 a^2 y_1^2 \left(a^2+b^2\right)+2 a y_2 y_1 (a-b) \left(2 a^2+a b+b^2\right)+y_2^2 \left(a^2+b^2\right) \left(3 a^2+2 a b+b^2\right)}\right)\right\lvert^2, \\
m_3^2 &=&\bigg| \left(a^2\left(y_2+2 z\right)+a b \left(y_1-y_2\right)+b^2 z\right. \notag \\
&  & +\left.\sqrt{2 a^2 y_1^2 \left(a^2+b^2\right)+2 a y_2 y_1 (a-b) \left(2 a^2+a b+b^2\right)+y_2^2 \left(a^2+b^2\right) \left(3 a^2+2 a b+b^2\right)}\right) \bigg|^2.   
\end{eqnarray}

With the values of the parameters of our best-fit point of the Eq. \eqref{ec:bestpara}, we obtain the results of the neutrino sector that are shown in the Table \ref{table:neutrinos_value}, together with the experimental values in the range of $1\sigma$ and $3\sigma$, whose experimental data were taken from \cite{deSalas:2020pgw}. In the Table \ref{table:neutrinos_value}, we can see that the difference of the square of the neutrino masses ($\Delta m_{21}^2$, $\Delta m_{31}^2$) and the of solar and reactor mixing neutrinos ($\sin^2\theta_{12}^{(l)}$, $\sin^2\theta_{13}^{(l)}$) are in the range of $1\sigma$, while the angle of the atmospheric neutrino ($\sin^2\theta_{23}^{(l)}$) is in the range of $2\sigma$ and the CP violation phase ($\delta_{\text{CP}}$) is in the range of $3\sigma$.

\begin{table}[tp]
\resizebox{16.8cm}{!}{
\begin{tabular}{c|c|cccccc}
\toprule[0.13em] Observable & range & $\Delta m_{21}^{2}$ [$10^{-5}$eV$^{2}$]
& $\Delta m_{31}^{2}$ [$10^{-3}$eV$^{2}$] & $\sin^2\theta^{(l)}_{12}/10^{-1}$
& $\sin^2\theta^{(l)}_{13}/10^{-3}$ & $\sin^2\theta^{(l)}_{23}/10^{-1}$ & $%
\delta^{(l)}_{\text{CP}} (^{\circ })$ \\ \hline
Experimental & $1\sigma$ & $7.50_{-0.20}^{+0.22}$ & $2.55_{-0.03}^{+0.02}$ & 
$3.18\pm 0.16$ & $2.200_{-0.062}^{+0.069}$ & $5.74\pm 0.14$ & $%
194_{-22}^{+24}$ \\ 
Value & $3\sigma$ & $6.94-8.14$ & $2.47-2.63 $ & $2.71-3.69$ & $2.000-2.405$
& $4.34-6.10$ & $128-359$ \\ \hline
Fit & $1\sigma-3\sigma$ & $7.498$ & $2.54$ & $3.02$ & $2.191$ & $5.93$ & $246.4$\\ \hline
\end{tabular}
}
\caption{Model predictions for the scenario of normal order (NO) neutrino
mass. The experimental values are taken from Ref. \protect\cite{deSalas:2020pgw}}
\label{table:neutrinos_value}
\end{table}

\begin{figure}[]
\centering
\subfigure[]{
\includegraphics[scale=0.35]{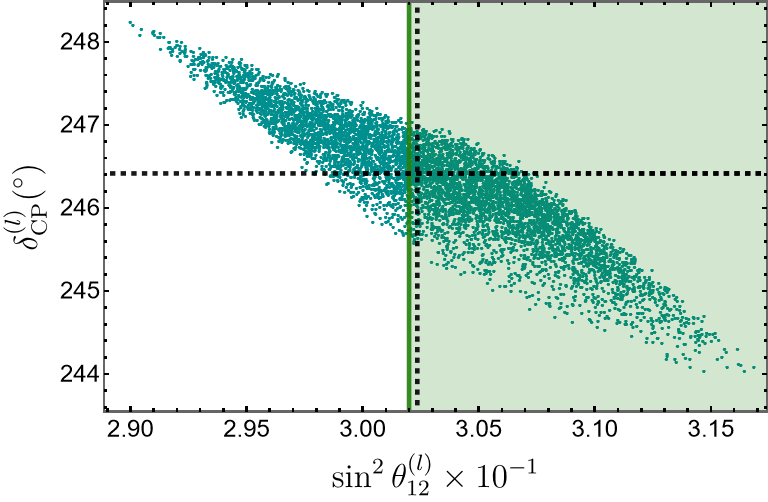}
} \quad
\subfigure[]{
\includegraphics[scale=0.35]{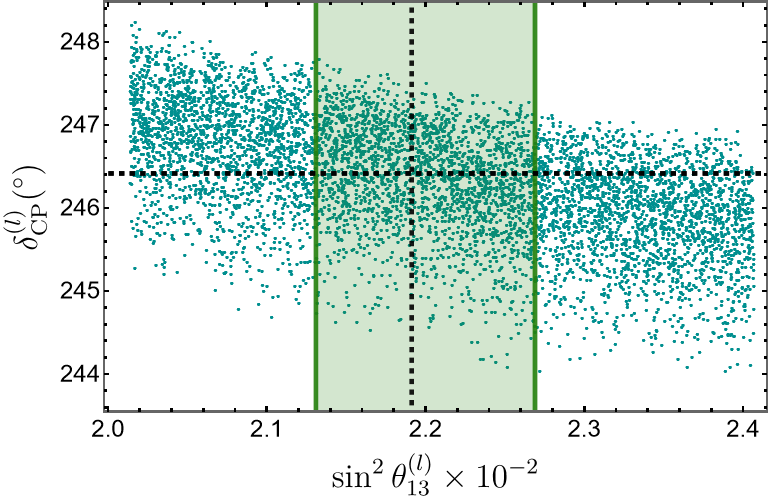}
}\quad
\subfigure[]{
\includegraphics[scale=0.35]{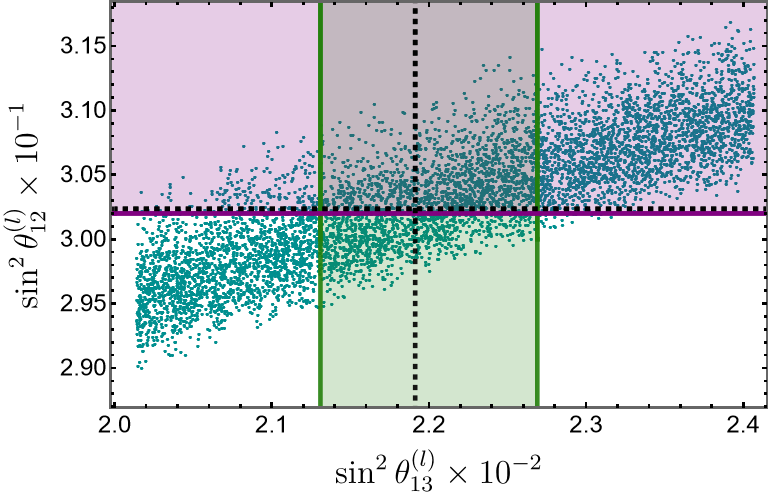}
}
\subfigure[]{
\includegraphics[scale=0.35]{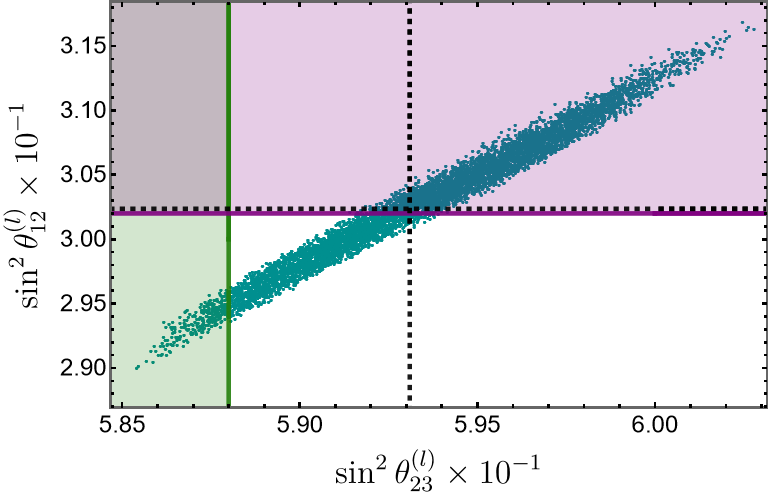}
}
\subfigure[]{
\includegraphics[scale=0.35]{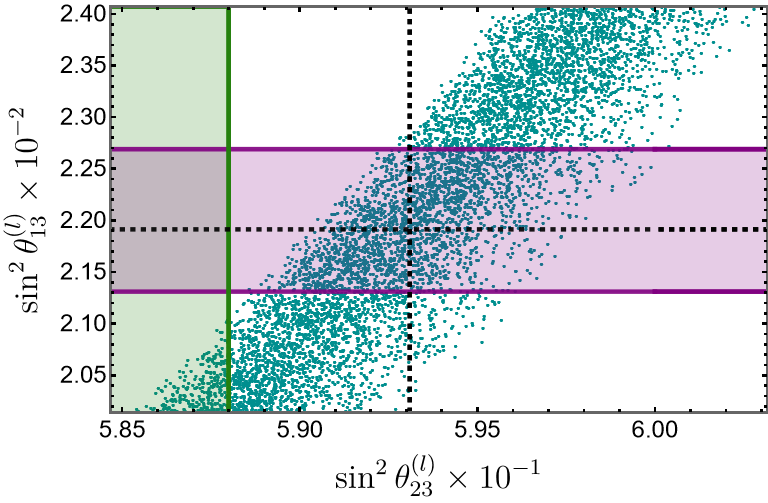}
}
\subfigure[]{
\includegraphics[scale=0.35]{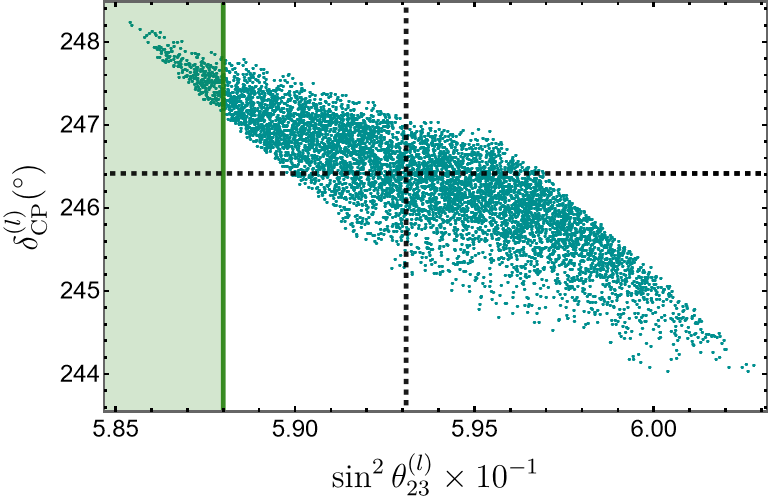}
}
\caption{Correlation between the mixing angles of the neutrino sector and the CP violation phase obtained with our model. The green and purple bands represent the $1\sigma$ range in the experimental values, while the dotted line (black) represents the best-fit point of our model.}
\label{fig:neutrino-corr}
\end{figure}

Fig. \ref{fig:neutrino-corr} shows the correlation between the leptonic Dirac CP violating phase and the neutrino mixing angles as well as the correlations among the leptonic mixing angles, where the green and purple background fringes represent the $1\sigma$ range of the experimental values and the black bands the dotted lines represent our best-fit point for each observable. In Fig. \ref{fig:neutrino-corr}, we see that for the mixing angles, we can get values in the $1\sigma$ range, while for the CP violating phase, we obtain values up to $3\sigma$, where each lepton sector observable is obtained in the following range of values: $0.290\leq \sin^2\theta^{(l)}_{12}\leq 0.317$, $0.0201\leq \sin^2\theta^{(l)}_{13}\leq 0.0241$, $0.584\leq \sin^2\theta^{(l)}_{23}\leq 0.603$ and $244^\circ\leq \delta_{\text{CP}}\leq 248^\circ$.

\section{\label{scalarsector-331}Scalar potential for the $SU(3)_{L}$ triplets.}\label{scalar}
\lhead[\thepage]{\thesection. Scalar potential $SU(3)_{L}$ triplets}

In this section we will discuss the low energy scalar potential as well as the resulting scalar mass spectrum. As previously mentioned, the discrete groups are spontaneously broken at energy scale much larger than the scale of spontaneoys breaking of the $SU(3)_C\times SU(3)_L\times U(1)_X\times U(1)_{L_g}$ symmetry, then implying that the mixing angles between the gauge singlet scalar fields and the $SU(3)_L$ scalar triplets are strongly suppressed by the ratios between their VEVs (in the scenario of quartic scalar couplings of the same order of magnitude), as follows from the method of method of recursive expansion proposed in \cite{Grimus:2000vj}. Because of this reason, the singlet scalar fields do not have a relevant impact in the electroweak precision observables since they do not couple with the SM gauge bosons and their mixing angles with the neutral components of the $SU(3)_L$ scalar triplets are very small. Therefore, under the aforementioned considerations, we restrict of our analysis of the scalar spectrum to the one resulting from the scalar potential built from $SU(3)_L$ triplets of our model, which are the relevant scalar degrees of freedom at energies corresponding to the scale were the $SU(3)_C\times SU(3)_L\times U(1)_X\times U(1)_{L_g}$ symmetry is spontaneously broken. It is worth mentioning that the most general renormalizable potential invariant under $S_4$ which we can write with the four triplets of the Eqs. (\ref{triplet}) is given by
\begin{eqnarray}\label{Vhiggs}
V &=&-\mu _{\chi }^{2}(\chi ^{\dagger }\chi ) + \mu
_{\eta _{1}}^{2}\eta _{1}^{\dagger }\eta _{1}-\mu _{\eta _{2}}^{2}\eta
_{2}^{\dagger }\eta _{2} -\mu _{\rho }^{2}(\rho
^{\dagger }\rho )+ \varkappa~ \eta _{2}\rho \chi +\lambda_{1}(\chi ^{\dagger }\chi )(\chi ^{\dagger }\chi )+\lambda_{2}(\eta ^{\dagger }\eta )_{\mathbf{1}}(\eta ^{\dagger }\eta )_{\mathbf{1}}
\notag \\
&&+\lambda _{3}(\eta ^{\dagger }\eta )_{\mathbf{1}^{\prime }}(\eta ^{\dagger
}\eta )_{\mathbf{1}^{\prime }}+\lambda _{4}(\eta ^{\dagger }\eta )_{\mathbf{2%
}}(\eta ^{\dagger }\eta )_{\mathbf{2}} +\lambda _{5}(\eta ^{\dagger }\eta )_{\mathbf{1}}(\chi ^{\dagger }\chi)+\lambda _{6}\left[ (\eta ^{\dagger }\chi )(\chi ^{\dagger }\eta )\right] _{%
\mathbf{1}}\\
&&+\lambda _{7}(\rho ^{\dagger }\rho )^{2}+\lambda _{8}(\eta
^{\dagger }\eta )_{\mathbf{1}}(\rho ^{\dagger }\rho )+\lambda _{9}(\chi
^{\dagger }\chi )(\rho ^{\dagger }\rho )+h.c,  \notag
\end{eqnarray}
where $\eta_{1}$ is an inert $SU(3)_L$ scalar triplet, the $\mu$’s are mass parameters and $\varkappa$ is the trilinear scalar coupling, has dimensions of mass, while $\lambda$’s are the quartic dimensionless couplings. 


In the de Born level analysis of the mass spectrum of the Higgs bosons and the physical basis of scalar particles, we must construct the scalar mass matrices of the model. Then substituting the equations of constraintst Eq.~(\ref{constraints}), and Eq.~\eqref{triplet} into the scalar potential Eq.~\eqref{Vhiggs}, the square mass matrices are determined, calculating the second derivatives of the potential
\begin{equation}
\left(M^2_{\phi^{\pm}}\right)_{ij}=\left.\frac{\partial V}{\partial \phi^+_i \partial \phi_j^-}\right\vert_{\left\langle \phi^{\pm}\right\rangle }, \quad \quad
\left(M^2_{\phi}\right)_{ij}=\left.\frac{\partial V}{\partial \phi_i \partial \phi_j}\right\vert_{\left\langle \phi\right\rangle },
\end{equation}

where $\phi^{\pm}=\chi_2^{+},\rho_1^{+},\rho_3^{+},\eta_{21}^{+},\eta_{22}^{+} ,\chi_2^{-},\rho_1^{-},\rho_3^{-},\eta_{21}^{-},\eta_{22}^{-}$, for charged fields and $\phi =$ $\xi_\chi,\xi_\rho,\xi_{\eta_1},\xi_{\eta_2},\zeta_{\chi},$ $\zeta_{\rho},$ $\zeta_{\eta_1 },$ $\zeta_{\eta_2},$ $\chi_1^0,\eta_{31}^0,\eta_{32}^0$, for neutral scalar fields. Due to the symmetry of our models, the matrices of the CP odd and CP even sectors contain two diagonal blocks, a situation similar to that presented in other works on 3-3-1 models \cite{CarcamoHernandez:2019iwh}, however, our results differ in higher dimension matrices due to the extra intert field.
\noindent
In the charged sector, we can obtain a mass squared matrix in the basis $\left(\eta_{21}^{\pm},\eta_{22}^{\pm},\rho_1^{\pm},\rho_3^{\pm},\chi_2^{\pm}\right)$ of the form.
\begin{equation}
M^2_{\phi^{\pm}}=\left(
\begin{array}{ccccc}
\mu_{\eta_1}^2 +\left(\lambda-\lambda_4\right)v_{\eta}^2+\frac{\lambda_8}{2} v_{\rho}^2+\frac{\lambda_5}{2} v_{\chi}^2& 0 & 0 & 0 & 0 \\
 0 & \frac{A v_{\rho } v_{\chi }}{\sqrt{2} v_{\eta _2}} & \frac{A v_{\chi }}{\sqrt{2}} & 0 &
   0 \\
 0 & \frac{A v_{\chi }}{\sqrt{2}} & \frac{A v_{\chi } v_{\eta _2}}{\sqrt{2} v_{\rho }} & 0 &
   0 \\
 0 & 0 & 0 & \frac{A v_{\chi } v_{\eta _2}}{\sqrt{2} v_{\rho }} & \frac{A v_{\eta
   _2}}{\sqrt{2}} \\
 0 & 0 & 0 & \frac{A v_{\eta _2}}{\sqrt{2}} & \frac{A v_{\rho } v_{\eta _2}}{\sqrt{2}
   v_{\chi }} \\
\end{array}
\right)
\end{equation}
\noindent
In sector CP odd, the square matrices in the pseudoscalar neutral basis $\left(\zeta_{\eta_1},\zeta_{\eta_2},\zeta_{\rho},\zeta_{\chi}\right)$
\begin{equation}
M^2_{\zeta}=\left(
\begin{array}{cccc}
\mu _{\eta _1}^2+\left(\lambda _2-2 \lambda _3-\lambda _4\right) v_{\eta _2}^2+\frac{\lambda_8}{2}v_{\rho}^2+\frac{\lambda_5}{2}v_{\chi}^2 & 0 & 0 & 0 \\
 0 & \frac{A v_{\rho } v_{\chi }}{\sqrt{2} v_{\eta
   _2}} & \frac{A v_{\chi }}{\sqrt{2}} & \frac{A
   v_{\rho }}{\sqrt{2}} \\
 0 & \frac{A v_{\chi }}{\sqrt{2}} & \frac{A
   v_{\chi } v_{\eta _2}}{\sqrt{2} v_{\rho }} &
   \frac{A v_{\eta _2}}{\sqrt{2}} \\
 0 & \frac{A v_{\rho }}{\sqrt{2}} & \frac{A
   v_{\eta _2}}{\sqrt{2}} & \frac{A v_{\rho }
   v_{\eta _2}}{\sqrt{2} v_{\chi }} \\
\end{array}
\right)
\end{equation}
and neutral scalar complex at the basis $\left(\text{Im}\eta_{31}^0,\text{Im}\eta_{32}^0,\text{Im}\chi_{1}^0 \right)$
\begin{equation}
M_{\phi_{\text{Im}}}^2=\left(
\begin{array}{ccc}
 2 \mu _{\eta _1}^2+2 \left(\lambda _2-\lambda
   _4\right) v_{\eta _2}^2+\lambda _8 v_{\rho
   }^2+\left(\lambda _5-\lambda _6\right) v_{\chi
   }^2 & 0 & 0 \\
 0 &
   \frac{\sqrt{2} A v_{\rho
   }v_{\chi }}{v_{\eta _2}}-\lambda _6 v_{\chi }^2&
   \lambda _6 v_{\eta _2} v_{\chi }-\sqrt{2} A
   v_{\rho } \\
 0 & \lambda _6 v_{\eta _2} v_{\chi }-\sqrt{2} A
   v_{\rho } & \frac{\sqrt{2} A
    v_{\eta _2} v_{\rho }}{v_{\chi }}-\lambda _6 v_{\eta
   _2}^2 \\
\end{array}
\right)
\end{equation}
\noindent
In the sector CP even, the square mass matrices in the scalar neutral basis $\left(\xi_{\eta_1},\xi_{\eta_2},\xi_{\rho},\xi_{\chi}\right) $
\begin{equation}
M^2_\xi=\left(
\begin{array}{cccc}
\mu _{\eta _1}^2+\left(\lambda _2+\lambda _4\right)
   v_{\eta _2}^2+\frac{\lambda_8}{2}v_{\rho}^2+\frac{\lambda _5}{2} v_{\chi}^2 & 0 & 0 & 0 \\
 0 &
 \frac{A v_{\rho } v_{\chi }}{\sqrt{2} v_{\eta _2}^2}
 +2\left(\lambda_2+\lambda_4 \right)v_{\eta _2}^2
 & \lambda _8 v_{\eta _2}
   v_{\rho }-\frac{A v_{\chi }}{\sqrt{2}} &
   \lambda _5 v_{\eta _2} v_{\chi }-\frac{A
   v_{\rho }}{\sqrt{2}} \\
 0 & \lambda _8 v_{\eta _2} v_{\rho }-\frac{A
   v_{\chi }}{\sqrt{2}} & \frac{A v_{\eta _2}
   v_{\chi }}{\sqrt{2} v_{\rho }}+2 \lambda _7
   v_{\rho }^2 & \lambda _9 v_{\rho } v_{\chi
   }-\frac{A v_{\eta _2}}{\sqrt{2}} \\
 0 & \lambda _5 v_{\eta _2} v_{\chi }-\frac{A
   v_{\rho }}{\sqrt{2}} & \lambda _9 v_{\rho }
   v_{\chi }-\frac{A v_{\eta _2}}{\sqrt{2}} &
   \frac{A v_{\eta _2} v_{\rho }}{\sqrt{2} v_{\chi
   }}+2 \lambda _1 v_{\chi }^2 \\
\end{array}
\right)
\end{equation}
and neutral scalar complex at the basis $\left(\text{Re}\eta_{31}^0,\text{Re}\eta_{32}^0,\text{Re}\chi_{1}^0 \right)$
\begin{equation}
M_{\phi_{\text{Re}}}^2=\left(
\begin{array}{ccc}
 2 \mu _{\eta _1}^2+2 \left(\lambda _2-\lambda
   _4\right) v_{\eta _2}^2+\lambda _8 v_{\rho
   }^2+\left(\lambda _5+\lambda _6\right) v_{\chi
   }^2 & 0 & 0 \\
 0 &  \frac{\sqrt{2} A v_{\rho
   }v_{\chi }}{v_{\eta _2}}+\lambda _6 v_{\chi }^2 &
   \sqrt{2} A v_{\rho }+\lambda _6 v_{\eta _2}
   v_{\chi } \\
 0 & \sqrt{2} A v_{\rho }+\lambda _6 v_{\eta _2}
   v_{\chi } &  \frac{\sqrt{2} A
   v_{\eta _2}v_{\rho }}{v_{\chi }}+\lambda _6 v_{\eta
   _2}^2 \\
\end{array}
\right)
\end{equation}
Furthermore, the  minimization conditions of the scalar potential yield the following relations
\begin{eqnarray}
\nonumber \mu_{\chi}^2 &=&-\frac{\varkappa}{\sqrt{2}} \frac{v_{\eta _2} v_{\rho }}{v_{\chi}}
+\frac{\lambda_5}{2}v_{\eta _2}^2 
+\frac{\lambda_9}{2}v_{\rho }^2 
+\lambda_1 v_{\chi}^2,\\
\label{constraints} \mu_{\eta_2}^2 &=& -
\frac{\varkappa}{\sqrt{2}} \frac{v_{\rho } v_{\chi }}{v_{\eta _2}}+\frac{\lambda_8}{2} v_{\rho}^2 +\frac{\lambda_5}{2} v_{\chi}^2 +\left(\lambda_2+\lambda_4\right)  v_{\eta _2}^2, \\
\nonumber \mu_{\rho}^2 &=& -
\frac{\varkappa}{\sqrt{2}} \frac{v_{\eta _2} v_{\chi }}{v_{\rho }}+\frac{\lambda_8}{2}v_{\eta_2}^2+\frac{\lambda_9}{2}v_{\chi }^2+\lambda_7 v_{\rho}^2.
\end{eqnarray}

\begin{figure}
\subfigure[]{\includegraphics[scale=0.3]{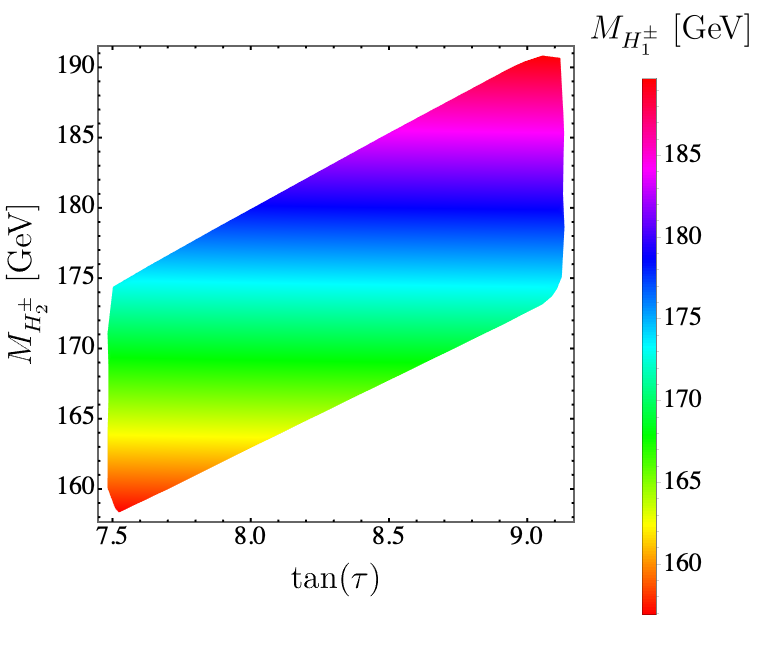}}
\subfigure[]{\includegraphics[scale=0.3]{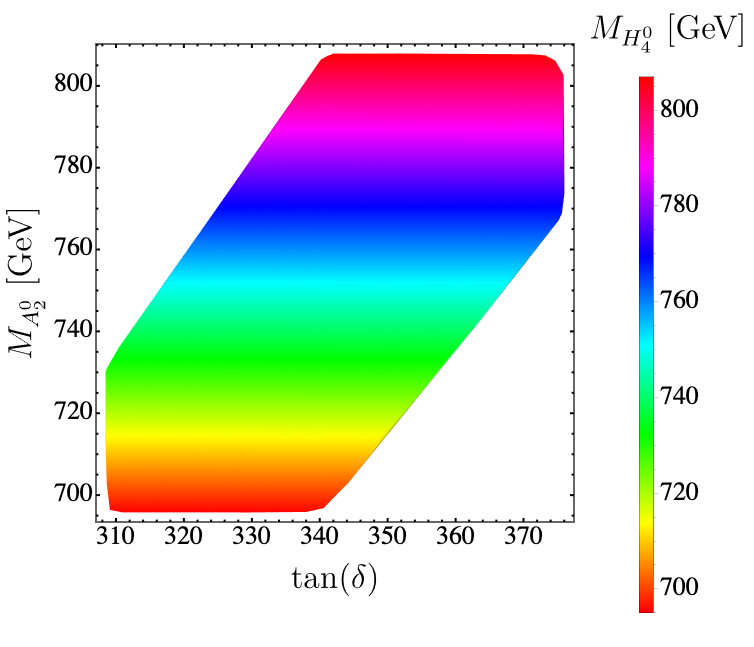}}
\caption{Correlations between mixing angles and the masses of the physical charged scalar, neutral scalar/pseudoscalar fields.}
\label{mangles}
\end{figure}

After spontaneous symmetry breaking, the Higgs mass spectrum comes from the diagonalization of the squared mass matrices. The mixing angles\footnote{The symbol $\beta$ is used in the scalar potential for the $SU(3)_{L}$ triplets as one of the mixing angles, it is a symbol different from the definition of the operator electric charge.} for the physical eigenstates are:
\begin{eqnarray} \label{mixingangles}
\tan\alpha &=& \frac{v_{\eta_2}}{v_{\chi}}, \quad \tan\beta \ =\ \frac{v_{\rho}}{v_{\chi}}, \quad \tan\tau \ =\ \frac{v_{\eta_2}}{v_{\rho}}, \quad \tan\delta = \frac{v_{\chi}}{v_{\rho} \sin\tau}, \quad \tan2\vartheta = \frac{2\sqrt{2} \lambda_8}{\varkappa v_\chi}\frac{v_{\eta_2}^2 v_\rho^2}{v_\rho^2-v_{\eta_2}^2 },
\end{eqnarray}  
The Fig.~\ref{mangles}, presents correlation plots demonstrating the relationships between mixing angles and physical scalar masses, these correlation plots have been obtained by varying the point of best fit of the potential sector parameters around $10\%$. These plots highlight specific correlations, such as the correlation between the $\tau$ angle and charged fields, and the correlation between the $\delta$ angle and charged and pseudo-scalar fields. These correlations provide invaluable insights into the interactions among scalar fields and enhance our understanding of particle properties and the relationships between their masses and mixing angles. Moreover, similar correlations were observed for other mixing angles. The correlation analyses of the mixing angles offer valuable information regarding the underlying theoretical structure and relationships within the model.

We find that the charged sector is composed of two Goldstone bosons and three massive charged scalars.
\begin{eqnarray}
M_{G_{1}^{\pm }}^{2} &=& M_{G_{2}^{\pm }}^{2}=0, \\
M_{H_{1}^{\pm }}^{2} &=&\frac{\varkappa v_{\eta _{2}}v_{\rho }}{\sqrt{2}v_{\chi }}%
+\frac{\varkappa v_{\eta _{2}}v_{\chi }}{\sqrt{2}v_{\rho }} ,\\
M_{H_{2}^{\pm }}^{2} &=&\frac{\varkappa v_{\eta _{2}}v_{\chi }}{\sqrt{2}v_{\rho }}%
+\frac{\varkappa v_{\rho }v_{\chi }}{\sqrt{2}v_{\eta _{2}}}, \\
M_{H_{3}^{\pm }}^{2} &=&\mu_{\eta_1}^{2}+\left( \lambda _{2}-\lambda _{4}\right)
v_{\eta _{2}}^{2}+\frac{\lambda _{8}v_{\rho }^{2}}{2}+\frac{\lambda
_{5}v_{\chi }^{2}}{2}.
\end{eqnarray}

The Goldstone bosons come only from mixing between $\rho_3^{\pm}$ and $\eta_{22}^\pm$, through the angle $\tau$, while $\chi_2^\pm$ is a massive field charged, the other two bulk fields correspond to the blending of $\rho_1^\pm$ and the charged component of the scalar inert $\eta_{21}^\pm$ by blending angle $\beta$, i.e,
\begin{equation}
\begin{array}{lll}
G_{1}^{\pm }=\cos \tau \ \rho _{3}^{\pm }-\sin \tau \ \eta _{22}^{\pm }, & 
& G_{2}^{\pm }=\sin \tau \ \rho _{3}^{\pm }+\cos \tau \ \eta _{22}^{\pm },
\\ 
H_{1}^{\pm }=\cos \beta \ \rho _{1}^{\pm }-\sin \beta \ \eta _{21}^{\pm }, & 
& H_{2}^{\pm }=\sin \beta \ \rho _{1}^{\pm }+\cos \beta \ \eta _{21}^{\pm },
\\ 
H_{3}^{\pm }=\chi _{2}^{\pm }. &  & 
\end{array}
\end{equation}

The physical mass eigenvalues of the CP odd scalars $A_ 1^0, \ A_2^0$ and the Goldstone bosons $G_1^0, \ G_2^0$ can be written as:
\begin{eqnarray}
M_{G_{1}^{0}}^{2} &=&M_{G_{2}^{0}}^{2} \ = \ 0, \\
M_{A_{1}^{0}}^{2} &=&\frac{\varkappa v_{\eta _{2}}v_{\rho }}{\sqrt{2}v_{\chi }}+%
\frac{\varkappa v_{\eta_2} v_{\chi}}{\sqrt{2} v_{\rho}} +\frac{\varkappa v_{\rho} v_{\chi}}{\sqrt{2} v_{\eta_2}}, \\
M_{A_{2}^{0}}^{2} &=&\mu _{\eta_1}^{2}+v_{\eta _{2}}^{2}\left( \lambda_{2}-2\lambda _{3}-\lambda_{4}\right) +\frac{\lambda_{8}v_{\rho }^{2}}{2}+%
\frac{\lambda _{5}v_{\chi }^{2}}{2}.
\end{eqnarray}

\noindent
We have the following relationship between the original physical eigenstates:
\begin{equation}
\begin{array}{lll}
G_{1}^{0}=\cos \tau \ \zeta _{\rho }-\cos \delta \cos \tau \ \zeta _{\eta_1}-\sin \tau \ \zeta _{\eta _{2}}, &  & G_{2}^{0}=\cos \delta \ \zeta _{\rho
}+\sin \delta \ \zeta _{\eta _{1}}, \\ 
A_{1}^{0}=\sin \tau \ \zeta _{\rho }-\cos \delta \sin \tau \ \zeta _{\eta
_{1}}+\cos \tau \ \zeta _{\eta _{2}}, &  & A_{2}^{0}=\zeta _{\chi },%
\end{array}
\end{equation}
where consider the following limit $v_\chi\gg v_{\rho},v_{\eta_2}$.
\noindent
The masses of the light and heavy eigenstates for CP even scalars are given as:
\begin{eqnarray}
M^2_h  &=& \varkappa \frac{\left( v_{\eta_2}^2 + v_\rho^2\right) v_\chi}{2\sqrt{2}v_{\eta_2} v_\rho } - \frac{1}{2\sqrt{2}v_{\eta_2} v_\rho }  \sqrt{\varkappa^2\left( v_{\eta_2}^2 - v_\rho^2\right)^2 v_\chi^2 +8 \lambda_8^2 v_{\eta_2}^4  v_\rho^4} ,\\
M_{H_1^0}^2 &=&  \varkappa \frac{\left( v_{\eta_2}^2 + v_\rho^2\right) v_\chi}{2\sqrt{2}v_{\eta_2} v_\rho } +\frac{1}{2\sqrt{2}v_{\eta_2} v_\rho }  \sqrt{\varkappa^2\left( v_{\eta_2}^2 - v_\rho^2\right)^2 v_\chi^2 +8 \lambda_8^2v_{\eta_2}^4  v_\rho^4} ,\\
M_{H_2^0}^2&=& \mu _{\eta _1}^2+\left(\lambda _2+\lambda _4\right)
   v_{\eta _2}^2+\frac{\lambda_8}{2}v_{\rho}^2+\frac{\lambda _5}{2} v_{\chi}^2,\\
M_{H_3^0}^2 &=&  2 \lambda _1 v_{\chi }^2.
\end{eqnarray}

The lighter mass eigenstate $h$ is identified as the SM Higgs boson. The two mass eigenstates $h$ and $H_1^0$ are related with the $\xi_{\eta_2}$ and $\xi_\rho$ fields through the rotation angle $\vartheta$ as:
 \begin{eqnarray}
h&\simeq &  \xi_{\eta_2}\cos\vartheta -  \xi_{\rho}  \sin\vartheta ,\\
 H_1^0&\simeq & \xi_{\eta_2}\sin\vartheta +  \xi_{\rho}  \cos\vartheta, 
\end{eqnarray}
while the heavier fields are related as $H_2^0 \simeq \xi_{\eta_1}$ and $H_3^0 \simeq \xi_\chi$.

Finally, for the pseudoscalar and scalar neutral complex fields, we have composed the mixture of the Imaginary and Real parts of $\eta_{31}^0,\ \eta_{32}^0,\ \chi_{1}^0$, respectively,
\begin{eqnarray}
M_{G_{3}^{0}}^{2} &=&M_{G_{4}^{0}}^{2}\ = \ 0 \\
M_{A_{3}^{0}}^{2} &=& \sqrt{2}\varkappa \left(
\frac{v_{\eta _{2}}v_{\rho }}{v_{\chi }}+
\frac{v_{\rho }v_{\chi }}{v_{\eta _{2}}}
\right)
-\lambda _{6}\left(
v_{\chi }^{2}+v_{\eta _{2}}^{2}\right),  \\
M_{A_{4}^{0}}^{2} &=&2\mu _{\eta_1}^{2}
+2\left(\lambda_2-\lambda_4 \right)v_{\eta _{2}}^{2}
+\lambda_8 v_\rho^2 + \left(\lambda_5-\lambda_6 \right) v^2_\chi,\\
M_{H_{4}^{0}}^{2} &=& \sqrt{2}\varkappa\left(
\frac{v_{\eta _{2}}v_{\rho }}{v_{\chi }}+
\frac{v_{\rho }v_{\chi }}{v_{\eta _{2}}}
\right)
+\lambda _{6}\left(
v_{\chi }^{2}+v_{\eta _{2}}^{2}\right) , \\
M_{H_{5}^{0}}^{2} &=&2\mu _{\eta_1}^{2}+2\left(\lambda _{2}-\lambda_{4}\right) v_{\eta _{2}}^{2}+\lambda _{8}v_{\rho }^{2}+\left(\lambda _{5}+\lambda _{6}\right) v_{\chi }^{2}.
\end{eqnarray}

In the physical eigenstates, there are two Goldstone bosons and one pseudoscalar massive boson, and a scalar from the mixture of the complex neutral part of $\eta_1$ and $\eta_2$, while,
\begin{equation}
\begin{array}{lll}
G_{3}^{0}=\sin \alpha\ \text{Im}\eta_{31}^0-\cos \alpha \ \text{Im}\eta _{32}^{0}, &  & G_{4}^{0}=-\sin \alpha \ \text{Re}\eta_{31}^{0}+\cos \alpha \ \text{Re}\eta_{32}^{0}, \\ 
A_{3}^{0}=\cos \alpha \ \text{Im}\eta _{31}^{0}+\sin \alpha \ \text{Im}\eta_{32}^{0}, &  & H_{4}^{0}=\cos \alpha \ \text{Re}\eta _{31}^{0}+\sin\alpha \ \text{Re}\eta_{32}^{0}, \\ 
A_{4}^{0}=\text{Im}\chi_{1}^{0}, &  & H_{5}^{0}=\text{Re}\chi_{1}^{0}.
\end{array}
\end{equation}

\begin{figure}
\centering
\subfigure[]{\includegraphics[scale=0.4]{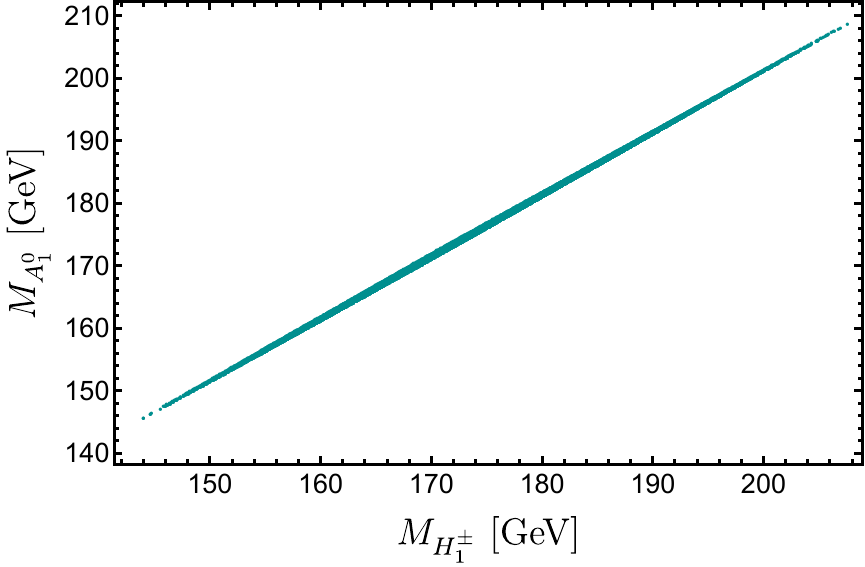}}
\subfigure[]{\includegraphics[scale=0.4]{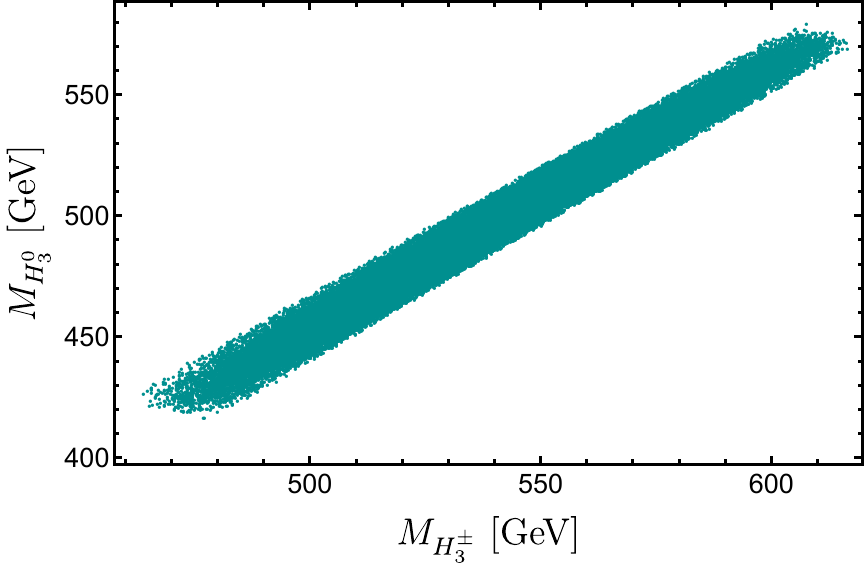}}
\caption{Correlation plot between the pseudoscalar neutral, scalar neutral, and scalar charged masses.}
\label{fig:scalar-sector}
\end{figure}

In our model, the physical scalar masses can also be expressed 
in terms of the scalar mixing angles, as shown in table \ref{tab:massanglesmix}, where the light scalar field $h$, is identified as the SM-like Higgs boson, additionally six charged fields $\left(H_1^{\pm}, H_2^{\pm},H_3^{\pm} \right)$, five CP even $\left(H_1^0,H_2^0,H_3^0,H_4^0, H_5^0 \right)$ and four CP odd fields $\left(A_1^0,A_2^0,A_3^0, A_4^0 \right)$. Fig.~\ref{fig:scalar-sector}, displays 
the correlation between the charged and neutral scalar masses. Fig.~\ref{fig:scalar-sector} \textcolor{red}{(a)}, shows a linear correlation between of the masses between the charged field and the pseudoscalar neutral field, $H_1^{\pm}$ and $A_1^0$ respectively, in Fig.~\ref{fig:scalar-sector} \textcolor{red}{(b)}, a linear correlation between the masses of the pseudoscalar and scalar neutral field, $A_3^0$ and $H_3^0$ respectively. The charged Goldstone bosons $\left(G_1^{\pm},G_2^{\pm} \right)$ are related to the longitudinal components of the $W^{\pm}$ and $W'^{\pm}$ gauge bosons respectively; while the neutral Goldstone bosons $\left(G_1^0,G_2^0,G_3^0,G_4^0\right)$ are associated to the longitudinal components of the $Z$, $Z'$, $K^0$ and $K'^0$ gauge bosons.

\begin{table}[h]
    \centering
    \begin{tabular}{|c|c|}
        \hline
        \text{Scalar} & \text{Masses} \\
        \hline
        $M_{G_{1}^{\pm }}^{2}, \, M_{G_{2}^{\pm }}^{2}$ & $0$ \\
        $M_{H_1^{\pm}}^{2}$ & $\frac{\varkappa}{\sqrt{2}}a v_{\eta_2}  \csc (\beta ) \sec (\beta )$ \\
        $M_{H_2^{\pm}}^{2}$ & $\frac{\varkappa}{\sqrt{2}} v_\chi \csc (\tau ) \sec (\tau )$ \\
        $M_{H_3^{\pm}}^{2}$ & $\mu_{\eta_1}^{2} + v_\chi^2 \left( \frac{\lambda_5}{2} + \left(\lambda_2 - \lambda_4 \right) \tan^2(\alpha ) + \frac{\lambda_8}{2} \tan^2(\beta) \right)$ \\
        \hline
        $M_{G_{1}^{0}}^{2}, \, M_{G_{2}^{0}}^{2}$ & $0$ \\
        $M_{A_{1}^{0}}^{2}$ & $\frac{\varkappa}{\sqrt{2}} \csc (\tau ) \left( v_{\eta_2} \cot (\delta ) + v_\chi \sec (\tau ) \right)$ \\
        $M_{A_{2}^{0}}^{2}$ & $\mu_{\eta_1}^{2} + v_\chi^2 \left( \frac{\lambda_5}{2} + \left( \lambda_2 - 2\lambda_3 - \lambda_4 \right) \tan^2(\alpha ) + \frac{\lambda_8}{2} \tan^2(\beta) \right)$ \\
        \hline
        $M_{h}^{2}$ & $\frac{\varkappa }{2\sqrt{2}} v_\chi \csc (\tau ) \sec (\tau ) \left( 1 + \sec (2 \vartheta ) \cos (2 \tau ) \right)$ \\
        $M_{H_1^0}^{2}$ & $\frac{\varkappa }{2\sqrt{2}} v_\chi \csc (\tau ) \sec (\tau ) \left( 1 - \sec (2 \vartheta ) \cos (2 \tau ) \right)$ \\
        $M_{H_2^0}^{2}$ & $\mu_{\eta_1}^{2} + v_\chi^2 \left( \frac{\lambda_5}{2} + \left( \lambda_2 + \lambda_4 \right) \tan^2(\alpha ) + \frac{\lambda_8}{2} \tan^2(\beta) \right)$ \\
        $M_{H_3^0}^{2}$ & $2 \lambda_1 v_{\chi }^2$ \\
        \hline
        $M_{G_{3}^{0}}^{2}, \, M_{G_{4}^{0}}^{2}$ & $0$ \\
        $M_{A_{3}^{0}}^{2}$ & $v_{\eta_2} \csc^2(\alpha) \left( \sqrt{2} \varkappa \tan(\alpha) - \lambda_6 v_{\eta_2} \right)$ \\
        $M_{A_{4}^{0}}^{2}$ & $2 \mu_{\eta_1}^{2} + v_\chi^2 \left( \lambda_5 - \lambda_6 + 2 \left( \lambda_2 - \lambda_4 \right) \tan^2(\alpha) + \lambda_8 \tan^2(\beta) \right)$ \\
        $M_{H_{4}^{0}}^{2}$ & $v_{\eta_2} \csc^2(\alpha) \left( \sqrt{2} \varkappa \tan(\alpha) + \lambda_6 v_{\eta_2} \right)$ \\
        $M_{H_{5}^{0}}^{2}$ & $2 \mu_{\eta_1}^{2} + v_\chi^2 \left( \lambda_5 + \lambda_6 + 2 \left( \lambda_2 - \lambda_4 \right) \tan^2(\alpha) + \lambda_8 \tan^2(\beta) \right)$ \\
        \hline
    \end{tabular}
    \caption{Physical mass spectrum of the scalars in terms of mixing angles.}
    \label{tab:massanglesmix}
\end{table}

\begin{figure}
\centering
\includegraphics[scale=0.35]{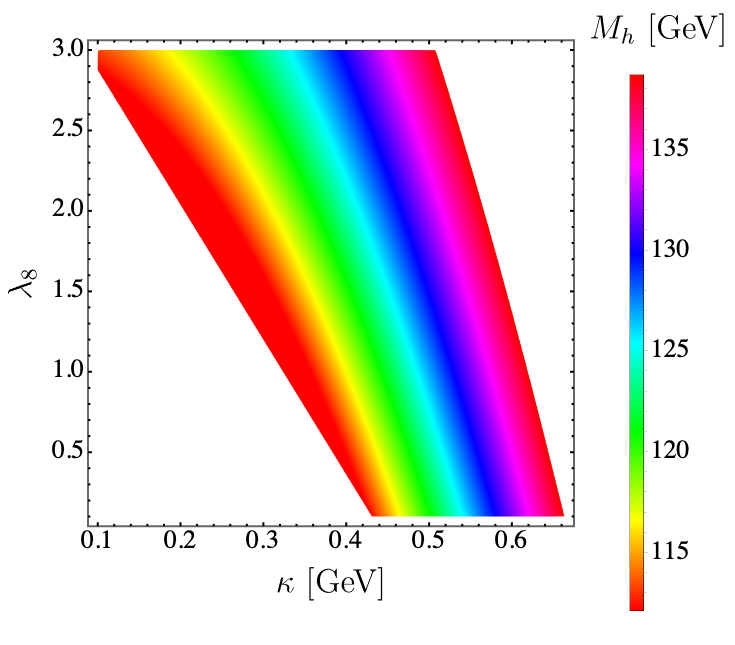}
\caption{Correlation color plot between the trilinear scalar parameter $\varkappa$ and the quartic scalar coupling $\lambda_8$ consistent with the SM-like Higgs mass.}
\label{fig:kappavslambda8}
\end{figure}

In analyzing the scalar sector of the model, 
it should be noted that in the potential of Eq.~\eqref{Vhiggs}, the quartic coupling parameters must meet the following constraint $\left\vert \lambda_i \right\vert < 4\pi$ due to perturbative unitarity. Taking this into account we can successfully accommodate the $125$ GeV mass for the SM like Higgs boson found at the LHC for 
the following VEV values:
\begin{equation}\label{eq:vevsmodel}
v_{\chi }\simeq 9.994\ \text{TeV}, \quad  v_{\eta _{2}}\simeq 244.2\ \text{GeV}, \quad v_{\rho }\simeq 29.54\ \text{GeV},
\end{equation}
which yield a mass $m_h=125.387\ \text{GeV}$ for the SM like Higgs boson. Note that the Higgs SM boson mass depends on the trilinear coupling $\varkappa$ and the quartic coupling $\lambda_8$, which is lower than its upper limit of $4\pi$ arising from perturbativity. 
Fig. \ref{fig:kappavslambda8} displays the correlation between the trilinear scalar parameter $\varkappa$ and the quartic scalar coupling $\lambda_8$ consistent with the experimental values of the SM like Higgs boson mass. Here we have used the numerical values of the VEVs of Eq.~\eqref{eq:vevsmodel} 
The new scalar fields introduce corrections to the phenomenological processes that arise in this model and provide more precise determinations of phenomena such as, for instance $K$ meson oscillations. In the SM-type two-photon Higgs decay constraints, where extra-charged scalar fields induce one loop level corrections to the Higgs diphoton decay. 
These phenomenological processes are studied in more detail in the following sections.

\subsection{The scalar potential for a $S_4$ doublet and triplet}
\label{ScalarS4doubletandtriplet}
The scalar potential for a $S_4$ doublet $\eta$ is given by
\begin{equation} \label{VD}
V_D=-g_{\eta}^2 \left( \eta \eta^*\right)_\textbf{1}+k_1 \left( \eta \eta^*\right)_\textbf{1}\left( \eta \eta^*\right)_\textbf{1} +k_2 \left( \eta \eta^*\right)_{\textbf{1}^{\prime}}\left( \eta \eta^*\right)_{\textbf{1}^{\prime}} + k_3   \left( \eta \eta^*\right)_\textbf{2} \left( \eta \eta^*\right)_\textbf{2}.
\end{equation} 

The components of this expression are characterized by the presence of four unconstrained parameters: a bilinear coupling and three quartic couplings. From the scalar potential minimization: 
\begin{equation}
    \begin{aligned}
        \frac{\partial  \langle V_D \rangle}{\partial v_{\eta_2}} = & \  -2 g_{\eta}^2 v_{\eta_2} +4 k_1 v_{\eta_2}^3+4 k_3 v_{\eta_2}^3 = 0,
    \end{aligned}
\end{equation}
we obtain the parameter $g_{\eta}$ as a function of the other three parameters, i.e.
\begin{equation}
    \begin{aligned}
        g_{\eta}^2 = & \ 2 \left(k_1+k_3 \right) v_{\eta_2}^2,
    \end{aligned}
\end{equation}
with $k_1, \ k_3 \in \mathbb{R}$.
On the other hand, the condition of having a global minimum:
\begin{equation}
    \begin{aligned}
        \frac{\partial^2 \langle V_D \rangle}{\partial v_{\eta_2}^2}  = & \  -g_{\eta}^2 + 6 v_{\eta_{2}}^{2} (k_1 + k_3) > 0,
    \end{aligned}
\end{equation}
gives rise to the followingfollowing inequality:
\begin{equation}
    \begin{aligned}
        g_{\eta}^2 < & \ 6 v_{\eta_{2}}^{2} (k_1 + k_3),
    \end{aligned}
\end{equation}.

This result suggests that the VEV pattern of the $S_4$ doublet $\eta$ as presented in Eq. (\ref{veveta}), is consistent with a global minimum of the scalar potential of Eq. (\ref{VD}), in a large parameter space region.

\noindent
The relevant terms determining the VEV directions of any $S_4$ scalar triplet are:
\begin{eqnarray}
V_T &= & -g_{\Omega}^2 \left( \Omega \Omega^*\right)_\textbf{1}+k_1 \left( \Omega \Omega^*\right)_\textbf{1}\left( \Omega \Omega^*\right)_\textbf{1} +k_2 \left( \Omega \Omega^*\right)_{\textbf{3}}\left( \Omega \Omega^*\right)_{\textbf{3}}+ k_3 \left( \Omega \Omega^*\right)_{\textbf{3}^{\prime}}\left( \Omega \Omega^*\right)_{\textbf{3}^{\prime}} \\
\nonumber & &+ k_4 \left( \Omega \Omega^*\right)_\textbf{2} \left( \Omega \Omega^*\right)_\textbf{2}
+H.c.
\label{VT}
\end{eqnarray} 
\noindent
where $\Omega=S_e,S_{\mu},S_{\tau},\Phi,\xi$.

The part of the scalar potential for each $S_4$ scalar triplet has five free parameters: one bilinear and
four quartic couplings. The minimization conditions of the scalar potential for a $S_4$ triplet yield the
following relations:

\begin{eqnarray} \label{vevS4}
\nonumber \frac{\partial \langle V_T \rangle}{\partial v_{\Omega_1}}   & = & -2g_{\Omega}^2 v_{\Omega_1} +8 k_2 v_{\Omega_1}  \left(v_{\Omega_2}^2 +v_{\Omega_3}^2 \right) + 4k_1 v_{\Omega_1} \left( v_{\Omega_1}^2 + v_{\Omega_2}^2 + v_{\Omega_3}^2 \right) - \frac{4}{3} k_4 v_{\Omega_1} \left( -2v_{\Omega_1}^2 + v_{\Omega_2}^2 + v_{\Omega_3}^2 \right)  \\
 \nonumber & = & 0, \\
 \nonumber & & \\
\frac{\partial \langle V_T \rangle}{\partial v_{\Omega_2}}   & = & -2g_{\Omega}^2 v_{\Omega_2} + 8 k_2 v_{\Omega_2} \left( v_{\Omega_1}^2 + v_{\Omega_3}^2  \right)+ 4 k_1 v_{\Omega_2} \left( v_{\Omega_1}^2 + v_{\Omega_2}^2 + v_{\Omega_3}^2 \right) +2k_4  v_{\Omega_2}  \left( v_{\Omega_2}^2 - v_{\Omega_3}^2\right) \\
\nonumber & & + \frac{2}{3} k_4 v_{\Omega_2} \left(-2v_{\Omega_1}^2 +v_{\Omega_2}^2 + v_{\Omega_3}^2 \right) \\
 \nonumber & = & 0, \\
\nonumber  & & \\
\nonumber \frac{\partial \langle V_T \rangle}{\partial v_{\Omega_3}}   & = & -2g_{\Omega}^2 v_{\Omega_3}
+8 k_2 v_{\Omega_3}   \left( v_{\Omega_1}^2 + v_{\Omega_2}^2 \right) + 4k_1 v_{\Omega_3} \left( v_{\Omega_1}^2 + v_{\Omega_2}^2 + v_{\Omega_3}^2 \right)- 2k_4  v_{\Omega_3}  \left( v_{\Omega_2}^2 - v_{\Omega_3}^2\right) \\
\nonumber & & + \frac{2}{3} k_4v_{\Omega_3} \left(-2v_{\Omega_1}^2 +v_{\Omega_2}^2 + v_{\Omega_3}^2 \right) \\
 \nonumber & = & 0.
\end{eqnarray}
whereas the condition of having a global minimum gives rise to the following inequalities:
\begin{equation}
\begin{aligned}
\frac{\partial^2 \langle V_T \rangle}{\partial v_{\Omega_1}^2} = & \  -2 g_{\Omega}^2 + 8 k_{2} \left( v_{\Omega_{2}}^{2} + v_{\Omega_{3}}^{2} \right) + 4 k_{1} \left( 3 v_{\Omega_{1}}^{2} + v_{\Omega_{2}}^{2} + v_{\Omega_{3}}^{2} \right) + \frac{4}{3} k_{4} \left( 6 v_{\Omega_{1}}^{2} - v_{\Omega_{2}}^{2}- v_{\Omega_{3}}^{2} \right) > 0, \\
 & \\
\frac{\partial^2 \langle V_T \rangle}{\partial v_{\Omega_2}^2} = & -2 g_{\Omega}^2 + 8 k_{2} \left( v_{\Omega_{1}}^{2} + v_{\Omega_{3}}^{2} \right) + 4 k_{1} \left( v_{\Omega_{1}}^{2} + 3 v_{\Omega_{2}}^{2} + v_{\Omega_{3}}^{2} \right) + 2 k_{4} \left( 3 v_{\Omega_{2}}^{2} - v_{\Omega_{3}}^{2} \right) + \\
& - \frac{4}{3} k_{4} \left(  v_{\Omega_{1}}^{2} -6 v_{\Omega_{2}}^{2} + v_{\Omega_{3}}^{2} \right) > 0, \\
 & \\
\frac{\partial^2 \langle V_T \rangle}{\partial v_{\Omega_3}^2} = & -2 g_{\Omega}^2 + 8 k_{2} \left( v_{\Omega_{1}}^{2} + v_{\Omega_{2}}^{2} \right) + 4 k_{1} \left( v_{\Omega_{1}}^{2} + v_{\Omega_{2}}^{2} + 3 v_{\Omega_{3}}^{2} \right) - 2 k_{4} \left( v_{\Omega_{2}}^{2} - 3 v_{\Omega_{3}}^{2} \right) \\ 
& - \frac{4}{3} k_{4} \left(  v_{\Omega_{1}}^{2} + v_{\Omega_{2}}^{2}-6 v_{\Omega_{3}}^{2} \right) > 0,
\end{aligned}
\end{equation}
\noindent
From the scalar potential minimization equations. From the first minimization condition we get:
\begin{eqnarray}
g_{S_k}^2 & = & \frac{\left(3k_1 + 2k_4 \right)}{3}v_{S_k}^2, \\
g_{\Phi}^2 & = &\frac{2  \left(2 e^{4 i \theta } \left(6 \left(k_1+k_2\right)+k_4\right) r_1^4+4 e^{2 i\theta } \left(3 k_1+6 k_2-k_4\right) r_1^2+3 k_1+2 k_4\right)}{3+6 e^{2 i \theta } r_1^2} v_\Phi^2,\\
g_{\xi}^2&=& \frac{2  \left(2 k_4 \left(r_2^2-1\right){}^2+3 k_1 \left(r_2^2+2\right){}^2+12 k_2 \left(2
   r_2^2+1\right)\right)}{3 \left(r_2^2+2\right)} v_\xi^2,
\end{eqnarray}
where $k=e,\mu,\tau$.
This shows that the VEV configuration of the $S_4$ triplet $S_e$
given in the equations \eqref{eq:vev-lep} and \eqref{eq:vev-lep2}, is in accordance with the scalar potential minimization condition of Eq. \eqref{vevS4}. The other $S_4$ triplets in our model are also consistent with the scalar potential minimization conditions, which can be demonstrated using the same procedure described in this section.
These results show that the VEV directions for the $S_4$ triplets $S_e$, $S_{\mu}$, $S_{\tau}$, $\Phi$, $\xi$ are consistent with a global minimum of the scalar potential for a large region of parameter space.
\section{Meson mixings}\label{meson}
\lhead[\thepage]{\thesection. Meson mixings}

In this section we discuss the implications of our model in
the Flavour Changing Neutral Current (FCNC) interactions in the down type
quark sector. These FCNC down type quark Yukawa interactions produce $K^{0}-%
\bar{K}^{0}$, $B_{d}^{0}-\bar{B}_{d}^{0}$ and $B_{s}^{0}-\bar{B}_{s}^{0}$
meson oscillations, whose corresponding effective Hamiltonians are: 
\begin{equation}
\mathcal{H}_{eff}^{\left( K\right) }\mathcal{=}\sum_{j=1}^{3}\kappa
_{j}^{\left( K\right) }\left( \mu \right) \mathcal{O}_{j}^{\left( K\right)
}\left( \mu \right) ,
\end{equation}%
\begin{equation}
\mathcal{H}_{eff}^{\left( B_{d}\right) }\mathcal{=}\sum_{j=1}^{3}\kappa
_{j}^{\left( B_{d}\right) }\left( \mu \right) \mathcal{O}_{j}^{\left(
B_{d}\right) }\left( \mu \right) ,
\end{equation}%
\begin{equation}
\mathcal{H}_{eff}^{\left( B_{s}\right) }\mathcal{=}\sum_{j=1}^{3}\kappa
_{j}^{\left( B_{s}\right) }\left( \mu \right) \mathcal{O}_{j}^{\left(
B_{s}\right) }\left( \mu \right) ,
\end{equation}

where: 
\begin{eqnarray}
\mathcal{O}_{1}^{\left( K\right) } &=&\left( \overline{s}_{R}d_{L}\right)
\left( \overline{s}_{R}d_{L}\right) ,\hspace{0.7cm}\hspace{0.7cm}\mathcal{O}%
_{2}^{\left( K\right) }=\left( \overline{s}_{L}d_{R}\right) \left( \overline{%
s}_{L}d_{R}\right) ,\hspace{0.7cm}\hspace{0.7cm}\mathcal{O}_{3}^{\left(
K\right) }=\left( \overline{s}_{R}d_{L}\right) \left( \overline{s}%
_{L}d_{R}\right) ,  \label{op3f} \\
\mathcal{O}_{1}^{\left( B_{d}\right) } &=&\left( \overline{d}%
_{R}b_{L}\right) \left( \overline{d}_{R}b_{L}\right) ,\hspace{0.7cm}\hspace{%
0.7cm}\mathcal{O}_{2}^{\left( B_{d}\right) }=\left( \overline{d}%
_{L}b_{R}\right) \left( \overline{d}_{L}b_{R}\right) ,\hspace{0.7cm}\hspace{%
0.7cm}\mathcal{O}_{3}^{\left( B_{d}\right) }=\left( \overline{d}%
_{R}b_{L}\right) \left( \overline{d}_{L}b_{R}\right) , \\
\mathcal{O}_{1}^{\left( B_{s}\right) } &=&\left( \overline{s}%
_{R}b_{L}\right) \left( \overline{s}_{R}b_{L}\right) ,\hspace{0.7cm}\hspace{%
0.7cm}\mathcal{O}_{2}^{\left( B_{s}\right) }=\left( \overline{s}%
_{L}b_{R}\right) \left( \overline{s}_{L}b_{R}\right) ,\hspace{0.7cm}\hspace{%
0.7cm}\mathcal{O}_{3}^{\left( B_{s}\right) }=\left( \overline{s}%
_{R}b_{L}\right) \left( \overline{s}_{L}b_{R}\right) ,
\end{eqnarray}

and the Wilson coefficients take the form: 
\begin{eqnarray}
\kappa _{1}^{\left( K\right) } &=&\frac{x_{h\overline{s}_{R}d_{L}}^{2}}{%
m_{h}^{2}}+\sum_{m=1}^{5}\sum_{n=1}^{4}\left( \frac{x_{H_{m}^{0}\overline{s}_{R}d_{L}}^{2}}{%
m_{H_{m}^{0}}^{2}}-\frac{x_{A_{n}^{0}\overline{s}_{R}d_{L}}^{2}}{m_{A_{n}^{0}}^{2}}%
,\right) \\
\kappa _{2}^{\left( K\right) } &=&\frac{x_{h\overline{s}_{L}d_{R}}^{2}}{%
m_{h}^{2}}+\sum_{m=1}^{5}\sum_{n=1}^{4}\left( \frac{x_{H_{m}^{0}\overline{s}_{L}d_{R}}^{2}}{%
m_{H_{m}^{0}}^{2}}-\frac{x_{A_{n}^{0}\overline{s}_{L}d_{R}}^{2}}{m_{A_{n}^{0}}^{2}}%
\right) ,\hspace{0.7cm}\hspace{0.7cm} \\
\kappa _{3}^{\left( K\right) } &=&\frac{x_{h\overline{s}_{R}d_{L}}x_{h%
\overline{s}_{L}d_{R}}}{m_{h}^{2}}+\sum_{m=1}^{5}\sum_{n=1}^{4}\left( \frac{x_{H_{m}^{0}%
\overline{s}_{R}d_{L}}x_{H_{m}^{0}\overline{s}_{L}d_{R}}}{m_{H_{m}^{0}}^{2}}-\frac{%
x_{A_{n}^{0}\overline{s}_{R}d_{L}}x_{A_{n}^{0}\overline{s}_{L}d_{R}}}{m_{A_{n}^{0}}^{2}}%
\right) ,
\end{eqnarray}%
\begin{eqnarray}
\kappa _{1}^{\left( B_{d}\right) } &=&\frac{x_{h\overline{d}_{R}b_{L}}^{2}}{%
m_{h}^{2}}+\sum_{m=1}^{5}\sum_{n=1}^{4}\left( \frac{x_{H_{m}^{0}\overline{d}_{R}b_{L}}^{2}}{%
m_{H_{m}^{0}}^{2}}-\frac{x_{A_{n}^{0}\overline{d}_{R}b_{L}}^{2}}{m_{A_{n}^{0}}^{2}}%
\right) , \\
\kappa _{2}^{\left( B_{d}\right) } &=&\frac{x_{h\overline{d}_{L}b_{R}}^{2}}{%
m_{h}^{2}}+\sum_{m=1}^{5}\sum_{n=1}^{4}\left( \frac{x_{H_{m}^{0}\overline{d}_{L}b_{R}}^{2}}{%
m_{H_{m}^{0}}^{2}}-\frac{x_{A_{n}^{0}\overline{d}_{L}b_{R}}^{2}}{m_{A_{n}^{0}}^{2}}%
\right) , \\
\kappa _{3}^{\left( B_{d}\right) } &=&\frac{x_{h\overline{d}_{R}b_{L}}x_{h%
\overline{d}_{L}b_{R}}}{m_{h}^{2}}+\sum_{m=1}^{5}\sum_{n=1}^{4}\left( \frac{x_{H_{m}^{0}%
\overline{d}_{R}b_{L}}x_{H_{m}^{0}\overline{d}_{L}b_{R}}}{m_{H_{m}^{0}}^{2}}-\frac{%
x_{A_{n}^{0}\overline{d}_{R}b_{L}}x_{A_{n}^{0}\overline{d}_{L}b_{R}}}{m_{A_{n}^{0}}^{2}}%
\right) ,
\end{eqnarray}%
\begin{eqnarray}
\kappa _{1}^{\left( B_{s}\right) } &=&\frac{x_{h\overline{s}_{R}b_{L}}^{2}}{%
m_{h}^{2}}+\sum_{m=1}^{5}\sum_{n=1}^{4}\left( \frac{x_{H_{m}^{0}\overline{s}_{R}b_{L}}^{2}}{%
m_{H_{m}^{0}}^{2}}-\frac{x_{A_{n}^{0}\overline{s}_{R}b_{L}}^{2}}{m_{A_{n}^{0}}^{2}}%
\right) , \\
\kappa _{2}^{\left( B_{s}\right) } &=&\frac{x_{h\overline{s}_{L}b_{R}}^{2}}{%
m_{h}^{2}}+\sum_{m=1}^{5}\sum_{n=1}^{4}\left( \frac{x_{H_{m}^{0}\overline{s}_{L}b_{R}}^{2}}{%
m_{H_{m}^{0}}^{2}}-\frac{x_{A_{n}^{0}\overline{s}_{L}b_{R}}^{2}}{m_{A_{n}^{0}}^{2}}%
\right) , \\
\kappa _{3}^{\left( B_{s}\right) } &=&\frac{x_{h\overline{s}_{R}b_{L}}x_{h%
\overline{s}_{L}b_{R}}}{m_{h}^{2}}+\sum_{m=1}^{5}\sum_{n=1}^{4}\left( \frac{x_{H_{m}^{0}%
\overline{s}_{R}b_{L}}x_{H_{m}^{0}\overline{s}_{L}b_{R}}}{m_{H_{m}^{0}}^{2}}-\frac{%
x_{A_{n}^{0}\overline{s}_{R}b_{L}}x_{A_{n}^{0}\overline{s}_{L}b_{R}}}{m_{A_{n}^{0}}^{2}}%
\right) ,
\end{eqnarray}%
where the $x_{h_a\overline{d}_{cL}d_{fR}}$ effective parameters are the couplings of the physical scalar and pseudoscalar fields $h_a$ with the $d_{c,f}$-type quarks, with $h_a=h,\ H_m^0,\ A_n^0$ and $d_{c,f}=d,\; s,\; b$. Furthermore, we have used the notation of section \ref{scalar} for the physical
scalars, assuming $h$ is the lightest of the CP-even ones and
corresponds to the SM Higgs.
The $K-\bar{K}$, $B_{d}^{0}-\bar{B}_{d}^{0}$ and $B_{s}^{0}-\bar{B}_{s}^{0}$%
\ meson mass splittings read: 
\begin{equation}
\Delta m_{K}=\Delta m_{K}^{\left( SM\right) }+\Delta m_{K}^{\left( NP\right)
},\hspace{1cm}\Delta m_{B_{d}}=\Delta m_{B_{d}}^{\left( SM\right) }+\Delta
m_{B_{d}}^{\left( NP\right) },\hspace{1cm}\Delta m_{B_{s}}=\Delta
m_{B_{s}}^{\left( SM\right) }+\Delta m_{B_{s}}^{\left( NP\right) },
\label{Deltam}
\end{equation}

\begin{figure}[tbp]
\centering
\subfigure[] {\includegraphics[scale=0.35]{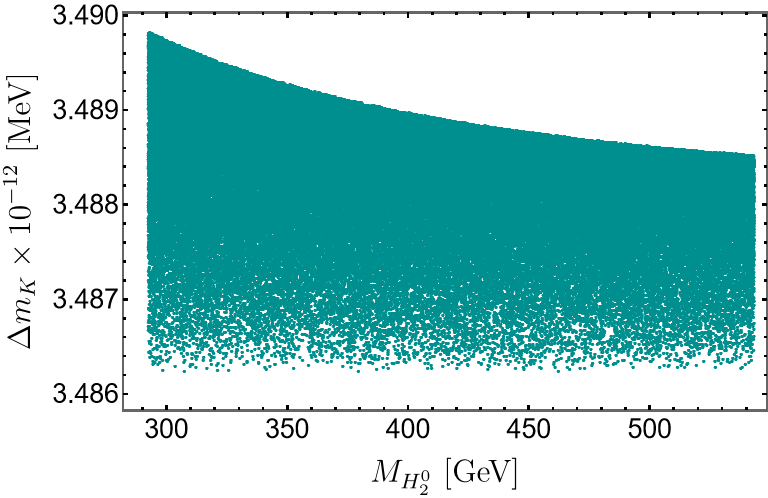}}
\quad \subfigure[]{%
\includegraphics[scale=0.35]{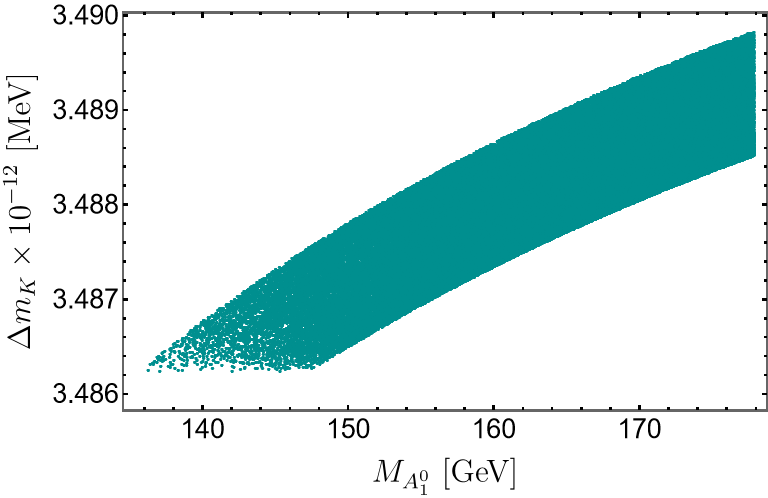}} 
\caption{Correlation a) between the $\Delta m_{K}$ mass splitting and the lightest CP even scalar mass $m_{H_2^0}$, b) between the $\Delta m_{K}$ mass splitting and the lightest CP odd scalar mass $m_{A_1^0}$.}
\label{fig:mesonmixing}
\end{figure}

where $\Delta m_{K}^{\left( SM\right) }$, $\Delta m_{B_{d}}^{\left(
SM\right) }$ and $\Delta m_{B_{s}}^{\left( SM\right) }$ correspond to the SM
contributions, while $\Delta m_{K}^{\left( NP\right) }$, $\Delta
m_{B_{d}}^{\left( NP\right) }$ and $\Delta m_{B_{s}}^{\left( NP\right) }$
are due to new physics effects. Our model predicts the following new physics
contributions for the $K-\bar{K}$, $B_{d}^{0}-\bar{B}_{d}^{0}$ and $%
B_{s}^{0}-\bar{B}_{s}^{0}$ meson mass differences: 
\begin{eqnarray}
\Delta m_{K}^{\left( NP\right) }&\simeq&\frac{8}{3}f_{K}^{2}\eta _{K}B_{K}m_{K}%
\left[ r_{2}^{\left( K\right) }\kappa _{3}^{\left( K\right) }+r_{1}^{\left(
K\right) }\left( \kappa _{1}^{\left( K\right) }+\kappa _{2}^{\left( K\right)
}\right) \right]~,\label{eq:mK} \\
\Delta m_{B_{d}}^{\left( NP\right) }&\simeq&\frac{8}{3}f_{B_{d}}^{2}\eta
_{B_{d}}B_{B_{d}}m_{B_{d}}\left[ r_{2}^{\left( B_{d}\right) }\kappa
_{3}^{\left( B_{d}\right) }+r_{1}^{\left( B_{d}\right) }\left( \kappa
_{1}^{\left( B_{d}\right) }+\kappa _{2}^{\left( B_{d}\right) }\right) \right]~, \label{eq:mBd}\\
\Delta m_{B_{s}}^{\left( NP\right) }&\simeq&\frac{8}{3}f_{B_{s}}^{2}\eta
_{B_{s}}B_{B_{s}}m_{B_{s}}\left[ r_{2}^{\left( B_{s}\right) }\kappa
_{3}^{\left( B_{s}\right) }+r_{1}^{\left( B_{s}\right) }\left( \kappa
_{1}^{\left( B_{s}\right) }+\kappa _{2}^{\left( B_{s}\right) }\right) \right]~.\label{eq:mBs}
\end{eqnarray}

Since the contribution arising from the flavor changing down type quark interaction involving the $Z^{\prime}$ gauge boson exchange is very small and subleading, the main contributions to the meson mass differences is due to the virtual exchange of additional scalar and pseudosalar fields participating in the flavor violating Yukawa interactions of the model under consideration. Using the following numerical values of the meson parameters \cite%
{Dedes:2002er,Aranda:2012bv,Khalil:2013ixa,Queiroz:2016gif,Buras:2016dxz,Ferreira:2017tvy,NguyenTuan:2020xls}
: 
\begin{eqnarray}
\left(\Delta m_{K}\right)_{\exp }&=&\left( 3.484\pm 0.006\right) \times
10^{-12}\, \mathrm{{MeV},\hspace{1.5cm}\left( \Delta m_{K}\right)_{\text{SM}}=3.483\times 10^{-12}\, {MeV}}  \notag \\
f_{K} &=&155.7\, \mathrm{{MeV},\hspace{1.5cm} B_{K}=0.85,\hspace{1.5cm}\eta_{K}=0.57,}  \notag \\
r_{1}^{\left( K\right) } &=&-9.3,\hspace{1.5cm}r_{2}^{\left(K\right) }=30.6,%
\hspace{1.5cm}m_{K}=\left(497.611\pm 0.013\right)\, \mathrm{{MeV},}
\end{eqnarray}%
\begin{eqnarray}
\left( \Delta m_{B_{d}}\right) _{\exp } &=&\left(3.334\pm 0.013\right)
\times 10^{-10}\, \mathrm{{MeV},\hspace{1.5cm}\left( \Delta m_{B_{d}}\right)
_{SM}=\left(3.653\pm 0.037\pm 0.019\right)\times 10^{-10}\, {MeV},}  \notag
\\
f_{B_{d}} &=&188\, \mathrm{{MeV},\hspace{1.5cm}B_{B_{d}}=1.26,\hspace{1.5cm}%
\eta _{B_{d}}=0.55,}  \notag \\
r_{1}^{\left( B_{d}\right) } &=&-0.52,\hspace{1.5cm}r_{2}^{\left(
B_{d}\right) }=0.88,\hspace{1.5cm}m_{B_{d}}=\left(5279.65\pm 0.12\right)\,%
\mathrm{{MeV},}
\end{eqnarray}%
\begin{eqnarray}
\left( \Delta m_{B_{s}}\right) _{\exp } &=&\left(1.1683\pm 0.0013\right)
\times 10^{-8}\, \mathrm{{MeV},\hspace{1.5cm}\left( \Delta m_{B_{s}}\right)
_{SM}=\left(1.1577\pm 0.022\pm 0.051\right) \times 10^{-8}\, {MeV},}  \notag
\\
f_{B_{s}} &=&225\, \mathrm{{MeV},\hspace{1.5cm}B_{B_{s}}=1.33,\hspace{1.5cm}%
\eta _{B_{s}}=0.55,}  \notag \\
r_{1}^{\left( B_{s}\right) } &=&-0.52,\hspace{1.5cm}r_{2}^{\left(
B_{s}\right) }=0.88,\hspace{1.5cm}m_{B_{s}}=\left(5366.9\pm 0.12\right)\, 
\mathrm{{MeV},}
\end{eqnarray}

Fig. \ref{fig:mesonmixing} \textcolor{red}{(a)} and Fig. \ref{fig:mesonmixing} \textcolor{red}{(b)} show the correlations of the mass splitting $\Delta m_K$ with the mass of the lightest CP-even and CP-odd scale $m_ {H_2^0}$ and $m_ {A_1^0}$, respectively. In our numerical analysis, for the sake of simplicity, we have set the couplings of flavor-changing Yukawa neutral interactions that produce $(K^0- \overline{K}^0)$ mixings to be equal to $10^{-6}$. In addition, we have varied the masses around 20\% of their best fit-point values obtained in the analysis of the scalar sector shown in the plots of Fig. \ref{fig:scalar-sector}. As indicated in Fig. \ref{fig:mesonmixing}, our model can successfully accommodate the experimental constraints arising from $(K^0-\overline{K}^ 0) $ meson oscillations for the above specified range of parameter space. We have numerically verified that in the range of masses described above, the values obtained for the mass splittings $\Delta m_{B_d}$ and $\Delta m_{B_s}$ are consistent with the experimental data on meson oscillations for flavor violating Yukawa couplings equal to $10^{ -4}$ and $2.5\times 10^{ -4}$, respectively.

\section{Higgs di-photon decay rate}\label{di-photon}
\lhead[\thepage]{\thesection. Higgs di-photon decay rate}

In order to study the implications of our model in the decay of the $125\  \mathrm{GeV}$ Higgs into a photon pair, one introduces the Higgs diphoton signal strength $R_{\gamma \gamma}$, which is defined as \cite{Bhattacharyya:2014oka}:
\begin{equation}
R_{\gamma \gamma}=\frac{\sigma(p p \rightarrow h) \Gamma(h \rightarrow \gamma \gamma)}{\sigma(p p \rightarrow h)_{\text{SM}} \Gamma(h \rightarrow \gamma \gamma)_{\text{SM}}} \simeq a_{h t t}^2 \frac{\Gamma(h \rightarrow \gamma \gamma)}{\Gamma(h \rightarrow \gamma \gamma)_{\text{SM}}} .
\end{equation}

That Higgs diphoton signal strength, normalizes the $\gamma \gamma$ signal predicted by our model in relation to the one given by the SM. Here we have used the fact that in our model, single Higgs production is also dominated by gluon fusion as in the Standard Model. In the 3-3-1 model $\sigma\left( pp \rightarrow h\right)=a_{htt}^2 \sigma\left( pp \rightarrow h\right)_{\text{SM}}$, so $R_{\gamma \gamma}$ reduces to the ratio of branching ratios.\\
The decay rate for the $h\rightarrow \gamma \gamma$ process takes the form \cite{Bhattacharyya:2014oka,Logan:2014jla,Hernandez:2021uxx}:
\begin{equation}\label{diphoton}
\Gamma(h \rightarrow \gamma \gamma)=\frac{\alpha_{\text{em}}^2 m_h^3}{256 \pi^3 v^2}\left|\sum_f a_{h f f} N_C Q_f^2 F_{1 / 2}\left(\varrho_f\right)+a_{h W W} F_1\left(\varrho_W\right)+\sum_{k} \frac{C_{h H_k^{\pm} H_k^{\mp} }v}{2 m_{H_k^{\pm}}^2} F_0\left(\varrho_{H_k^{\pm}}\right)\right|^2
\end{equation}

where $\alpha_{\text{em}}$ is the fine structure constant, $N_C$ is the color factor ($N_C=3$ for quarks and $N_C=1$ for leptons) and $Q_f$ is the electric charge of the fermion in the loop. From the fermion-loop contributions we only consider the dominant top quark term. The $\varrho_i$ are the mass ratios $\varrho_i=4 M_i^2/m_h^2$ with $M_i=m_f, M_W,M_{H_k^{\pm}}$ with $k=1,2,3$. Furthermore, $C_{h H_k^{\pm} H_k^{\mp}}$ is the trilinear coupling between the SM-like Higgs and a pair of charged Higges, whereas $a_{h t t}$ and $a_{h W W}$ are the deviation factors from the SM Higgs-top quark coupling and the SM Higgs-W gauge boson coupling, respectively (in the SM these factors are unity). Such deviation factors are close to unity in our model, which is a consequence of the numerical analysis of its scalar, Yukawa and gauge sectors.

The form factors for the contributions from spin-$0$, $1/2$ and $1$ particles are:
\begin{eqnarray}
F_0(\varrho) & =& -\varrho(1-\varrho f(\varrho)),\\
F_{1 / 2}(\varrho) & =& 2\varrho(1+(1-\varrho) f(\varrho)), \\
F_1(\varrho) & =& -\left(2+3\varrho+3\varrho \left(2-\varrho\right) f(\varrho)\right),
\end{eqnarray}
with
\begin{equation}
f(\varrho)= \begin{cases}\arcsin ^2 \sqrt{\varrho^{-1}} & \text { for } \varrho \geq 1 \\ -\frac{1}{4}\left[\ln \left(\frac{1+\sqrt{1-\varrho}}{1-\sqrt{1-\varrho}}\right)-i\pi\right] ^2& \text { for } \varrho<1\end{cases}
\end{equation}

\begin{figure} 
\centering
\includegraphics[scale=0.55]{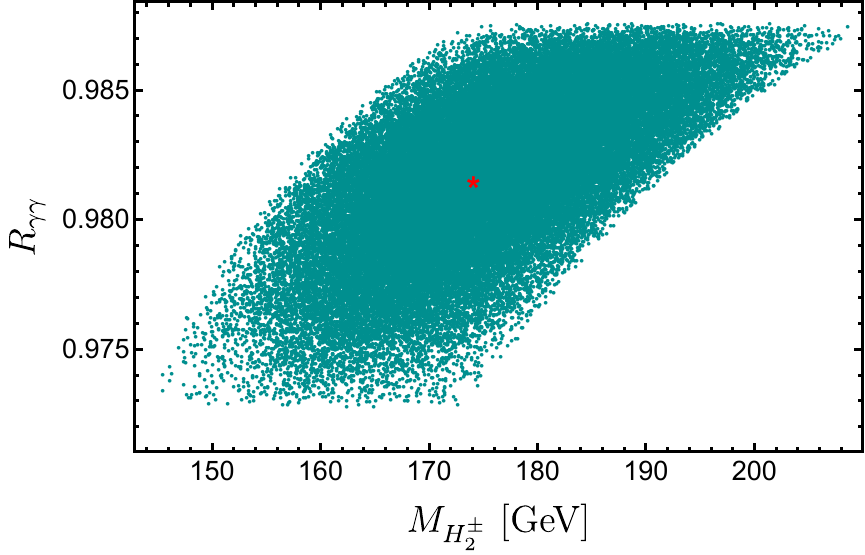}
\caption{Correlation of the Higgs di-photon signal strength with the charged scalar mass. The red star point corresponds to the best fit for $R_{\gamma \gamma}$ (see Table \ref{tab2photon}).}
\label{2photon}
\end{figure}

Table \ref{tab2photon} displays the best-fit values of the $R_{\gamma \gamma}$ ratio in comparison to the best-fit signals measured in CMS \cite{10.1007/978-981-19-2354-8_33}  and ATLAS \cite{ATLAS:2022tnm}. In this analysis, the electrically charged scalar fields play a key role in determining the value of the ratio, while the other fields have an indirect impact through the parameter space involving the VEV (vacuum expectation values) and the trilinear scalar coupling $A$, as well as some $\lambda_i$. From our numerical analysis, it follows that our model favors a Higgs diphoton decay rate lower than the SM expectation but inside the $3\sigma$ experimentally allowed range. The correlation of the Higgs diphoton signal strength with the charged scalar mass $M_{H_2^{\pm}}$ is shown in Fig. \ref{2photon}, which indicates that our model successfully accommodates the current Higgs diphoton decay rate constraints. Additionally, it should be noted that the correlation with $M_{H_1^{\pm}}$ is similar; however, the correlation is weaker than for $M_{H_3^{\pm}}$.
\begin{table}[H]
\centering
\begin{tabular}{lccc}
 \hline\hline
& Model value  & CMS & ATLAS \\ \hline
$R_{\gamma \gamma}$ & $0.982 \pm 0.08$ & $1.02_{-0.09}^{+0.11}$ & $1.04_{-0.09}^{+0.10} $  \\ 
\hline\hline
\end{tabular}
\caption{The best fit for the ratio of Higgs boson diphoton decay obtained from the model indicates a lower Higgs decay rate into two photons compared to the expectation of the Standard Model in ATLAS \cite{ATLAS:2022tnm} and CMS \cite{10.1007/978-981-19-2354-8_33} collaboration.  However, this value still falls within the $1\sigma$ experimentally allowed range.}
\label{tab2photon}
\end{table}

\section{Oblique $T$, $S$ and $U$ parameters}\label{oblique}
\lhead[\thepage]{\thesection. Oblique $T$, $S$ and $U$ parameters}

The parameters $S$, $T$, and $U$ basically quantify the corrections to the two-point functions of gauge bosons through loop diagrams. In our case, where there are three $SU(3)_L$ scalar triplets that introduce new scalar particles, which lead to new Higgs-mediated contributions to the self-energies of gauge bosons through loop diagrams. Based on references \cite{Peskin:1991sw,Altarelli:1990zd,Barbieri:2004qk}, the parameters $S$, $T$, and $U$ can be defined as follows:
\begin{eqnarray}
 T &=& \frac{1}{\alpha_{\text{em}} M_W} \left. \left[ \Pi_{11}\left(q^2\right) - \Pi_{33} \left( q^2\right)\right] \right|_{q=0} \\
 S & = &-\left.\frac{4  c_W s_W}{ \alpha_{\text{em}}} \frac{d}{d q^2} \Pi_{30}\left(q^2\right)\right|_{q^2=0},\\
 U &= & \frac{4 s_W }{\alpha_{\text{em}} } \left. \frac{d}{d q^2}\left[\Pi_{11}\left(q^2\right)-\Pi_{33}\left(q^2\right)\right]\right|_{q^2=0} ,
 \end{eqnarray}
with $s_W=\sin \theta_W$ and $c_W=\cos\theta_W$, where $\theta_W$ is the electroweak mixing angle, the quantity $\Pi_{ij} \left(q \right)$ is defined in terms of the vacuum-polarization tensors
\begin{equation}
\Pi_{ij} ^{\mu \nu} \left(q^2\right) = g^{\mu \nu} \Pi_{ij}\left(q^2\right)-i  q^{\mu} q^{\nu}  \Delta_{ij}\left(q^2\right),
\end{equation}
where $i,j=0,1,3$ for the $B$, $W_1$ and $W_3$ bosons respectively, or possibly $i,j=W,Z,\gamma$. If the new physics enters at the TeV scale, the effect of the theory will be well-described by an expansion up to linear order in $q^2$ for $\Pi_{i j} \left(q^2\right)$ as presented in reference \cite{Peskin:1991sw}. 

For our 331 model, the scalar fields arising from the $SU(3)_L$ triplets $\eta_k$ ($k=1,2$) and $\rho$ provide the dominant contributions to the 
new physics values of the 
$T$, $S$, and $U$ oblique parameters, as they couple with the $W$ and $Z$, and then we must take into account the scalar mixing angles. 
We can calculate the parameters considering 
that the low energy effective field theory below the scale of spontaneos breaking of the $SU(3)_L\times U(1)_X\times U(1)_{L_g}$ symmetry corresponds to a three Higgs doublet model (3HDM), where the three Higgs doublets arise from the $\eta_1$, $\eta_2$ and $\rho$ $SU(3)_L$ scalar triplets. Then, following these considerations, in the above described low energy limit scenario, the leading contributions to the oblique $T$, $S$ and $U$ parameters take the form \cite{Long:1999bny,Grimus:2007if,Grimus:2008nb,CarcamoHernandez:2015smi,Long:2018dun,CarcamoHernandez:2022vjk,Batra:2022arl}:
\begin{eqnarray}
T & \simeq & t_0 \left[ 
\sum_{a=1}^{2}\sum_{k=1}^2\left[\left(R_C\right)_{a k}\right]^2 m_{H_k^{ \pm}}^2
+\sum_{a=1}^{2} \sum_{i=1}^2 \sum_{j=1}^2\left[ \left(R_H\right)_{a i} \left(R_A\right)_{a j}\right]^2 F\left(m_{H_i^0}^2, m_{A_j^0}^2\right)
\right. ,\\
\nonumber & & \left.- \sum_{a=1}^{2}\sum_{i=1}^2 \sum_{k=1}^2 \left\lbrace  \left[\left(R_H\right)_{a i}\left(R_C\right)_{a k}\right]^2 F\left(m_{H_i^0}^2, m_{H_k^{ \pm}}^2\right)+
\left[\left(R_A\right)_{a i}\left(R_C\right)_{a k}\right]^2 F\left(m_{A_i^0}^2, m_{H_k^{ \pm}}^2\right)\right\rbrace \right]  \\
\nonumber &&\\
S & \simeq & \frac{1}{12 \pi} \sum_{i=1}^2 \sum_{j=1}^2 \sum_{k=1}^2 \left[\left(R_H\right)_{k i}\left(R_A\right)_{k j}\right]^2 K\left(m_{H_i^0}^2, m_{A_j^0}^2, m_{H_k^{ \pm}}^2\right) ,\\
\nonumber &&\\
U &\simeq & -S+\sum_{a=1}^{2}\sum_{i=1}^2 \sum_{k=1}^2  \left\lbrace \left[ \left(R_A\right)_{a i}\left(R_C\right)_{a k}\right]^2 G\left(m_{A_i^0}^2, m_{H_k^{ \pm}}^2\right)  +\left[\left(R_H\right)_{a i}\left(R_C\right)_{a k}\right]^2 G\left(m_{H_i^0}^2, m_{H_k^{ \pm}}^2\right) \right\rbrace ,\qquad
\end{eqnarray}
where $t_0=\left( 16\pi^2 v^2 \alpha_{\text{em}}\left(M_Z\right) \right)^{-1}$ and $R_C$, $R_H$, $R_A$ are the mixing matrices for the charged scalar fields, neutral scalar and pseudoscalars, respectively presented in the Sec. \ref{scalarsector-331}. Furthermore, the following loop functions $F\left(m_1^2, m_2^2\right)$, $G\left(m_1^2, m_2^2\right)$ and $K\left(m_1^2, m_2^2, m_3^2\right)$ were introduced in \cite{CarcamoHernandez:2022vjk}:
\begin{eqnarray}
F\left(m_1^2, m_2^2\right) &=& \frac{m_1^2 m_2^2}{m_1^2-m_2^2} \ln \left(\frac{m_1^2}{m_2^2}\right),\\
G\left(m_1^2, m_2^2\right)&=&\frac{-5 m_1^6+27 m_1^4 m_2^2-27 m_1^2 m_2^4+6\left(m_1^6-3 m_1^4 m_2^2\right) \ln \left(\frac{m_1^2}{m_2^2}\right)+5 m_2^6}{6\left(m_1^2-m_2^2\right)^3},\\
K\left(m_1^2, m_2^2, m_3^2\right) &= & \frac{1}{\left(m_2^2-m_1^2\right)^3}\left\{m_1^4\left(3 m_2^2-m_1^2\right) \ln \left(\frac{m_1^2}{m_3^2}\right)-m_2^4\left(3 m_1^2-m_2^2\right) \ln \left(\frac{m_2^2}{m_3^2}\right)\right. \\
\nonumber & & \left.-\frac{1}{6}\left[27 m_1^2 m_2^2\left(m_1^2-m_2^2\right)+5\left(m_2^6-m_1^6\right)\right]\right\}.
\end{eqnarray}

Besides that, the experimental limits for $S$, $T$, and $U$ are given in ref \cite{Workman:2022ynf}:
\begin{eqnarray}
T_{\text{exp}} &= & 0.03 \pm 0.12\\
S_{\text{exp}}  &=& -0.02 \pm 0.1\\
U_{\text{exp}}  &= & 0.01 \pm 0.11
\end{eqnarray}

\begin{figure} 
\centering
\subfigure[]{\includegraphics[scale=.5]{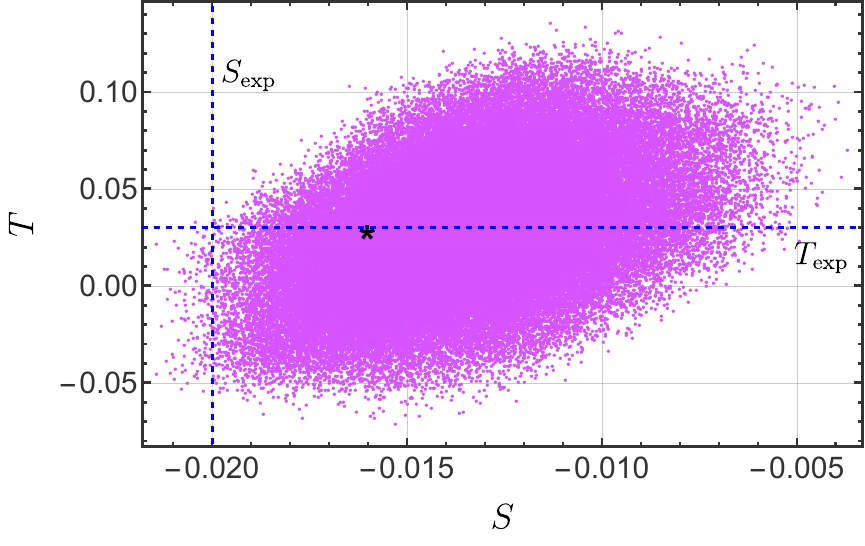}}
\subfigure[]{\includegraphics[scale=.5]{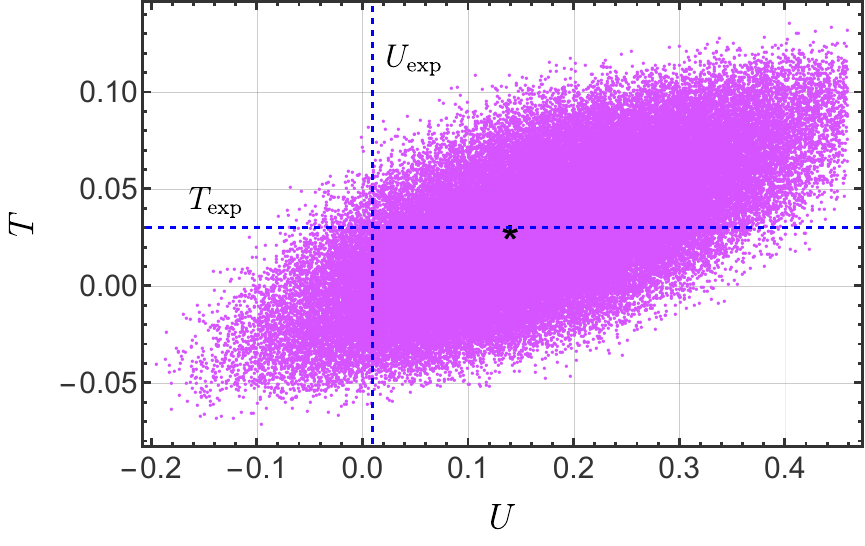}}   
\subfigure[]{\includegraphics[scale=.5]{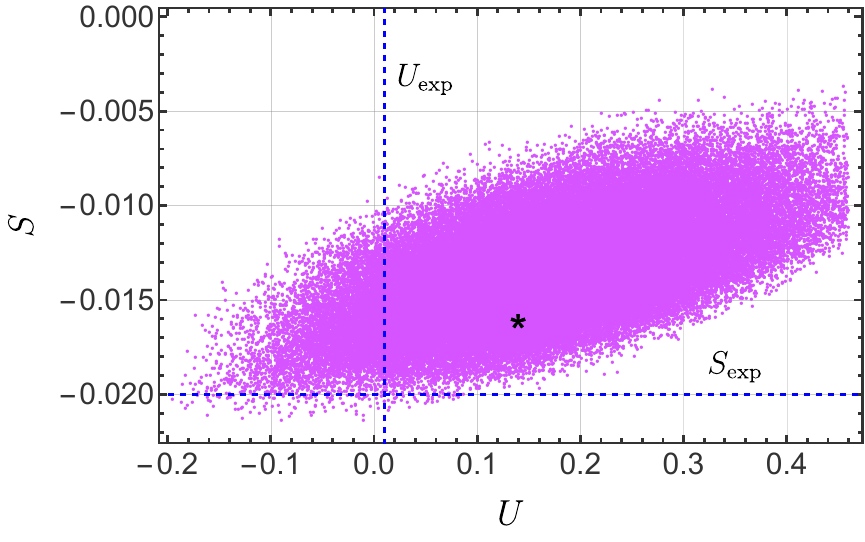}}
\caption{Correlation between the oblique parameters, the blue dashed line represents the central value of the Ref.~\cite{Workman:2022ynf} and the star corresponds to the best fit of this model.}
\label{oblicue}
\end{figure}

From the numerical analysis, the $S_4$ flavored $331$ model has restricted parameters because the determination of the new physics from $S$, $T$, and $U$ is determined by the physical masses of the model, also within the limit indicated by the oblique parameters present some correlation shown in Fig.~\ref{oblicue}, we see that the evolution of parameter space is adjusted within the $1 \sigma$ experimental range.
The figures \ref{oblicue} \textcolor{red}{(a)} and \textcolor{red}{(c)} of dispersion involving the $U$ parameter is produced with values larger than the central one, despite this, the values of $U$ fit within the range of $ 1 \sigma$ and the statistical discrepancy is minimal. In the case of the $S$ value, due to the large uncertainty value, the value fits more naturally, as shown in the figure \ref{oblicue} \textcolor{red}{(b)}.

Our analysis shows that our model allows a successfull fit for the oblique $S$, $T$ and $U$ parameters, consistent with their current experimental limits.
The obtained best fit point values for the oblique $S$, $T$ and $U$ parameters in our model are: 
\begin{eqnarray}
T &= & 0.029 \pm 0.009,\\
S &= & -0.016\pm 0.006, \\
U &= & 0.14\pm 0.04.
\end{eqnarray}

Our results also suggest that the model favors a larger value for $U$ within statistical uncertainty. While the values for $S$ and $T$ fit within the relative error $0.2$ and $0.1$, respectively.

\newpage
$\ $
\thispagestyle{empty} 
\let\cleardoublepage\clearpage
\chapter{Conclusions}\label{cap.conclu}
\markboth{CONCLUSIONS}{CONCLUSIONS}
\lhead[\thepage]{\thesection. Conclusions}
We proposed three viable low-scale seesaw models in which the SM symmetry and particle content of the fermionic and scalar sectors were extended. We obtained values for the fermion masses and mixing in agreement with the experimental data, where the smallest neutrino masses were obtained by implementing a seesaw mechanism in the three models. We also studied the phenomenological aspects predicted by each model, analyzing from the fermionic and scalar dark matter candidates, the leptonic flavor-violating processes, flavor-changing neutral current interactions, and exploring the corrections given to the oblique parameters.\\

In the first model, we extended the particle content by adding two scalar fields and six neutral leptons, and supplemented the SM symmetry global $U(1)_X$ symmetry, whose spontaneous breaking produces a preserved $Z_2$ symmetry, allowing us to obtain scalar and fermionic dark matter candidates, focusing on the study of the fermionic dark matter candidate. The six additional neutral leptons allowed the successful implementation of a radiative inverse seesaw mechanism to generate active neutrino masses, where the Majorana submatrix corresponding to the lepton number violating  $\mu$ parameter arises at one loop level. This radiative inverse seesaw model allowed us to successfully reproduce the experimental values for the observed relic density and to comply with the constraints arising from dark matter direct detection. In turn, the implications of the model in lepton flavour violating (LFV) processes were studied, finding the model successfully complies with the constraints arising from these processes and predicts charged LFV decays within the reach of forthcoming experiments.\\

In the second model, extensions to the 3HDM and 4HDM theories were constructed, where in the case of 3HDM, an inert scalar singlet was added, whereas in the case of 4HDM, an inert scalar doublet was included. Besides that, several singlet scalar fields were included in both theories in order to have a predictive pattern of lepton masses and mixings. Besides that, the SM gauge symmetry was supplemented by the inclusion of the preserved $Z_2$ and spontaneously broken $S_4\times Z_4$ symmetry. In these theories, the preserved $Z_2$ symmetry ensures the stabiity of the dark matter candidate as well as the radiative nature of the one loop level seesaw mechanism that produces the tiny active neutrino masses, and for the charged fermions, the masses were generated through symmetry breaking where at first glance, the model seemed to have several parameters, however, due to the symmetry of the model it was reduced to a few effective parameters, being able to obtain values for the masses and mixing in agreement with the experimental data in the range of $1\sigma$. The extra scalars in our model provide radiative corrections to the oblique parameters, where due to the presence of the scalar inert doublet, in the case 4HDM is less restrictive than 3HDM. Furthermore, flavor-changing neutral current interactions mediated by CP even scalars and CP odd scalars give rise to $(K^0-\overline{K}^0)$ and $(B_{d,s}^0-\overline{B}_{d,s}^0)$ meson oscillations, whose experimental constraints are successfully fulfilled for an appropiate region of parameter space. Also, scalar dark matter candidates were studied in both cases, the inert singlet in the case of 3HDM and the inert doublet in the case of 4HDM, obtaining values for the relic density in agreement with the experimental data, where the inert singlet provides strong constraints, while the doublet must have a mass of at least $10$ TeV, to be consistent with the experimental data.\\

In the third model, we considered a theory based on the SU(3)CxSU(3)LxU(1)X gauge symmetry, which was supplemented by the $U(1)_{L_g}\times S_4\times Z_4\times Z_4^{\prime}\times Z_2$, where the quark and lepton sectors were analyzed, implementing an inverse seesaw mechanism to generate the masses of the light active neutrinos, while the masses of the quark sector were obtained by spontaneous symmetry breaking, allowing to obtain correlations between the mixing angles and the Jarlskog invariant for the quark sector. Regarding the lepton sector, our model predicts a diagonal SM charged lepton mass matrix, thus implying that the leptonic mixing only arises from the neutrino sector, where correlations between the leptonic mixing angles and the leptonic CP violating phase were obtained. We have found that the considered model successfully reproduces the experimental values for the physical observables of both quark and lepton sectors within the 1sigma experimentally allowed range. Furthermore, the consequences of the model in meson oscillations produced by flavor-changing interactions were analyzed as well, finding that the resulting experimental constraints were successfully fulfilled for an appropriate region of the parameter space. The charged scalars of our model provide the new physical contribution to the Higgs diphoton decay rate, where our proposed model favors a $R_{\gamma\gamma}$ value lower than the SM expectation, and within the $3\sigma$ experimentally allowed range measured by the CMS and ATLAS collaborations. The extra scalar fields of our model induce radiative corrections to the oblique $S$, $T$, and $U$ parameters, where the numerical analysis yields correlations between these parameters and, in addition, their obtained values are within the $1 \sigma$ experimentally allowed range.

\let\cleardoublepage\cleardoublepage 


\cleardoublepage
\addcontentsline{toc}{chapter}{Bibliography}
\lhead[\thepage]{Bibliography}
\bibliographystyle{utphys}
\bibliography{jmarchant.bib} 

\appendix
\clearpage

\lhead[\thepage]{APPENDIX \thechapter. \rightmark}
\rhead[APPENDIX \thechapter. \leftmark]{\thepage}
\newpage
$\ $
\thispagestyle{empty} 

\chapter{Symmetry groups}\label{anex.A}
\markboth{SYMMETRY GROUPS}{SYMMETRY GROUPS}

\section{Discrete symmetry groups.}

Imagine having an object with a specific geometric shape to which we want to apply a transformation that preserves the original shape. The set of all transformations that can be performed to maintain the original shape is called the “Symmetry Group''. The smallest group, in which no action is taken, is called the “Identity'' and includes the element $e$.\\

\noindent
Before examining some examples of symmetry groups, let us first discuss the fundamental axioms of groups:\\
\textbf{Group Axioms}
\begin{itemize}
\item A group is a set of elements $a,b,\ldots$ which can be combined together with $ab$ inside the set.
\item $(ab)c=a(bc)$.
\item The element $e$ satisfies $ae=ea=a$ for all $a$.
\item For each element “$a$'' there is an element $a^{-1}$ which satisfies $aa^{-1}=a^{-1}a=e$.
\end{itemize}

\section{Group $Z_n$}

The group $Z_n$ is an Abelian group that represents the rotation group of a regular N-polygon and is a subgroup of SO(2). For instance, $Z_2$ represents the permutation of two elements, $Z_3$ corresponds to an equilateral triangle, $Z_4$ to a square, $Z_5$ to a pentagon, and so on. Where the elements of the object are:
\begin{equation}
\left\lbrace e,a,a^2,\ldots,a^{n-1}\right\rbrace,
\end{equation}
\noindent
where $a^n=e$ and $e$ is the identity.\\

\noindent
The group $Z_n$ can be represented as discrete rotations whose generator corresponds to the rotation~\cite{Ishimori:2010au},
\begin{equation}
Z_N=e^{2\pi i/N}.
\end{equation}

\section{Group $S_N$}

The group $S_N$ is the group of all possible permutations among the $N$ objects $x_i$ $(i=1,\dotsc, N)$ of order $N!$. Next, we will see the case for $N=4$, which corresponds to the full symmetry group of the tetrahedron or the rotation symmetry of a cube.

\subsection{Group $S_4$}
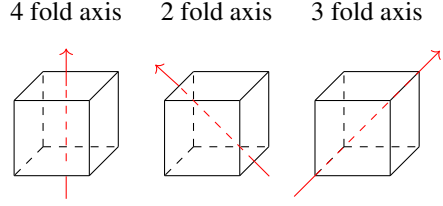
\begin{figure}
\centering
\begin{tikzpicture}
  \draw[dashed] (0,0,0) -- (1,0,0);
  \draw[dashed] (0,0,0) -- (0,0,1);
  \draw[dashed] (0,0,0) -- (0,1,0);
  \draw (1,0,0) -- (1,1,0);
  \draw (1,0,0) -- (1,0,1);
  \draw (0,0,1) -- (0,1,1);
  \draw (0,0,1) -- (1,0,1);
  \draw (1,1,1) -- (0,1,1);
  \draw (1,1,1) -- (1,1,0);
  \draw (1,1,1) -- (1,0,1);
  \draw (0,1,0) -- (0,1,1);
  \draw (0,1,0) -- (1,1,0);

  \draw[dashed,red] (0.5,0,0.5) -- (0.5,1,0.5);
  \draw[->,red] (0.5,1,0.5) -- (0.5,1.5,0.5);
  \draw[red] (0.5,-0.2,0.5) -- (0.5,-0.5,0.5);
  \node at (0.5,2,0.5) {4 fold axis};
  \draw[dashed] (2,0,0) -- (3,0,0);
  \draw[dashed] (2,0,0) -- (2,0,1);
  \draw[dashed] (2,0,0) -- (2,1,0);
  \draw (3,0,0) -- (3,1,0);
  \draw (3,0,0) -- (3,0,1);
  \draw (2,0,1) -- (2,1,1);
  \draw (2,0,1) -- (3,0,1);
  \draw (3,1,1) -- (2,1,1);
  \draw (3,1,1) -- (3,1,0);
  \draw (3,1,1) -- (3,0,1);
  \draw (2,1,0) -- (2,1,1);
  \draw (2,1,0) -- (3,1,0);

  \draw[dashed,red] (3,0,0.5) -- (2,1,0.5);
  \draw[->,red] (2,1,0.5) -- (1.8,1.4,0.8);
  \draw[red] (3,0,0.5) -- (3.2,-0.2,0.5);
  
  \node at (2.5,2,0.5) {2 fold axis};

  \draw[dashed] (4,0,0) -- (5,0,0);
  \draw[dashed] (4,0,0) -- (4,0,1);
  \draw[dashed] (4,0,0) -- (4,1,0);
  \draw (5,0,0) -- (5,1,0);
  \draw (5,0,0) -- (5,0,1);
  \draw (4,0,1) -- (4,1,1);
  \draw (4,0,1) -- (5,0,1);
  \draw (5,1,1) -- (4,1,1);
  \draw (5,1,1) -- (5,1,0);
  \draw (5,1,1) -- (5,0,1);
  \draw (4,1,0) -- (4,1,1);
  \draw (4,1,0) -- (5,1,0);

  \draw[dashed,red] (4,0,1) -- (5,1,0);
  \draw[->,red] (5,1,0) -- (5.2,1.2,-0.2);
  \draw[red] (4,0,1) -- (3.8,-0.2,1.2);
  \node at (4.5,2,0.5) {3 fold axis};
\end{tikzpicture}
\caption{The figure shows the $S_4$ symmetry of the cube with its different fold axis.}
\label{fig:rotationS4}
\end{figure}

\noindent
$S_4$ are all permutations among the four objects $\left(x_1,x_2,x_3,x_4\right)$~\cite{Ishimori:2010au}:
\begin{equation}
\left( x_1,x_2,x_3,x_4\right),\quad \rightarrow \quad \left( x_i,x_j,x_k,x_l\right)
\end{equation}

whose order is $4!=24$, therefore there are 24 possible elements or permutations. In $S_4$, we have three types of rotations, which can be seen in figure \ref{fig:rotationS4}, therefore, we have three generators commonly called $S$, $T$ and $U$, where the generator $S$ is associated with the transposition of the objects $1$ and $2$ leaving the other two objects fixed, the generator $T$ is related to the cycle $(123)$, leaving the fourth object fixed and the generator $U$ is associated with the transposition of the objects $3$ and $4$, whose matrix representations are:
\begin{equation}
S= \begin{pmatrix}
1 & 0 & 0\\
0 & -1 & 0\\
0 & 0 & -1
\end{pmatrix},\qquad T= \begin{pmatrix}
0 & 1 & 0\\
0 & 0 & 1\\
1 & 0 & 0
\end{pmatrix},\qquad U= \begin{pmatrix}
1 & 0 & 0\\
0 & 0 & 1\\
0 & 1 & 0
\end{pmatrix}.
\end{equation}

\noindent
The operators $S$, $T$ and $U$ comply with the following rules:
\begin{equation}
S^2=T^3= U^2= \left(ST\right)^3=\left(SU\right)^2= \left(TU\right)^2=\left(STU\right)^4=e
\label{eq:s4-e}
\end{equation}

where $e$ is the identity and the Eq.~\eqref{eq:s4-e} is known as ``presentation''. Therefore, according to this ``presentation'', we can obtain all the elements of the group by combining the products of the generators.
\begin{align}
a_1&:e & a_2&:S & a_3&:ST & a_4&:ST^2 \notag\\
b_1&:TST^2 & b_2&:T^2ST & b_3&:U & b_4&:T \notag\\
c_1&:T^2 & c_2&:TUT^2 & c_3&:T^2UT & c_4&:TSUT \\
d_1&:T^2SUT^2 & d_2&:TU & d_3&:T^2U & d_4&:SUST \notag\\
e_1&:STU & e_2&:SU & e_3&:STUT & e_4&:ST^2UT^2 \notag\\
f_1&:TSUS & f_2&:T^2SUS & f_3&:SUS^{-1} & f_4&:T^2U \notag
\end{align}

Furthermore, we can classify the elements of the group according to their class, where the elements of the same class must comply with $a^h= e$, so, according to the equation \eqref{eq:s4-e}, we will have five conjugate classes:
\begin{align}
C_1 &: \left\lbrace a_1 \right\rbrace, &h=1, \notag\\
C_3 &: \left\lbrace a_2, a_3, a_4 \right\rbrace, &h=2, \notag\\
C_6 &: \left\lbrace d_1, d_2, e_1, e_4, f_1, f_3 \right\rbrace, &h=2, \\
C_8 &: \left\lbrace b_1, b_2, b_3, b_4, c_1, c_2, c_3, c_4 \right\rbrace, &h=3, \notag\\
C_{6'} &: \left\lbrace d_3, d_4, e_2, e_3, f_2, f_4 \right\rbrace, &h=4. \notag
\end{align}
\noindent
Since we have five conjugate classes, we have five irreducibles in the group $S_4$, described as: $\mathbf{1_{1},1_{2},2,3_{1},3_{2}}$.

\subsection{The product rules of the $S_4$ discrete group}\label{S4}

The $S_{4}$ is the smallest non abelian group having doublet, triplet and singlet irreducible representations. As mentioned above, $S_{4}$ is the group of permutations of four objects, which includes five irreducible representations, i.e., $\mathbf{1_{1},1_{2},2,3_{1},3_{2}}$ fulfilling the following tensor product rules \cite{Ishimori:2010au}: 
\begin{align}
\begin{pmatrix}
a_{1} \\ 
a_{2}%
\end{pmatrix}%
_{\mathbf{2}}\otimes 
\begin{pmatrix}
b_{1} \\ 
b_{2}%
\end{pmatrix}%
_{\mathbf{2}}& =(a_{1}b_{1}+a_{2}b_{2})_{\mathbf{1}_{1}}\oplus
(-a_{1}b_{2}+a_{2}b_{1})_{\mathbf{1}_{2}}\oplus 
\begin{pmatrix}
a_{1}b_{2}+a_{2}b_{1} \\ 
a_{1}b_{1}-a_{2}b_{2}%
\end{pmatrix}%
_{\mathbf{2}\ ,} \\
\begin{pmatrix}
a_{1} \\ 
a_{2}%
\end{pmatrix}%
_{\mathbf{2}}\otimes 
\begin{pmatrix}
b_{1} \\ 
b_{2} \\ 
b_{3}%
\end{pmatrix}%
_{\mathbf{3}_{1}}& =%
\begin{pmatrix}
a_{2}b_{1} \\ 
-\frac{1}{2}(\sqrt{3}a_{1}b_{2}+a_{2}b_{2}) \\ 
\frac{1}{2}(\sqrt{3}a_{1}b_{3}-a_{2}b_{3})%
\end{pmatrix}%
_{\mathbf{3}_{1}}\oplus 
\begin{pmatrix}
a_{1}b_{1} \\ 
\frac{1}{2}(\sqrt{3}a_{2}b_{2}-a_{1}b_{2}) \\ 
-\frac{1}{2}(\sqrt{3}a_{2}b_{3}+a_{1}b_{3})%
\end{pmatrix}%
_{\mathbf{3}_{2}\ ,} \\
\begin{pmatrix}
a_{1} \\ 
a_{2}%
\end{pmatrix}%
_{\mathbf{2}}\otimes 
\begin{pmatrix}
b_{1} \\ 
b_{2} \\ 
b_{3}%
\end{pmatrix}%
_{\mathbf{3}_{2}}& =%
\begin{pmatrix}
a_{1}b_{1} \\ 
\frac{1}{2}(\sqrt{3}a_{2}b_{2}-a_{1}b_{2}) \\ 
-\frac{1}{2}(\sqrt{3}a_{2}b_{3}+a_{1}b_{3})%
\end{pmatrix}%
_{\mathbf{3}_{1}}\oplus 
\begin{pmatrix}
a_{2}b_{1} \\ 
-\frac{1}{2}(\sqrt{3}a_{1}b_{2}+a_{2}b_{2}) \\ 
\frac{1}{2}(\sqrt{3}a_{1}b_{3}-a_{2}b_{3})%
\end{pmatrix}%
_{\mathbf{3}_{2}\ ,} \\
\begin{pmatrix}
a_{1} \\ 
a_{2} \\ 
a_{3}%
\end{pmatrix}%
_{\mathbf{3}_{1}}\otimes 
\begin{pmatrix}
b_{1} \\ 
b_{2} \\ 
b_{3}%
\end{pmatrix}%
_{\mathbf{3}_{1}}& =(a_{1}b_{1}+a_{2}b_{2}+a_{3}b_{3})_{\mathbf{1}%
_{1}}\oplus 
\begin{pmatrix}
\frac{1}{\sqrt{2}}(a_{2}b_{2}-a_{3}b_{3}) \\ 
\frac{1}{\sqrt{6}}(-2a_{1}b_{1}+a_{2}b_{2}+a_{3}b_{3})%
\end{pmatrix}%
_{\mathbf{2}}  \notag \\
& \ \oplus 
\begin{pmatrix}
a_{2}b_{3}+a_{3}b_{2} \\ 
a_{1}b_{3}+a_{3}b_{1} \\ 
a_{1}b_{2}+a_{2}b_{1}%
\end{pmatrix}%
_{\mathbf{3}_{1}}\oplus 
\begin{pmatrix}
a_{3}b_{2}-a_{2}b_{3} \\ 
a_{1}b_{3}-a_{3}b_{1} \\ 
a_{2}b_{1}-a_{1}b_{2}%
\end{pmatrix}%
_{\mathbf{3}_{2}\ ,} \\
\begin{pmatrix}
a_{1} \\ 
a_{2} \\ 
a_{3}%
\end{pmatrix}%
_{\mathbf{3}_{2}}\otimes 
\begin{pmatrix}
b_{1} \\ 
b_{2} \\ 
b_{3}%
\end{pmatrix}%
_{\mathbf{3}_{2}}& =(a_{1}b_{1}+a_{2}b_{2}+a_{3}b_{3})_{\mathbf{1}%
_{1}}\oplus 
\begin{pmatrix}
\frac{1}{\sqrt{2}}(a_{2}b_{2}-a_{3}b_{3}) \\ 
\frac{1}{\sqrt{6}}(-2a_{1}b_{1}+a_{2}b_{2}+a_{3}b_{3})%
\end{pmatrix}%
_{\mathbf{2}}  \notag \\
& \ \oplus 
\begin{pmatrix}
a_{2}b_{3}+a_{3}b_{2} \\ 
a_{1}b_{3}+a_{3}b_{1} \\ 
a_{1}b_{2}+a_{2}b_{1}%
\end{pmatrix}%
_{\mathbf{3}_{1}}\oplus 
\begin{pmatrix}
a_{3}b_{2}-a_{2}b_{3} \\ 
a_{1}b_{3}-a_{3}b_{1} \\ 
a_{2}b_{1}-a_{1}b_{2}%
\end{pmatrix}%
_{\mathbf{3}_{2}\ ,} \\
\begin{pmatrix}
a_{1} \\ 
a_{2} \\ 
a_{3}%
\end{pmatrix}%
_{\mathbf{3}_{1}}\otimes 
\begin{pmatrix}
b_{1} \\ 
b_{2} \\ 
b_{3}%
\end{pmatrix}%
_{\mathbf{3}_{2}}& =(a_{1}b_{1}+a_{2}b_{2}+a_{3}b_{3})_{\mathbf{1}%
_{2}}\oplus 
\begin{pmatrix}
\frac{1}{\sqrt{6}}(2a_{1}b_{1}-a_{2}b_{2}-a_{3}b_{3}) \\ 
\frac{1}{\sqrt{2}}(a_{2}b_{2}-a_{3}b_{3})%
\end{pmatrix}%
_{\mathbf{2}}  \notag \\
& \ \oplus 
\begin{pmatrix}
a_{3}b_{2}-a_{2}b_{3} \\ 
a_{1}b_{3}-a_{3}b_{1} \\ 
a_{2}b_{1}-a_{1}b_{2}%
\end{pmatrix}%
_{\mathbf{3}_{1}}\oplus 
\begin{pmatrix}
a_{2}b_{3}+a_{3}b_{2} \\ 
a_{1}b_{3}+a_{3}b_{1} \\ 
a_{1}b_{2}+a_{2}b_{1}%
\end{pmatrix}%
_{\mathbf{3}_{2}\ .}
\end{align}

\chapter{Quark sector 3HDMS4 model}\label{quarks-app}
\markboth{QUARK SECTOR 3HDMS4 MODEL}{QUARK SECTOR 3HDMS4 MODEL}

We would like to give more details about the method to diagonalize the quark
mass matrix which has been assumed to be complex. Then, let us start from
the $\hat{ \mathbf{M}}_{q}=\mathbf{O}^{T}_{q}\mathbf{\bar{m}}_{q}\mathbf{O}%
_{q}$ where the $\mathbf{\bar{m}}_{q}$ mass matrix is written in the
following way 
\begin{equation}
\mathbf{\bar{m}}_{q}= \vert g_{q}\vert \mathbf{1}_{3\times 3}+\overbrace{ 
\begin{pmatrix}
\vert A_{q}\vert-\vert g_{q}\vert & \vert b_{q}\vert & 0 \\ 
\vert b_{q}\vert & \vert B_{q}\vert-\vert g_{q}\vert & \vert C_{q}\vert \\ 
0 & \vert C_{q}\vert & 0%
\end{pmatrix}
}^{\tilde{m}_{q}}.
\end{equation}

To diagonalize $\mathbf{\bar{m}}_{q}$, we just focus in $\mathbf{\tilde{m}}%
_{q}$. This means, once the latter is diagonalized, the former one will be too.
Given the above decomposition, one obtains $\tilde{\hat{ \mathbf{M}}}_{q}=%
\mathbf{O}^{T}_{q}\mathbf{\tilde{m}}_{q}\mathbf{O}_{q}=\text{diag}%
.(\mu_{q_{1}}, \mu_{q_{2}}, \mu_{q_{3}})$ where $\mu_{q_{i}}=m_{q_{i}}-\vert g_{q}\vert $
with $i=1,2,3$. Then, $\mathbf{O}_{q}$ is built by means of the eigenvectors $%
X_{q_{i}}$, which are given by
\begin{equation}
X_{q_{i}}=\frac{1}{N_{q_{i}}}%
\begin{pmatrix}
\vert b_{q}\vert \vert C_{q}\vert \\ 
\left[\mu_{q_{i}}-\left(\vert A_{q}\vert-\vert g_{q}\vert \right)\right]\vert C_{q}\vert \\ 
\left[\mu_{q_{i}}-\left(\vert A_{q}\vert-\vert g_{q}\vert\right)\right]\left[ \mu_{q_{i}}-%
\left(\vert B_{q}\vert -\vert g_{q}\vert\right)\right]-\vert b_{q}\vert^{2}%
\end{pmatrix}%
.
\end{equation}
with $N_{q_{i}}$ being the normalization factors. Due to the orthogonality
condition $\mathbf{O}^{T}_{q} \mathbf{O}_{q}=\mathbf{1}=\mathbf{O}_{q}%
\mathbf{O}^{T}_{q}$, then one can obtain the normalization factors.

In addition, some parameters can be fixed by using the following invariants 
\begin{equation}
\text{tr}\left( \tilde{\hat{ \mathbf{M}}}_{q}\right),\qquad \frac{1}{2}\left[%
\text{tr}\left( \tilde{\hat{ \mathbf{M}}}_{q} \tilde{\hat{ \mathbf{M}}}%
_{q}\right)- \left\{\text{tr}\left( \tilde{\hat{ \mathbf{M}}}%
_{q}\right)\right\}^{2}\right],\qquad \text{det}\left( \tilde{\hat{ \mathbf{M%
}}}_{q}\right).
\end{equation}
where tr and det stand for the trace and determinant respectively. As a result, one gets 
\begin{eqnarray}
b_{q}&=&\sqrt{\frac{\left(\vert A_{q}\vert-m_{q_{1}}\right)\left(m_{q_{2}}-\vert A_{q}\vert \right)%
\left(m_{q_{3}}-\vert A_{q}\vert\right)}{\vert g_{q}\vert-\vert A_{q}\vert}};  \notag \\
\vert C_{q}\vert&=&\sqrt{\frac{\left(\vert g_{q}\vert -m_{q_{1}}\right)%
\left(\vert g_{q}\vert -m_{q_{2}}\right)\left(m_{q_{3}}-\vert g_{q}\vert\right)}{\vert g_{q}\vert-\vert A_{q}\vert}}; 
\notag \\
B_{q}&=& m_{q_{3}}-m_{q_{2}}+m_{q_{1}}-\vert g_{q}\vert-\vert A_{q}\vert~,
\end{eqnarray}
As notices, there is
is a hierarchy among the free parameters, this is, $%
m_{q_{3}}>\vert g_{q}\vert >m_{q_{2}}>\vert A_{q}\vert >m_{q_{1}}$ in order to have real parameters.
Finally, $\mathbf{O}_{q}=\left(X_{q_{1}}, X_{q_{2}}, X_{q_{3}}\right)$,
this is written explicitly as 
\begin{equation}
\mathbf{O}_{q}=%
\begin{pmatrix}
\sqrt{\frac{\left(\vert g_{q}\vert -m_{q_{1}}\right)\left(m_{q_{2}}-\vert A_{q}\vert\right)
\left(m_{q_{3}}-\vert A_{q}\vert \right)}{\mathcal{M}_{q_{1}}}} & \sqrt{\frac{
\left(\vert g_{q}\vert -m_{q_{2}}\right)\left(m_{q_{3}}-\vert A_{q}\vert \right)\left(
\vert A_{q}\vert -m_{q_{1}}\right)}{\mathcal{M}_{q_{2}}}} & \sqrt{\frac{
\left(m_{q_{3}}-\vert g_{q}\vert\right)\left(m_{q_{2}}-\vert A_{q}\vert\right)\left(
\vert A_{q}\vert -m_{q_{1}}\right)}{\mathcal{M}_{q_{3}}}} \\ 
-\sqrt{\frac{\left(\vert g_{q}\vert-\vert A_{q}\vert \right)\left(\vert g_{q}\vert-m_{q_{1}}\right)
\left(\vert A_{q}\vert-m_{q_{1}}\right)}{\mathcal{M}_{q_{1}}}} & \sqrt{\frac{
\left(\vert g_{q}\vert-\vert A_{q}\vert \right)\left(\vert g_{q}\vert-m_{q_{2}}\right)\left(m_{q_{2}}-\vert A_{q}\vert 
\right)}{\mathcal{M}_{q_{2}}}} & \sqrt{\frac{\left(\vert g_{q}\vert-\vert A_{q}\vert\right)
\left(m_{q_{3}}-\vert g_{q}\vert \right)\left(m_{q_{3}}-\vert A_{q}\vert \right)}{\mathcal{M}%
_{q_{3}} }} \\ 
\sqrt{\frac{\left(\vert g_{q}\vert-m_{q_{2}}\right)\left(m_{q_{3}}-\vert g_{q}\vert\right)\left(
\vert A_{q}\vert-m_{q_{1}}\right)}{\mathcal{M}_{q_{1}}}} & -\sqrt{\frac{
\left(\vert g_{q}\vert-m_{q_{1}}\right)\left(m_{q_{2}}-\vert A_{q}\vert\right)
\left(m_{q_{3}}-\vert g_{q}\vert\right)}{\mathcal{M}_{q_{2}}}} & \sqrt{\frac{
\left(\vert g_{q}\vert-m_{q_{1}}\right)\left(\vert g_{q}\vert-m_{q_{2}}\right)
\left(m_{q_{3}}-\vert A_{q}\vert \right)}{\mathcal{M}_{q_{3}}}}%
\end{pmatrix}%
\end{equation}
with 
\begin{eqnarray}
\mathcal{M}_{q_{1}}&=&\left(\vert g_{q}\vert -\vert A_{q}\vert \right)\left(m_{q_{2}}-m_{q_{1}}%
\right)\left(m_{q_{3}}-m_{q_{1}}\right)  \notag \\
\mathcal{M}_{q_{2}}&=&\left(\vert g_{q}\vert -\vert A_{q}\vert \right)\left(m_{q_{2}}-m_{q_{1}}%
\right)\left(m_{q_{3}}-m_{q_{2}}\right)  \notag \\
\mathcal{M}_{q_{3}}&=&\left(\vert g_{q}\vert -\vert A_{q}\vert \right)\left(m_{q_{3}}-m_{q_{1}}%
\right)\left(m_{q_{3}}-m_{q_{2}}\right).
\end{eqnarray}

Therefore, the mixing matrix that takes places in the CKM mixing matrix is
given by $\mathbf{U}_{q}=\mathbf{U}_{\pi/4}\mathbf{P}_{q}\mathbf{O}_{q}$
with $q=u,d$, then $\mathbf{V}_{CKM}=\mathbf{U}^{\dagger}_{u}\mathbf{U}_{d}=%
\mathbf{O}^{T}_{u}\mathbf{\bar{P}_{q}} \mathbf{O}_{d}$ where $\mathbf{\bar{P}%
_{q}}=\mathbf{P}^{\dagger}_{u}\mathbf{P}_{d}=\text{diag.}\left( e^{i\bar{\eta}_{q_{1}}}, e^{i\bar{\eta}_{q_{2}}}, e^{i\bar{\eta}_{q_{3}}}
\right)$ with $\bar{\eta}_{q_{i}}=\eta_{d_{i}}-\eta_{u_{i}}$.

For the up and down quark sector, the orthogonal real matrices are 
\begin{eqnarray}
\mathbf{O}_{u}&=& 
\begin{pmatrix}
\sqrt{\frac{\left(\vert g_{u}\vert -m_{u}\right)\left(m_{c}-\vert A_{u}\vert \right)
\left(m_{t}-\vert A_{u}\vert \right)}{\mathcal{M}_{u}}} & \sqrt{\frac{
\left(\vert g_{u}\vert -m_{c}\right)\left(m_{t}-\vert A_{u}\vert \right)\left( \vert A_{q}\vert-m_{u}\right)}{%
\mathcal{M}_{c}}} & \sqrt{\frac{ \left(m_{t}-\vert g_{u}\vert \right)\left(m_{c}-\vert A_{u}\vert %
\right)\left( \vert A_{u}\vert -m_{u}\right)}{\mathcal{M}_{t}}} \\ 
-\sqrt{\frac{\left(\vert g_{u}\vert -\vert A_{u}\vert \right)\left(\vert g_{u}\vert -m_{u}\right)
\left(\vert A_{u}\vert -m_{u}\right)}{\mathcal{M}_{u}}} &  \sqrt{\frac{
\left(\vert g_{u}\vert -\vert A_{u}\vert \right)\left(\vert g_{u}\vert -m_{c}\right)\left(m_{c}-\vert A_{u}\vert  \right)}{%
\mathcal{M}_{c}}} & \sqrt{\frac{\left(\vert g_{u}\vert -\vert A_{u}\vert\right)
\left(m_{t}-\vert g_{u}\vert \right)\left(m_{t}-\vert A_{u}\vert \right)}{\mathcal{M}_{t} }} \\ 
\sqrt{\frac{\left(\vert g_{u}\vert -m_{c}\right)\left(m_{t}-\vert g_{u}\vert \right)\left(
\vert A_{u}\vert -m_{u}\right)}{\mathcal{M}_{u}}} & -\sqrt{\frac{ \left(\vert g_{u}\vert-m_{u}%
\right)\left(m_{c}-\vert A_{u}\vert \right) \left(m_{t}-\vert g_{u}\vert \right)}{\mathcal{M}_{c}}}
& \sqrt{\frac{ \left(\vert g_{u}\vert -m_{u}\right)\left(\vert g_{u}\vert -m_{c}\right)
\left(m_{t}-\vert A_{u}\vert \right)}{\mathcal{M}_{t}}}%
\end{pmatrix}%
;  \notag \\
\mathbf{O}_{d}&=& 
\begin{pmatrix}
\sqrt{\frac{\left(\vert g_{d}\vert -m_{d}\right)\left(m_{s}-\vert A_{d}\vert \right)
\left(m_{b}-\vert A_{d}\vert \right)}{\mathcal{M}_{d}}} & \sqrt{\frac{
\left(\vert g_{d}\vert -m_{s}\right)\left(m_{b}-\vert A_{d}\vert \right)\left( \vert A_{d}\vert -m_{d}\right)}{%
\mathcal{M}_{s}}} & \sqrt{\frac{ \left(m_{b}-\vert g_{d}\vert \right)\left(m_{s}-\vert A_{d}\vert%
\right)\left( \vert A_{d}\vert -m_{d}\right)}{\mathcal{M}_{b}}} \\ 
-\sqrt{\frac{\left(\vert g_{d}\vert -\vert A_{d}\vert \right)\left(\vert g_{d}\vert -m_{d}\right)
\left(\vert A_{d}\vert -m_{d}\right)}{\mathcal{M}_{d}}} &  \sqrt{\frac{
\left(\vert g_{d}\vert -\vert A_{d}\vert \right)\left(\vert g_{d}\vert-m_{s}\right)\left(m_{s}-\vert A_{d}\vert  \right)}{%
\mathcal{M}_{s}}} & \sqrt{\frac{\left(\vert g_{d}\vert -\vert A_{d}\vert \right)
\left(m_{b}-\vert g_{d}\vert \right)\left(m_{b}-\vert A_{d}\vert \right)}{\mathcal{M}_{b} }} \\ 
\sqrt{\frac{\left(\vert g_{d}\vert -m_{s}\right)\left(m_{b}-\vert g_{d}\vert \right)\left(
\vert A_{d}\vert -m_{d}\right)}{\mathcal{M}_{d}}} & -\sqrt{\frac{ \left(\vert g_{d}\vert-m_{d}%
\right)\left(m_{s}-\vert A_{d}\vert \right) \left(m_{b}-\vert g_{d}\vert \right)}{\mathcal{M}_{s}}}
& \sqrt{\frac{ \left(\vert g_{d}\vert -m_{d}\right)\left(\vert g_{d}\vert-m_{s}\right)
\left(m_{b}-\vert A_{d}\vert\right)}{\mathcal{M}_{b}}}%
\end{pmatrix}%
,
\end{eqnarray}
\begin{eqnarray}
\mathcal{M}_{u}&=&\left(\vert g_{u}\vert -\vert A_{u}\vert \right)\left(m_{c}-m_{u}\right)%
\left(m_{t}-m_{u}\right),\quad \mathcal{M}_{c}=\left(\vert g_{u}\vert -\vert A_{u}\vert \right)%
\left(m_{c}-m_{u}\right)\left(m_{t}-m_{c}\right);\nn\\
\mathcal{M}%
_{t}&=&\left(\vert g_{u}\vert -\vert A_{u}\vert \right)
\left(m_{t}-m_{u}\right)\left(m_{t}-m_{c}%
\right),\quad
\mathcal{M}_{d}=\left(\vert g_{d}\vert -\vert A_{d}\vert \right)\left(m_{s}-m_{d}\right)%
\left(m_{b}-m_{d}\right);\nn\\
\mathcal{M}_{s}&=&\left(\vert g_{d}\vert -\vert A_{d}\vert \right)%
\left(m_{s}-m_{d}\right)\left(m_{b}-m_{s}\right),\quad \mathcal{M}%
_{b}=\left(\vert g_{d}\vert -\vert A_{d}\vert \right)\left(m_{b}-m_{d}\right)\left(m_{b}-m_{s}%
\right).  \notag
\end{eqnarray}
\noindent
Having written the above expressions, we calculate the CKM matrix elements
which are given as 
\begin{eqnarray}
\big| \mathbf{V}^{ud}_{CKM}\big|&=& \big| \left(\mathbf{O}_{u}\right)_{11}\left(\mathbf{O}%
_{d}\right)_{11}+\left(\mathbf{O}_{u}\right)_{21}\left(\mathbf{O}%
_{d}\right)_{21}e^{i\bar{\alpha}_{q}}+\left(\mathbf{O}_{u}\right)_{31}\left(\mathbf{O}%
_{d}\right)_{31} e^{i\bar{\beta}_{q}}\big| ;  \notag \\
\big| \mathbf{V}^{us}_{CKM}\big|&=& \big| \left(\mathbf{O}_{u}\right)_{11}\left(\mathbf{O}%
_{d}\right)_{12}+\left(\mathbf{O}_{u}\right)_{21}\left(\mathbf{O}%
_{d}\right)_{22}e^{i\bar{\alpha}_{q}}+\left(\mathbf{O}_{u}\right)_{31}\left(\mathbf{O}%
_{d}\right)_{32} e^{i\bar{\beta}_{q}}\big|;  \notag \\
\big|\mathbf{V}^{ub}_{CKM}\big|&=& \big|\left(\mathbf{O}_{u}\right)_{11}\left(\mathbf{O}%
_{d}\right)_{13}+\left(\mathbf{O}_{u}\right)_{21}\left(\mathbf{O}%
_{d}\right)_{23}e^{i\bar{\alpha}_{q}}+\left(\mathbf{O}_{u}\right)_{31}\left(\mathbf{O}%
_{d}\right)_{33}e^{i\bar{\beta}_{q}}\big|;  \notag \\
\big|\mathbf{V}^{cd}_{CKM}\big|&=&\big| \left(\mathbf{O}_{u}\right)_{12}\left(\mathbf{O}%
_{d}\right)_{11}+\left(\mathbf{O}_{u}\right)_{22}\left(\mathbf{O}%
_{d}\right)_{21}e^{i\bar{\alpha}_{q}}+\left(\mathbf{O}_{u}\right)_{32}\left(\mathbf{O}%
_{d}\right)_{31}e^{i\bar{\beta}_{q}}\big|;  \notag \\
\big|\mathbf{V}^{cs}_{CKM}\big|&=& \big|\left(\mathbf{O}_{u}\right)_{12}\left(\mathbf{O}%
_{d}\right)_{12}+\left(\mathbf{O}_{u}\right)_{22}\left(\mathbf{O}%
_{d}\right)_{22}e^{i\bar{\alpha}_{q}}+\left(\mathbf{O}_{u}\right)_{32}\left(\mathbf{O}%
_{d}\right)_{32}e^{i\bar{\beta}_{q}}\big|;  \notag \\
\big|\mathbf{V}^{cb}_{CKM}\big|&=&\big| \left(\mathbf{O}_{u}\right)_{12}\left(\mathbf{O}%
_{d}\right)_{13}+\left(\mathbf{O}_{u}\right)_{22}\left(\mathbf{O}%
_{d}\right)_{23}e^{i\bar{\alpha}_{q}}+\left(\mathbf{O}_{u}\right)_{32}\left(\mathbf{O}%
_{d}\right)_{33}e^{i\bar{\beta}_{q}}\big|;  \notag \\
\big|\mathbf{V}^{td}_{CKM}\big|&=&\big| \left(\mathbf{O}_{u}\right)_{13}\left(\mathbf{O}%
_{d}\right)_{11}+\left(\mathbf{O}_{u}\right)_{23}\left(\mathbf{O}%
_{d}\right)_{21}e^{i\bar{\alpha}_{q}}+\left(\mathbf{O}_{u}\right)_{33}\left(\mathbf{O}%
_{d}\right)_{31}e^{i\bar{\beta}_{q}}\big|;  \notag \\
\big|\mathbf{V}^{ts}_{CKM}\big|&=& \big|\left(\mathbf{O}_{u}\right)_{13}\left(\mathbf{O}%
_{d}\right)_{12}+\left(\mathbf{O}_{u}\right)_{23}\left(\mathbf{O}%
_{d}\right)_{22}e^{i\bar{\alpha}_{q}}+\left(\mathbf{O}_{u}\right)_{33}\left(\mathbf{O}%
_{d}\right)_{32}e^{i\bar{\beta}_{q}}\big|;  \notag \\
\big|\mathbf{V}^{tb}_{CKM}\big|&=& \big|\left(\mathbf{O}_{u}\right)_{13}\left(\mathbf{O}%
_{d}\right)_{13}+\left(\mathbf{O}_{u}\right)_{23}\left(\mathbf{O}%
_{d}\right)_{23}e^{i\bar{\alpha}_{q}}+\left(\mathbf{O}_{u}\right)_{33}\left(\mathbf{O}%
_{d}\right)_{33}e^{i\bar{\beta}_{q}}\big|,
\end{eqnarray}
where $\bar{\alpha}_{q}=\bar{\eta}_{q_{2}}-\bar{\eta}_{q_{1}}$ and  $\bar{\beta}_{q}=\bar{\eta}_{q_{3}}-\bar{\eta}_{q_{1}}$. As we already commented, there are two effective phases that are relevant in the CKM matrix.

Now, let us show  that the well known Gatto-Sartori-Tonin relations can
be obtained in this model. To do this, we make some approximations on the
orthogonal matrix, $\mathbf{O}_{q}$. As can be noticed, the $\vert g_{q}\vert $ and $%
\vert A_{q}\vert $ free parameters could take two limiting values, this is, $\vert g_{q}\vert =m_{q_{3}}
$ and $\vert A_{q}\vert =m_{q_{2}}$ or $\vert g_{q}\vert =m_{q_{2}}$ and $\vert A_{q}\vert =m_{q_{1}}$.
Actually, some combination between those limiting cases could be considered, but
the CKM mixings can not be reproduced as one can check. Nonetheless, there
is a region in the parameters space where the quark mixing angles are fitted
quite well. With the following values $\vert g_{q}\vert =m_{q_{3}}-m_{q_{2}}$ and $%
\vert A_{q}\vert =2m_{q_{1}} $, one obtains
\begin{equation}
\mathbf{O}_{q}\approx 
\begin{pmatrix}
1-\frac{1}{2}\frac{m_{q_{1}}}{m_{q_{2}}} & \sqrt{\frac{m_{q_{1}}}{m_{q_{2}}}
} & \sqrt{\frac{m_{q_{1}}}{m_{q_{3}}}\frac{m_{q_{2}}}{m_{q_{3}}}\frac{
m_{q_{2}}}{m_{q_{3}}}} \\ 
-\sqrt{\frac{m_{q_{1}}}{m_{q_{2}}}} & 1-\frac{m_{q_{1}}}{m_{q_{2}}}& \sqrt{\frac{m_{q_{2}}}{m_{q_{3}}}} \\ 
\sqrt{\frac{m_{q_{2}}}{m_{q_{3}}}\frac{m_{q_{1}}}{m_{q_{2}}} } & -\sqrt{\frac{
m_{q_{2}}}{m_{q_{3}}}} & 1-\frac{m_{q_{2}}}{m_{q_{3}}}%
\end{pmatrix}
.
\end{equation}

One has to keep in mind that $q=u,d$. As a result, the following relations
are obtained 
\begin{eqnarray}
\big|\mathbf{V}^{us}_{CKM}\big| &\approx& \bigg|\sqrt{\frac{m_{d}}{m_{s}}} -\sqrt{%
\frac{m_{u}}{m_{c}}} e^{i\bar{\alpha}_{q}} \bigg|  \notag \\
\big|\mathbf{V}^{ub}_{CKM}\big| &\approx& \bigg|\frac{m_{s}}{m_{b}}\sqrt{\frac{ m_{d}}{m_{b}}}%
+\sqrt{\frac{m_{u}}{m_{c}}}\left(\sqrt{\frac{%
m_{c}}{m_{t}}}e^{i\bar{\beta}_{q}}-\sqrt{\frac{m_{s}}{m_{b}}}e^{i\bar{\alpha}_{q}}\right) \bigg| \notag \\
\big| \mathbf{V}^{cb}_{CKM}\big| &\approx& \bigg|\sqrt{\frac{m_{s}}{m_{b}}}e^{i\bar{\alpha}_{q}}- \sqrt{\frac{m_{c}}{m_{t}}}e^{i\bar{\beta}_{q}} \bigg| \notag \\
\big| \mathbf{V}^{td}_{CKM}\big| &\approx& \bigg| \frac{m_{c}}{m_{t}}\sqrt{\frac{ m_{u}}{m_{t}}}%
+\sqrt{\frac{m_{d}}{m_{s}}}\left(\sqrt{\frac{m_{s}}{m_{b}}} e^{i\bar{\beta}_{q}}-\sqrt{\frac{m_{c}}{m_{t}}}e^{i\bar{\alpha}_{q}}\right)\bigg|.
\end{eqnarray}

\end{document}